\documentclass[pdflatex,sn-mathphys-num]{sn-jnl}

\usepackage{soul}
\usepackage{graphicx}
\usepackage{epsf}
\usepackage{epsfig}
\usepackage[usenames]{color}
\usepackage{times}
\usepackage{multirow}
\usepackage{slashed}
\usepackage{amsmath,amssymb,amsfonts}
\usepackage{amsthm}
\usepackage{comment}
\usepackage{mathrsfs,bm}

\usepackage[caption=false]{subfig}
\DeclareMathOperator\arctanh{arctanh}
\DeclareMathOperator\sech{sech}

\usepackage{tikz}
\usepackage{cancel}
\usepackage{url}
\usepackage{pdflscape}

\usepackage[title]{appendix}
\usepackage{xcolor}
\usepackage{textcomp}
\usepackage{manyfoot}
\usepackage{booktabs}
\usepackage{algorithm}
\usepackage{algorithmicx}
\usepackage{algpseudocode}
\usepackage{listings}
\usepackage{etoolbox}
\usepackage{color}
\usepackage{xifthen}
\usepackage{stackengine}
\usepackage{booktabs}
\usepackage{array}
\usepackage{bigfoot}

\newcommand{\norm}[1]{\left\|#1\right\|}

\newcommand{\w}[1]{{#1}}
\newcommand{\be}{\begin{equation}}
\newcommand{\ee}{\end{equation}}
\newcommand{\bea}{\begin{eqnarray}}
\newcommand{\eea}{\end{eqnarray}}
\newcommand{\bal}{\begin{align}}
\newcommand{\eal}{\end{align}}
\newcommand{\nn}{\nonumber}

\newcommand{\bkps}[2]{\langle \w{#1}, \w{#2} \rangle_{\w{G}}}
\newcommand{\Gram}{{\w{G}}}
\newcommand{\GramE}{{\w{G}_\text{E}}}
\newcommand{\GramHp}{{\w{G}^{}_{\text{H}^p}}}
\newcommand{\ol}[1]{\overline{#1}}

\newcommand{\diag}[1]{\operatorname{diag}{#1}}

\usepackage{mathtools, stmaryrd}
\usepackage{xparse} 
\DeclarePairedDelimiterX{\Iintv}[1]{\llbracket}{\rrbracket}{\iintvargs{#1}}
\NewDocumentCommand{\iintvargs}{>{\SplitArgument{1}{,}}m}
{\iintvargsaux#1} 
\NewDocumentCommand{\iintvargsaux}{mm} {#1\mkern1.5mu,\mkern1.5mu#2}

\font\tenscr=rsfs10 scaled1100
\font\sevenscr=rsfs7 
\font\fivescr=rsfs5 
\skewchar\tenscr='177
\skewchar\sevenscr='177
\skewchar\fivescr='177
\newfam\scrfam
\textfont\scrfam=\tenscr
\scriptfont\scrfam=\sevenscr
\scriptscriptfont\scrfam=\fivescr

\def\scri{{\fam\scrfam I}}

\newcommand\sizefigmediumsmall{0.45}
\newcommand\sizefigmedium{0.5}

\makeatletter
\newrobustcmd{\fixappendix}{
  \patchcmd{\l@section}{1.5em}{7em}{}{}
  \patchcmd{\l@subsection}{2.3em}{7em}{}{}
}
\makeatother
\setstackEOL{\\}
\setstackgap{L}{\normalbaselineskip}
\strutlongstacks{T}

\begin{document}

\title{Quasi-normal mode expansions of black hole perturbations: a hyperboloidal Keldysh's approach}

\author*[1,2,3]{J. Besson}\email{jeremy.besson@aei.mpg.de}

\author[1]{J.L. Jaramillo}\email{Jose-Luis.Jaramillo-Martin@ube.fr}

\affil*[1]{Institut de Math\'ematiques de Bourgogne UMR 5584,
  Universit\'e Bourgogne Europe, CNRS, F-21000 Dijon, \country{France}}


\affil[2]{Albert-Einstein-Institut, Max-Planck-Institut für Gravitationsphysik, Callinstraße 38, 30167 Hannover, \country{Germany}}

\affil[3]{Leibniz Universit\"at Hannover, 30167 Hannover, \country{Germany}}


\abstract{We study  quasinormal mode expansions 
by adopting a Keldysh scheme for the spectral
construction of asymptotic resonant expansions. Quasinormal
modes are first cast in terms of a non-selfadjoint  problem by
adopting, in a black hole perturbation setting, a spacetime hyperboloidal approach.
Then the Keldysh expansion of the resolvent, built on bi-orthogonal systems,
provides a spectral version of Lax-Phillips expansions
on scattering resonances.
We clarify the role of scalar product structures in the
Keldysh setting~\cite{Gasperin:2021kfv}, that prove non-necessary to construct
the resonant expansions (in particular the quasinormal mode time-series
at null infinity), but are required to define the (constant) excitation coefficients
in the bulk resonant expansion.
We demonstrate the efficiency and accuracy of the Keldysh spectral approach
to (non-selfadjoint) dynamics,
even beyond its limits of validity, in particular recovering
Schwarzschild black hole late power-law tails.
We also study early dynamics by exploring i) the existence of an earliest time of validity
of the resonant expansion and ii) the interplay between overtones extracted with the
Keldysh scheme and regularity. Specifically, we address convergence aspects of the series
and, on the other hand, we implement non-modal analysis tools, namely
assessing $H^p$-Sobolev dynamical transient growths and constructing $H^p$-pseudospectra.
Finally, we apply the Keldysh scheme to calculate ``second-order'' quasinormal modes and
complement the qualitative study of overtone distribution 
by presenting the Weyl law for the counting of quasinormal modes in black holes with different (flat, De Sitter,
anti-De Sitter) spacetime asymptotics.

}


\maketitle

\tableofcontents

\section{Introduction: quasi-normal (resonant) expansions and black hole ringdown}
\label{s:intro}
Resonant or quasi-normal mode (QNM) expansions of scattered fields
play a key role in the description of open dissipative systems.
They have been used systematically in the physics literature ---(at least)
since Gamow's discussion of the $\alpha$
decay \cite{Gamow28}--- to describe the propagation of a linear field
on a given background in terms of a superposition of
damped oscillations, where the associated frequencies and time decay scales
are characteristic properties of the background.
From a mathematical perspective, they admit a sound treatment in the
Lax-Phillips and  Vainberg scattering theory \cite{LaxPhi89,Vainb73}.
In this mathematical setting, the resonant (or QNM) complex frequencies $\omega_n$'s
are characterised in terms of the poles of
the meromorphic extension of the resolvent (essentially the Green function) of the wave equation.
For concreteness, denoting formally the scattered field as $u(t,x)$
satisfying the following initial value problem of a linear wave equation (written in first-order Schr\"odinger form)
\bea
\label{e:wave_eq_1storder_u}
\displaystyle\left\{
\begin{array}{l}
 \partial_t u  = i L  u \ , \\
 u(t=0,x)=u_0(x) \ ,
 \end{array}
 \right.
\eea
subject to `outgoing' boundary conditions, the solution $u(t,x)$  can be expanded 
as an asymptotic series of damped sinusoids
(cf. e.g. \cite{TanZwo00,zworski2017mathematical,dyatlov2019mathematical})
\bea
\label{e:resonant_expansion}
u(t,x) \sim \sum_n {\cal A}_n(x) e^{i\omega_nt} \ , 
\eea
where the complex frequencies $\omega_n$'s are the poles of the meromorphic extension  of the resolvent of the
infinitesimal time generator $L$,
i.e. $R_L(\omega) = (L - \omega)^{-1}$, and the functions ${\cal A}_n(x)$ are obtained by acting
with  such resolvent  $R_L(\omega_n)$ on the initial data $u_0(x)$, that is
\bea
\label{e:coeff_asymptotic_functions}
{\cal A}_n(x) = R_L(\omega_n)(u_0) \ .
\eea
The series (\ref{e:resonant_expansion}) is an asymptotic one, in particular a non-convergent series  in the generic case.

\subsection{Normal modes in the conservative case}
\label{s:normal_modes}
In contrast with this non-conservative (dissipative) situation above, the notion of normal
modes in conservative systems provides an orthonormal basis where the 
solution $u(t,x)$ to the linear dynamics can be expanded. Specifically,
given the initial value problem (\ref{e:wave_eq_1storder_u})
with $L=H$ the selfadjoint  time generator of the dynamics,
acting in a Hilbert space ${\cal H}$ with scalar product\footnote{\label{fn:bracket}We use the notation
  $\langle \cdot, \cdot \rangle_{_G}$ for the scalar product $G:{\cal H}\times  {\cal H}\to \mathbb{C}$,
  that is $G(v, w)=\langle v, w\rangle_{_G}, \forall v, w\in{\cal H}$. We
  reserve the notation
  $\langle \cdot, \cdot \rangle$ for the dual pairing $\alpha(v)=  \langle \alpha, v\rangle$
  for $v\in {\cal H}$, $\alpha\in{\cal H}^*$. This choice permits to follow the notation in~\cite{MenMol03,BeyLatRot12},
  while still being consistent with the notation we have used in~\cite{Jaramillo:2020tuu,Gasperin:2021kfv}.}
$\langle \cdot, \cdot \rangle_{_G}$,
its (normalised) eigenfunctions
$\hat{v}_n$  provide an orthonormal (Hilbert) basis\footnote{We assume here a discrete spectrum for simplicity.}
such that the evolution $u(t,x)$ can be written as a convergent series
\bea
\label{e:normal_modes}
u(t,x) = \sum_{n=0}^\infty a_n \hat{v}_n(x) e^{i\omega_nt} \ ,
\eea
where 
\bea
\label{e:coeff_normal_modes}
 a_n = \langle\hat{v}_n, u_0\rangle_{_G} \  , \ \ \hbox{with} \  \ H \hat{v}_n = \omega_n\hat{v}_n \ ,
 \eea
 with $\omega_n$ real  and $\langle\hat{v}_n, \hat{v}_m\rangle_{_G}= \delta_{nm}$.
  Note that, in contrast with the prescription (\ref{e:resonant_expansion}) and (\ref{e:coeff_asymptotic_functions})
  for the dissipative case, the determination of the expansion (frequencies $\omega_n$'s
  and expansion coefficients $a_n$'s) in the conservative case  
  reduces to a spectral problem. This spectral nature is at the basis of the powerful
  character of the expansion (\ref{e:normal_modes}) and ultimately relies on the
  validity of the spectral theorem for selfadjoint (more generally, `normal') operators.

\subsection{Dissipative case: approaches to `completeness and orthogonality' of QNMs}
  In the non-selfadjoint (non-normal) case such a spectral theorem is absent and,
  therefore, no straightforward extension of the spectral approach underlying
  the normal mode expansion (\ref{e:normal_modes}) is available.
However, the formal comparison between the conservative and dissipative cases has prompted
long-standing efforts in the physics literature to rewrite the asymptotic expansion
(\ref{e:resonant_expansion}) in a form more akin to the series (\ref{e:normal_modes}),
in particular in terms of a {\em spectral problem} with generalised
eigenfunctions $v_n$ subject to QNM `outgoing boundary conditions' that are,
in general, non-normalisable. A considerable effort has been devoted 
to identify appropriate notions of `completeness'
and `orthogonality' of the set of such generalised eigenfunctions  $v_n$'s in this QNM dissipative setting,
leading to different prescriptions in the spirit of Eq. (\ref{e:coeff_normal_modes})
for determining the corresponding analogues of the (`excitation'~\cite{Nollert:1998ys}) coefficients $a_n$.

Specifically, in the gravitational setting and strongly motivated in recent times by the analysis in
(linear) perturbation theory of the
ringdown phase of binary black hole mergers, a large body of literature is available.
As indicated above, various approaches involving different notions of completeness and
orthogonality have been introduced in the literature,  not always easily mutually
comparable or just simply not compatible (see
for instance~\cite{ChiLeuSue95} and \cite{Nollert:1998ys} and references
therein\footnote{For a discussion of these completeness and orthogonality relations
  in other physical settings, with a special emphasis in optics, see
  e.g.~\cite{LeuLiuYou94,ChiLeuMaa98,LalYanVynSauHug18}.}).
Although rigorous notions of QNM completeness can be developed for certain potentials,
as in the case of the P\"oschl-Teller potential studied by Beyer~\cite{Beyer:1998nu},
in contrast with the conservative (self-adjoint) case no appropriate general and
sound notion of completeness for the expansion (\ref{e:resonant_expansion_coefficients})
is available for generic potentials, as plainly discussed in~\cite{Warnick:2022hnc}.
As commented above, the roots of this fact can be traced to the loss of spectral theorem
in the non-selfadjoint (more precisely, `non-normal') case.

\subsection{The Keldysh approach to QNM resonant expansions}
\label{s:Keldysh_intro}
In the present work we do not dwell in the discussion above about completeness and
orthogonality of the set of QNM functions $v_n$. We rather focus on the
study of a systematic approach to render the resonant (Lax-Phillips) expansion
(\ref{e:resonant_expansion}) in terms of a proper spectral problem.

An underlying problem of many of the attempts mentioned above to cast resonant
expansions in terms of a spectral problem is that the considered `generalised
eigenfunctions' are not normalisable, in particular they do not belong a
well-controlled Banach space. This hinders the very definition of the QNM
frequencies as `proper eigenvalues' of the operator $L$. In contrast with
this situation, in those cases in which the operator $L$ can be defined 
on a Hilbert space ${\cal H}$ (more generally on a Banach space) and QNM frequencies $\omega_n$'s
can be characterised as proper eigenvalues of $L$,  i.e. the $\omega_n$ values
are (discrete) complex numbers in the point spectrum of $L$ ---so the corresponding
eigenfunctions are indeed normalisable--- then a proper spectral approach
can be devised for the resonant expansion (\ref{e:resonant_expansion}).
This is based on the so-called Keldysh expansion of the resolvent $R_L(\omega)$
in terms of the eigenvalue problem of the operator $L$ and its transpose\footnote{In
  \cite{Gasperin:2021kfv} we have discussed the Keldysh expansion in terms of the spectral problem
  of $L$ and its adjoint $L^\dagger$
  \bea
\label{e:right-left_eigenvalues_dagger}
L \hat{v}_n = \omega_n \hat{v}_n \ \ , \ \ L^\dagger \hat{w}_n = \bar{\omega}_n \hat{w}_n \ ,
\eea
 with $L^\dagger$ defined with respect to a given scalar product $\langle \cdot, \cdot \rangle_{_G}$.
As we will discuss below, in spite of the interest of such formulation in terms of a scalar product,
such a discussion can be traced to a more fundamental underlying result relying solely on
`dual pairing' notions, and therefore formulated in terms of $L^t$ rather than $L^\dagger$.}
operator $L^t$ 
\bea
\label{e:right-left_eigenvalues}
L v_n = \omega_n v_n \ \ , \ \ L^t \alpha_n = \omega_n \alpha_n \ ,
\eea
where $v_n$ and $\alpha_n$ are usually referred to as right- and left-eigenfunctions of $L$
and, also, as modes and comodes, respectively (note that if $v_n$ belong to an linear (Banach) space
${\cal H}$, then $\alpha_n$ belong to a dual space ${\cal H}^*$).
Crucially, they are normalisable (in the norm
of the corresponding Banach space ${\cal H}$). As we will see below,
the spectral problem (\ref{e:right-left_eigenvalues}) permits to expand
the resolvent $R_L(\omega)$ in terms of $v_n$ and $\alpha_n$, in such a way
that (\ref{e:resonant_expansion}) can be rewritten as
\bea
\label{e:resonant_expansion_coefficients}
u(t,x) \sim \sum_n \big(a_n v_n(x)\big) e^{i\omega_nt} \ , 
\eea
where the QNM frequencies $\omega_n$'s are now proper eigenvalues of $L$, their
corresponding QNM functions are the associated (normalisable) eigenfunctions $v_n$'s and the
expansion coefficients $a_n$'s are obtained from the `action' of the comode $\alpha_n$
onto the initial data $u_0$  in an  expression that parallels
(see details later) the 
projection of $u_0$ onto $\hat{v}_n$ in expression (\ref{e:coeff_normal_modes})
---note that in the selfadjoint (more generally `normal') case `modes' and `comodes' do
coincide.

An important point in the previous discussion is that different norms can be envisaged to
measure de `size' of modes and comodes, depending of the specific aspect we are studying.
This freedom impacts the normalization of the QNMs $v_n$ and the value
of the coefficients $a_n$. What remains invariant however is the product
``$a_n v_n(x)$'', that provides a spectral reconstruction of 
the function $a_n(x)$ in the QNM resonant expansion (\ref{e:resonant_expansion}), that is
\bea
\label{e:coeff_asymptotic_functions_spectral}
{\cal A}_n(x) = a_n v_n(x) \ .
\eea
In essence, this expression provides a `spectral prescription' for the evaluation
of ${\cal A}_n(x)$, as an alternative to the action of the resolvent in (\ref{e:coeff_asymptotic_functions}).
However, it is a remarkable fact that this change of perspective translates into
a powerful and efficient scheme to construct QNM expansions.

\subsubsection{The present work: a hyperboloidal Keldysh approach to scattering and QNMs.}
\label{s:Keldysh_hyper_intro}
In this work we adopt the (spectral) Keldysh approach to QNM (asymptotic) expansions sketched
above, revisiting and extending the discussion presented in~\cite{Gasperin:2021kfv}.

A necessary condition to apply such a Keldysh expansion is that the 
time generator $L$ must be a properly defined non-selfadjoint operator acting in
Hilbert (Banach) spaces. There are different manners of fulfilling
this condition. A successful approach, used systematically in the calculation of
QNMs in different physical and mathematical settings, is the so-called `complex scaling' method
(see e.g. \cite{simon1978resonances,moiseyev2011,zworski2017mathematical}). Here we rather adopt a geometrical approach,
  akin to the discussion
of spacetime causality and propagation properties in general relativity, namely the so-called
hyperboloidal approach to scattering. In this scheme,  spacetime is foliated by
constant time spacelike `hyperboloidal hypersurfaces' that asymptotically reach the spacetime
regions attained by null rays, namely such hyperboloidal slices are transverse
at regular cuts to future null infinity $\scri^+$ at large distances and to the event horizon
`inner boundary' in the case of black hole spacetimes. Very importantly, QNM eigenfunctions
become then normalisable, in stark contrast with QNM functions defined on `Cauchy slices'.

In particular, this hyperboloidal procedure provides a geometrical implementation of the outgoing boundary
conditions entering in the construction of QNMs, since the characteristics of the
associated wave equations (along the light cones) become `outgoing' at $\scri^+$ and the
event horizon, so no causal degree of freedom can enter the integration domain from the
boundary. At an analytical level, when combined with a (coordinate) compactification of the hyperboloidal
slices, this approach recasts the boundary conditions
into the (bulk) operator, the latter becoming `singular' in the sense that its principal
part appears multiplied by a function vanishing at the boundaries\footnote{We thank
  Juan A. Valiente-Kroon for pointing out the methodological similarity with the strategy
  followed in  Melrose's `geometric scattering' theory \cite{melrose1995geometric}.}. In concrete terms,
enforcing the outgoing boundaries conditions translates into enforcing appropriate (enhanced)
regularity of the QNM functions.

In summary, adopting a hyperboloidal approach permits to characterise QNMs
as (proper) eigenvalues of a well-defined non-selfadjoint operator, with (normalisable)
eigenfunctions belonging to an appropriate Hilbert space.
Such an approach to QNM in the BH setting has been pioneered by 
Warnick in \cite{Warnick:2013hba} and Ansorg \& Macedo
in~\cite{Ansorg:2016ztf} and then further developed in subsequent works
(cf.~\cite{PanossoMacedo:2018hab,Gajic:2019oem,Gajic:2019qdd,galkowski2021outgoing,Jaramillo:2020tuu,Joykutty:2021fgj,Joykutty:2024ctv,gajic2024quasinormal,Warnick:2024,PanossoMacedo:2024nkw} and references therein). In this non-selfadjoint setting, it is natural to
consider the Keldysh expansion of the resolvent $R_L(\omega)$ of $L$.
Exploiting this fact, Ref.~\cite{Gasperin:2021kfv} proposes precisely
an approach to BH QNM resonant expansions built on the Keldysh expansion.
In this work we revisit such Keldysh approach to QNM expansions
focusing on the following points:
\begin{itemize}
\item[i)] {\em Keldysh approach to QNM expansions: independence of the scalar product}.
  We refine and extend the spectral approach in~\cite{Gasperin:2021kfv} 
  for the construction of the version (\ref{e:resonant_expansion_coefficients})
  of the asymptotic QNM resonant expansions (\ref{e:resonant_expansion}),
  in particular stressing the fact ---not sufficiently discussed in~\cite{Gasperin:2021kfv}---
  that such expansion
  is independent of the chosen scalar product, depending only on the transpose $L^t$
  of $L$, rather than on its adjoint $L^\dagger$.

\item[ii)] {\em Keldysh QNM expansions in BH scattering: an accurate and efficient prescription}.
  We demonstrate numerically the remarkable accuracy of such Keldysh expansions
  in the BH setting, even with non-convergent 
  asymptotic series and, most unexpectedly, when applying the
  Keldysh prescription beyond its domain of validity by including not only QNMs but also
  discrete approximations of the continuous `branch cut' contribution.    
\end{itemize}
Regarding its relation with previous works, this Keldysh QNM expansion can be
seen, on the one hand, as a generalisation of the efficient spectral QNM expansions
introduced by Ansorg \& Macedo in~\cite{Ansorg:2016ztf}. Indeed, the scheme
presented in~\cite{Ansorg:2016ztf} makes use of
a discrete version of the Wronskian to construct the Green function (resolvent)
that limits its application essentially to $1+1$
problems, whereas the Keldysh expansion permits a priori to extend the analysis
to (odd) space dimensions\footnote{\label{footnote:Huygens}The `a priori' requirement of an odd (space)
dimensionality is related to Huygens' principle~\cite{belger1997survey,baker2003mathematical}.
In odd spatial dimensions, data for reconstructing the solution at a spacetime point $p$
are required only at the intersection of the past null cone of $p$ and
the initial Cauchy surface, whereas in even dimensions data with support in the interior
of the null cone are also required, entailing the appearance of tails.
The latter spoil the treatment of resonant expansions in Lax-Phillips
theory (notice e.g. the odd space dimensionality requirements in theorems in
\cite{dyatlov2019mathematical}).
In our case, in principle such a restriction also applies. However, the 
(unexpected) recovery of tails in Schwarzschild in our Keldysh approach
(cf. section \ref{s:S_tails}), suggests that the space even-dimensional case could be
actually successfully handled. We have nevertheless preferred
to be conservative on this point and kept the ``(odd)'' spatial dimensions requirement.}.
On the other hand, this Keldysh expansion connects
with the QNM expansions discussed by Joykutty in \cite{Joykutty:2021fgj,Joykutty:2024ctv}
in the BH scattering context, being precisely defined in terms of modes
and comodes of the operator $L$. In this sense, following the  suggestions
in~\cite{Joykutty:2021fgj,Joykutty:2024ctv},
and emphasising the absence of a fundamental role of a (definite-positive)
scalar product in the Keldysh expansion ---since the latter ultimately depends only on `transpose' (dual pairing) and not on `adjoint' (scalar product)
notions--- it is tantalising to consider the relation between Keldysh QNM expansions
and those QNM expansions proposed and discussed in \cite{Green:2022htq}.
Finally, it is worthwhile to note that, in the finite-rank (matrix) case,
this discussion in terms of modes $v_n$'s and comodes $\alpha_n$'s reduces
to the use of so-called bi-orthonormal bases, the key differences playing a role
in the present infinite-dimensional case, especially in the
discussion of QNM-expansion's convergence\footnote{\label{footnote:bi-orthogonal}For the reader familiar
with bi-orthonormal bases in the finite-rank (matrix) case, the right-eigenfunctions $v_n$'s
and the left-eigenfunctions $\alpha_n$'s in Eq. (\ref{e:right-left_eigenvalues_dagger})
form a 'bi-orthogonal system'.
In such matrix case, the Keldysh approach essentially reduces to the use of bi-orthogonal
bases (see e.g. \cite{moiseyev2011}). Note, though, that the Keldysh scheme provides a systematic
treatment of the non-diagonalisable case
for dealing with the Jordan blocks in terms of so-called `associated vectors' \cite{MenMol03}, namely
`Jordan chains'
(such a case is treated, to our knowledge, as an exceptional one in the standard bi-orthogonal treatment
\cite{moiseyev2011}). The Keldysh discussion also allows to treat systematically a broader class
of spectral problems going beyond the
standard (generalised) one (and occurring naturally, for instance, in optics),
including pencil operators with a non-linear dependence in the spectral parameter.
However, the key difference occurs in the infinite-dimensional case (precisely the
case we discuss), where convergence issues in the QNM series show up impacting
the QNM completeness that is generically lost. The Keldysh approach  clarifies the
generic asymptotic nature of the QNM series and crucially provides
a control on the loss of convergence of the QNM resonant expansion through the
error function $E_{N_{\mathrm{QNM}}}$ in Eq. (\ref{e:u_Keldysh_v7}).}.

\bigskip

The plan of the article is as follows. In section \ref{s:Keldysh} we revisit
the Keldysh  expansion of the resolvent of a non-selfadjoint operator and
apply it to the asymptotic resonant expansions in QNMs of a scattered field.
As an application of the versatility of the Keldysh expansion we sketch its
application to the problem of so-called ``non-linear QNMs''.
In section \ref{s:Time-domain_evolution} we illustrate the performance of the
Keldysh QNM expansion by comparing, in a proof-of-principle spirit, the (numerical) implementation
of the hyperboloidal time-evolution of a Gaussian testbed initial data in different spacetime
asymptotics with the corresponding Keldysh QNM resonant expansions,
demonstrating the remarkably performant spectral re-construction of the time-domain signal.
We present in section \ref{s:regularity_excitation_coefficients} a short discussion of some
regularity aspects
of the problem in terms of $H^p$-Sobolev norms and their implication in the role that the scalar product
plays the qualitative control of their excitation coefficients. In section \ref{s:phys-struc}
we present some physical and structural implications of the Keldysh analysis, encompassing
the late and early behaviour of the Keldysh expansion (namely tails in Schwarzschild, initial
time of validity of the expansion, transients...), convergence issues of the asymptotic series
and QNM Weyl asymptotics. In section \ref{s:conclusions} we present our conclusions and perspectives.
Finally, the main text  is complemented with three appendices covering the most
technical aspects of the discussion.

\section{Keldysh resonant expansions }
\label{s:Keldysh}
In this section we revisit the Keldysh QNM expansion introduced
in \cite{Gasperin:2021kfv}, but performing a crucial shift in the
argument and construction: whereas in \cite{Gasperin:2021kfv} a central
role is endowed to the notion of scalar product in a given
Hilbert space, here we will dwell on a more {\em primitive} version only
involving the notion of a Banach space ${\cal H}$ and its dual ${\cal H}^*$.

In Ref. \cite{Gasperin:2021kfv} the use of Hilbert spaces and the associated
adjoint operator $L^\dagger$ of a given operator $L$ were fully
justified, since that article focuses on the different implications
of the choice of a given scalar product in the discussion of BH QNM
instability. However, in the specific context of QNM Keldysh expansions,
such emphasis on the additional scalar product structure actually may
eclipse the key underlying structures actually responsible
of the expansion. In this section we provide such more general
and, simultaneously, more basic account of QNM Keldysh expansions
constructed on the basis of the transpose $L^t$ (and not the adjoint $L^\dagger$)
operator. The connection with the scalar product can be made
at a later stage.

\subsection{Keldysh expansion of the resolvent}
\label{s:Keldysh_resolvent}
The construction is based on the notion of right- and left-eigenvectors,
or modes and comodes, as defined in Eq. (\ref{e:right-left_eigenvalues}),
that we rewrite as
\bea
\label{e:right-left_eigenvalues_v2}
\left(L - \omega_n \; \mathrm{I}\right) v_n = 0 \ \ , \ \ \left(L^t - \omega_n \; \mathrm{I}\right) \alpha_n = 0 \ \ , \ \
v_n\in {\cal H}, \alpha\in{\cal H}^* \ .
\eea
In order to fully illustrate the generality of the procedure
we consider a more general (in general non-linear) eigenvalue problem.

\medskip

Following closely \cite{BeyLatRot12} (see also \cite{MenMol03,Beyn12,nicolet2023physically}), let us consider the application
\bea
  \label{e:F_function}
  \begin{array}{rccc}
  F:  &\Omega &\longrightarrow& {\cal L}({\cal H}, {\cal K}) \\
  &\omega &\mapsto&F(\omega) \ ,
  \end{array}
\eea
where $\Omega\in\mathbb{C}$ is a complex domain,
${\cal L}({\cal H}, {\cal K})$ is the space of
linear operators\footnote{The discussion in \cite{BeyLatRot12}
  deals with bounded operators. For the non-bounded case see
  \cite{yakubov1993completeness,markus2012introduction}.} from the
(complex) Banach space ${\cal H}$ into the Banach space ${\cal K}$.
For all $\omega\in\Omega$, we assume the operator
\bea
\label{e:F-lambda_operator}
F(\omega): {\cal H}\longrightarrow{\cal K} \ ,
\eea
to be Fredholm of index $0$. Defining the resolvent set $\rho(F)$ of $F$ as the
subset of $\Omega$ where $F(\omega)$ is invertible, with inverse $(F(\omega))^{-1}$,
we write the `resolvent application' $F^{-1}$ as 
\bea
\label{e:Finv_function}
\begin{array}{rccc}
F^{-1}:  &\rho(F)\in\Omega &\longrightarrow & \cal{L}(\cal{K}, \cal{H}) \\
         &\omega           &\mapsto         &(F(\omega))^{-1} \ .
\end{array} 
\eea
Under the conditions above, the spectrum $\sigma(F)$ of $F$, namely $\sigma(F) = \Omega\backslash\rho(F)$,
is a discrete subset of $\Omega$ and the resolvent $F^{-1}$ is a meromorphic function.

\medskip

{\em Example 1}.
To fix and illustrate the points above, we note that in the case of the eigenvalue problem (\ref{e:right-left_eigenvalues_v2}),
the function $F$ is defined just by $F(\omega) = L - \omega \mathrm{I}$. Therefore the function
$F^{-1}$ is the standard resolvent $R_L(\omega)$ of $L$, that is,
$F^{-1}= \left(L - \omega \mathrm{I}\right)^{-1}=R_L(\omega)$. Under these assumptions, it follows that $R_L(\omega)$
is meromorphic in $\omega$.

\medskip

We proceed now to discuss the Keldysh expansion. We consider the spaces ${\cal H}^*$ and ${\cal K}^*$,
respectively the dual spaces  of ${\cal H}$ and ${\cal K}$, and the transpose application $F^t(\omega)$ of
$F(\omega)$
\bea
\label{e:F-lambda_operator_transpose}
F(\omega)^t: {\cal K}^*\longrightarrow{\cal H}^* \ ,
\eea
defined by duality $\big(F(\omega)^t(\alpha)\big)(v):= \alpha\big(F(\omega)(v)\big)$ for all $v\in {\cal H}$ and all $\alpha\in {\cal K}^*$.
Using the notation $\langle \cdot, \cdot\rangle$ for the dual pairing
(cf. footnote \ref{fn:bracket})
we rewrite these relations as
$\langle F(\omega)^t(\alpha), v\rangle = \langle \alpha, F(\omega)(v)\rangle$. We take now
a bounded subdomain $\Omega_o\subset \Omega$ and 
consider the `eigenvalue problems'  associated with $F(\omega)$ and $F(\omega)^t$, for
$\omega\in\Omega_o$, namely the characterisation of their respective kernels. 
Under the assumptions above (see details in \cite{Beyn12}) eigenvalues $\omega_n$ are isolated
and we can write~\cite{MenMol03,Beyn12,nicolet2023physically} the eigenvalue problems\footnote{Formally we can write these eigenvalue problems in the  right- and left-eigenvector notation of matrices, namely
\bea
\label{e:general_eigenvalue_problems}
F(\omega_n)v_n = 0 \ \ , \ \  \alpha^t_nF(\omega_n) = 0 \ \ , \ \ \hbox{with} \  v_n\in {\cal H}, \  \alpha_n\in {\cal K}^*  \ ,
\eea
where $\alpha^t_n$ is the `row' vector transpose to the `column' vector $\alpha_n$.
Here $v_n$ and $\alpha_n$ are referred to, respectively, as  the right- and left-eigenvectors of $F(\omega_n)$.
} as
\bea
\label{e:general_eigenvalue_problems}
F(\omega_n)v_n = 0 \ \ , \ \  F(\omega_n)^t \alpha_n = 0 \ \ , \ \ \hbox{with} \  v_n\in {\cal H}, \ \alpha_n\in {\cal K}^* \ .
\eea
We assume in addition, for simplicity, that the $\omega_n$'s are non-degenerate (simple). Then, using
the operator $F'(\omega) = \frac{dF}{d\omega} \in {\cal L}({\cal H}, {\cal K})$, obtained by deriving $F$ with respect
to the spectral parameter $\omega$, we consider $\tilde{v}_n$ and  $\tilde{\alpha}_n$ 
satisfying the following (relative) normalisation
\begin{eqnarray}
  \label{e:normalization}
  \langle \tilde{\alpha}_n, F'(\omega_n)(\tilde{v}_n)\rangle = 1 \ .
\end{eqnarray}
With these elements we can write the Keldysh expansion of the resolvent application $F^{-1}$,
 evaluated at $\omega\in\Omega_o\backslash\sigma(F)$, as
follows~\cite{MenMol03,Beyn12}
\bea
\label{e:gen_resolvent_F_normalized}
F^{-1}(\omega) = \sum_{\omega_n\in\Omega_o}
\frac{\langle \tilde{\alpha}_n, \cdot\rangle}{\omega - \omega_n} \tilde{v}_n + H(\omega) \ \ , \ \
\hbox{with} \ \langle \tilde{\alpha}_n, F'(\omega_n)(\tilde{v}_n)\rangle = 1  \ . 
\eea
where $ H(\omega)\in {\cal L}({\cal H}, {\cal K})$ is holomorphic in the domain $\Omega_o$.

We can incorporate the normalisation (\ref{e:normalization}) into the expression
of the resolvent as follows. Given $\alpha_n$ and $v_n$ satisfying the eigenvalue
problem (\ref{e:general_eigenvalue_problems}) but not subject to any particular normalisation,
then the modes $\tilde{v}_n$ and comodes $\tilde{\alpha}_n$ defined as
\bea
\label{e:rescaling_alpha}
\tilde{v}_n = v_n  \ \ , \ \
\tilde{\alpha}_n &=& \frac{1}{\langle \alpha_n, F'(\omega_n)(v_n)\rangle} \alpha_n \ ,
\eea
satisfy
\bea
\langle \tilde{\alpha}_n, F'(\omega_n)(\tilde{v}_n)\rangle = \Big\langle \frac{1}{\langle \alpha_n, F'(\omega_n)(v_n)\rangle}
\alpha_n, F'(\omega_n)(v_n)\Big\rangle = 1 \ ,
\eea
and we can write
\bea
\label{e:gen_resolvent_F_non-normalized}
F^{-1}(\omega) =
\sum_{\omega_n\in\Omega_o}
\frac{\langle \alpha_n, \cdot\rangle }{\langle \alpha_n, F'(\omega_n)(v_n)\rangle}
\frac{v_n}{\omega - \omega_n} + H(\omega)
\ \ , \ \ \omega\in\Omega_o\setminus \sigma(L) \ . 
\eea
This expression of $F^{-1}(\omega)$ has the virtue of making
explicit the weight $1/\langle\alpha_n, F'(\omega_n)(v_n)\rangle$
entering in the structure of the resolvent. Note in particular that this expression
(\ref{e:gen_resolvent_F_non-normalized})
is invariant under arbitrary rescalings
of $v_n\in{\cal H}$ and $\alpha_n\in{\cal H}^*$ so, in contrast with $\tilde{v}_n$ and $\tilde{\alpha}_n$
in expression (\ref{e:gen_resolvent_F_normalized}) of $F^{-1}(\omega)$,
$v_n$ and $\alpha_n$ are not subject to any given normalization.

\medskip

{\em Example 2.} We apply the Keldysh construction of the resolvent to the case of the eigenvalue problem
(\ref{e:right-left_eigenvalues_v2}), with $F(\omega) = L - \omega \mathrm{I}$ as discussed
in the Example 1. Taking into account that $F'(\omega)=-\mathrm{I}$, we find
as normalisation $\langle \tilde{\alpha}_n, \tilde{v}_n\rangle = -1$. We can then write 
\bea
\label{e:resolvent_L_Keldysh}
R_L(\omega) &=& (L-\omega\mathrm{I})^{-1} =  \sum_{\omega_n\in\Omega_o}
\frac{\langle \tilde{\alpha}_n, \cdot\rangle}{\omega - \omega_n} \tilde{v}_n + H(\omega) \nn \\
&=&
\sum_{\omega_n\in\Omega_o}
\frac{\langle \alpha_n, \cdot\rangle}{\langle \alpha_n, F'(\omega_n)(v_n)\rangle} \frac{v_n}{\omega - \omega_n}  + H(\omega) \nn \\
&=&  \sum_{\omega_n\in\Omega_o}
\frac{\langle \alpha_n, \cdot\rangle}{\langle \alpha_n, v_n\rangle} \frac{v_n}{\omega_n - \omega}  + H(\omega) \nn \\
&=&
\sum_{\omega_n\in\Omega_o}
\frac{\langle \alpha_n, \cdot\rangle}{\omega_n - \omega} v_n  + H(\omega)
\ \ , \ \ \omega\in\Omega_o\setminus \sigma(L) \ ,
\eea
where in the third line we have used $F'(\omega)=-\mathrm{I}$ and in the fourth we have chosen
$\langle \alpha_n, v_n\rangle = 1$. For concreteness, we will make use of the last expression
of $R_L(\omega)$ in
later sections, valid for modes $v_n$ and  comodes $\alpha_n$ satisfying
\bea
\label{e:resolvent_L_Keldysh_modes_comodes}
L v_n = \omega_n v_n \ \ , \ \ L^t \alpha_n = \omega_n \alpha_n \ \ , \ \ \hbox{with} \ \ 
\langle \alpha_n, v_n\rangle = 1 \ .
\eea

The resolvent in
the (bounded) $\Omega_o$ region is then expressed as a finite sum of poles (assuming that the
$\omega_n$'s do not accumulate in $\Omega_o$) plus an
analytical (holomorphic) operator function $H(\omega)$.  This finite sum
will play a key role in the assessment of the infinite sum given by the
QNM expansion (\ref{e:resonant_expansion_coefficients})
as an asymptotic series and not (in general) a convergent series,
in contrast with the selfadjoint (normal) case in (\ref{e:normal_modes}),
as we will see in the following subsection.

\subsection{Keldysh asymptotic QNM resonant expansions}
\label{s:QNM_aymptotic_expansions}
Let us apply the Keldysh construction of the resolvent to the partial differential equation (PDE)
wave problem in Schr\"odinger form, defined in Eq. (\ref{e:wave_eq_1storder_u}),
that we rewrite as 
\begin{equation}
\label{e:wave_eq_1storder_u_tau}
\displaystyle\left\{
\begin{array}{l}
 \partial_\tau u  = i L  u \ , \\
 u(\tau=0,x)=u_0(x) \ \ , \ \ ||u_0||<\infty \ ,
 \end{array}
 \right. 
\end{equation}
denoting by $\tau$ the hyperboloidal time parameter (cf. appendix \ref{a:Hyperboloidal approach}
for a review of the hyperboloidal approach) and using an appropriate
norm $||\cdot||$ in the  Banach space of initial data.
Following closely~\cite{Gasperin:2021kfv}, we apply a Laplace transform to
solve (\ref{e:wave_eq_1storder_u_tau}). Considering 
  $\mathrm{Re}(s)>0$
\begin{equation}
 \label{e:Laplace_transform_1st_u}
       u(s;x) := \big({\cal L} u\big)(s;x) =\int_0^\infty e^{-s\tau}
       u(\tau,x)d\tau \ ,
\end{equation}
and applying it to (\ref{e:wave_eq_1storder_u_tau}), we get
\bea
s\;u(s;x) - u(\tau=0,   x) = i L u(s; x) \ .
\eea
Dropping the explicit $s$-dependence and introducing $u(\tau\!=\!0,x)=u_0(x)$
from~(\ref{e:wave_eq_1storder_u_tau}), we write
\bea
\label{e:non-homogenous_s}
\big(L + is\big)u(s;x) = i u_o(x) \ ,
\eea
To solve this non-homogeneous equation, we need the 
expression for the resolvent $\big(L + is\big)^{-1}= R_L(-is)$ of $L$,
namely
\bea
\label{e:u_Laplace}
  u(s;x) = i(L +is)^{-1} u_0(x) =  i R_L(-is) u_0 \ . 
\eea
This is the point in which the above-discussed Keldysh's expansion
of the resolvent enters into scene.
Using the relation $s=i\omega$, we have  $R_L(-is) = R_L(\omega)$, for  $\omega\in\Omega_o\setminus \sigma(L)$, so
\bea
\label{e:u_QNM_deriv0}
  u(s;x) = i R_L(-is) u_0 = i R_L(\omega) u_0  \ \ , \ \ \omega\in\Omega_o\setminus \sigma(L) \ ,
\eea
and employing expression (\ref{e:resolvent_L_Keldysh}) for the resolvent
with modes and comodes in (\ref{e:resolvent_L_Keldysh_modes_comodes}), we write
\bea
\label{e:u_QNM_deriv1}
 u(s;x) &=& i \sum_{\omega_n\in\Omega_o}
 \frac{\langle\alpha_n, u_0\rangle}{\omega_n - \omega} v_n  + iH(\omega)(u_0) \ \ , \ \ \omega\in\Omega_o\setminus \sigma(L) \nn \\
  &=& \sum_{s_n\in\Omega_o}
 \frac{\langle\alpha_n, u_0\rangle}{s - s_n} v_n  + i\tilde{H}(s)(u_0) \ \ , \ \ s\in i\Omega_o\setminus \sigma(L) \ ,
 \eea
 with $\tilde{H}(s) = H(-is)$ an holomorphic function (and relative normalization
 $\langle \alpha_n, v_n\rangle = 1$).

 \medskip
 
  The time-domain scattered field $u(\tau,x)$ is obtained with
  the inverse Laplace transform 
  \bea
  \label{e:Laplace_inverse}
 u(\tau,x) = \frac{1}{2\pi i }\int_{c-i\infty}^{c+i\infty} e^{s\tau}u(s;x) ds \ ,
 \eea
 with $c\in \mathbb{R}^+$. Considering bounded domains $\Omega$ containing
 $[c-iR, c+iR]$ with $R\in\mathbb{R}^+$,  we can then write (under the hypothesis
 of convergence of this limit) 
      \bea
         \label{e:u_Keldysh_v1}
       u(\tau,x) &=& \lim_{R\to\infty}\frac{1}{2\pi i }\int_{c-iR}^{c+iR}  e^{s\tau} u(s;x) ds   \nn
       \\&=&
       \lim_{R\to\infty}\frac{1}{2\pi i }\int_{c-iR}^{c+iR} e^{s\tau}
       \Big(\sum_{s_n\in\Omega_o}
 \frac{\langle\alpha_n, u_0\rangle}{s - s_n} v_n  + i\tilde{H}(s)(u_0)\Big) ds \ .
       \eea
       Taking a contour $C$ in the $s\text{-}\mathbb{C}$ complex plane composed by the interval
       $[c-iR, c+iR]$  closed on the left half-plane by a semi-circle $S$ centered at $c+i0$ and of radius $R$,
       i.e. $C = [c-iR, c+iR]\cup S$,
       we denote  by $\Omega_R$ the domain bounded by $C$ in $s\text{-}\mathbb{C}$.
       Under the hypotheses in section \ref{s:Keldysh_resolvent}, the number of $L$-eigenvalues
       $s_n\in\Omega_R$
       (poles in the Keldysh expansion of the resolvent $R_L(-is)$) 
       is finite and we can interchange the (finite) sum and the integral 
       \bea
       \label{e:u_Keldysh_v2}
        u(\tau,x) =
       \lim_{R\to\infty} \Bigg(\sum_{s_n\in\Omega_R} \frac{1}{2\pi i }
       \oint_C e^{s\tau}   \frac{\langle\alpha_n, u_0\rangle}{s - s_n} v_n ds +
       \frac{1}{2\pi  }
       \oint_C e^{s\tau} \tilde{H}(s)(u_0) ds\Bigg) \nn\\
        - \lim_{R\to\infty} \Bigg(\sum_{s_n\in\Omega_R} \frac{1}{2\pi i }
       \int_S e^{s\tau}   \frac{\langle\alpha_n, u_0\rangle}{s - s_n} v_n ds +
       \frac{1}{2\pi  }
       \int_S e^{s\tau} \tilde{H}(s)(u_0) ds\Bigg) \ .
       \eea
       The integral of the analytic function $e^{s\tau} \tilde{H}(s)(u_0)$ along the contour $C$  vanishes.
       For large enough $R$, the first integral along the  semi-circle $S$ also vanishes.
       On the contrary, the second integral along the
       semi-circle $S$ does not in general vanish, depending on the particular function $\tilde{H}(s)$.
       Such last term then produces in general a term $C_R(\tau; u_0)$. Then we write
 \bea
 \label{e:u_Keldysh_v3}
 u(\tau,x)  &=& 
 \lim_{R\to\infty}\Big(  \sum_{s_n\in\Omega_R}
 e^{s_n\tau} \langle\alpha_n, u_0\rangle v_n + C_R(\tau; u_0)\Big) \nn \\
 &=& \lim_{R\to\infty}\Big(  \sum_{\omega_n\in\Omega_R}
 e^{i\omega_n\tau} \langle\alpha_n, u_0\rangle v_n + C_R(\tau; u_0)\Big) \ ,
 \eea
 by applying the Cauchy theorem,
 where in the second line we have used the Fourier rather than the Laplace spectral parameter.  
 The limit in expression (\ref{e:u_Keldysh_v3}) does not necessarily exist, 
 in stark contrast with `normal' (in particular selfadjoint)
 operators where the spectral theorem guarantees it.
 But, although such a resonant expansion for $u(\tau,x)$ cannot in general be written as a
 convergent series, as in the selfadjoint (normal) case in Eq. (\ref{e:normal_modes}),
 a proper notion of asymptotic QNM resonant expansion does exist. We denote the latter formally as
 \bea
 \label{e:first_QNM_expansion}
  u(\tau,x)  \sim   \sum_n
 e^{i\omega_n\tau} \langle\alpha_n, u_0\rangle v_n \ \ , \ \ \langle \alpha_n, v_n\rangle = 1 \ .
 \eea
 The meaning of such asymptotic expression is the following. Given a (bounded) domain
 $\Omega$ in the complex plane, we can always write the exact solution to the
 evolution problem (\ref{e:wave_eq_1storder_u_tau}) as
   \begin{equation}
   \label{e:u_Keldysh_v6}
   u(\tau,x) = \sum_{\omega_n\in\Omega} e^{i\omega_{n}\tau} \langle\alpha_n, u_0\rangle v_n(x)
    + E_\Omega(\tau;u_0)(x) \ ,
   \end{equation}
   where note that, under our assumptions, the number of terms in the sum is finite.
   The key point is that the Keldysh expansion (\ref{e:u_Keldysh_v3}) permits to
   find a fine bound of the error $E_\Omega(\tau;u_0)(x)$ made when approximating $u(\tau,x)$
   by the finite sum. Specifically, choosing an appropriate norm and
   defining $a_\Omega= \max\{\mathrm{Im}(\omega), \omega\in \Omega\}$ then, from the
   structure of the integrand  $e^{s\tau}\tilde{H}(s)(u_0)$ in the last term in
   (\ref{e:u_Keldysh_v2}), we can estimate the norm of $E_\Omega(\tau;u_0)(x)$ as   
\bea
   \label{e:bound_E}
   ||E_\Omega(\tau;S)|| \leq C(a_\Omega,L) e^{-a_\Omega\tau}||u_0|| \ ,
   \eea
   where  $C(a_\Omega,L)$ is a constant 
   that depends on  $a_\Omega$ and the evolution operator $L$ but, and this is a key point,
   not on the initial data $u_0$. In those cases where a finite number of QNMs are located
   in the region $\Omega$ (this will be the case in the black hole potentials we
   will consider later), we can count the QNMs in $\Omega$ with $n\in\{0, \ldots, N_{\mathrm{QNM}}\}$. 
   We can then emphasize the number of QNMs employed in the approximation
   of the evolution field $u(\tau,x)$, rather than the considered region $\Omega$ in
   the complex plane containing those $N_\mathrm{QNM}+1$ modes, and rewrite
   \begin{eqnarray}
   \label{e:u_Keldysh_v7}
   u(\tau,x) &=& \sum_{n=0}^{N_{\mathrm{QNM}}} e^{i\omega_{n}\tau}  \langle\alpha_n, u_0\rangle v_n(x) + E_{N_{\mathrm{QNM}}}(\tau;u_0) \nn \\
   \hbox{with } &&||E_{N_{\mathrm{QNM}}}(\tau;u_0)|| \leq C(N_{\mathrm{QNM}}, L) e^{-a_{_{N_{\mathrm{QNM}}}}\tau}||u_0|| \ ,
   \end{eqnarray}
   with $a_\Omega$ cast as $a_{_{N_{\mathrm{QNM}}}}$ to emphasise the role of QNMs.
   In summary, we can write 
   \bea
   \label{e:Keldysh_QNM_expansion_u_an}
   u(\tau,x) \sim \sum_n e^{i\omega_{n}\tau} a_n v_n(x) \ \ \ , \ \ \ \hbox{with} \
   a_n = \langle\alpha_n, u_0\rangle \ , \ \ \langle\alpha_n, v_n\rangle=1 \ .
   \eea
   These expressions provide the QNM resonant expansion of the propagating field in terms
   of its initial data. For completeness, we provide the expression of $a_n$ when we do not
   impose $\langle \alpha_n, v_n\rangle=1$ in (\ref{e:resolvent_L_Keldysh}).
   Repeating the steps from  Eq. (\ref{e:u_QNM_deriv0})
   by inserting in it the expression of  $R_L(-is)$
   in the third line of Eq. (\ref{e:resolvent_L_Keldysh}), we finally get for the
   coefficient $a_n$ the expression
   \bea
   \label{e:a_n_general}
   a_n = \frac{\langle\alpha_n, u_0\rangle}{\langle\alpha_n, v_n\rangle} \ ,
   \eea
   that is, we can write the general form of the QNM expansion of the solution
   \bea
   \label{e:Keldysh_QNM_expansion_u_scale_invariant}
   u(\tau,x) \sim \sum_n e^{i\omega_{n}\tau} \frac{\langle\alpha_n, u_0\rangle}{\langle\alpha_n,
     v_n\rangle} v_n(x) \ ,
   \eea
   where no normalisation is imposed on $\alpha_n$ and $v_n$ and, actually, the expression
   is explicitly invariant under (independent) rescalings of $\alpha_n$ and $v_n$.

   \medskip

\noindent   {\em Remarks.} We recapitulate some important points in the construction: 

   \begin{itemize}
   \item[i)] {\em Generalisation of the Ansorg \& Macedo QNM expansions.}
     The Keldysh QNM expansion (\ref{e:Keldysh_QNM_expansion_u_an}) generalises, to arbitrary
     dimension and in a general formalism valid for `arbitrary'\footnote{There are, of course,
       restrictions on $L$. In particular here we are strongly using its Fredholm character,
       for the discreteness of the eigenvalues. In all the applications discussed in this
       manuscript the assumption on the non-degeneracy of $\omega_n$ eigenvalues holds.
       The extension to non-simple eigenvalues, but still a diagonalisable operator $L$, does not
       require any change in the reasoning. In the case of non-diagonalisable operators
       $L$, the extension is straightforward by 
       resorting (see footnote \ref{footnote:bi-orthogonal}) to the general expression of the
       Keldysh expansion
       of the resolvent that takes into the account the `associated vectors' in Jordan blocks,
       leading to the general Lax-Phillips resonant expressions~\cite{dyatlov2019mathematical}.
       As we have pointed out above, we will not need this
       in the present study and we will present it elsewhere.} operators $L$,
     the one-dimensional QNM  expansions introduced and (for the first time)
     implemented in the BH context by Ansorg \& Macedo in ~\cite{Ansorg:2016ztf}
     (see also \cite{Ammon:2016fru,PanossoMacedo:2018hab}).

   \item[ii)] {\em No fundamental role of scalar product in QNM resonant expansions.}
     The Keldysh expansion does not make use of scalar product structures, but only
     of the notion of duality and the associated transpose operator $L^t$, rather than
     the adjoint $L^\dagger$.

   \item[iii)] {\em No intrinsic `excitation coefficients' $a_n$ in the Keldysh expansion.}
     In the absence of a scalar product (or more generally a norm in ${\cal H}$), only the product
     ``$a_n v(x)$'' is well-defined. This can be easily seen in expressions
     (\ref{e:Keldysh_QNM_expansion_u_an}) and  (\ref{e:Keldysh_QNM_expansion_u_scale_invariant}).
     Indeed, if the norm of
     $v(x)$ is not fixed,
     we can rescale it by a factor $f\neq 0$. For concreteness, considering expression
     (\ref{e:Keldysh_QNM_expansion_u_an}) for  $u(\tau,x)$,
     from the constraint
     $\langle\alpha_n, v_n\rangle=1$ we must rescale $\alpha_n$ by $1/f$,
     so $a_n$ is also rescaled by $1/f$, leaving the product ``$a_n v(x)$''
     unchanged\footnote{The fact that only this combination ``$a_n v(x)$'' is relevant in
       the structural aspects of the QNM expansion was already remarked in Ansorg \& Macedo~\cite{Ansorg:2016ztf}.}. The same conclusion follows,
     for arbitrary rescalings of $v_n$ and $\alpha_n$, from the scale
     invariant expression (\ref{e:Keldysh_QNM_expansion_u_scale_invariant}).
     
     \item[iv)] {\em Expansion coefficients $a_n$ and choice of scalar product.} 
       In contrast with the point above, scalar products (and more generally norms)
       provide an additional structure permitting to fix the norm of QNMs
       $v_n$'s and, consequently, the associated coefficients $a_n$. This is the
       setting adopted in~\cite{Gasperin:2021kfv} and we provide now the connection with
       this latter work.
\bea
\label{e:QNM_scalar_product}
u(\tau, x) \sim \sum_n a_n \hat{v}_n(x) e^{i\omega_n \tau} \ ,
\eea

     \item[v)] {\em General $F(\omega)=0$ problems: the ``generalised eigenvalue problem'' case.} The standard
       eigenvalue problem (\ref{e:right-left_eigenvalues_v2}) is not the only  spectral problem $F(\omega)=0$
       relevant in the context of QNM expansions. Another one particularly important in the BH setting,
       that we will discuss later, is given by the so-called generalised eigenvalue.

       \medskip
       
       We start by considering the evolution problem
       \begin{equation}
\label{e:wave_eq_1storder_u_tau_generalised}
\displaystyle\left\{
\begin{array}{l}
 B \partial_\tau u  = i L  u \ , \\
 u(\tau=0,x)=u_0(x) \ \ , \ \ ||u_0||<\infty \ .
 \end{array}
 \right. 
\end{equation}
       Taking the Laplace transform, the analogue of Eq. (\ref{e:non-homogenous_s}) is
       the non-homogeneous equation
   \bea
\label{e:non-homogenous_s_B}
\big(L + isB\big)u(s;x) = i Bu_o(x) \ .
\eea
The homogeneous part, $\big(L + isB\big)u(s;x) = 0$, leads to the generalised eigenvalue problem
\bea
L u(\omega;x) = \omega B u(\omega;x) = 0 \ \ \longleftrightarrow \big(L - \omega B\big)u(\omega;x) \ ,
\eea
where we have used $s=i\omega$, that we can rewrite as
\bea
\label{e:F_generalised_eigenvalue}
F(\omega) u(\omega;x) = 0 \ \ , \ \ F(\omega) = L - \omega B \ .
\eea
Following exactly the procedure followed in section \ref{s:QNM_aymptotic_expansions}, we can write
\bea
u(\omega;x) = i F^{-1}(\omega) (B u_o) \ ,
\eea
and then use Eq. (\ref{e:gen_resolvent_F_non-normalized}) with
\bea
F'(\omega) = -B \ ,
\eea
to write
\bea
\label{e:gen_resolvent_F_non-normalized_B}
F^{-1}(\omega) =
\sum_{\omega_n\in\Omega_o}
\frac{\langle \alpha_n, \cdot\rangle }{\langle \alpha_n, B v_n\rangle}
\frac{v_n}{\omega_n - \omega} + H(\omega)
\ \ , \ \ \omega\in\Omega_o\setminus \sigma(L) \ . 
\eea
Finally, taking the inverse Laplace transform  (\ref{e:Laplace_inverse}), we
get the QNM resonant expansion 
\bea
u(\tau,x) \sim  \sum_n e^{i\omega_{n}\tau} \frac{\langle\alpha_n,B u_0\rangle}{\langle\alpha_n,
  B v_n\rangle} v_n(x)
\eea
where no particular normalisation is assumed for $\alpha_n(x)$ and $v_n(x)$.
Alternatively, and in particular, one can adopt $\langle\alpha_n, B v_n\rangle=1$
in parallel with (\ref{e:first_QNM_expansion})
and (\ref{e:Keldysh_QNM_expansion_u_an}), to write
\bea
u(\tau,x) \sim \sum_n e^{i\omega_{n}\tau} \langle\alpha_n,B u_0\rangle v_n(x) \ \ , \ \ \langle\alpha_n,
  B v_n\rangle = 1 \ ,
  \eea
  by adapting the normalisation to the structure of the generalised eigenvalue problem.

   \end{itemize}

\subsection{``Second-order'' QNMs in general relativity: a Keldysh-approach first analysis }
\label{s:2nd-order}
The Keldysh expansion for the resolvent has a straightforward application in
so-called `second-order QNMs'. The latter emerge in 
second-order perturbation theory of general relativity and has received much
attention recently~(see e.g. \cite{zlochower2003mode,nakano2007second,london2014modeling,baibhav2023agnostic,bourg2025quadratic} and references therein).

In our non-selfadjoint setting, general relativity second-order perturbation theory
has been considered in the context of pseudo-resonances in~\cite{Jaramillo:2022kuv}. 
We adopt the notation there and, specifically, 
following~\cite{Isaac68a,CunPriMon80,Miller:2016hjv,Pound:2015fma}
the metric $g_{ab}$ is expanded in a small parameter $\epsilon$ as
\bea
\label{e:pert_metric}
 g_{ab} = g^{(0)}_{ab} + \epsilon h^{(1)}_{ab} +  \epsilon^2 h^{(2)}_{ab} + O(\epsilon^3) \ ,
 \eea
and, in the proper gauges, the perturbed (vacuum) Einstein equations write \cite{Pound:2015fma}
 \bea
 \label{e:GR_pert}
 \delta G_{ab} \cdot h^{(1)} &=& 0 \nn \\
 \delta G_{ab} \cdot h^{(2)} &=& \delta^2G_{ab}[h^{(1)}, h^{(1)}] \ .
 \eea
This expansion provides a hierarchy of equations that share the  
linearised Einstein tensor $\delta G_{ab}$ evaluated on $g^{(0)}_{ab}$ in the left-hand-side,
acting linearly on first and second-order perturbations, $h^{(1)}_{ab}$
and $h^{(2)}_{ab}$ respectively. On the right-hand-side, the first-order equation
has no source, whereas the second-order perturbation has a source
quadratic in $h^{(1)}_{ab}$.

Following~\cite{Jaramillo:2022kuv}, and in the spirit of Eqs. (\ref{e:pert_metric}) and
(\ref{e:GR_pert}), we write
 \bea
 u(\tau, x) = u^{(1)}(\tau, x)  + \epsilon  u^{(2)}(\tau, x) + O(\epsilon^2) \ ,
 \eea
 for the perturbative expansion the master function for black hole perturbations
 and we assume that a gauge exists where its dynamical corresponding
 can be perturbatively written as
\bea
 \label{e:close-limit_2ndorder}
  \left(\partial_\tau -  i L\right) u^{(1)} &=& 0 \nn \\
 \left(\partial_\tau -  i L\right) u^{(2)} &=& S(\tau,x; u^{(1)}) \ ,
 \eea
 where  $S(\tau,x; u^{(1)})$ is a quadratic expression in $u^{(1)}$. The first equation
 in (\ref{e:close-limit_2ndorder}) is exactly (\ref{e:right-left_eigenvalues_v2}) and
 therefore, as
 shown in section \ref{s:Keldysh_resolvent}, its solution $u^{(1)}(\tau, x)$
 admits the Keldysh QNM expansion (\ref{e:Keldysh_QNM_expansion_u_an}).
 Regarding $u^{(2)}(\tau, x)$, taking the Laplace transform in the
 second equation of (\ref{e:close-limit_2ndorder}) and rearranging, we
 can write the analogue of Eq. (\ref{e:u_Laplace}) as
 \bea
\label{e:u2_Laplace}
u^{(2)}(s;x) = i(L +is)^{-1} \left(u^{(2)}_0(x) + S(s,x; u^{(1)}) \right)
= i R_L(-is)  \left(u^{(2)}_0(x) + S(s,x; u^{(1)}) \right) \ .
\eea
Before applying expression (\ref{e:resolvent_L_Keldysh}) for $R_L(-is)$, we
need a model for $S(s,x; u^{(1)})$. Taking into account its quadratic
dependence in $u^{(1)}$, we consider the quadratic expression in QNMs\footnote{This is a just a formal
expression,
sufficient for the present illustration purpose. A more faithful one would also involve
spatial and time derivatives (the latter making appear the QNM frequencies $\omega_n$'s)
of the QNM eigenfunctions $v_n(x)$'s.}
\bea
\label{S_tau}
S(\tau,x; u^{(1)}) &\sim& \sum_{k\ell} S_{k\ell}(\tau,x; u^{(1)})
= \sum_{k\ell} a_{k\ell} \left(v_k(x)e^{i\omega_k\tau}\right)\left(v_\ell(x)e^{i\omega_\ell\tau}\right) H(\tau) \nn\\
&=&  \sum_{k\ell} a_{k\ell} v_k(x)v_\ell(x)  e^{i(\omega_k+\omega_\ell)\tau} H(\tau) \ ,
\eea
where $a_{k\ell}$ are constants and $H(\tau)$ is the Heaviside function
(with $H(\tau)=1, \forall \tau\geq 0$ and $H(\tau)=0, \forall \tau < 0$. Taking the
Laplace transform we get (remember the relation $s=i\omega$)
\bea
\label{S_s}
  S(s, x; u^{(1)}) \sim  \sum_{k\ell} \frac{a_{k\ell} v_k(x)v_\ell(x)}{s-(s_k+s_\ell)} \ .
  \eea
  Inserting this expression into Eq. (\ref{e:u2_Laplace}) and proceeding as in
  section \ref{s:QNM_aymptotic_expansions}, we finally can write
  \bea
  u(\tau, x) \sim
  \sum_n\biggl(\bigl(a_n^{(1)} + \epsilon(a_n^{(2)} - \sum_{k\ell}a_{n,k\ell}^{(2)}) \bigr)e^{i\omega_n\tau}
  + \epsilon  \sum_{k\ell}a_{n,k\ell}^{(2)}e^{i(\omega_k+\omega_\ell)\tau} \biggr) v_n(x)
  \eea
  with
  \bea
  a_n^{(1)} &=& \langle \alpha_n, u_0^{(1)}\rangle \nn \\
  a_n^{(2)} &=& \langle \alpha_n, u_0^{(2)}\rangle \nn \\
  a_{n,k\ell}^{(2)}  &=& i \frac{a_{k\ell}}{\omega_n-(\omega_k+\omega_\ell)} \langle \alpha_n, v_k v_\ell\rangle \ ,
  \eea
  where the initial data is written $u_0(x)= u_0^{(1)}(x) + \epsilon u_0^{(2)}(x)$. Expressions
  usually employed in gravitational wave ringdown context are obtained by evaluating
  this expression at $x=x_{\scri^+}$. Note that only new sum frequencies $\omega_k+\omega_\ell$
  appear at second order, but the
  excitation coefficients of the fundamental QNM frequencies get also modified.

   \section{Hyperboloidal evolution in black hole backgrounds: time-domain versus spectral QNM expansions}
   \label{s:Time-domain_evolution}
   In this section we present some exploratory results on the comparison between the time-domain evolution
   and the spectral-domain QNM resonant
expansions in a class of scattering problems corresponding to the linear propagation of
scalar, electromagnetic and gravitational field perturbations on backgrounds given
by (the domain of outer of communication of) stationary  BH spacetimes,
by adopting the hyperboloidal approach discussed in
\cite{zenginouglu2008hyperboloidal,Zenginoglu:2011jz,Warnick:2013hba,Ansorg:2016ztf,PanossoMacedo:2018hab,Jaramillo:2020tuu,PanossoMacedo:2020biw,panosso2024hyperboloidal,zenginouglu2024hyperbolic,macedo2024hyperboloidal}.
We follow the treatment in these references, in particular adopting the notation in \cite{Jaramillo:2020tuu}
(details are given in appendix \ref{a:Hyperboloidal approach}). Such a hyperboloidal
approach permits to cast the evolution problem in the form (\ref{e:wave_eq_1storder_u_tau}),
with $L$ non-selfadjoint, in such a way that the Keldysh approach to QNMs discussed
in section \ref{s:QNM_aymptotic_expansions} can be applied straightforwardly.

We adopt a ``proof-of-principle'' approach, restraining ourselves
to spherically symmetric spacetimes{\footnote{A very important point in our discussion, however, is the fact that
  the Keldysh scheme extends beyond
spherically black hole spacetimes to arbitrary (stationary) geometries, without spatial
symmetries and,  in addition, in arbitrary (odd) spatial dimensions (cf. footnote \ref{footnote:Huygens}).
It generalises in this sense the
``effective one-dimensional'' discussion by Ansorg \& Macedo in reference~\cite{Ansorg:2016ztf}.}. The basic starting
  equation can be taken then as the $1+1$-dimensional problem
  \bea
 \label{e:wave_equation_generic}
\left(\frac{\partial^2}{\partial \overline{t}^2} - \frac{\partial^2}{\partial \overline{x}^2}
+ \hat{V} \right)\phi=0 \ ,
\eea
where $\overline{x}\in]-\infty, \infty[$, subject to outgoing boundary conditions
  at $\overline{x}\to\pm\infty$ and 
 with  
initial data $\phi(\overline{t}=0, \overline{x}) = \phi_1(\overline{x})$
and $\partial_{\overline{t}} \phi(\overline{t}=0, \overline{x}) = \phi_2(\overline{x})$ in a Cauchy slice.
Adopting the hyperboloidal scheme, with a (coordinate) compactification in the
spatial hyperboloids
(cf. appendix \ref{a:Hyperboloidal approach}), this equation
is written in the  Schr\"odinger form (\ref{e:wave_eq_1storder_u_tau}), with $L$ given by
\bea
\label{e:L_L_1_L_2_main}
	 L = \frac{1}{i}
	\left(\begin{array}{c|c}
    0 & 1\\ \hline
    L_1 & L_2
	\end{array}\right) \ ,
        \eea
        where the explicit form of $L_1$ and $L_2$ are given in (\ref{e:L_1-L_2}) and $L$
        acts on
        \bea
        u = \begin{pmatrix} \phi\\ \psi \end{pmatrix}  \ \ , \  \ \hbox{with}  \ \psi = \partial_\tau \phi  \ .
        \eea    
        The $L_1$ and $L_2$ encode, respectively, the bulk content in the potential and the
        outgoing boundary conditions. In particular, $L_2$ is responsible of the
        non-selfadjoint character of $L$.

\medskip
        
        In this and the subsequent sections, we
        systematically study a reference toy-model and three spherically
        symmetric BH spacetimes with different asymptotics at null infinity, namely:
        \begin{itemize}

        \item[i)] The P\"oschl-Teller potential (details in appendix \ref{a:PT}).

        \item[ii)] The Schwarzschild BH, asymptotically flat case  (details in appendix \ref{a:S})

      \item[iii)] The Schwarzschild-de Sitter (S-dS) BH (details in appendix \ref{a:SdS}).

        \item[iv)] The Schwarzschild-Anti-de Sitter (S-AdS) BH (details in appendix \ref{a:SAdS}).
                
        \end{itemize}
        For better comparison among the four cases of study and in the mentioned
        ``proof-of-principle'' spirit, we discuss the time-domain evolution and frequency-domain
        spectral QNM expansions with the same Gaussian initial data, namely the
        one presented in appendix \ref{e:Gaussian_ID} (note however that the S-AdS case
        involves a slight adaptation). We have explored other families of initial data,
        always finding the same performance in both the time-domain and
        frequency-domain spectral treatments. For the sake of clarity, we restrain here
        to Gaussian initial data, leaving the systematic discussion of more
        generic initial data for future work \cite{BesJarPoo24}.

\subsection{Illustration of the hyperboloidal evolutions: qualitative proof of principle}\label{s:hyperboloidal evolutions}
In order to  illustrate the behaviour of the constructed numerical evolutions of the field $\phi$ in the hyperboloidal scheme, we present  in Fig. \ref{waveforms} the evolution in time $\tau$, for the four cases of study,  of the field evaluated at future null infinity (one endpoint of the grid). In the AdS case we rather plot the field at the horizon 
(after rescaling the field by $\sigma$, it only makes sense to show the waveform at the event horizon $\sigma=1$). That is, this is the field an observer at null infinity (or at the horizon in the AdS case) would measure.
  \begin{figure}[htp]
    \centering
    \subfloat[P\"oschl-Teller]{
      \includegraphics[clip,width=\sizefigmedium\columnwidth]{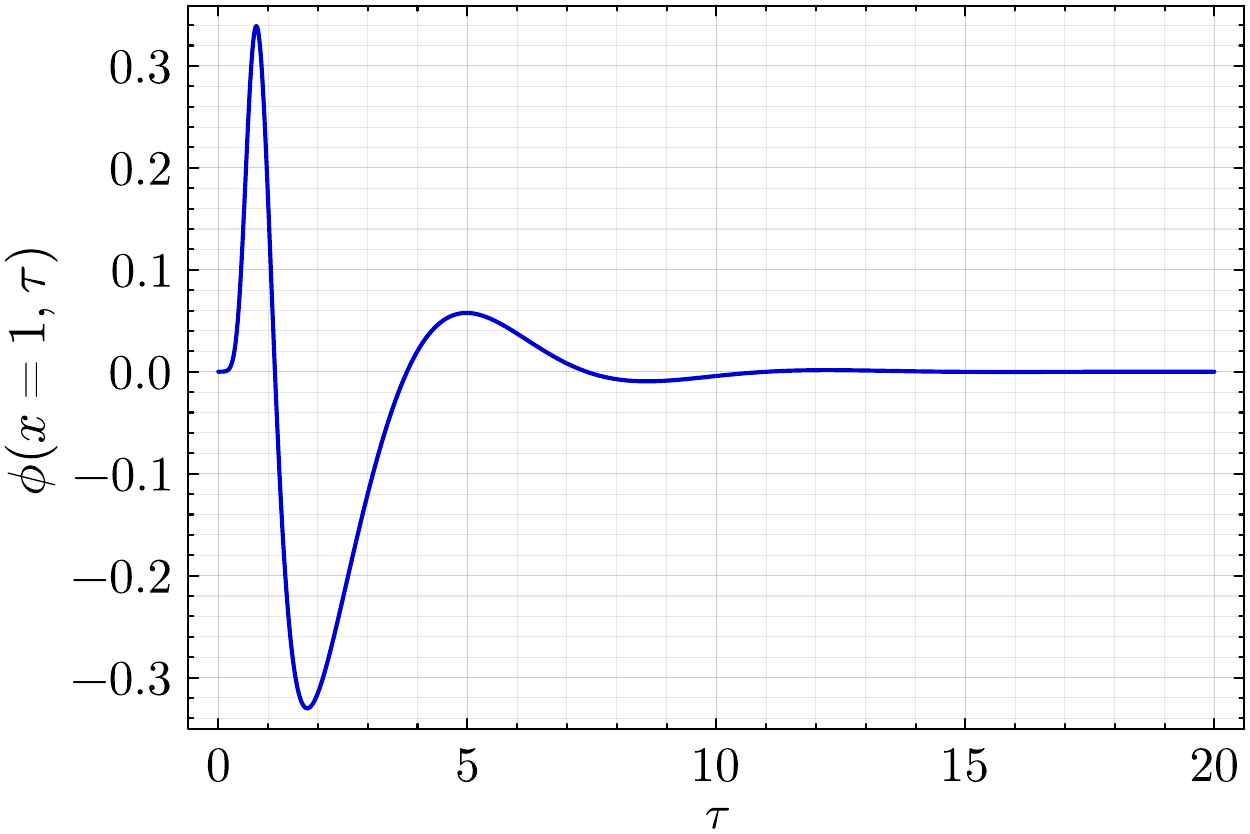}\label{waveforms:PT}
    }
    \subfloat[Schwarzschild-dS]{
      \includegraphics[clip,width=\sizefigmedium\columnwidth]{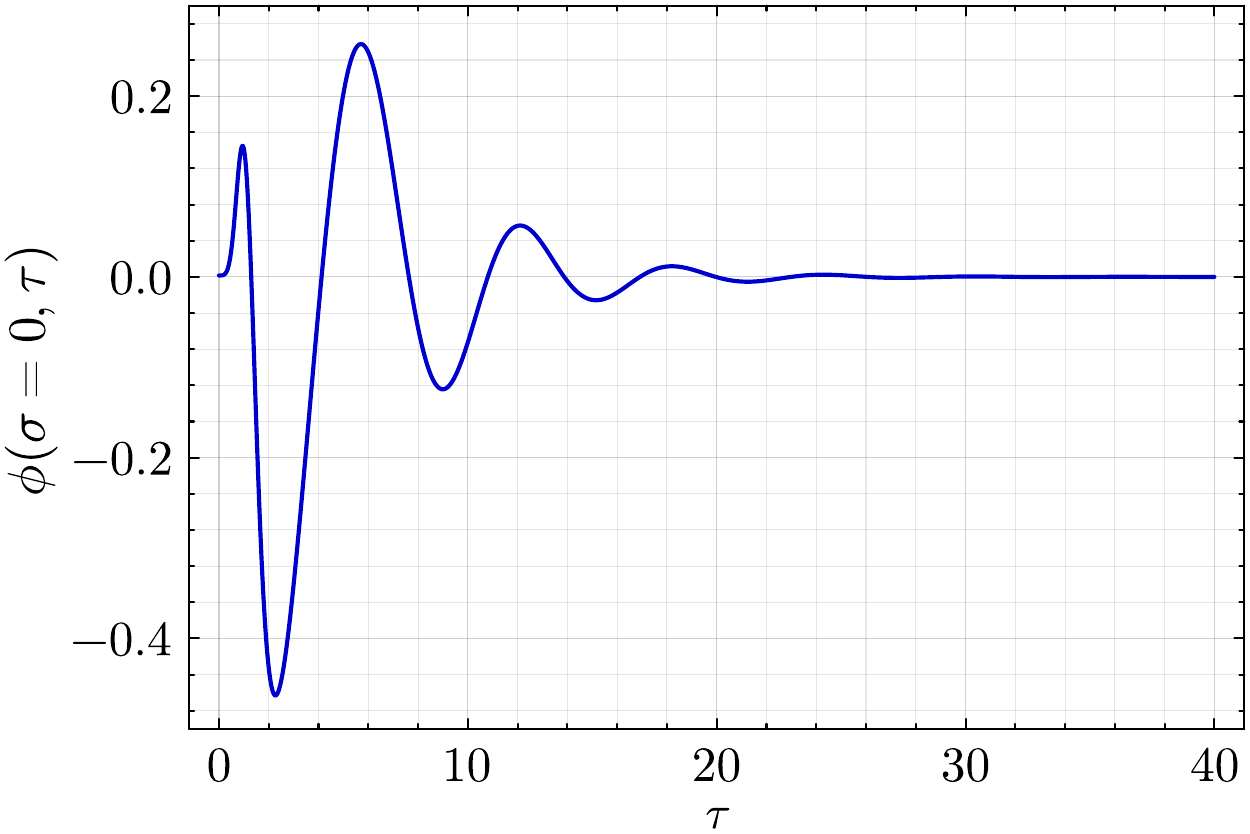}\label{waveforms:dS}
    }
  
    \subfloat[Schwarzschild]{
      \includegraphics[clip,width=\sizefigmedium\columnwidth]{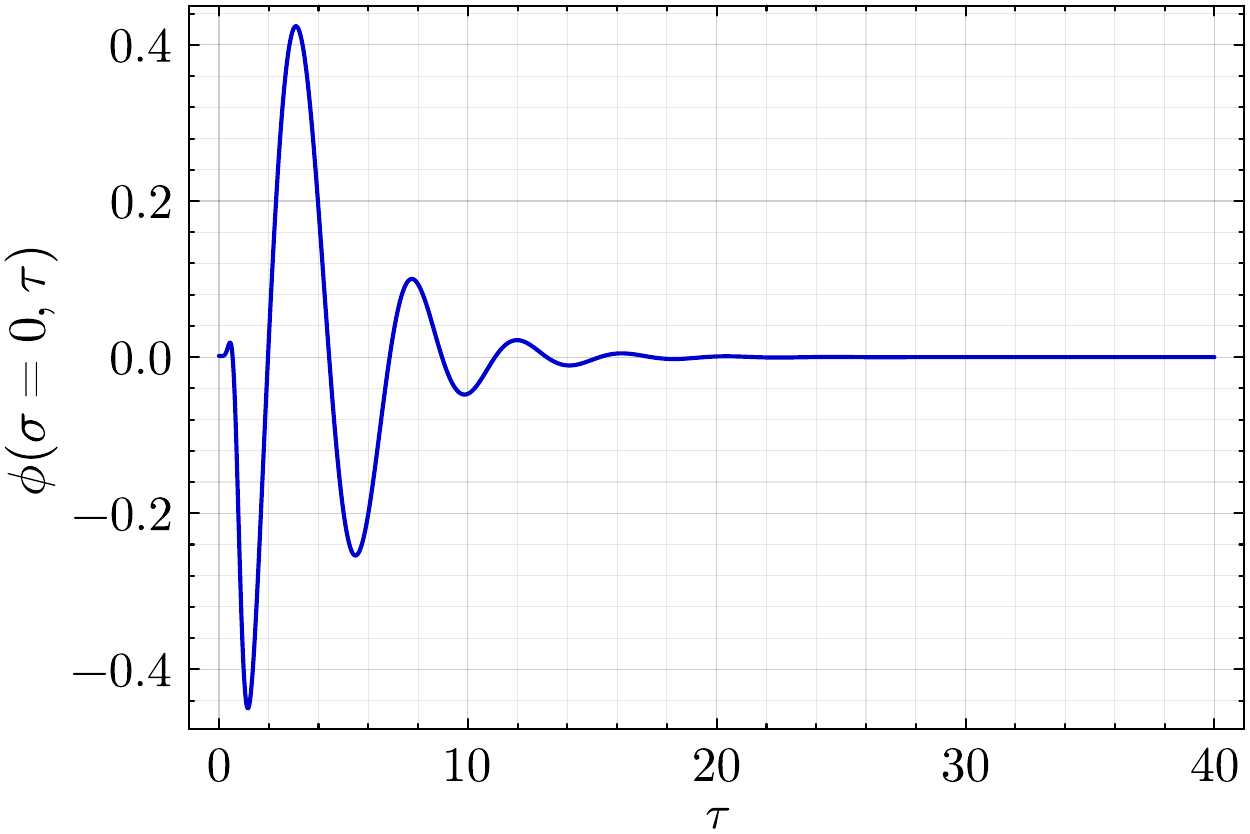}\label{waveforms:S}
    }
    \subfloat[Schwarzschild-AdS]{
      \includegraphics[clip,width=\sizefigmedium\columnwidth]{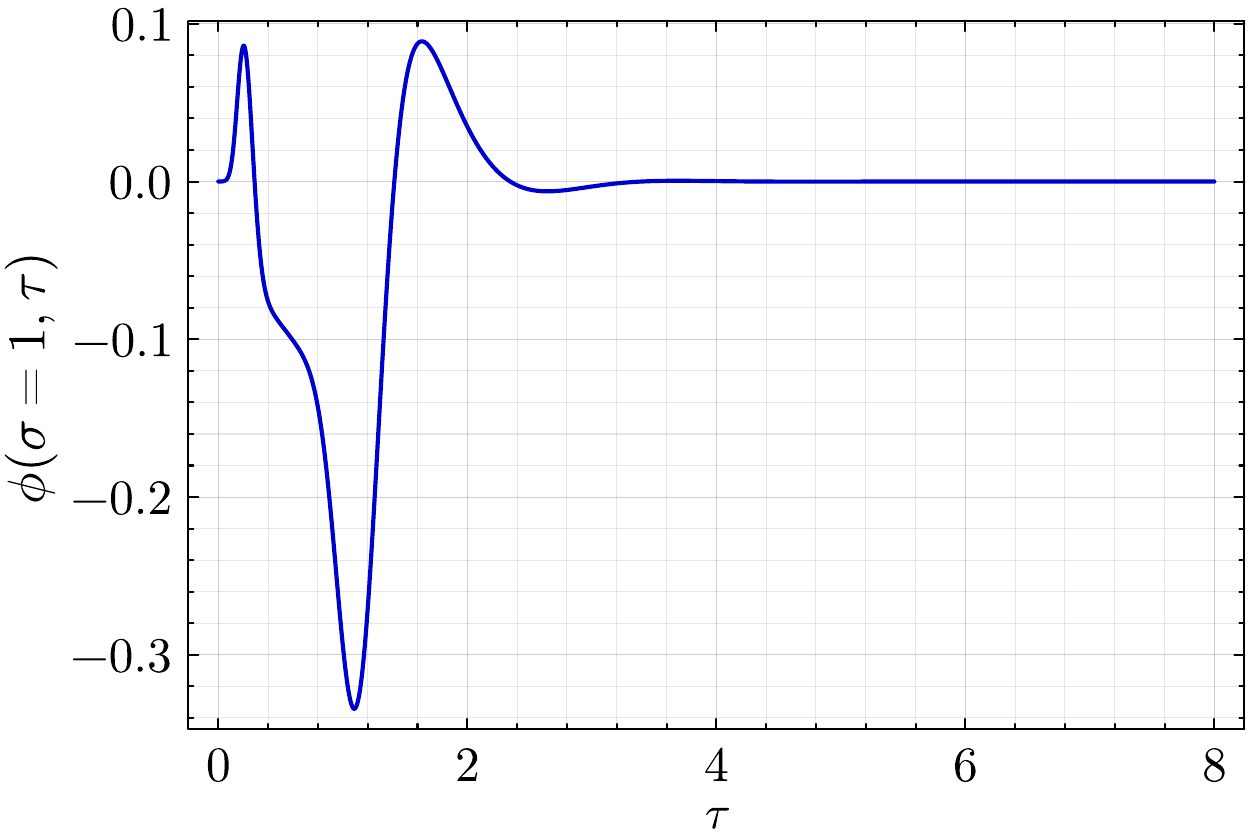}\label{waveforms:AdS}
    }
    \caption{Panels \ref{waveforms:PT}, \ref{waveforms:dS}, \ref{waveforms:S} and  \ref{waveforms:AdS} are waveforms at future null infinity and at the event horizon for the AdS case (timeseries at a boundary of the Chebyshev-Lobatto grid).}
    \label{waveforms}
  \end{figure}
  
  Since the Gaussian initial condition is initially very small at the boundary of the interval, the timeseries is initially flat and close to zero during a very short time, then it increases abruptly, oscillates until it reaches its peak and finally decreases,  the signal being dominated at late times by the fundamental mode in the absence of tails (notice the fundamental mode in the P\"oschl-Teller case is a constant function of $x$ illustrated later on panel \ref{eigenfunctions:PT}).

  We provide a different view of the field in the supplementary material (\verb|PT_evol.mp4| and \verb|SAdS_evol.mp4|) that aims in particular at verifying that the boundary conditions are well-implemented, namely, the field must be purely outgoing at the edges of the interval.
  
  The time-series in Fig. \ref{waveforms} (and specially the full simulations
  in the  supplementary material)
  are meant to get some qualitative intuition. Among the different details
  in such simulations, we make the following two comments:
\begin{itemize}
  \item [i)] \textit{P\"oschl-Teller case.} The field travels outwards in both directions and becomes apparently flat in the spatial direction (dominated by the fundamental mode) after $\tau\approx1.8$ (see \verb|PT_evol.mp4|), corresponding to the minimum of $\phi(x=1,\tau)$ according to Fig. \ref{waveforms:PT}. 
  \item [ii)] \textit{Schwarzschild-AdS case.} The time evolution on \verb|SAdS_evol.mp4| is obtained using the DAE scheme (see Eq. (\ref{DAE_sol}) in appendix \ref{a:method_lines}), in particular the initial condition is adapted from the Gaussian in appendix \ref{e:Gaussian_ID}, due to the rescaling $u_0=\sigma\widetilde{u_0}$. Qualitatively, the AdS boundary acts as a reflecting box at $\sigma=0$, the amplitude of the signal \ref{waveforms:AdS} relative to the maximum of the initial condition on \verb|SAdS_evol.mp4| is much higher compared to P\"oschl-Teller because it can only dissipate energy at the event horizon.
\end{itemize}

\subsection{Keldysh QNM expansion : cases of study}\label{Keldysh_expansion_cases_study}
After the previous first contact and qualitative illustration of the hyperboloidal time evolutions, we now proceed to a systematic study of the comparison of this `time-domain' evolutions with the `frequency-domain' evolutions provided by the
asymptotic Keldysh resonant expansions discussed in section \ref{s:Keldysh}. We perform this comparison for the four cases of study, starting with the numerical construction of spectral elements, i.e. the QNM frequencies and QNM functions, as eigenvalues
and eigenfunctions of the associated spectral problem, respectively. With these elements at hand, we proceed to the assessment of the Keldysh expansions by calculating the amplitude coefficients, by addressing the contribution of overtones to the waveform  and the presence of tails in the Schwarzschild case. The systematic
comparison of the time-domain and frequency-domain calculation will serve to assess and validate the latter and,
simultaneously, to provide a form of convergence test for the former.
In this section we focus on the time-series at boundaries $x_b$, calculating in particular
the coefficients  $\mathcal{A}_n(x_b)$, leaving the bulk discussion
and the excitation coefficient $a_n$'s for the next section.

\begin{figure}[htp]
  \centering
  \subfloat[P\"oschl-Teller]{
    \includegraphics[clip,width=\sizefigmedium\columnwidth]{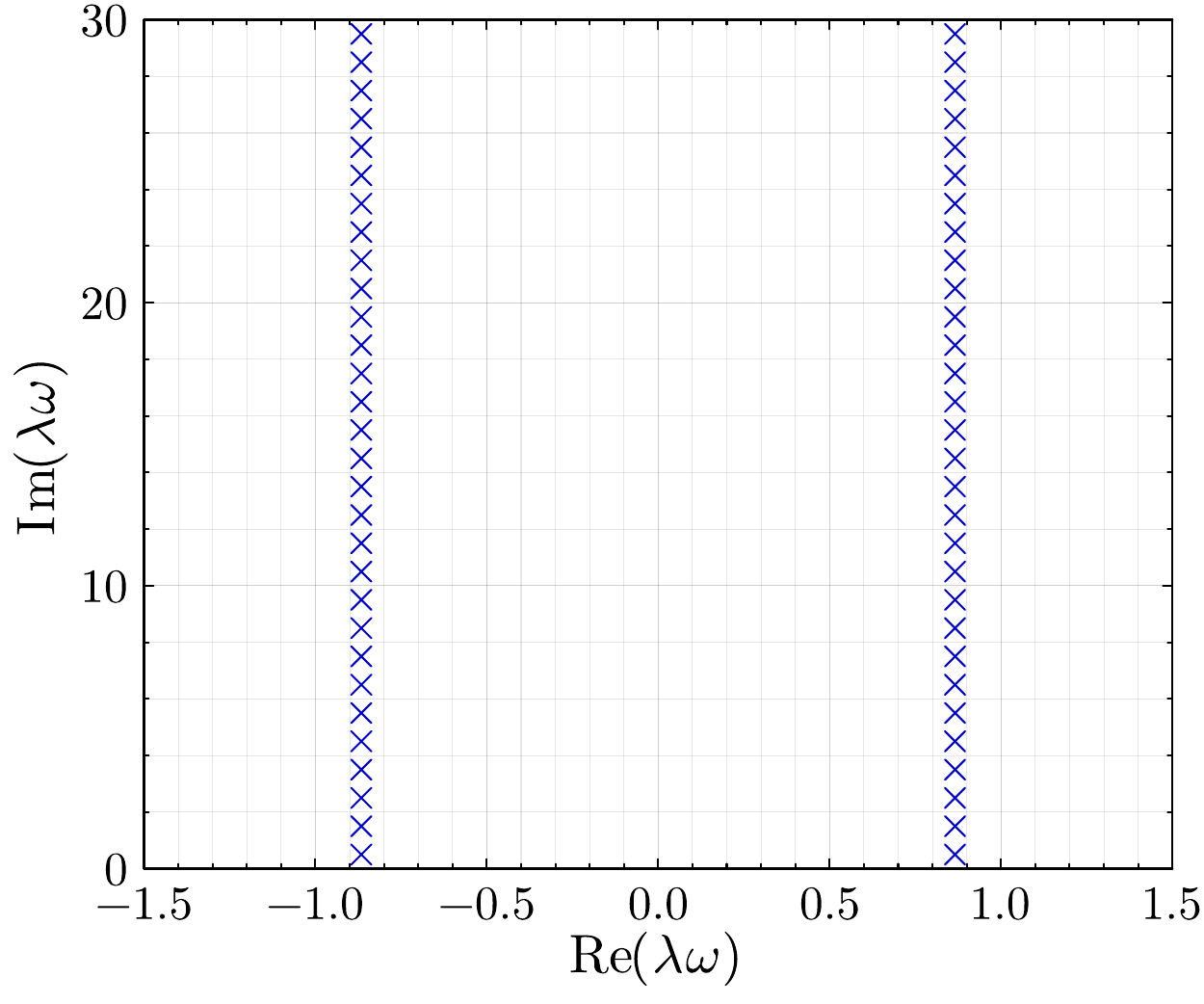}\label{spectres:PT}
  }
  \subfloat[Schwarzschild-dS]{
    \includegraphics[clip,width=\sizefigmedium\columnwidth]{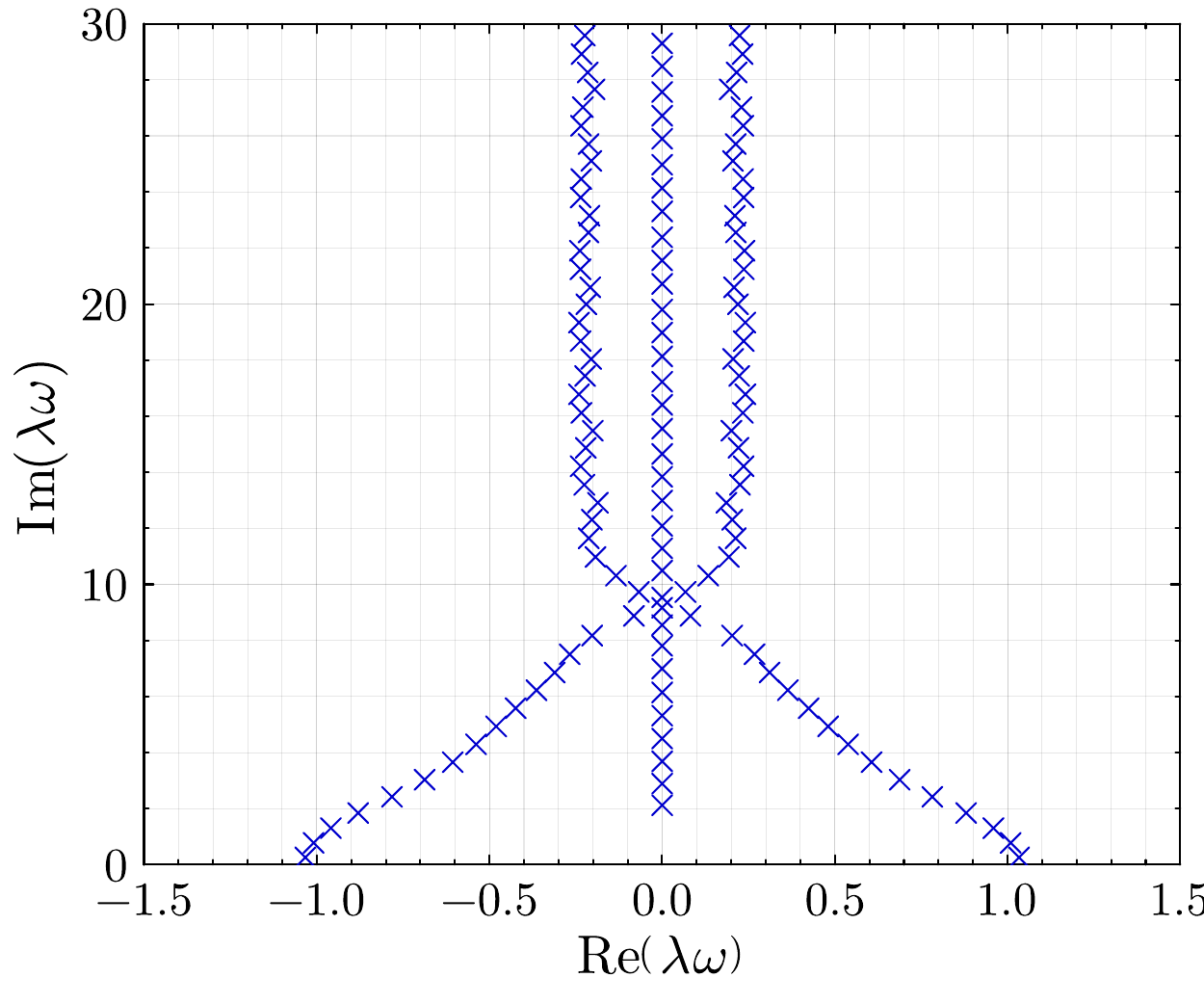}\label{spectres:SdS}
  }

  \subfloat[Schwarzschild]{
    \includegraphics[clip,width=\sizefigmedium\columnwidth]{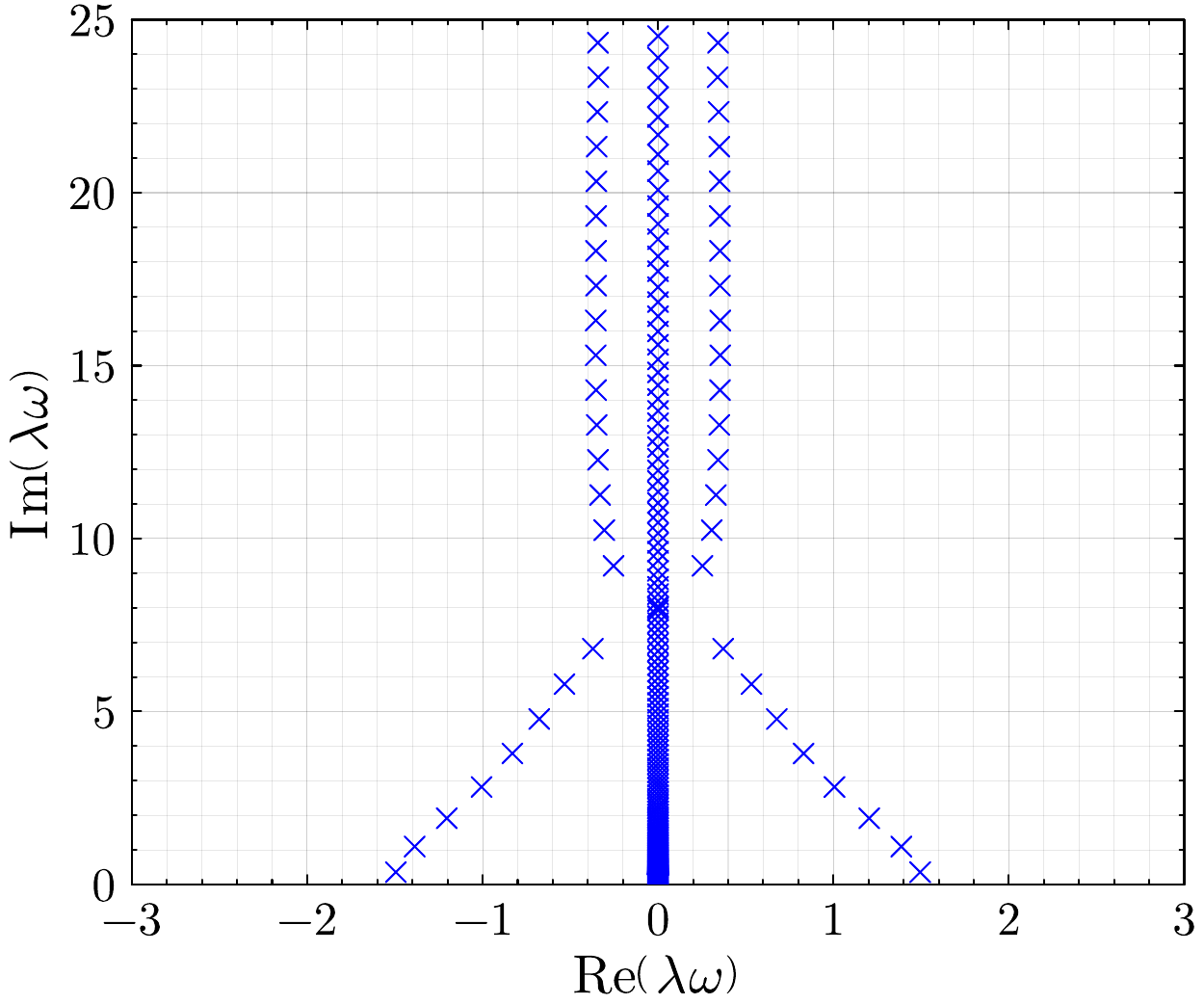}\label{spectres:S}
   }
   \subfloat[Schwarzschild-AdS]{
     \includegraphics[clip,width=\sizefigmedium\columnwidth]{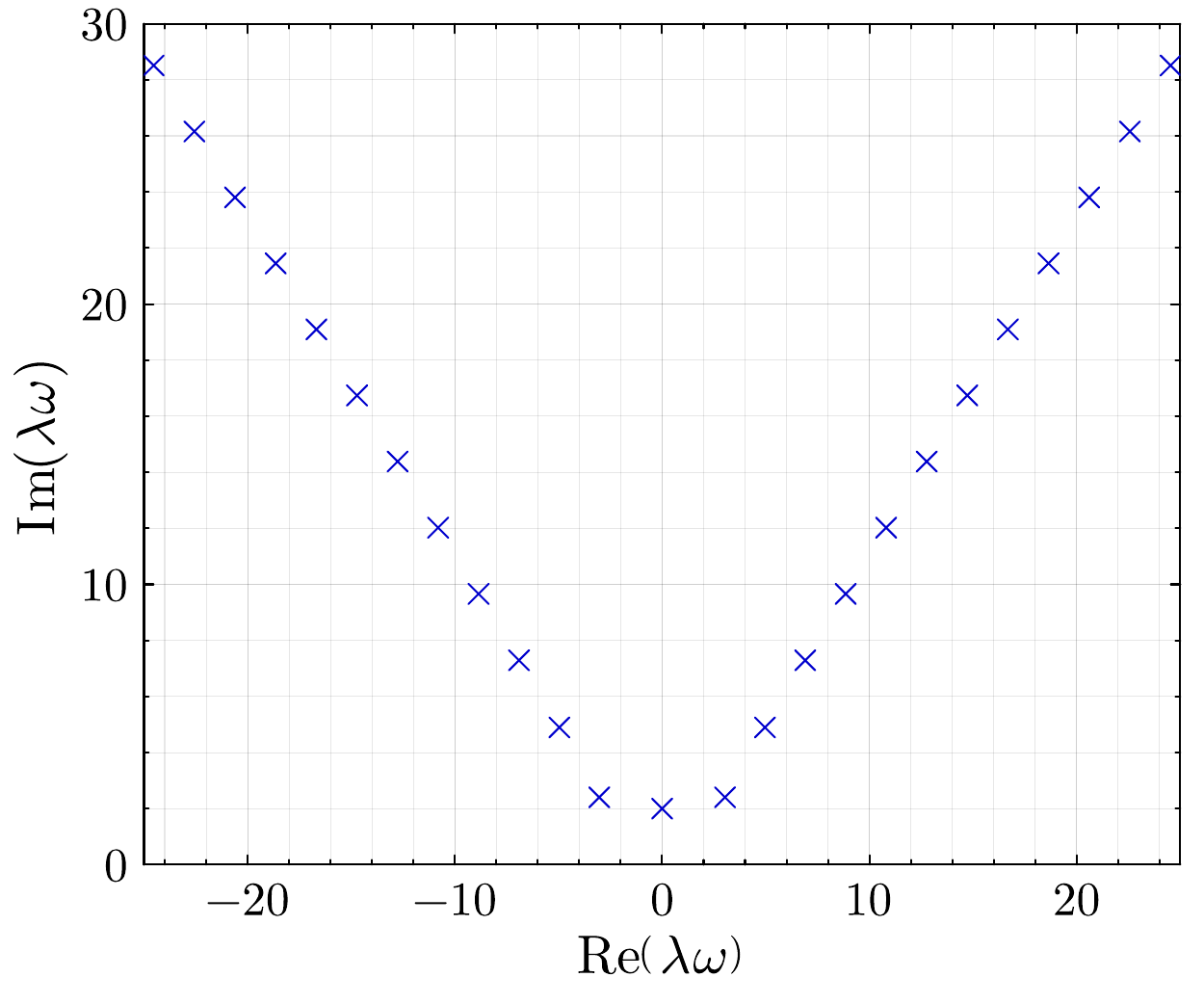}\label{spectres:SAdS}
   }

   \caption{Panels \ref{spectres:PT}, \ref{spectres:SdS}, \ref{spectres:S} and  \ref{spectres:SAdS} are spectra of the cases of study. These figures are made using $N=700, 600, 600$ and $400$ respectively.}
   \label{spectres}
 \end{figure}

 \begin{figure}[htp]
  \centering
  \subfloat[P\"oschl-Teller]{
    \includegraphics[clip,width=\sizefigmedium\columnwidth]{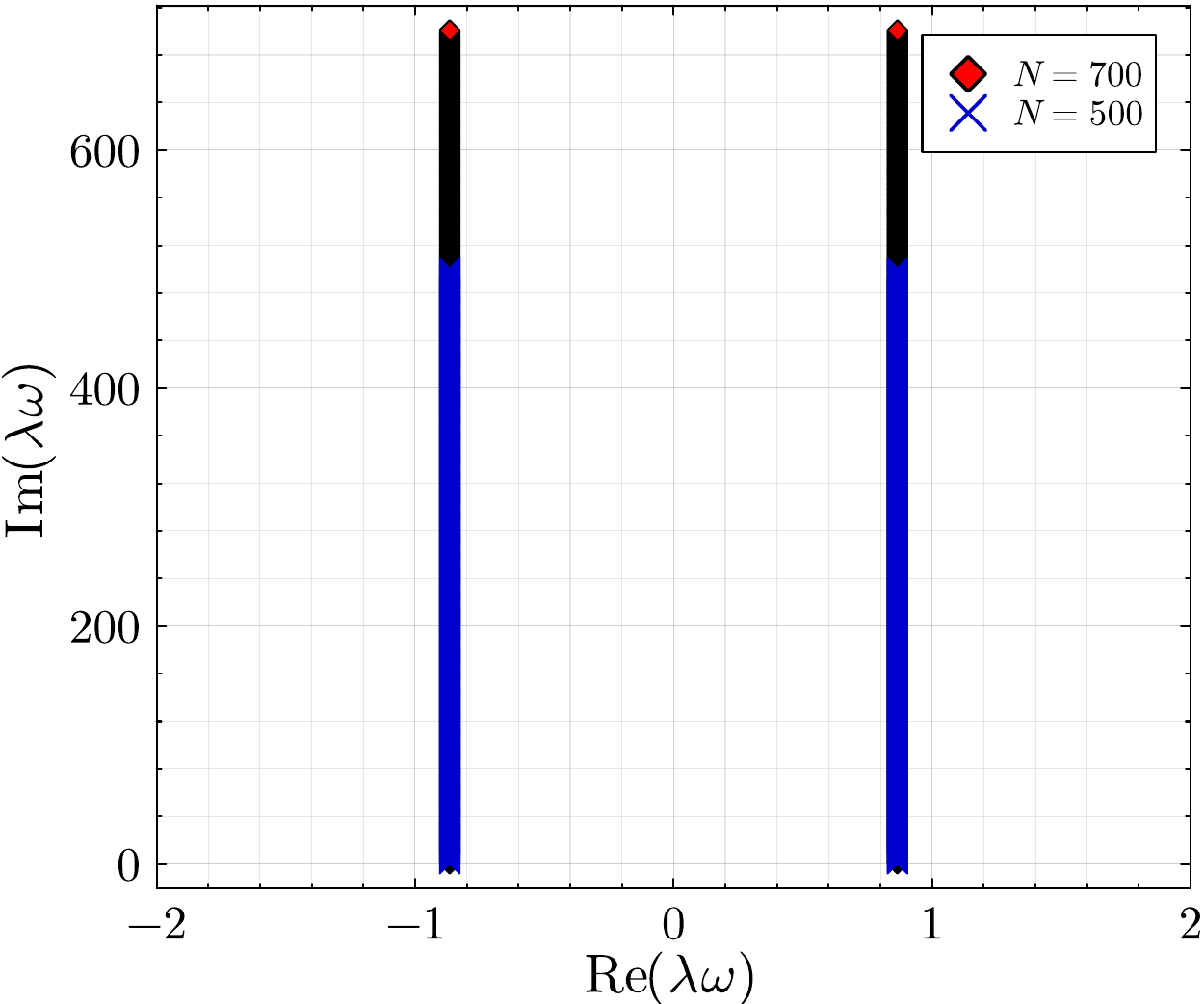}\label{spectresconv:PT}
  }
  \subfloat[Schwarzschild-dS]{
    \includegraphics[clip,width=\sizefigmedium\columnwidth]{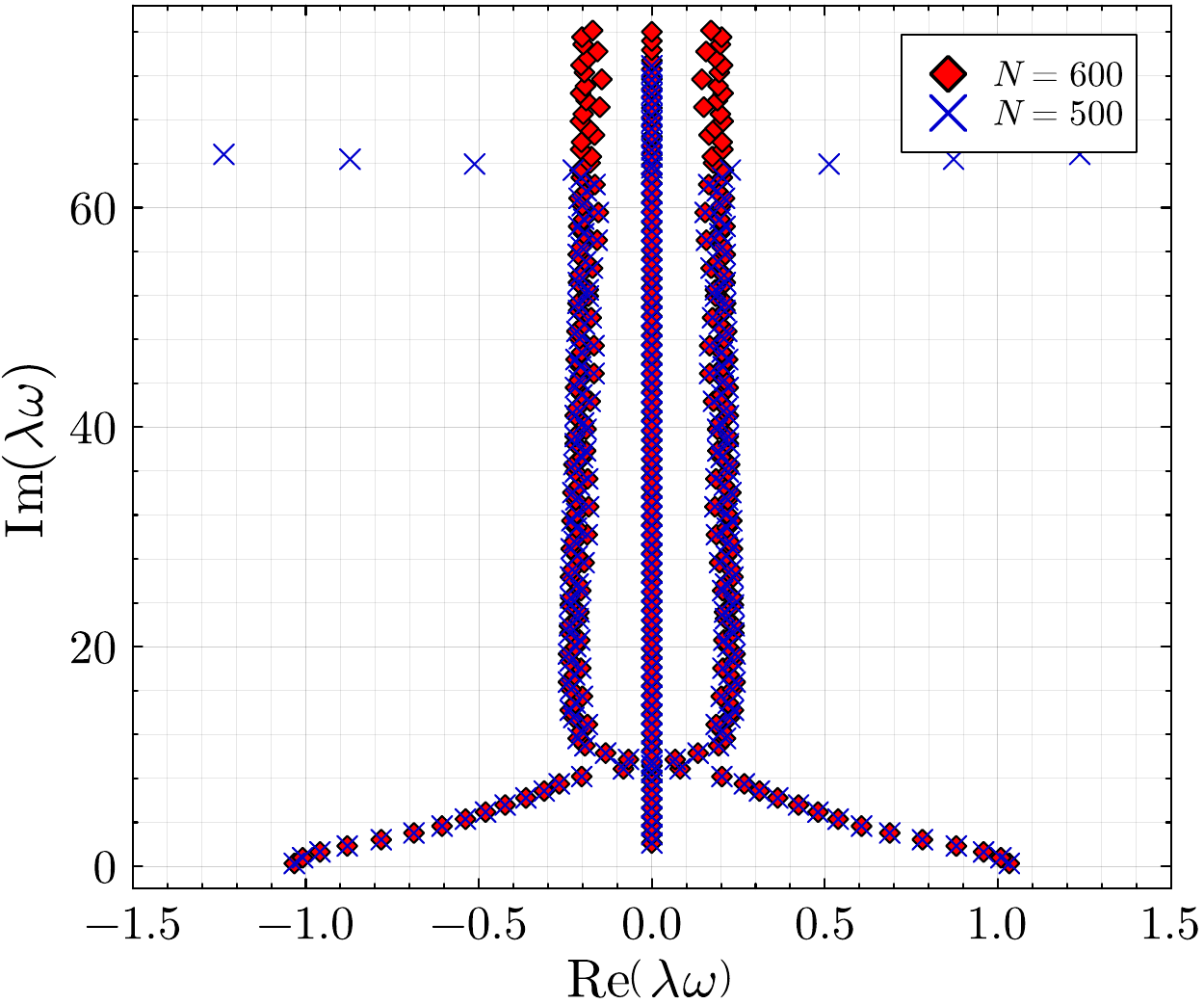}\label{spectresconv:SdS}
  }

  \subfloat[Schwarzschild]{
    \includegraphics[clip,width=\sizefigmedium\columnwidth]{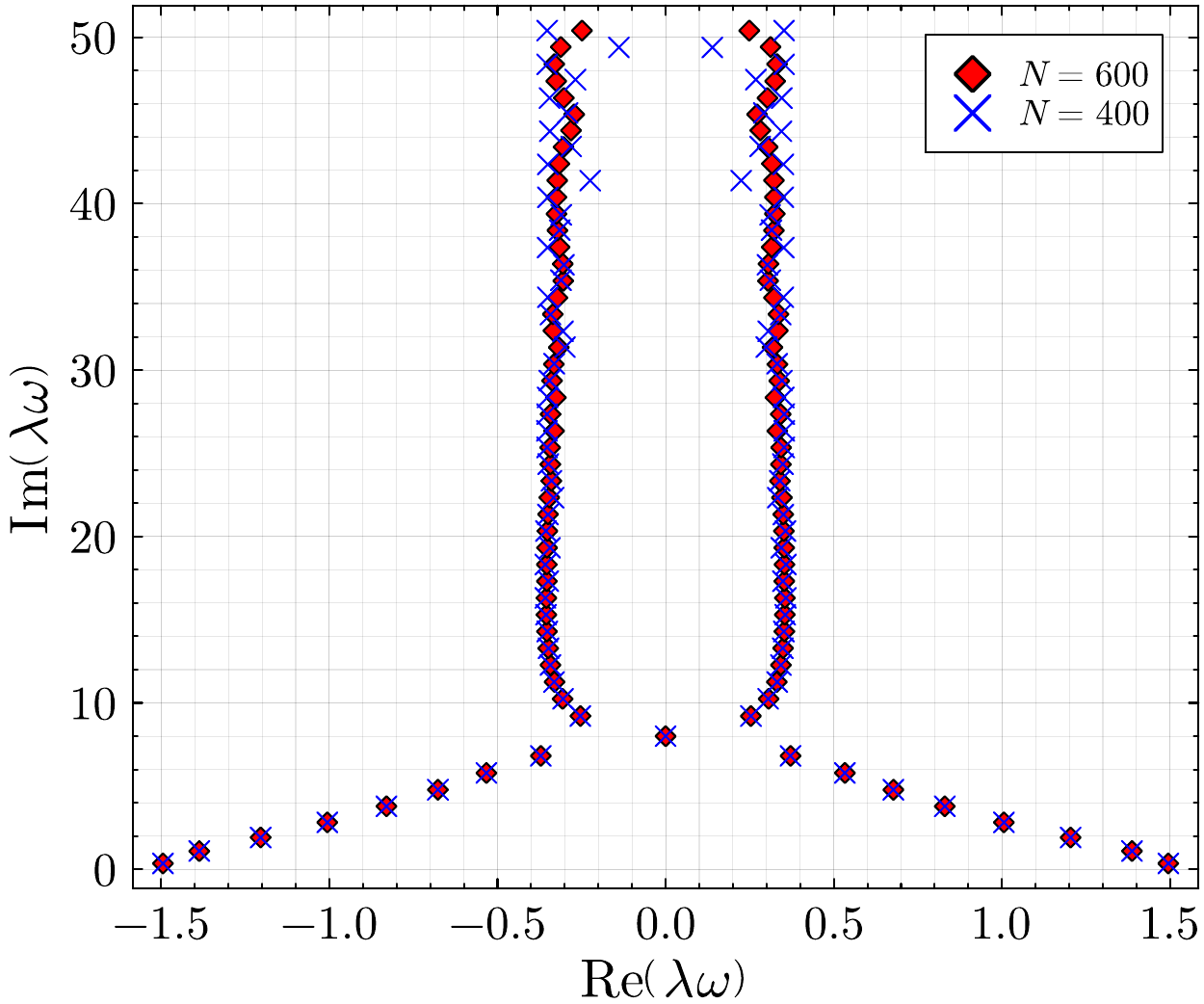}\label{spectresconv:S}
  }
  \subfloat[Schwarzschild-AdS]{
    \includegraphics[clip,width=\sizefigmedium\columnwidth]{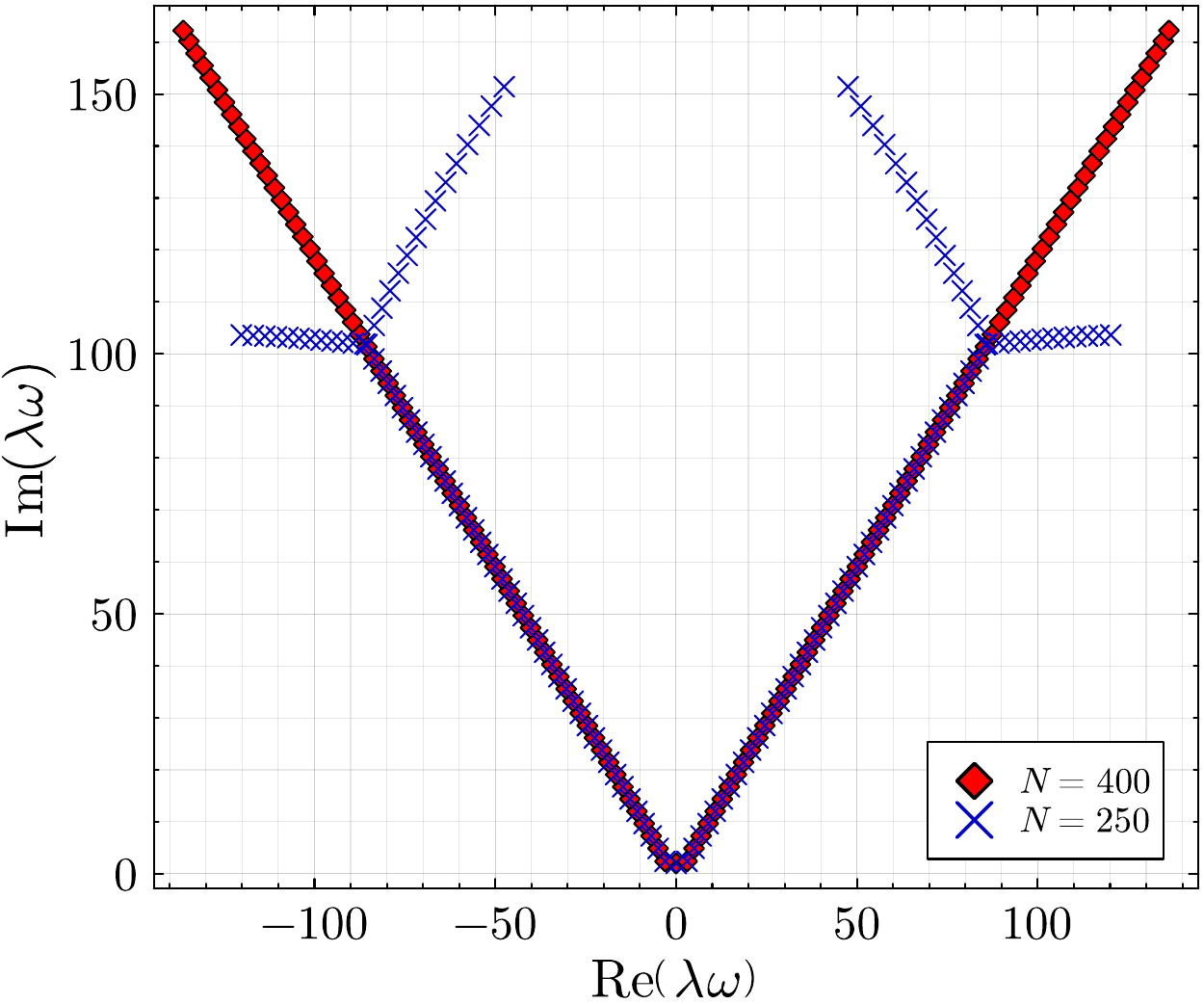}\label{spectresconv:AdS}
  }
  \caption{Panels \ref{spectresconv:PT}, \ref{spectresconv:SdS}, \ref{spectresconv:S} and \ref{spectresconv:AdS} are spectra of the cases of study for different values of $N$.}
  \label{spectresconv}
\end{figure}

\subsubsection{QNM spectral problem.} \label{s:QNM_spectral_problem} In a first step, we solve numerically the spectral problems in Eqs.
(\ref{e:right-left_eigenvalues}), obtaining the numerical approximations to the QNM frequencies
(eigenvalues) $\omega_n$, and the numerical approximations to the
right- and left-eigenvectors $v_n$ and $\alpha_n$, respectively.

As an illustration of the result, Figs.~\ref{spectres} and \ref{spectresconv} (the latter
addresses the convergence of the former, see below) provide a spectral follow-up to the time-domain
Fig. \ref{waveforms}, by presenting a view of the QNM spectra upon which we are going 
to construct our spectral discussion. Figure \ref{eigenfunctions} shows the first eigenfunctions for the four cases of study.

Some general comments are in order:
\begin{itemize}
\item[i)] {\em Labelling of QNMs}. Regarding the labelling of QNMs, given the particular
  structure in the complex plane of the studied QNM spectra,
each eigenvalue $\omega^\pm_n$ (in a given QNM branch $\omega^\pm_n$)
is labelled by $n$ and ordered by increasing
imaginary part $\mathrm{Im}{\left(\omega_n^\pm\right)}$.

\item[ii)] {\em General description of spectra}. P\"oschl-Teller and Schwarzschild panels, respectively panel \ref{spectres:PT}
and panel \ref{spectres:S},
simply recover the results in \cite{Jaramillo:2020tuu}. Even if they are not QNMs, note that in the Schwarzschild case we have kept the eigenvalues corresponding to the discretisation of the `branch cut'. They do not converge when $N$ increases, but we keep them in the discussion for later convenience. Regarding the asymptotically dS and AdS cases, there is a
dependence
on the choice of the cosmological constant $\Lambda$. Rather than a systematic study on the dependence on this parameter, and in the spirit of a `proof-of-principle' calculation, we choose some particular $\Lambda$. 
Specifically, in the asymptotically dS case, the high QNM  overtones in panel \ref{spectres:SdS} have a slightly oscillating behaviour that depends on the chosen cosmological constant
($\Lambda=0.07/M^2$ here). Choosing a higher cosmological constant, closer to the extremal limit $\Lambda_{\text{ext}}=1/(9M^2)$, where the event horizon $r_+$ and the cosmological $r_\Lambda$ radii
(see Eq. (\ref{e:f_SdS})) coalesce,
does reduce these oscillations, actually leading to a P\"oschl-Teller-like QNM structure at
extremal Schwarzschild-dS $\Lambda_{\text{ext}}=1/(9M^2)$. Generically speaking, we need to use a high grid size $N$ in the Schwarzschild and Schwarzschild-dS cases to capture the structure of the overtones, only revealed ``deep'' in the complex plane.

\item[iii)] {\em Convergence of the QNM frequencies}. The convergence of these  QNM $\omega_n$'s,
  cast as eigenvalues
  of the appropriate non-selfadjoint operator corresponding to each potential,
  has been studied in the literature (cf. e.g. \cite{Jaramillo:2020tuu,sarkar23}). For the purpose of the present discussion,
  we consider the straightforward (qualitative) test of assessing which $\omega_n$'s converge when $N$ increases. 
  Specifically, we calculate  $\omega_n$'s for different resolutions and keep only those that coincide when calculated with
  the different resolutions. For the sake of clarity in Fig. \ref{spectresconv}  we show the calculation 
with two different resolutions $N_1<N_2$, and we keep only those eigenvalues coinciding for both resolutions.
Since further increasing the
resolution in $N$ does not change the coefficients already stabilised, we take this as a criterion of convergence. A more systematic study of this point will be developed in \cite{BesBoyJar24}.

The Schwarzschild case is however particularly delicate, among our cases of study, a feature that
impacts the
Keldysh expansion we will discuss later.
The branch cut in panel \ref{spectres:S} is excluded from the convergence test in
panel \ref{spectresconv:S} since these eigenvalues  do not converge with $N$ (unlike the de Sitter modes in panel \ref{spectresconv:SdS}).
Their presence seems to heavily influence the actual Schwarzschild QNMs, even those that are not very high in the complex plane.
As a consequence, it becomes more subtle to compute a QNM expansion out of a truncated sum of modes that have converged. This will be discussed later in section \ref{s:conv_coeffs_keldysh}.

\end{itemize}

\begin{figure}[htp]
  \centering
  \subfloat[P\"oschl-Teller]{
    \includegraphics[clip,width=\sizefigmedium\columnwidth]{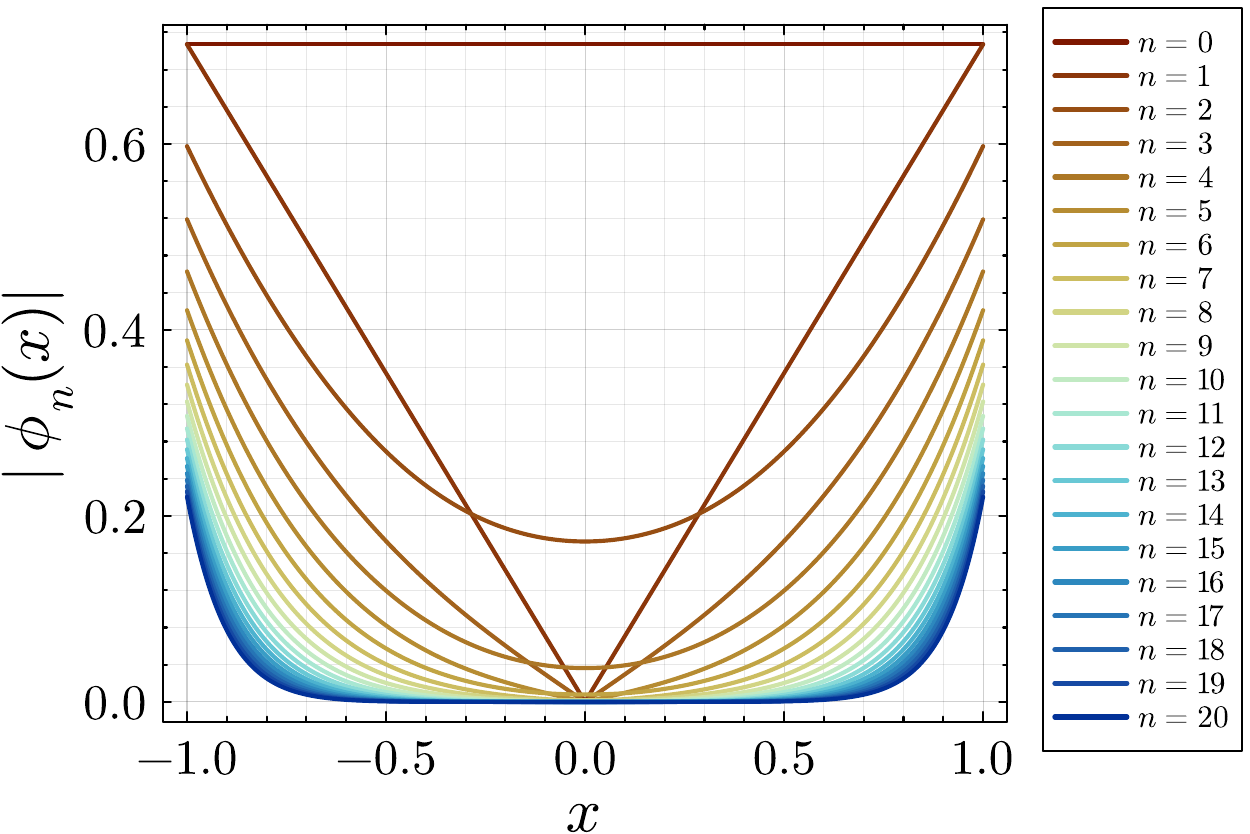}\label{eigenfunctions:PT}
  }
  \subfloat[Schwarzschild-dS]{
    \includegraphics[clip,width=\sizefigmedium\columnwidth]{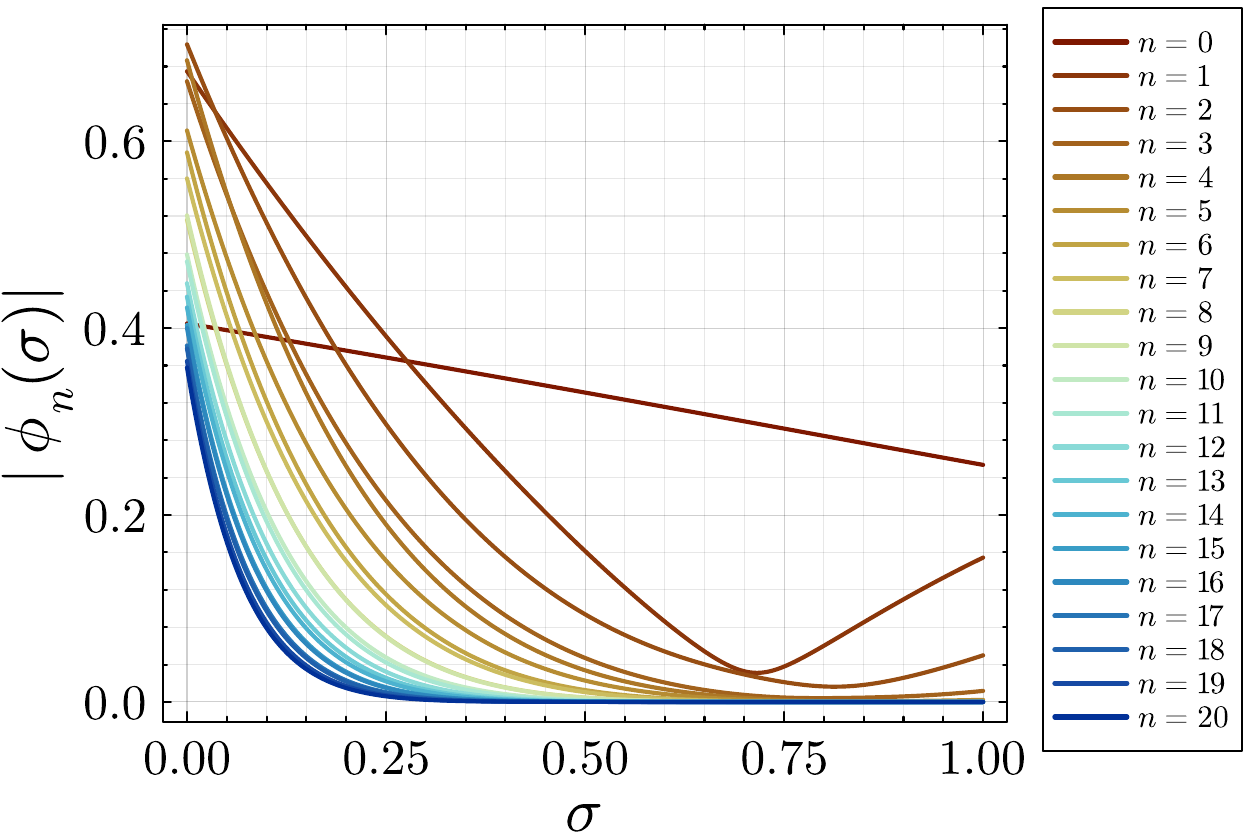}\label{eigenfunctions:SdS}
  }

  \subfloat[Schwarzschild]{
    \includegraphics[clip,width=\sizefigmedium\columnwidth]{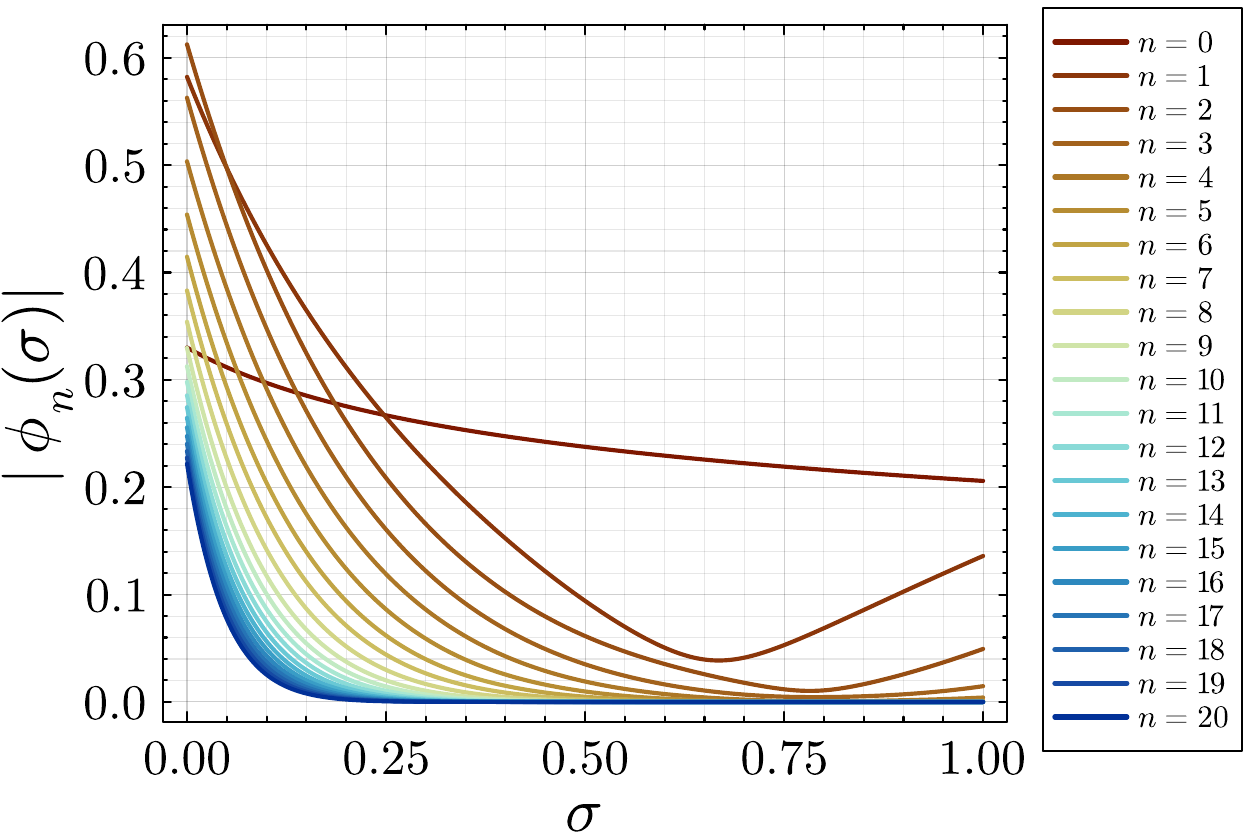}\label{eigenfunctions:S}
  }
  \subfloat[Schwarzschild-AdS]{
    \includegraphics[clip,width=\sizefigmedium\columnwidth]{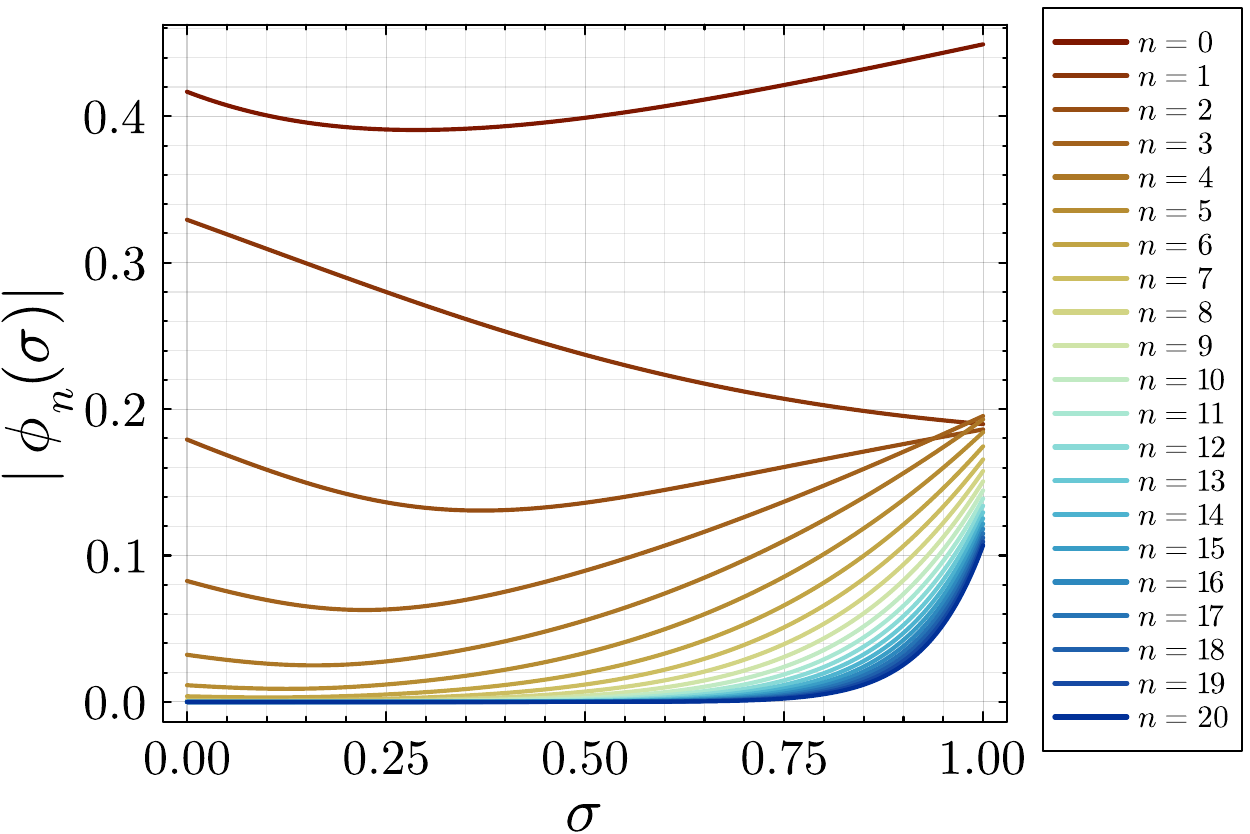}\label{eigenfunctions:AdS}
  }
  \caption{Panels \ref{eigenfunctions:PT}, \ref{eigenfunctions:SdS}, \ref{eigenfunctions:S} and \ref{eigenfunctions:AdS} show the first 21 eigenfunctions $\phi_n$ for each case of study. The eigenfunctions are normalized using the energy norm (see \ref{s:QNM_expansions_scalar_product} and \cite{Jaramillo:2020tuu}).}
  \label{eigenfunctions}
\end{figure}

\subsubsection{Calculation of the Keldysh expansion.}\label{s:calculation_keldysh} Once we have calculated numerically the spectral elements
$\omega_n$, $v_n$ and $\alpha_n$, and given our choice of `proof-of-principle' initial data $u_0$
(cf. appendix \ref{e:Gaussian_ID}),
we can make use of the expressions discussed in section
\ref{s:Keldysh} and summarised in appendix \ref{a:easy_access_expressions}.
Specifically, we implement expressions (\ref{e:Keldysh_QNM_expansion_conclusions}) and 
(\ref{e:Lax_Phillips_QNM_expansion_conclusions}) to construct the Keldysh QNM expansions
\bea
u(\tau,x) &\sim& \sum_n e^{i\omega_{n}\tau} a_n v_n(x)  = \sum_n{\cal A}_n(x)  e^{i\omega_{n}\tau} \ .
\eea
The individual contribution of each quasinormal mode in the Keldysh QNM expansion is therefore
given by $\mathcal{A}_n(x)e^{i\omega_n\tau}=a_n v_n(x)e^{i\omega_n\tau}$. The coefficients $\mathcal{A}_n(x)$ are ``agnostic'' to
the particular prescription in section \ref{s:Keldysh} to compute them
(either the use of the transpose $L^t$ or rather the adjoint $L ^\dagger$,
the chosen normalisation of $v_n$ and $\alpha_n$, et cetera), they only depend on the choice of slicing
and the compactified coordinate $x$.
Conversely, the coefficients $a_n$ are independent on $x$ but rely on the normalization of the eigenfunctions $v_n$,
and therefore on the choice of the scalar product. We will come back to this latter point below in
section \ref{s:QNM_expansions_scalar_product} and, at this point, we rather focus on presenting  
the coefficients ${\cal A}_n^\infty$ of the time series
\bea
\label{e:time_series_calculation}
u(\tau,x_{\scri^\infty})\sim \sum_n{\cal A}_n^\infty  e^{i\omega_{n}\tau} \ ,
\eea
that an observer at null infinity $\scri^\infty$ would observe, and that are independent of the
chosen hyperboloidal foliation\footnote{We lack a proof of the later statement, but it is consistent with uniqueness of the Lax-Phillips resonant expansion in (\ref{e:resonant_expansion}) and (\ref{e:coeff_asymptotic_functions}).}.
The interest of the Keldysh approach is that of providing a straight-forward spectral algorithm for calculating
the time series (\ref{e:time_series_calculation}) for given initial data $u_0(x)$:
\begin{itemize}

\item[i)] Solve the spectral problem: this step produces the family $\{\omega_n , v_n(x), \alpha_n(x); \forall n \in \mathbb{N}\}$.
  
\item[ii)] Calculate the coefficients $a_n$ of the Keldysh expansion: given $u_0(x)$, simply evaluate
  $a_n = \langle \alpha_n(x), u_0(x)\rangle$.

  \item[iii)] Calculate the Lax-Phillips-expansion coefficients ${\cal A}_n(x)$: evaluate\footnote{\label{note:indexing}A difficulty when it comes to the numerical implementation of this projection algorithm is the correct indexing of the families $\{v_n(x)\}_{n}$ and $\{\alpha_n(x)\}_{n}$ so that ${\cal A}_n(x)= a_n v_n(x)$ and $\langle \alpha_n(x), v_n(x)\rangle$ are evaluated correctly. Indeed, we require $v_n$ and $\alpha_n$ to be respectively eigenfunctions of $L$ and $L^t$ associated to the same eigenvalue $\omega_n$, this becomes an obvious difficulty if the spectral problems of $L$ and $L^t$ are solved separately and generate two families $\{v_n(x)\}_{n\in J_1}$ and $\{\alpha_n(x)\}_{n\in J_2}$ with different index sets $J_1$ and $J_2$. The appendix \ref{a:Keldysh_evol_op} presents a way to overcome this tedious numerical aspect if $L$ is diagonalisable.}  ${\cal A}_n(x)= a_n v_n(x)$.

\item[iv)] Determine ${\cal A}_n^\infty$: the coefficients of the time-series (\ref{e:time_series_calculation})
  are simply given by evaluating ${\cal A}_n(x)$ at null infinity, that is  ${\cal A}_n^\infty={\cal A}_n(x_{\scri^\infty})$.

\end{itemize}
Such coefficients $\mathcal{A}_n^\infty$ (corresponding to the considered initial data in (\ref{e:initial_data}), illustrated in Fig. \ref{u0} and giving rise to the time-evolutions in Figs. \ref{waveforms}
for the different spacetimes) are presented in
Fig.\ref{Aconv}, namely their moduli  $|\mathcal{A}_n^\infty|$.
Before proceeding to the comparison with the time-domain waveforms, we comment below on the convergence of these $\mathcal{A}_n^\infty$'s.
\begin{figure}[htp]
  \centering
  \subfloat[P\"oschl-Teller]{
    \includegraphics[clip,width=\sizefigmedium\columnwidth]{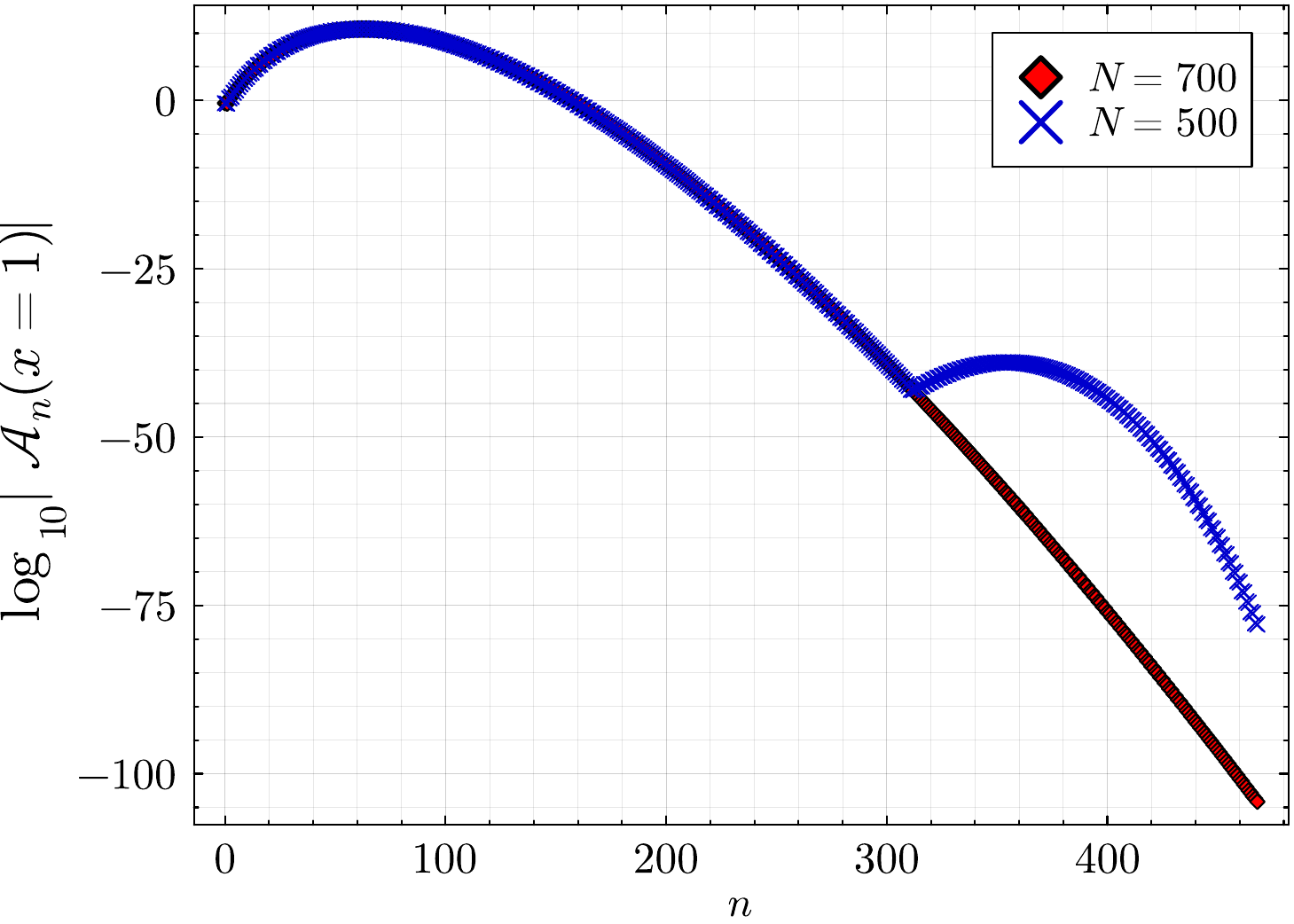}\label{Aconv:PT}
  }
  \subfloat[Schwarzschild-dS]{
    \includegraphics[clip,width=\sizefigmedium\columnwidth]{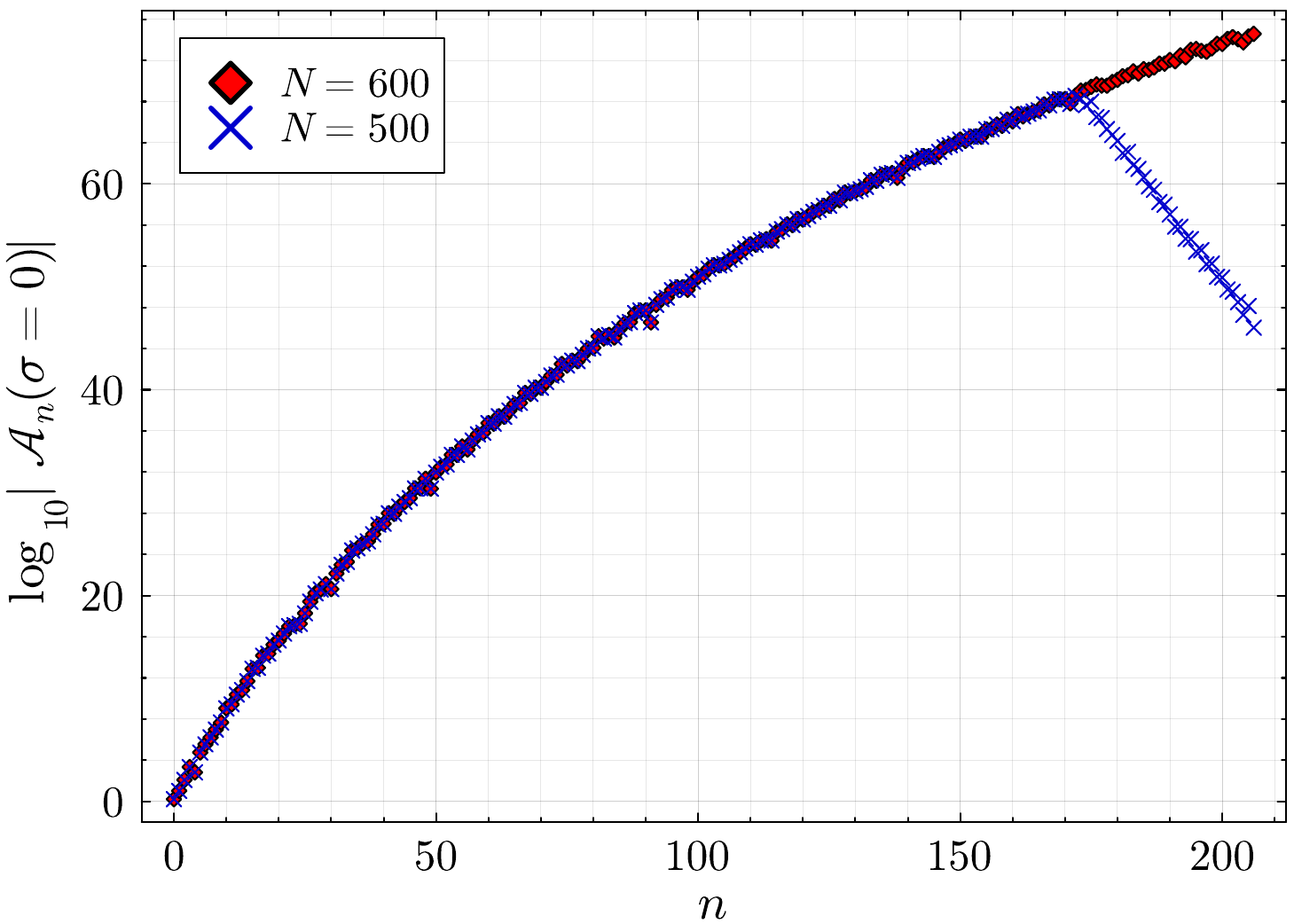}\label{Aconv:dS}
  }

  \subfloat[Schwarzschild]{
    \includegraphics[clip,width=\sizefigmedium\columnwidth]{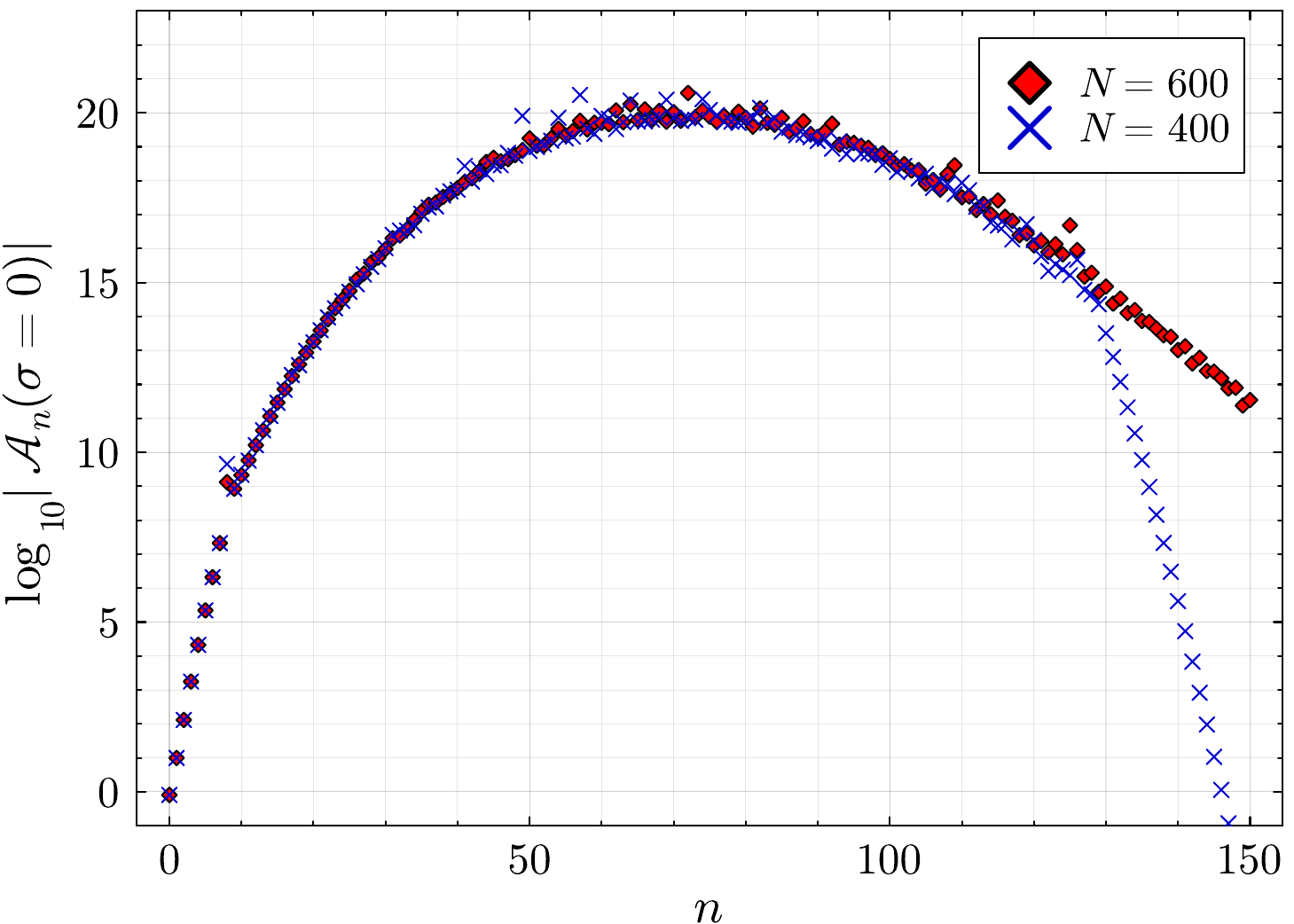}\label{Aconv:S}
  }
  \subfloat[Schwarzschild-AdS]{
    \includegraphics[clip,width=\sizefigmedium\columnwidth]{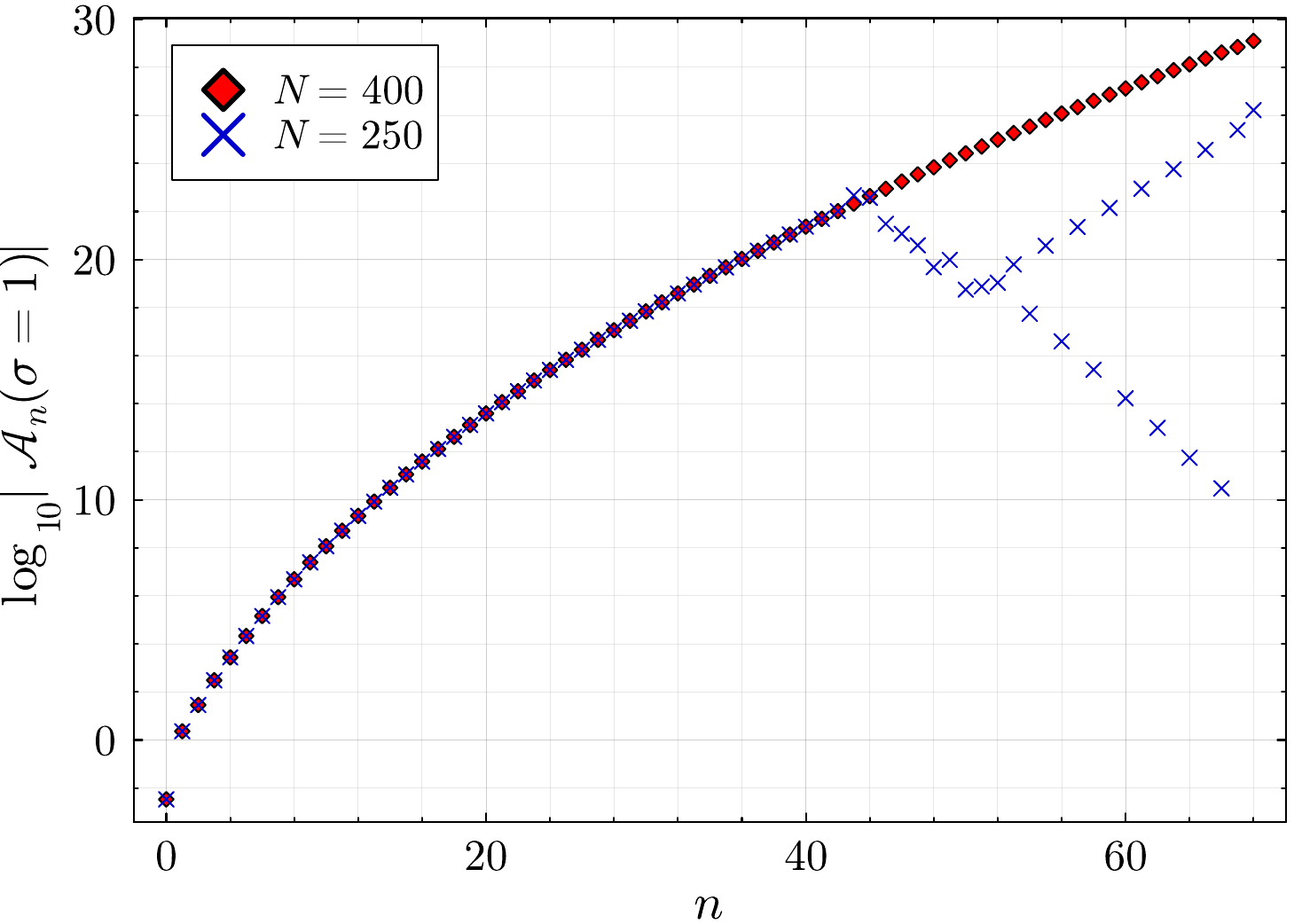}\label{Aconv:AdS}
  }
  \caption{Panels \ref{Aconv:PT}, \ref{Aconv:dS}, \ref{Aconv:S} and \ref{Aconv:AdS} show the modulus of the coefficients $\mathcal{A}_n$ at null infinity (or the event horizon in the Schw.-AdS case) for the cases of study. The modes are labelled by $n$ and ordered by increasing imaginary part.}
  \label{Aconv}
\end{figure}

\subsubsection{Convergence and growth of coefficients in the Keldysh expansion.}\label{s:conv_coeffs_keldysh} We proceed with the same methodology followed in
section \ref{s:QNM_spectral_problem}, when considering the convergence of QNM frequencies $\omega_n$'s.
We focus on $\mathcal{A}_n^\infty$ coefficients, although the same analysis can be done for
the $a_n$'s
for a given normalization of $v_n$ and $\alpha_n$ (see later in section \ref{s:QNM_expansions_scalar_product}).

The systematics of the  convergence of the  $\mathcal{A}_n^\infty$'s, as the
grid resolution $N$ increases, is apparent from Fig. \ref{Aconv}.
As with the QNM frequencies, there is always a clear threshold $n_{T}$ such that for
$n<n_T$ the $\mathcal{A}_n^\infty$'s corresponding to two resolutions $N_1$ and $N_2$ overlap and for
$n>n_T$ the coefficients split. Specifically, the splittings occur (for the  $N_1$ and $N_2$
in Fig.  \ref{Aconv}) at: i) $n_T\approx 311$ for
P\"oschl-Teller in panel \ref{Aconv:PT},  ii) $n_T\approx 172$ for Schwarzschild-dS in panel \ref{Aconv:dS},
iii) $n_T\approx 128$ for
Schwarzschild in panel \ref{Aconv:S},  i) $n_T\approx 43$ for
Schwarzschild-AdS in panel \ref{Aconv:AdS}.
An important point is that, as it was in the case of the QNM frequencies, the
assessment of the convergence of the $\mathcal{A}_n^\infty$'s for asymptotically flat Schwarzschild is
more delicate than in the other cases, as a consequence of the spurious eigenvalues corresponding to the discretised branch cut,
making the construction of the Keldysh resonant expansion more subtle.

We comment now on the growth of the coefficients  $\mathcal{A}_n^\infty$. This, of course, depends critically on the
chosen initial data $u_0$. Here we consider our `reference' Gaussian initial data in (\ref{e:initial_data}) and,
therefore, the discussion below is not meant to refer to the generic physical case.
This specific case rather provides an illustration of the involved concepts and tools.
The following discussion  is needed for the later comparison
with the time-domain results in subsection  \ref{s:time_vs_frequency_domain}. 

In the case of P\"oschl-Teller in panel \ref{Aconv:PT},  coefficients $\mathcal{A}_n^\infty$ reach a maximum
around $n\approx 125$ and then decrease. In the asymptotically de Sitter and Anti-de Sitter cases,
respectively in panel \ref{Aconv:dS} and panel \ref{Aconv:AdS},
we get a monotonic increasing profile but we cannot rule out the possibility that the coefficients decrease if the resolution of the grid is high enough to capture overtones higher in the complex plane.

Regarding the Schwarzschild case \ref{Aconv:S}, it exhibits a maximum and a decreasing (averaged)
trend as $n$ increases, as in P\"oschl-Teller,
before the $\mathcal{A}_n^\infty$'s at different resolutions split in two directions. However, considering
individual $\mathcal{A}_n^\infty$ coefficients, we observe the same type of fluctuations that we had with the spectrum \ref{spectresconv:S} and the individual amplitude coefficients of these high overtones
is more difficult to assess.

As commented above, no conclusions about realistic initial data should be drawn. However this test is quite
remarkable in the sense that it shows that coefficients $\mathcal{A}_n^\infty$ can be reliably calculated
for data containing very high overtones and that, in spite of this non-trivial behaviour in $n$
(even monotonically increasing, as in the Schwarzschild-dS and Schwarzschild-AdS cases), the convergence
properties of the resulting asymptotic series are surprisingly good. Indeed, as we will see 
in section \ref{s:time_vs_frequency_domain}, the comparison with the time-domain signal
indicates a very good behaviour of the QNM series, with
high overtones playing a key role in the accurate reconstruction at early times.
We comment further on these ``unexpectedly good'' convergence properties\footnote{Note
however that
  in Fig. \ref{Aconv}
  we are only showing the modulus of the $\mathcal{A}_n^\infty$ coefficients. For the good convergence it
  is crucial to take into account their complex nature and the associated interference phenomenon.
  This has been observed in \cite{Ansorg:2016ztf}, where the good convergence (starting from an initial
  time $\tau_o$) led the authors to propose
  a conjecture of a certain sense of `completeness' of the QNM (and tails). Under the light of these results of
  Ansorg and Macedo, the good convergence properties are not so `unexpected'.}
in section \ref{s:role_overtones} and they  will be
studied in detail in \cite{BesJarPoo24}.

\subsection{A first comparison between time and frequency domain evolutions}\label{s:time_vs_frequency_domain}
We attain in this section a central point of this work: the direct comparison between
the time-domain signal, constructed from the direct time integration of
Eq. (\ref{e:wave_eq_1storder_u_tau}), and the frequency-domain
spectral QNM expansion (\ref{e:u_Keldysh_v7}),
built for the initial data $u_0$ in (\ref{e:wave_eq_1storder_u_tau}).
As we have stressed, we stay at a ``proof of principle'' perspective,  providing the basic elements
of the comparison and building on the specific initial condition (\ref{e:initial_data})
employed in the
previous section \ref{Keldysh_expansion_cases_study} to explore the different black hole asymptotics.
A more detailed and extended analysis, in particular concerning larger classes
of initial data, is left for a future work.

We start from the asymptotic expansion (\ref{e:u_Keldysh_v7}) and
introduce the finite truncated QNM expansion
with the first $N_{\mathrm{QNM}}$ QNMs (for each branch $\omega_n^\pm$), that we denote
as $u^{\mathrm{QNM}}(\tau,x)$
\begin{equation}
  \label{e:u_truncated_sum}
  u^{\mathrm{QNM}}(\tau,x)=\sum_{n=0}^{N_{\mathrm{QNM}}} \mathcal{A}^\pm_n(x) e^{i\omega^\pm_n\tau} \ .
\end{equation}
In Fig. \ref{fit_log} it is presented the times-series corresponding to the
evaluation of $u(\tau,x)$ at null infinity (P\"oschl-Teller, Schwarzschild), the
cosmological horizon (Schwarzschild-dS) or the horizon (Schwarzschild-AdS) for
both the time-domain solutions (in blue)
presented in section \ref{Keldysh_expansion_cases_study}
and the corresponding truncated QNM expansions (in red), calculated
accordingly to the Keldysh prescription. Regarding the time-domain signal,
it corresponds exactly to the panels in Fig. \ref{waveforms}, but in a logarithmic plot.

\begin{figure}[htp]
  \centering
  \subfloat[P\"oschl-Teller]{
    \includegraphics[clip,width=\sizefigmedium\columnwidth]{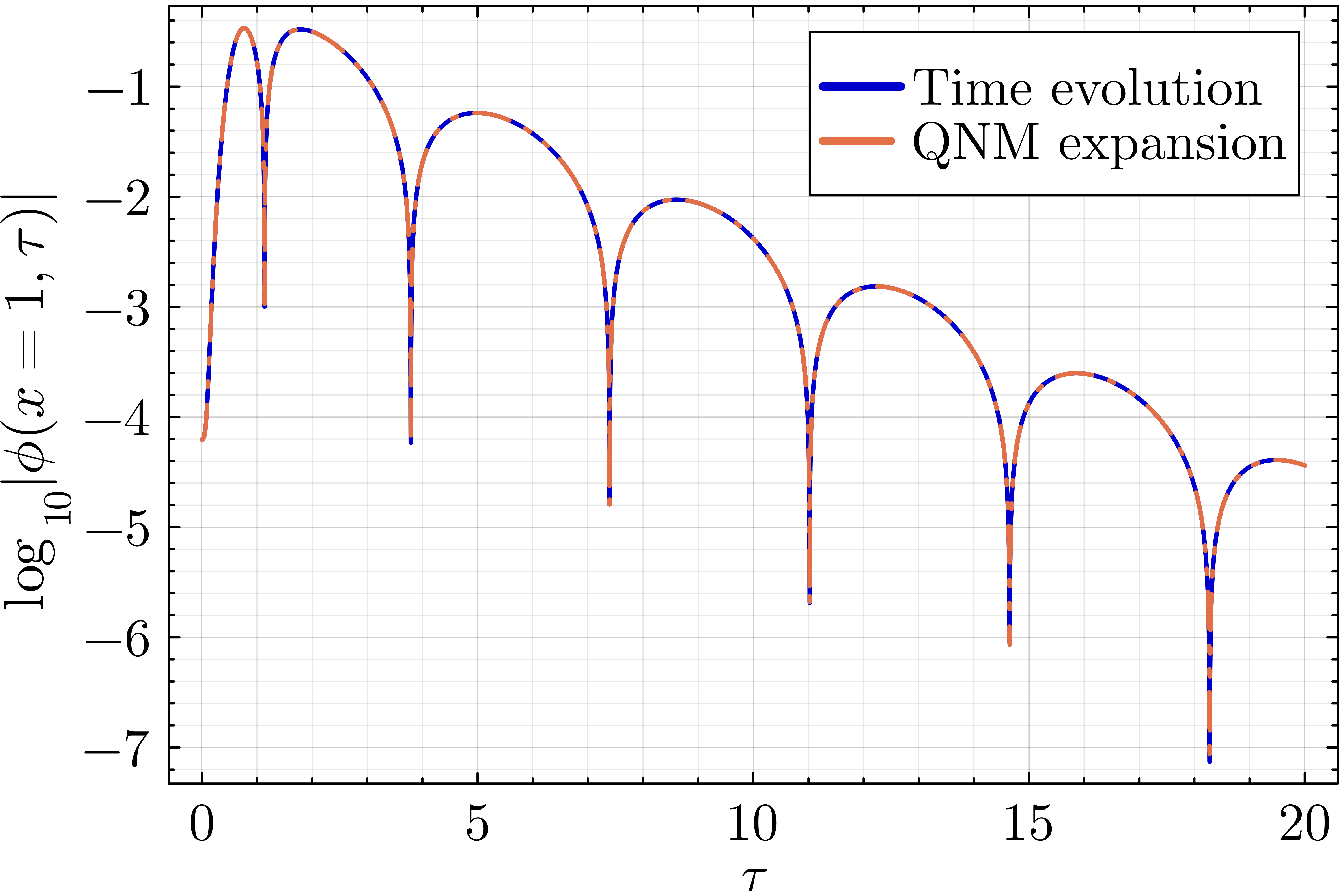}\label{fit_log:PT}
  }
  \subfloat[Schwarzschild-dS]{
    \includegraphics[clip,width=\sizefigmedium\columnwidth]{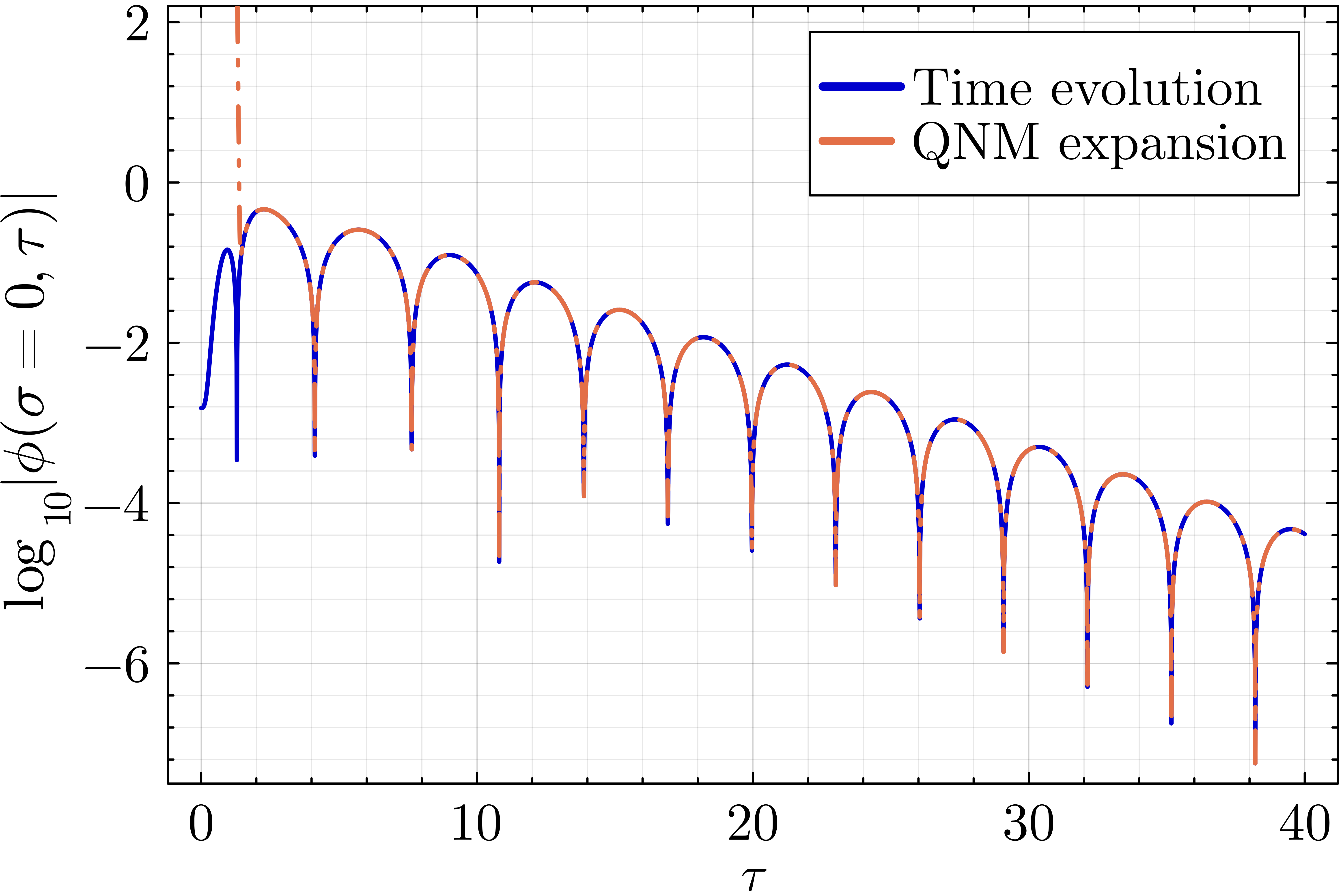}\label{fit_log:dS}
  }

  \subfloat[Schwarzschild]{
    \includegraphics[clip,width=\sizefigmedium\columnwidth]{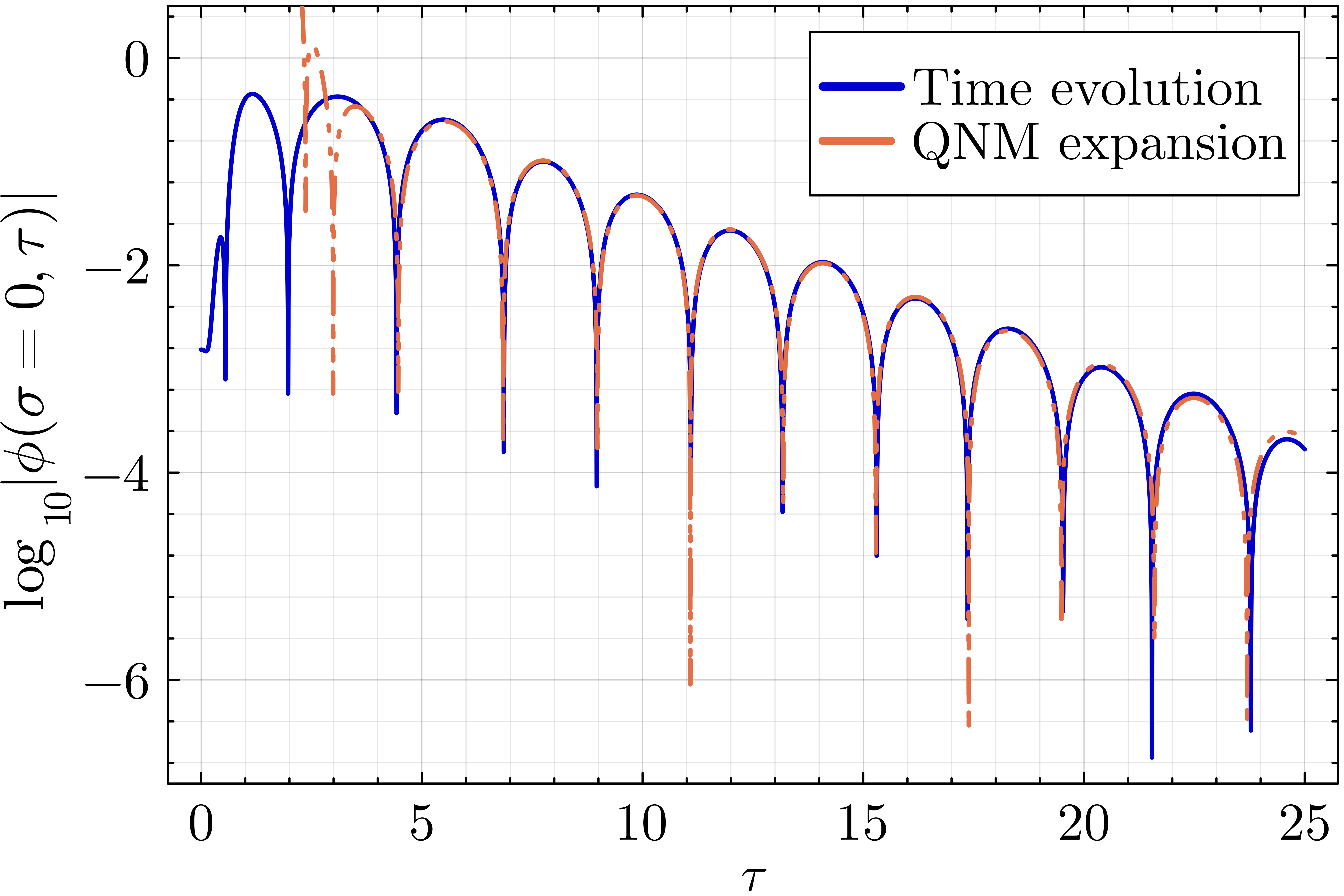}\label{fit_log:S}
  }
  \subfloat[Schwarzschild-AdS]{
    \includegraphics[clip,width=\sizefigmedium\columnwidth]{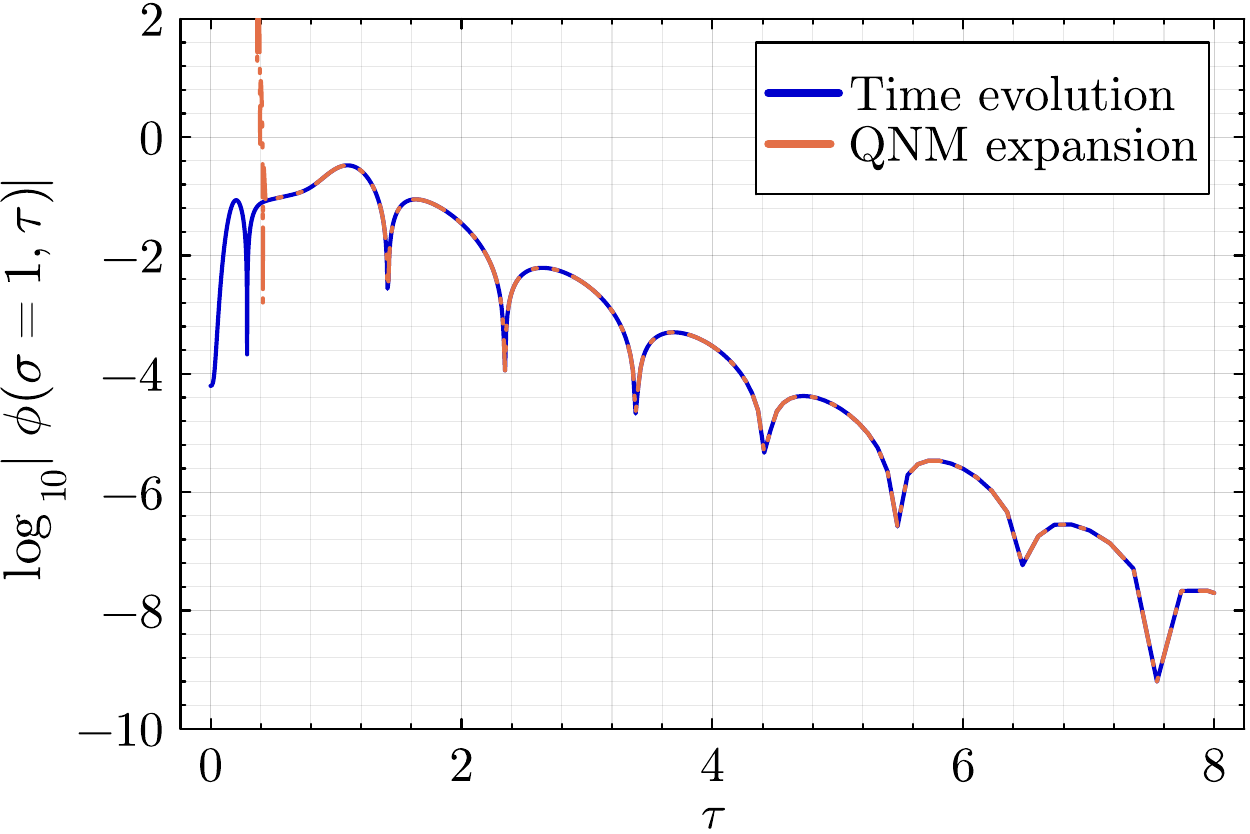}\label{fit_log:AdS}
  }
  \caption{Panels \ref{fit_log:PT}, \ref{fit_log:dS}, \ref{fit_log:S} and \ref{fit_log:AdS} show both the ODE/DAE solution and the numerical resonant expansion.}
  \label{fit_log}
\end{figure}

There are two parameters to control the Keldysh QNM expansions in Fig. \ref{fit_log}: on the one hand,
the number $N_{\mathrm{QNM}}$ determining the QNMs employed in each truncated QNM expansion (namely
$N_{\mathrm{QNM}}+1$ QNMs) and, on the other hand,
the size $N$ of the Chebyshev-Lobatto grids employed in the discrete approximations
of $u^{\mathrm{QNM}}$ (namely using $N+1$ Chebyshev-Lobatto collocation points).
Regarding $N_{\mathrm{QNM}}$ in  Fig. \ref{fit_log},  it is determined by the number of QNMs
whose coefficients ${\cal A}_n^\infty$ in Fig. \ref{Aconv} have already converged.
Regarding the grid resolution $N$, we choose it as the finest
grid used Fig. \ref{Aconv}. In Fig.  \ref{diff} we plot the absolute
difference between the time domain and the QNM time-series,
namely $\left|u(\tau, x_{\mathrm{boundary}})-u^{\mathrm{QNM}}(\tau, x_{\mathrm{boundary}})\right|$.

We comment below on the four cases studied :
\begin{itemize}

\item[i)] P\"oschl-Teller case ($N_{\mathrm{QNM}}=469$, $N=700$): early times in the times-series
  are very accurately described by the QNM expansion, the error always being smaller than
  the tolerance  in the time-domain evolution (here we refer to evolution with the
  method of lines in appendix \ref{a:method_lines}; results are even more accurate
  if using the spectral scheme in appendix \ref{a:Keldysh_evol_op}).
  The first points of the time-series \ref{diff:PT} even suggests that the error might be
  below $10^{-100}$, which is coherent with the value of the highest correct overtone on \ref{spectresconv:PT}.

\item[ii)] Schwarzschild-dS case ($N_{\mathrm{QNM}}=205$, $N=600$): for the chosen number $N_{\mathrm{QNM}}$ of QNMs,
  the early times of the signal presents a huge error which decreases quickly and gets below the numerical precision of the time-evolution solver. The oscillations appearing in Fig. \ref{diff}
  correspond to artefacts of the discretisation scheme of the time derivatives determined
  when choosing the time-evolution solver's algorithm (cf. appendix \ref{a:method_lines}).

\item[iii)] Schwarzschild-AdS ($N_{\mathrm{QNM}}=69$, $N=400$): very similar qualitative behaviour
  to the  Schwarzschild-dS case, although with a much faster decay that is captured with a significantly
  lower number of QNMs.
  
\item[iv)] Schwarzschild ($N_{\mathrm{QNM}}=30$, $N=600$): 
  unlike the previous three cases, the error in the case of Schwarzschild is not limited by the time-domain
  solver. We will comment this case in more detail below in subsection \ref{s:S_tails}, in particular when looking
  at late times when the signal is not dominated by QNMs but by tails.
  The agreement between the time-domain QNM signal and the truncated QNM expansion is nevertheless very good.

\end{itemize}
We would like to comment on two features in Figs. \ref{fit_log} and \ref{diff}. The first one concerns
fundamentally P\"oschl-Teller,  Schwarzschild-dS and Schwarzschild-AdS (but also Schwarzschild in a smaller
degree). Specifically, as it can be seen in Figs. \ref{fit_log} and \ref{diff}, the global agreement
between the time-domain and the QNM expansion signals is remarkable. The accurate agreement at late times
(or intermediate times in the case of Schwarzschild) is expected, since the signal is then controlled by slow-decaying QNMs.
More interesting is the fact that, as seen in  Fig. \ref{diff}, such accuracy is maintained during most of
the whole signal including quite early times and that this agreement can be
pushed to even earlier times by adding additional QNMs. We address this point,
concerning pointwise convergence of the time-series, in subsection \ref{s:convergence_fixed_x}.
The second feature concerns specifically Schwarzschild, namely the only case with a ``branch cut'',
and the tails at late times. Namely, power-law tails are unexpectedly well
recovered by blindly applying a Keldysh projection scheme. We address this point in subsection \ref{s:S_tails}.

\begin{figure}[htp]
  \centering
  \subfloat[P\"oschl-Teller]{
    \includegraphics[clip,width=\sizefigmedium\columnwidth]{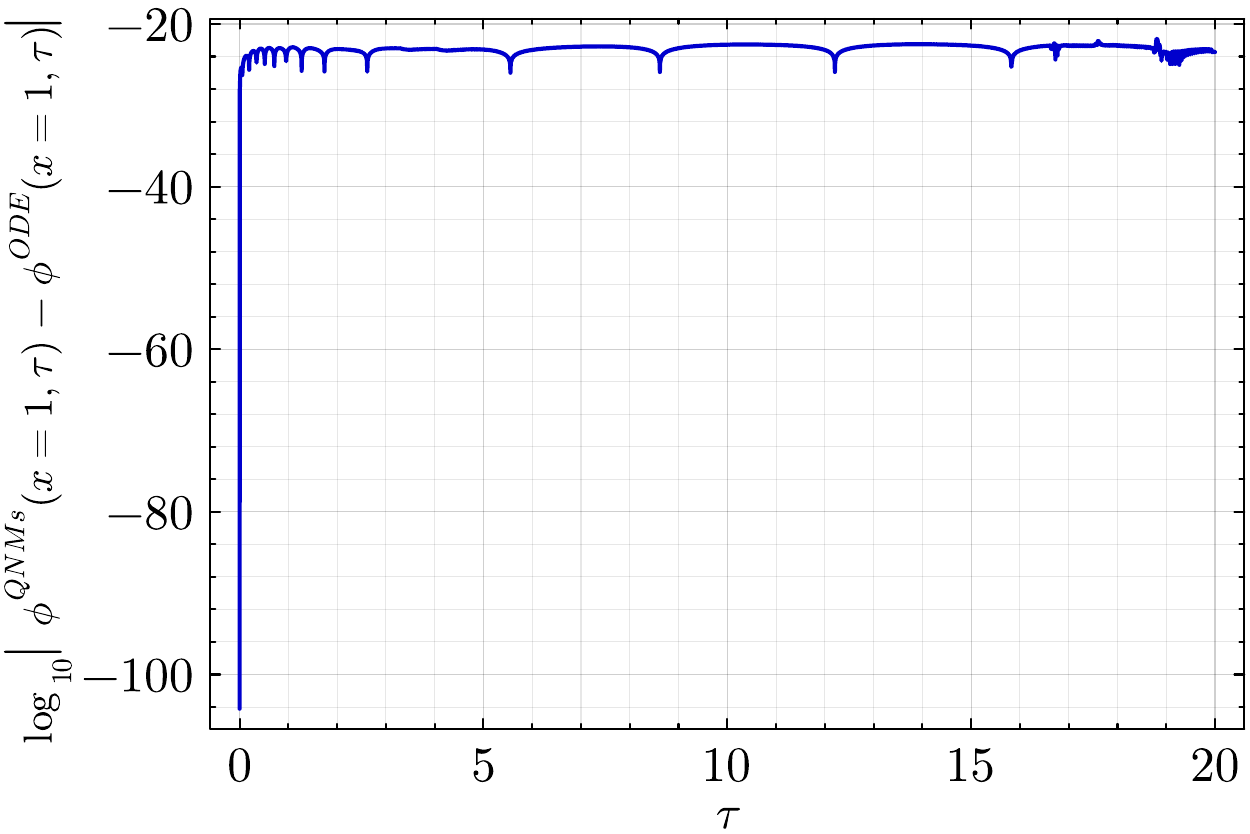}\label{diff:PT}
  }
  \subfloat[Schwarzschild-dS]{
    \includegraphics[clip,width=\sizefigmedium\columnwidth]{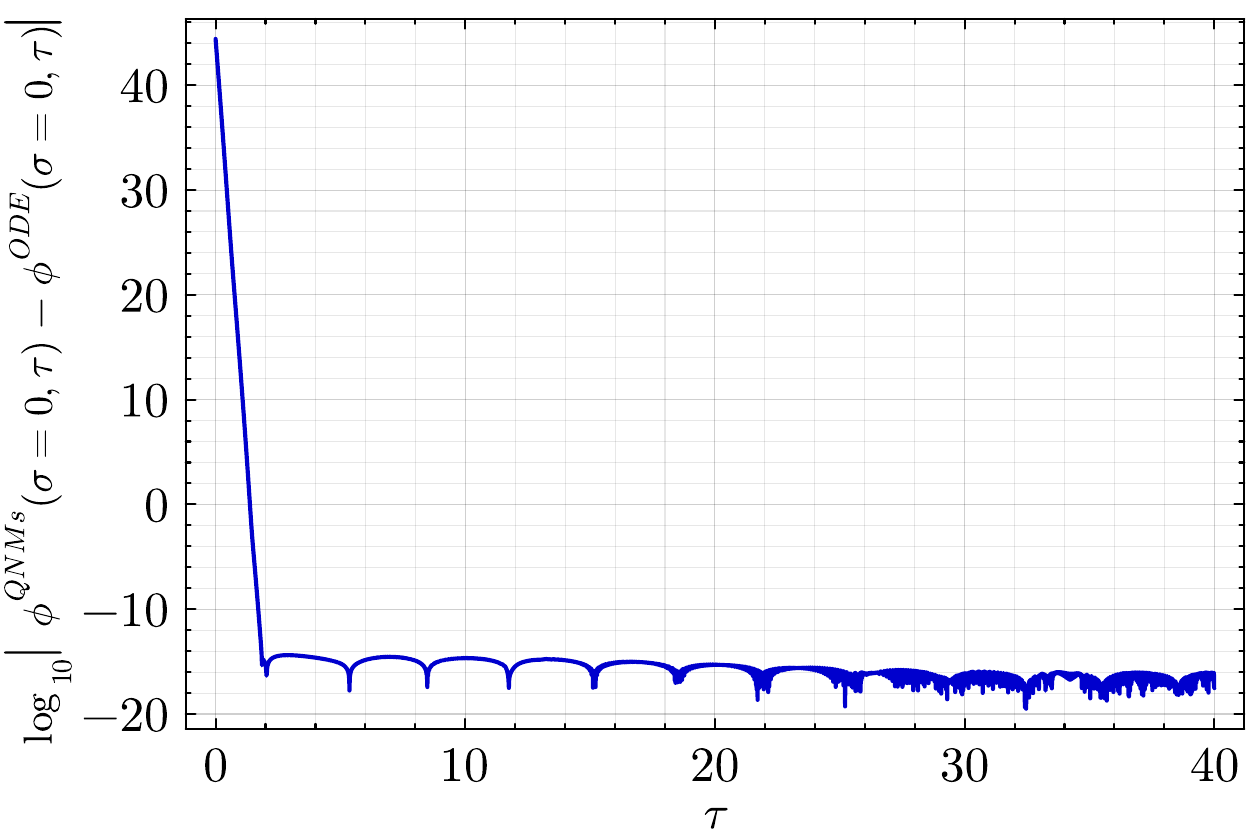}\label{diff:dS}
  }

  \subfloat[Schwarzschild]{
    \includegraphics[clip,width=\sizefigmedium\columnwidth]{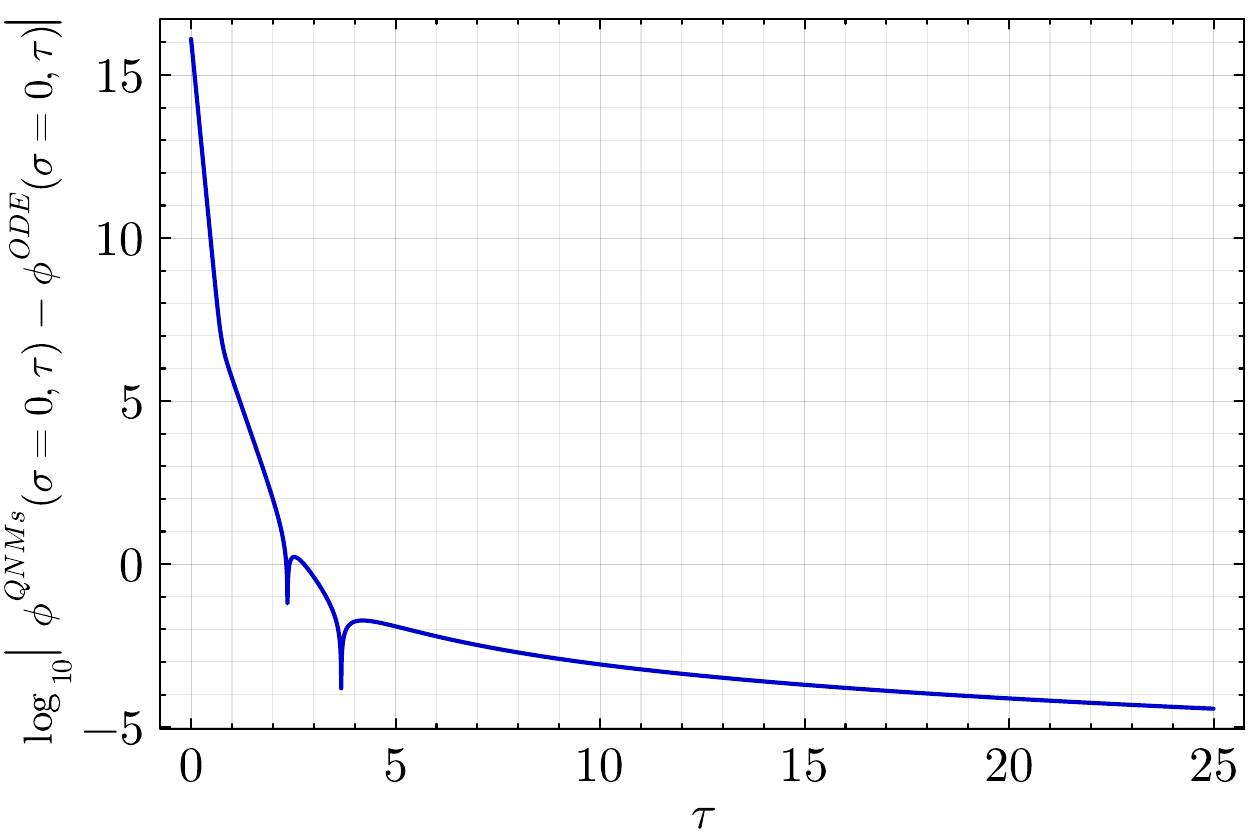}\label{diff:S}
  }
  \subfloat[Schwarzschild-AdS]{
    \includegraphics[clip,width=\sizefigmedium\columnwidth]{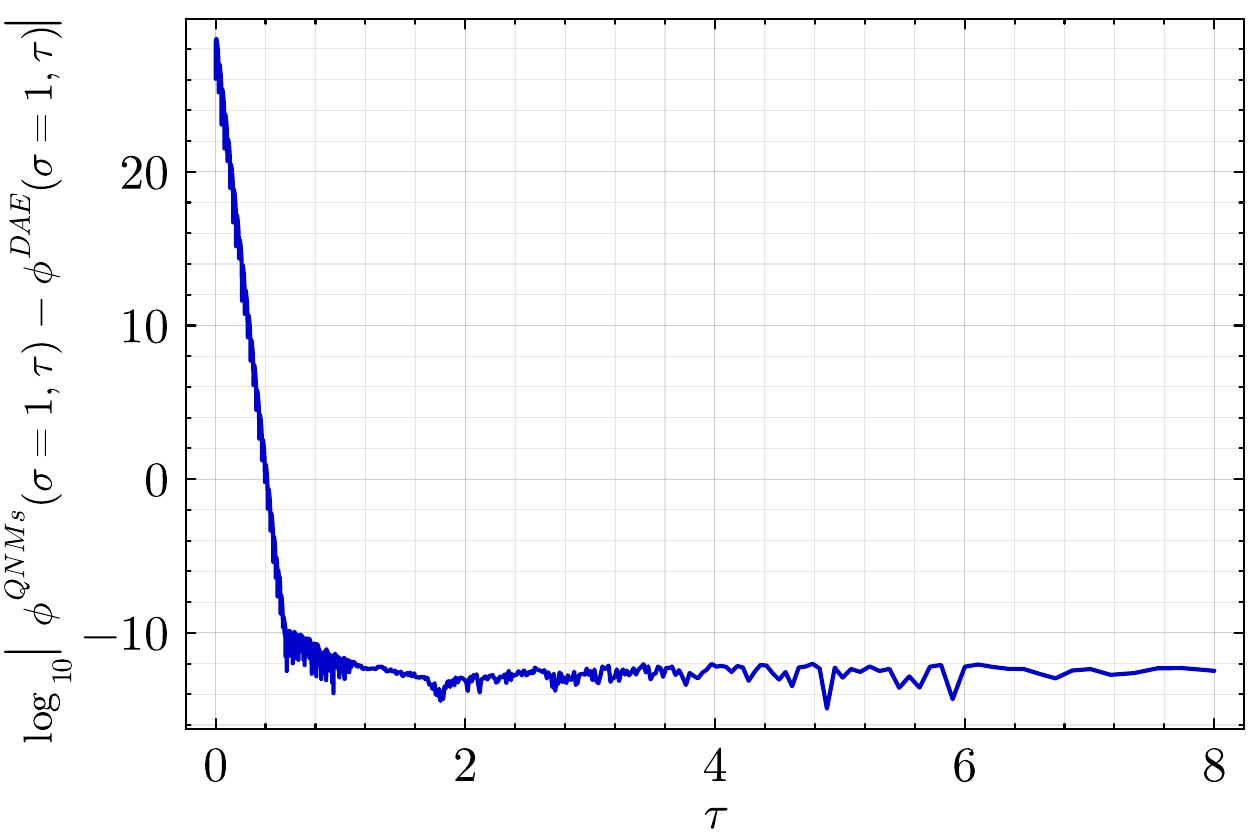}\label{diff:AdS}
  }
  \caption{Panels \ref{diff:PT}, \ref{diff:dS}, \ref{diff:S} and \ref{diff:AdS} show the difference between the ODE/DAE solution and the numerical resonant QNM expansion. The numerical precision of the solver (the tolerance) is $10^{-20}, 10^{-15}, 10^{-15}$ and $10^{-10}$ respectively for the 4 cases under study. We have limited control over the precision since the solver is a black box for us and it might start with a much high precision for $\tau$ small as the P\"oschl-Teller panel suggests.}
  \label{diff}
\end{figure}

\section{Scalar product and QNM excitation coefficients}\label{s:regularity_excitation_coefficients}

\subsection{QNM expansions: from Keldysh to a scalar product approach}\label{s:QNM_expansions_scalar_product} 
As discussed in section \ref{s:QNM_aymptotic_expansions}, scattered fields admit an expression in terms
of resonant expansions but
the latter  do not encode by themselves a meaningful notion of (``excitation'') coefficients
       $a_n$'s in the QNM expansion, either in the Lax-Phillips or in the Keldysh version. For the latter, a measure
       of ``large/small'' is needed. This is provided by the notion of scalar product (or more generally
       a norm), in accordance with the strategy adopted in \cite{Gasperin:2021kfv}.

       Accordingly,  in section \ref{s:calculation_keldysh} we have insisted on the ``agnostic nature''
       of the Keldysh QNM expansion and, in particular, of the ``excitation/expansion
       functions'' $\mathcal{A}_n(x)$, in the sense of not depending
on any additional (scalar product) structure, just on `dual pairing'. However, on the one hand,
from a physical perspective there can be  scenarios in which
a  QNM expansion of the form given in (\ref{e:QNM_scalar_product})
may be of interest\footnote{A physical setting in which such a QNM expansion is relevant is 
  in optical nanoresonators \cite{LalYanVyn17}, in particular when considering quantization
  schemes starting from coefficients $a_n$. In a gravitational context, having meaningful
  $a_n$ can provide the starting point for a (non-conservative) second quantization scheme where
  coefficients $a_n$ are promoted to operators. This might be of particular interest in
  AdS settings, that are very close in spirit to optical cavity problems. In this sense
  AdS/CFT may provide a gravitational setting where expression (\ref{e:QNM_scalar_product}) proves of interest.}.
Such a QNM expansion would be in the spirit of the normal mode expansion (\ref{e:normal_modes}), where
constant ``excitation/expansion coefficients'' $a_n$ are well-defined.
Writing such a QNM expansion (\ref{e:QNM_scalar_product})  amounts to have a physically well motivated
scalar product $\langle \cdot, \cdot \rangle_{_G}$.
The physical meaning of the corresponding coefficients $a_n$'s
would then be encoded in the physical content of the chosen scalar product. 
On the other hand, from a structural perspective, the choice of scalar product is intimately related
to the regularity properties of QNMs, known to be a key element in both the definition
and the instability problem of QNMs, as shown by
Warnick \cite{Warnick:2024,Warnick:2013hba,Boyanov:2024} (see also \cite{BesBoyJar24}).

With these physical and structural motivations, we provide now the connection between the
`dual pairing' Keldysh expansion in section \ref{s:QNM_aymptotic_expansions} and
the `scalar product' one in~\cite{Gasperin:2021kfv}.

\paragraph{From Keldysh expansion to excitation coefficients $a_n$'s.}
\label{s:connection_GasJar22}
       Given an operator $L:{\cal H}\to {\cal H}$, its transpose $L^t:{\cal H}^*\to {\cal H}^*$
       is defined by its action $L^t(\alpha)$ on $\alpha\in  {\cal H}^*$, with
       $\langle L^t\alpha, v\rangle = L^t(\alpha)(v) = \alpha(L v)=\langle \alpha, L v\rangle$, $\forall v\in {\cal H}$.
       If a scalar product $\langle \cdot, \cdot \rangle_{_G}$ is defined by $G:{\cal H}\times {\cal H}\to \mathbb{C}$,
       with $\langle v, w\rangle_{_G} = G(u, w)$, then the (formal) adjoint operator $L^\dagger$ is
       defined by $\langle v, L w\rangle_{_G} = \langle L^\dagger v, w\rangle_{_G}$, $\forall v, w \in{\cal H}$.
       We can then write 
       the corresponding systems of eigenvalue problems associated with these operators as
       \bea
       \label{e:eigen_L-Lt}
       L v_n = \omega_n v_n \ \ , \ \  L^t \alpha_n = \omega_n \alpha_n
       \ \ , \ \ v_n\in{\cal H}, \alpha_n\in{\cal H}^* \ ,
       \eea
       and 
       \bea
       \label{e:eigen_L-Ldagger}
       L v_n = \omega_n v_n \ \ , \ \  L^\dagger w_n = \ol{\omega}_n     w_n
       \ \ , \ \ v_n, w_n\in{\cal H} \ .
       \eea
       In order to relate these eigenvalue systems, let us note that
       the scalar product $\langle \cdot, \cdot \rangle_{_G}$ defines an application $\Phi_G: {\cal H}\to  {\cal H}^*$,
       with $\Phi_G(v)\in  {\cal H}^*$ for $v\in {\cal H}$, defined by
       $\Phi_G(v)(w) = G(v, w) = \langle v, w \rangle_{_G}$, $\forall w\in {\cal H}$.
       Then  the modes $\alpha_n\in {\cal H}^* $ and $w_n\in {\cal H}$,
       respectively in Eqs. (\ref{e:eigen_L-Lt}) and Eqs. (\ref{e:eigen_L-Ldagger}), are related as
       (cf. \ref{a:Keldysh_adjoint_transpose} for a justification in the finite-rank (matrix) case) 
       \bea
       \label{e:alpha_n-w_n}
       \alpha_n = \Phi_G(w_n) \ ,
       \eea
       and it holds
       \bea
        \label{e:scalar_product-dual}
       \langle w_n, v\rangle_{_G} = \langle \alpha_n, v\rangle, \ \ \forall v\in {\cal H} \ .
       \eea
       Applying the latter expression to rewrite $a_n$
       in (\ref{e:a_n_general}), we get
       \bea
       \label{e:coeff_an_scalar-dual_projections}
       a_n= \frac{\langle\alpha_n, u_0\rangle}{\langle\alpha_n, v_n\rangle} =
       \frac{\langle w_n, u_0\rangle_{_G}}{\langle w_n, v_n\rangle_{_G}} \ ,
       \eea
       and we can cast the asymptotic QNM resonant expansion (\ref{e:Keldysh_QNM_expansion_u_an})
       as (note $\tilde{v}_n(x) = v_n(x)$)
       \bea
       \label{e:Keldysh_scalar_product}
       u(\tau,x) &\sim& \sum_n e^{i\omega_{n}\tau} \frac{\langle w_n, u_0\rangle_{_G}}{\langle w_n, v_n\rangle_{_G}} v_n(x)
       \nn \\
       &=& \sum_n e^{i\omega_{n}\tau} \frac{||w_n||_{_G} ||v_n||_{_G}}{\langle w_n, v_n\rangle_{_G}}
       \Big\langle \frac{w_n}{||w_n||_{_G}}, u_0\Big\rangle_{_G} \frac{v_n(x)}{||v_n||_{_G}} \nn \\
        &=&\sum_n e^{i\omega_{n}\tau} \kappa_n \langle \hat{w}_n, u_0\rangle_{_G} \hat{v}_n(x) \ ,
       \eea
       with $\hat{v}_n(x)$ and $\hat{w}_n(x)$ the  modes and comodes normalised in the norm
       $||\cdot||_{_G}$ associated with the scalar product $\langle \cdot, \cdot\rangle_{_G}$,
       namely $||\hat{v}_n||_{_G} = ||\hat{w}_n||_{_G}= 1$, and with
       \bea
       \kappa_n = \frac{||w_n||_{_G} ||v_n||_{_G}}{\langle w_n, v_n\rangle_{_G}} \ ,
       \eea
       the condition number of the eigenvalue $\omega_n$ in the norm
       $||\cdot||_{_G}$. The expression (\ref{e:Keldysh_scalar_product})
       recovers exactly the Eq. (153) in \cite{Gasperin:2021kfv}, generalising the expression there
       for the ``energy scalar product''  $\langle \cdot, \cdot\rangle_{_E}$ to a general scalar product
       $\langle \cdot, \cdot\rangle_{_G}$.
       In particular, the QNM expansion (\ref{e:QNM_scalar_product})
       in section \ref{s:QNM_aymptotic_expansions} is recovered from Eq. (\ref{e:Keldysh_scalar_product})
       by defining the excitation coefficient as $a_n=\kappa_n \langle \hat{w}_n, u_0\rangle_{_G}$.
       
\subsection{Choice of scalar product: energetic and regularity aspects}
Having justified expression (\ref{e:QNM_scalar_product})  for the QNM expansion, we are left with the freedom to choose
the scalar product to normalise the eigenfunctions $v_n$ and compute $a_n$, something that translates into exploring
different scalar products to control the behaviour of the excitation coefficients, in particular
high overtones.
Given the dependence of the $a_n$ on the scalar product $\langle \cdot, \cdot \rangle_{_G}$ is
natural to denote then as $a_n^G$.

Whereas the goal in section \ref{s:conv_coeffs_keldysh} was to introduce and assess the coefficients of the ``time-series QNM expansion'' {\em at a fixed $x_o$}, namely
${\cal A}_n(x_o)$, with a focus on $x_o$ at  $\scri^+$ and the black hole event horizon,
we now consider the (asymptotic)
QNM expansion {\em at a fixed $\tau_o$} as a sum over the normalised eigenfunctions\footnote{The
resulting distinct QNM series {\em at a fixed $x_o$} and {\em at a fixed $\tau_o$} have, in particular,
different convergence properties, for which a first sketchy discussion is presented, respectively,
in sections \ref{s:convergence-fixed_tau} and \ref{s:convergence_fixed_x}.} 
$\hat v_n(x)=\frac{v_n(x)}{\norm{v_n}_{_{G}}}$: as we explained in section \ref{s:calculation_keldysh} we trade
function-coefficients ${\cal A}_n(x)$'s that depend on $x$ for constant-coefficients $a_n^G$'s that
depend on the scalar product, by means of the normalised  eigenfunctions
$\hat v_n(x)$.
The relation between the $a_n$'s and the norm  $\norm{v_n}_{_G}$ is given by $a^G_n= a_n \norm{v_n}_{_G}$.
Explicitly
\bea
\label{e:a_n_norm}
\!\!\!\!\!\!\!\!\!\!\!\!\!\!\!\!      {\cal A}_n(x) =a_n v_n(x) = a_n \norm{v_n}_{_G} \frac{v_n(x)}{\norm{v_n}_{_G}} =
      a_n \norm{v_n}_{_G} \hat{v}^G_n(x) = a^G_n  \hat{v}^G_n(x) \  .
\eea
In this section we focus on the study of  such $a^G_n$'s in two specific cases of
$\langle \cdot, \cdot \rangle_{_G}$:
\begin{itemize}
\item[i)]  Energy scalar product, $\langle \cdot, \cdot \rangle_{_E}$: the natural
  choice in the physical discussion of conservation/dissipation of
  energy~\cite{Driscoll:1996,Gasperin:2021kfv,Jaramillo:2022kuv,Carballo_2024,Chen_2024}
  as well as in QNM spectral instability~\cite{Jaramillo:2020tuu,Gasperin:2021kfv}.

\item[ii)] Sobolev spaces $H^p$,  $\langle \cdot, \cdot \rangle_{H^p}$:
  controlling the $L^2$-norm of the $p$-th spatial derivative, it is the natural choice in the discussion
  of regularity aspects
  in QNMs, e.g. in the QNM definition as an eigenvalue problem of a non-selfadjoint operator
  (cf.~\cite{Warnick:2013hba,Boyanov:2024,Warnick:2024} and references therein) or when
  assessing time-domain
  stability under ultraviolet perturbations~\cite{Gasperin:2021kfv,BesBoyJar24}.
\end{itemize}
We develop these two points below. Other scalar products are key in other problems,
as it is illustrated in the discussion of transient growths in the setting of superradiance~\cite{Carballo:2025ajx}.

\subsubsection{QNM expansion coefficients $a^E_n$ in the energy norm}
\label{s:a_n-scalarproducts}
The energy scalar product is, when applied to vectors, directly related to the total energy
contained in a slice $\Sigma_\tau$ at constant $\tau$
(when dealing with the induced `energy operator norm' the interpretation is more subtle;
see section 6.1 in \cite{Gasperin:2021kfv}), through
its associated norm~\cite{Jaramillo:2020tuu,Gasperin:2021kfv}
\bea
\label{e:energy_norm}
||u||^2_{_E} = \langle u, u \rangle_{_E}  = E_\tau = \int_{\Sigma_\tau} T_{ab} t^an^b d\Sigma_t \ ,
\eea
with $t^a$ the timelike Killing and $n^a$ the timelike normal to the
hyperboloidal slice $\Sigma_\tau$ at fixed $\tau$.
This leads to the expression of the energy scalar product  (cf. notation
in section \ref{a:hyperboloidal_scheme})
\begin{equation}\label{eq_eff_en_inner_product}
   \langle u_1, u_2\rangle_{_E}  =
   \Big\langle \begin{pmatrix} \phi_1 \\ \psi_1 \end{pmatrix}, \begin{pmatrix} \phi_2 \\ \psi_2 \end{pmatrix}
   \Big\rangle_{_E} =
  \frac{1}{2} \hspace{-1mm}\int_{a}^{b}\hspace{-1mm}
  (w(x)\bar{\psi}_1 \psi_2 +
  p(x)\partial_x\bar{\phi}_1\partial_x\phi_2 +
  \tilde{V}(x)\bar{\phi}_1\phi_2) dx \ .
 \end{equation}
As commented above, this scalar product is the natural one when discussing energetic considerations,
in particular when studying well-posedness of the PDE  initial data problem in
Eq. (\ref{e:wave_eq_1storder_u_tau}) or the conservative/dissipative nature of the
system. In the hyperboloidal setting it has been extensively employed in the
assessment of the QNM spectral instability under small (ultraviolet)
perturbations~\cite{Jaramillo:2020tuu,Jaramillo:2021tmt,Gasperin:2021kfv} (see reviews, see
\cite{destounis2024black,PanossoMacedo:2024nkw}), both through the direct
study of spacetime perturbations whose scale is controlled by the energy norm (\ref{e:energy_norm})
or by the construction of the pseudospectrum of the non-perturbed background.
The latter approach presents however a ``convergence'' problem that has been
pointed out in \cite{Boyanov:2024} and to which we will come back below in section
\ref{s:Hp-pseudospectrum}.

In this setting, coefficients $a_n^E$ calculated by using 
the energy norm are expected to encode information on the
energy contained in the $n$-th mode, although the precise manner
in which this is to be interpreted is not straightforward in this
non-orthogonal QNM setting and we do not develop further on this
point here. 
Figure \ref{aconv} presents the coefficients $a_n^E$ for the QNM expansion of our
Gaussian ``proof-of-principle'' test initial data $u_0$ in Eq. (\ref{e:initial_data})
in P\"oschl-Teller and in spherically symmetric black holes with
different spacetime asymptotics.
\begin{figure}[htp]
  \subfloat[P\"oschl-Teller]{
    \centering
    \includegraphics[clip,width=\sizefigmedium\columnwidth]{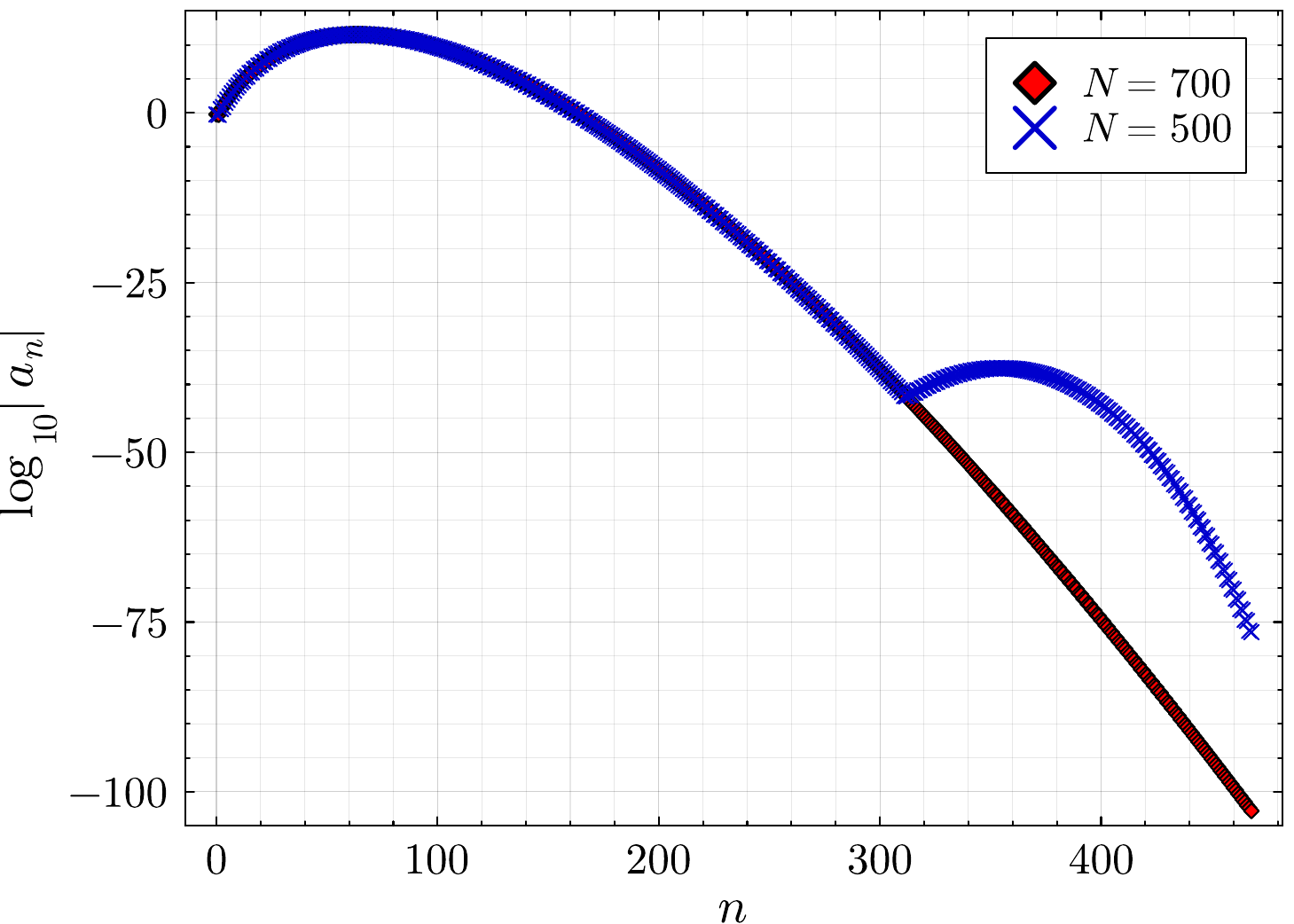}\label{aconv:PT}
  }
  \subfloat[Schwarzschild-dS]{
    \includegraphics[clip,width=\sizefigmedium\columnwidth]{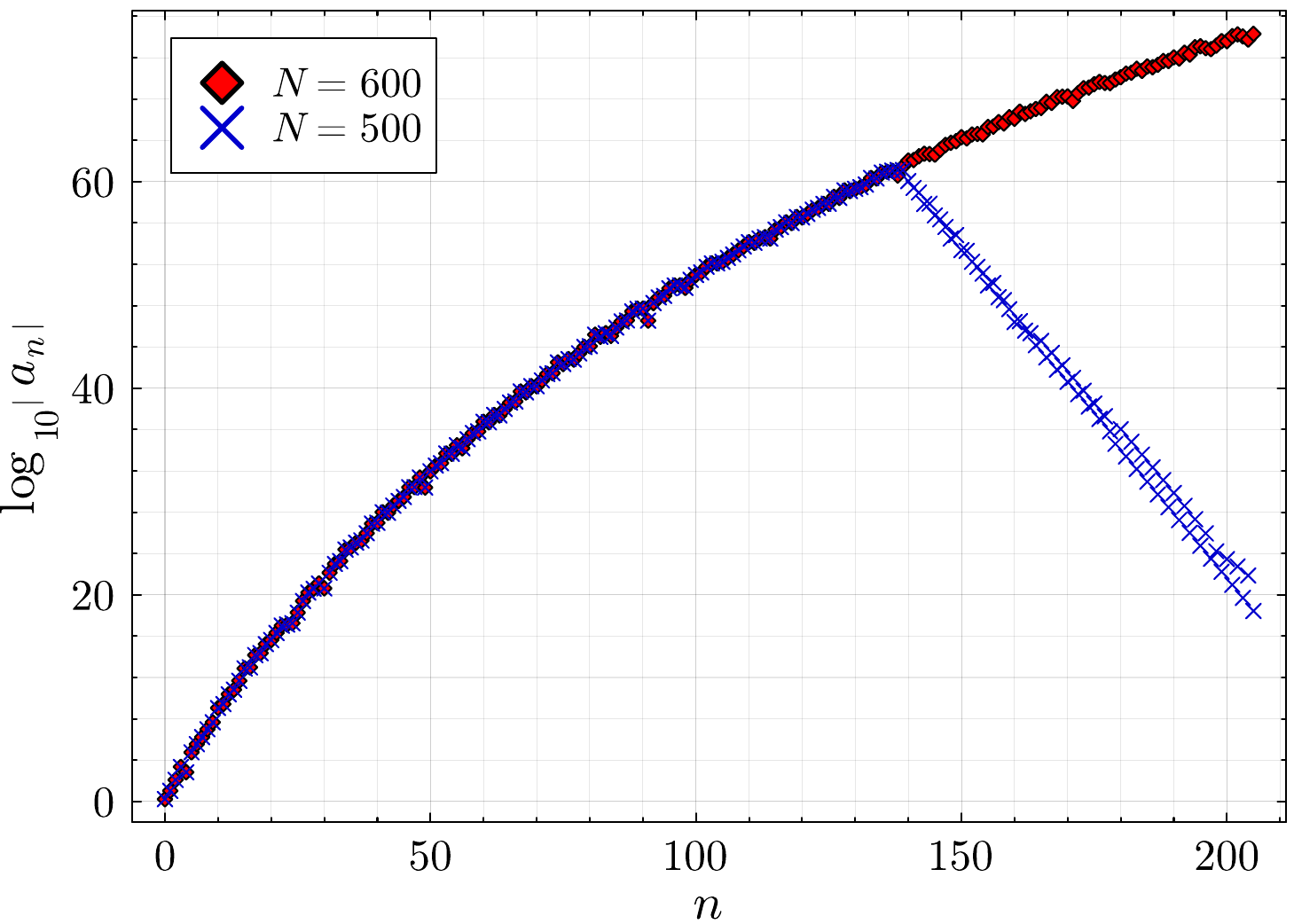}\label{aconv:dS}
  }

  \subfloat[Schwarzschild]{
    \includegraphics[clip,width=\sizefigmedium\columnwidth]{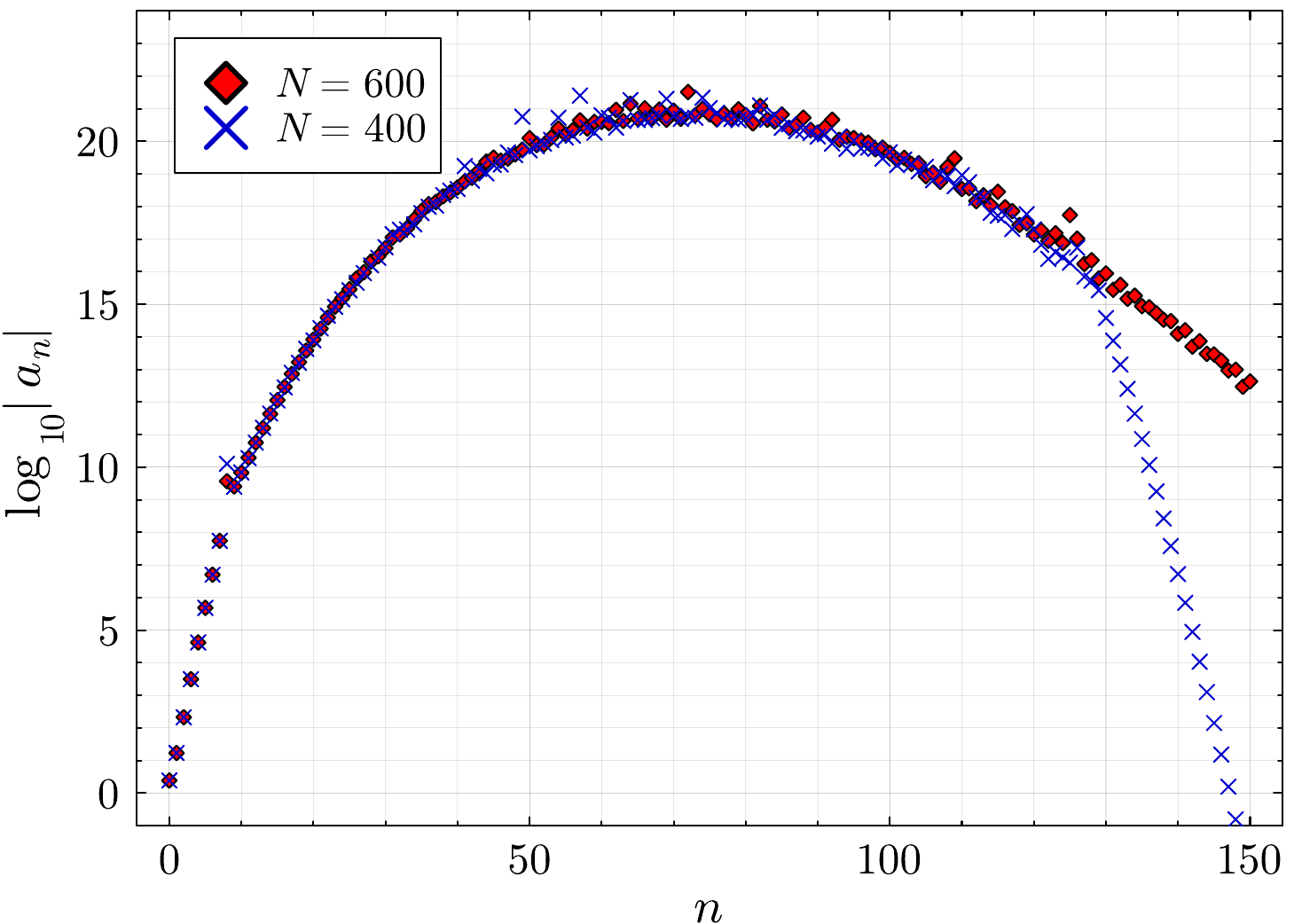}\label{aconv:S}
  }
  \subfloat[Schwarzschild-AdS]{
    \includegraphics[clip,width=\sizefigmedium\columnwidth]{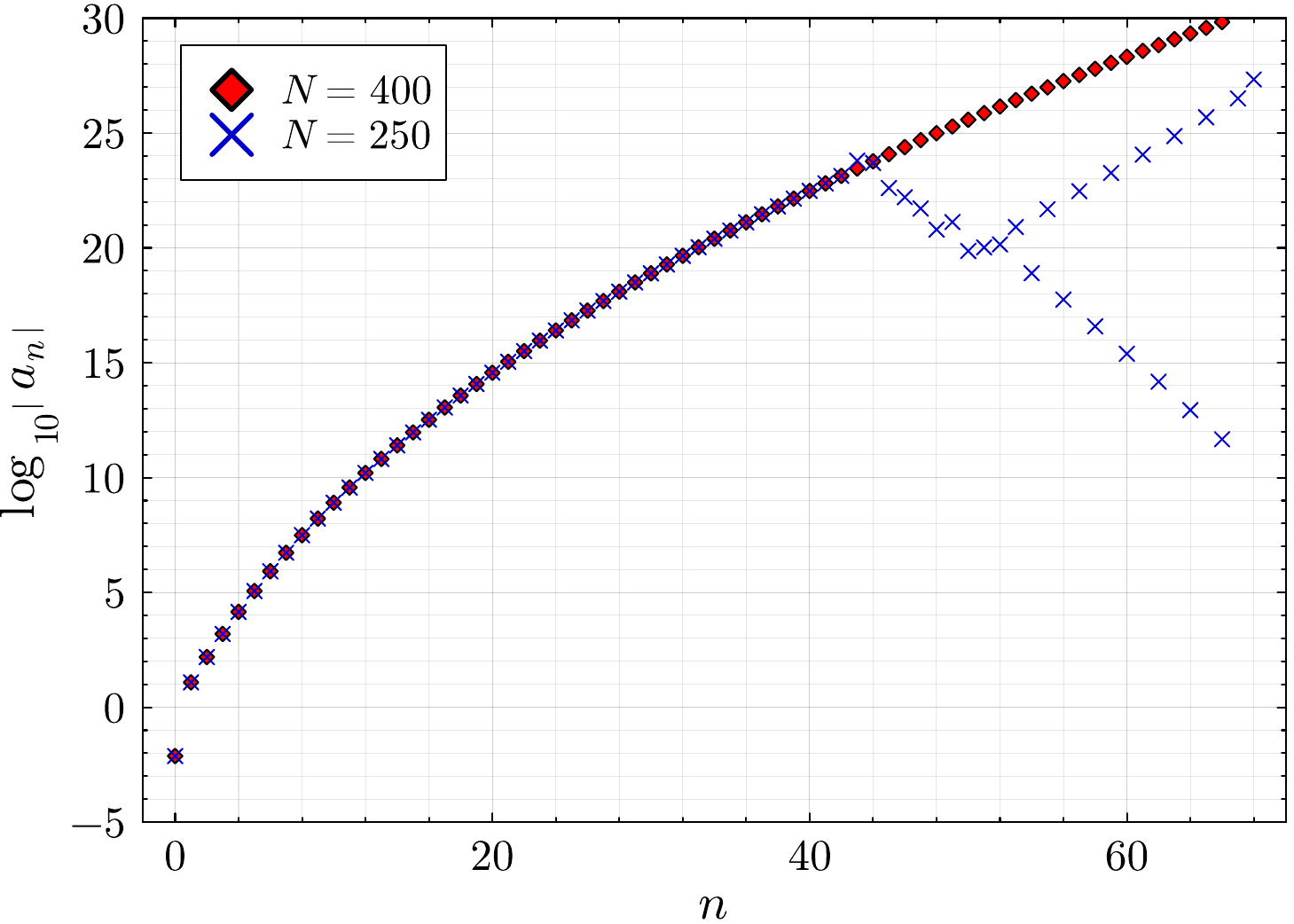}\label{aconv:AdS}
  }
  \caption{Panels \ref{aconv:PT}, \ref{aconv:dS}, \ref{aconv:S} and \ref{aconv:AdS} show the modulus of the coefficients $a_n$ corresponding to the energy norm. The modes are labelled by $n$ and ordered by increasing imaginary part.}
  \label{aconv}
\end{figure}

We note that, remarkably, the $a_n^E$ are very similar to the $\mathcal{A}^{\infty}_n$, upon comparison
between Figs \ref{Aconv} and \ref{aconv} we notice that their respective values
are only moved up slightly.
From the relation (\ref{e:a_n_norm}), this amounts to $|\hat{v}_n(1)|\sim 1$.
We explore below the behaviour of $a_n^G$'s when choosing rather a Sobolev norm,
focusing on the test-bed case provided by P\"oschl-Teller.

\subsubsection{Coefficients in the $H^p$ Sobolev scalar product}
\label{s:coeffs_Sobolev}
Sobolev $H^p$ scalar products and their associated norms are fundamental in the
discussion of regularity aspects of solutions to PDEs.
Here we will restraint ourselves to the discussion of $H^p$ spaces
in the P\"oschl-Teller case, leaving the actual black hole case
in different spacetimes asymptotics to a devoted study in \cite{BesBoyJar24}.

The main motivation
in our QNM setting comes, on the one hand, from the definition problem of QNMs
in terms of an eigenvalue problem and, on the other hand, from the assessment
of the possibility of time-domain  instabilities triggered by small-scale (ultraviolet)
perturbations and the related potential loss of regularity of the propagating solution.
Specifically:
\begin{itemize}

\item[i)] {\em Definition problem of QNMs}. A key ingredient in the characterisation
  of QNM frequencies $\omega_n$'s as eigenvalues of a non-selfadjoint operator $L$ is
  the appropriate identification of the Hilbert space in which the QNMs $v_n$'s live.
  This is not a free choice, but is actually determined from the regularity requirements
  to recover a discrete spectrum of QNMs: if we demand too much regularity  (e.g analyticity)
  no  eigenvalues are found, if we demand too little regularity (e.g $C^\infty$-smoothness)
  then a continuum of eigenvalues emerges. In the asymptotically AdS case (as well as
  in the asymptotically dS case, this including the P\"oschl-Teller case discussed here),
  Warnick~\cite{Warnick:2013hba} has identified $H^p$ Sobolev spaces as the ones
  providing the appropriate regularity to define QNMs as eigenvalues in a  band
  in the $\omega$-complex upper half-plane (in our convention, stable QNM frequencies are in the
  upper half-plane)  characterised by $0 \leq \mathrm{Im}(\omega) \lesssim  p\cdot \kappa$,
  i.e. a horizontal band of width
\bea
\label{e:QNM_band}
\Delta \omega_{H^p-\mathrm{QNM}} \sim p\cdot \kappa \ ,
\eea
where $\kappa$ is the surface gravity (namely the exponential decay rate of the
potential in the P\"oschl-Teller case). With the stationary BH QNM large-$n$ asymptotics,
this typically permits to define the first $p$ QNMs, but for higher overtones
the $H^p$ scheme fails (namely all points $\omega\in \mathbb{C}$ with
$\mathrm{Im}(\omega)\gtrsim \Delta \omega_{p\mathrm{-QNM}}$ are eigenvalues of $L$)
and the
definition of the $(p+1)$-th QNM requires to resort to the space $H^{p+1}$.
The energy scalar product discussed in section \ref{s:a_n-scalarproducts} simply
does not work beyond the fundamental QNM: it is appropriate to discuss physical
perturbations, but not to define QNM overtones as eigenvalues~\cite{BesBoyJar24}. $H^p$ scalar products,
on the contrary, provide an adequate structure for such first $p$ QNMs.
This justifies, from our perspective, the interest to assess the
associated coefficients $a_n^{H^p}$.

\item[ii)] {\em Time-domain stability of `ultraviolet' instabilities}. In ref. \cite{Jaramillo:2020tuu}
  it was introduced a non-selfadjoint spectral approach to the study of
  BH QNM spectral instabilities first identified in  
  \cite{Nollert:1996rf,Nollert:1998ys,Aguirregabiria:1996zy,Vishveshwara:1996jgz}.
  Such  BH QNM spectral instabilities correspond to large variations of the QNM
  frequencies $\omega_n$'s under `very small' perturbations $\epsilon \delta L$ induced 
  by (small-scale) perturbation of the background. A crucial point is that the
  size of such small perturbations 
  was measured with the `energy norm', namely $||\epsilon \delta L||_E\lesssim \epsilon$,
  something well justified when considering  physically induced perturbations.
  However, when considering the time-domain problem and 
  solutions $u(\tau, x)$ and $u^\epsilon(\tau, x)$ to the
  evolution problem (\ref{e:wave_eq_1storder_u}), respectively  with $L$ and
  $L+\epsilon \delta L$ as time infinitesimal generators, it holds
(cf. Eq. (179) in \cite{Gasperin:2021kfv})
  \begin{equation}
    \label{e:timedomain_stability}
    || u - u^\epsilon||_E \lesssim \epsilon C ||u_0||_E
  \end{equation}
  for some constant $C$. That is, the time-domain problem is stable in the
  energy-norm, although the corresponding spectral QNM problem is not. Since the latter is a low-regularity
  phenomenon,  it is necessary (cf.   \cite{Gasperin:2021kfv}, footnote below  Eq. (179)) 
  to take into account higher derivatives in the norm to get
  a better control of the regularity of $u^\epsilon(\tau, x)$ than the one provided
  by (\ref{e:timedomain_stability}). Therefore, the natural tool to 
  assess of the `small-scale/low regularity'
  (in)stability in the time-domain of this dissipative problem is the $H^p$-Sobolev norm, through
  $|| u - u^\epsilon||_{H^p}$. This is our second motivation to consider $H^p$ scalar products.

\end{itemize}

\paragraph{A $H^p$ scalar product.}
We consider the following $H^p$-like scalar product
constructed on the energy scalar product and reducing\footnote{The notation is slightly
  in tension with the fact that the energy scalar product is actually
  a classical $H^1$ scalar product. We prefer to keep the notation since it is
  more convenient to discuss associated pseudospectra. On the other hand, we notice
  that other choices for the $H^p$-weights can be envisaged, but
  the generic qualitative behaviour of the different objects we consider does not change.
  Our choice in Eq. (\ref{e:Hp_scalarproduct}) is a ``minimalistic'' one 
  including the energy-scalar product and recovering Warnick’s results in \cite{Warnick:2013hba}
  $H^p$-QNMs.}
to energy scalar product for $p=0$
\bea
\label{e:Hp_scalarproduct}
\langle u_1, u_2\rangle_{_{H^p}} = 
\Big\langle \begin{pmatrix} \phi_1 \\ \psi_1 \end{pmatrix}, \begin{pmatrix} \phi_2 \\ \psi_2 \end{pmatrix}
\Big\rangle_{H^p}
= \sum_{j=0}^p \Big\langle\begin{pmatrix} 
		\partial_x^j \phi_1 \\ 
		\partial_x^j \psi_1\end{pmatrix},
                \begin{pmatrix} 
		\partial_x^j \phi_2 \\ 
		\partial_x^j \psi_2
	\end{pmatrix}\Big\rangle_{E} \ ,
                \eea
and leading to the norm
\begin{equation}
  \label{e:Hp_norm}
	\norm{\begin{pmatrix} \phi \\ \psi \end{pmatrix}}_{H^p}^2 := 
	\sum_{j=0}^p \norm{ \begin{pmatrix} 
		\partial_x^j \phi \\ 
		\partial_x^j \psi
	\end{pmatrix}}_E^2 \ .
\end{equation}
The corresponding Gram matrix  $\GramHp$ (see appendix \ref{s:pseudospectral_methods})
is related to the  Gram matrix for the energy scalar $\GramE$ product through
\begin{equation}
  \label{e:Gram_Hp}
  \GramHp = \sum_{j=0}^p
	\left(\begin{array}{c|c} 
	   \mathbb{D}_N^{(j)} & 0\\ \hline
	   0 & \mathbb{D}_N^{(j)} 
	\end{array}\right)^t
	\GramE
	\left(\begin{array}{c|c} 
	   \mathbb{D}_N^{(j)} & 0\\ \hline
	   0 & \mathbb{D}_N^{(j)}
	\end{array}\right) \ ,
\end{equation}
that adds the energy norm of the function $u$ and its first $p$
spatial derivatives\footnote{We note that, since the eigenfunctions $v_n$ of $L$ in the P\"oschl-Teller
  case are polynomials, if $p$ is high enough the first excitation coefficients
  computed from $H^p$ and $H^{p+1}$ are identical.}. We have explored different choices
for the weights in the $H^p$-scalar product and all render the same qualitative results.
We have chosen the ones in (\ref{e:Hp_scalarproduct}) and (\ref{e:Gram_Hp})
provided by the energy scalar product, this yields cleaner results in the $H^p$-pseudospectrum
and $H^p$-transient growths (see section \ref{s:transients}).

\paragraph{QNM expansion coefficients $a_n^{H^p}$ in the $H^p$ norm.}
The norms  $H^p$ with increasing $p$ are correspondingly more sensitive to the small
scale of the functions.
The use of $H^p$ norms enhances coefficients
$a_n^{H^p}$ corresponding to QNM eigenfunctions with more small scale structure, consistently with
Eq. (\ref{e:a_n_norm}) and the corresponding larger $a_n^{H^p}$.
This behaviour can be appreciated in Fig.\ref{Hp_an}, where coefficients $a_n^{H^p}$ with
larger $n$ gets bigger as $p$ increases. In summary, the discussion and the results in this section
suggest that the use of $H^p$ norms can be of interest in the study and understanding
of small scale/regularity issues in QNM expansions. This will be developed elsewhere~\cite{BesBoyJar24}.

\begin{figure}[htp]
  \centering
  \includegraphics[clip,width=0.8\columnwidth]{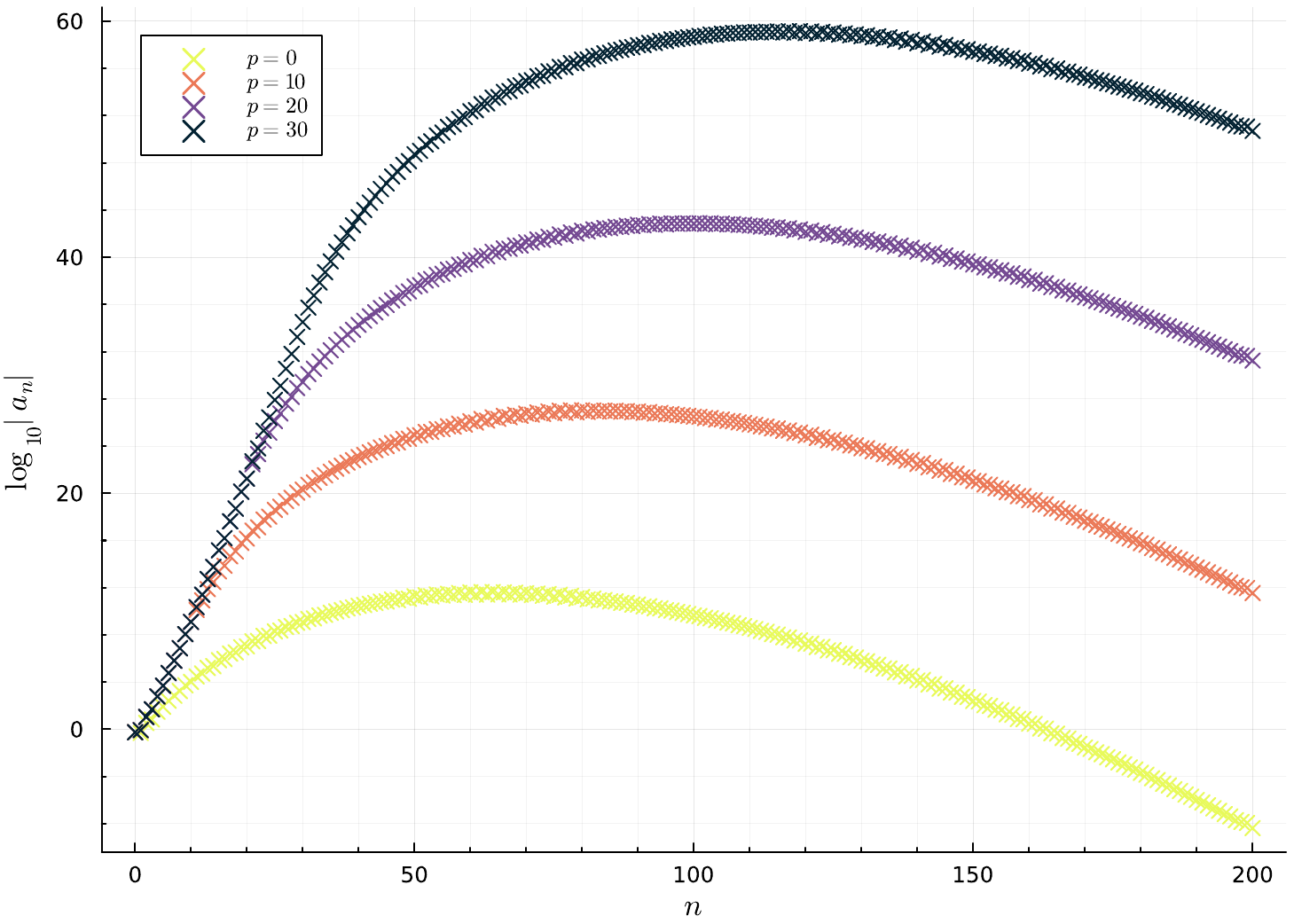}
  \caption{QNM expansion coefficients $a_n^{H^p}$, for different $H^p$ norms, as a function of $n$ and in the P\"oschl-Teller case.}
  \label{Hp_an}
\end{figure}

\section{Physical and structural implications}
\label{s:phys-struc}
After presenting the formalism and some technical aspects in the previous sections, we
discuss now the main results from a physical and structural perspective. We comment on:
\begin{itemize}
\item[a)] {\em Late dynamical behaviour}. Focus is placed on the unexpected
  good performance of the Keldysh expansion for the reconstruction of late tails in Schwarzschild.
\item[b)] {\em Early dynamical behaviour}. This discussion  includes:
  \begin{itemize}
  \item[i)] Early behaviour of the QNM expansion: this analysis involves the discussion
    of different kinds of convergence issues of the QNM series.

  \item[ii)] Non-modal transients growths: including a discussion of $H^p$-pseudospectra.

    \end{itemize}

\item[c)] {\em Black Hole QNM Weyl's law}. Study of the power-law asymptotics  of the
  QNM-frequency counting function
  in different spacetime asymptotics at null infinity.

\end{itemize}

\subsection{Late dynamics: Schwarzschild tails from Keldysh expansions}\label{s:S_tails}
We start by considering the late-time behaviour of the time-series obtained
by evaluating $\phi(\tau, x)$ at $x=x_{\scri^+}$.
A stark difference between the Schwarzschild case and the other three cases considered in
section \ref{s:Time-domain_evolution} is that its (asymptotically flat) null infinity is less
regular than its black hole horizon. Specifically, in the hyperboloidal scheme this translates into the
fact that the function $p(x)$ in $L_1$ (cf. Eqs. (\ref{e:L_1-L_2}) and (\ref{e:defs_in_L_1-L_2})) vanishes
quadratically at future null infinity, whereas at the horizon it vanishes linearly (as it does in all other three
cases at outer boundaries). As a consequence, the spectrum of the operator $L$ contains, apart from the discrete QNM eigenvalues,  a continuous part along the positive imaginary axis (known as ``branch-cut'' in the scattering resonance approach) that is responsible for a power-law tail of the waveform at late times. After discretisation, this branch-cut gives rise to (non-convergent) eigenvalues, as it can be seen in Figs. \ref{spectres} and \ref{spectresconv:S}, along the imaginary axis.

    When considering the spectral QNM decomposition of the scattered field, one of the assumptions in the
    discussion of the Keldysh expansion in section \ref{s:Keldysh} was the discreteness
    of the spectrum of $L$.
    As shown in section \ref{s:Time-domain_evolution},
    this has provided excellent results for the part of the signal
    dominated by QNMs. However, it also means that a priori it is not 
    a tool well adapted to study the tails, encoded
    in the continuum branch cut. In this setting it comes then as an unexpected result the fact that the
    naive application of the Keldysh scheme also to the branch cut (i.e. beyond its regime
    of validity) actually provides an excellent recovery
    of the power-law tail part of the signal.
    More specifically: {\em the straightforward application of the Keldysh scheme to the (non-convergent) eigenvalues corresponding to the discretisation of the branch cut does provide an accurate description of the late tails.}

    The latter is, in principle, an unexpected result that could be understood in terms
    of a Riemann sum approximation of the Bromwich integral to be calculated along the
    branch cut to account for the tails (see e.g. \cite{Ansorg:2016ztf}). However a simpler and more
    direct explanation is given in terms of the discussion presented
    in appendix \ref{a:Keldysh_evol_op}, were the dynamical evolution is obtained by applying the 
    discretised evolution operator $e^{i\tau L}$ on the initial data $u_0$\footnote{Given that
    (in the studied cases) the finite
    approximants $L^N$ are diagonalisable matrices, the complete evolution can be cast in terms
    of {\em all} the eigenvalues of $L^N$ by using a prescription that, crucially, turns out to be
    {\em exactly}  the Keldysh prescription one
    for the  calculation of  the QNM expansion coefficients.
    The difference with the Keldysh QNM expansion is that the sum in the full time-evolution
    is not restricted to QNM eigenvalues,
    but it includes all
    eigenvalues of $L^N$: when we only include QNMs and eigenvalues in the branch cut (disregarding all
    the other eigenvalues of $L^N$), we obtain an approximation to the total signal given by
    the superposition   of the QNM expansion {\em and}
    the tail. This justifies applying the Keldysh prescription to the branch cut eigenvalues to get the tail.}.
    In the following we comment on the main points regarding tails in this Keldysh approach:
    \begin{itemize}
      \item[i)] \textit{Polynomial tails from a Keldysh sum over ``branch-cut eigenvalues'': spectral
      separation of QNM and tails.}
      Fig. \ref{separating_QNM_and_tail_600} shows the calculation of the respective
      contributions of QNMs and branch-cut to the scattered field. Panel  \ref{separating_QNM_and_tail_600:waveform}
      shows the time-domain waveform (black) and the contributions of the QNM expansions (blue)
      and branch cut (orange). The corresponding eigenvalues from which QNM and tail signals are calculated
      (by applying  exactly the same Keldysh algorithm)
      are shown in panel \ref{separating_QNM_and_tail_600:eigen}, with the same code of colours.
      Finally panel \ref{separating_QNM_and_tail_600:waveform_log} shows a log-log plot
      where the power-law nature of the tails is apparent.

\begin{figure}[htp]
  \centering
  \subfloat[Waveform at future null infinity $\phi(\tau,\sigma)\vert_{\sigma=0}$]{
  \includegraphics[clip,width=0.4\columnwidth]{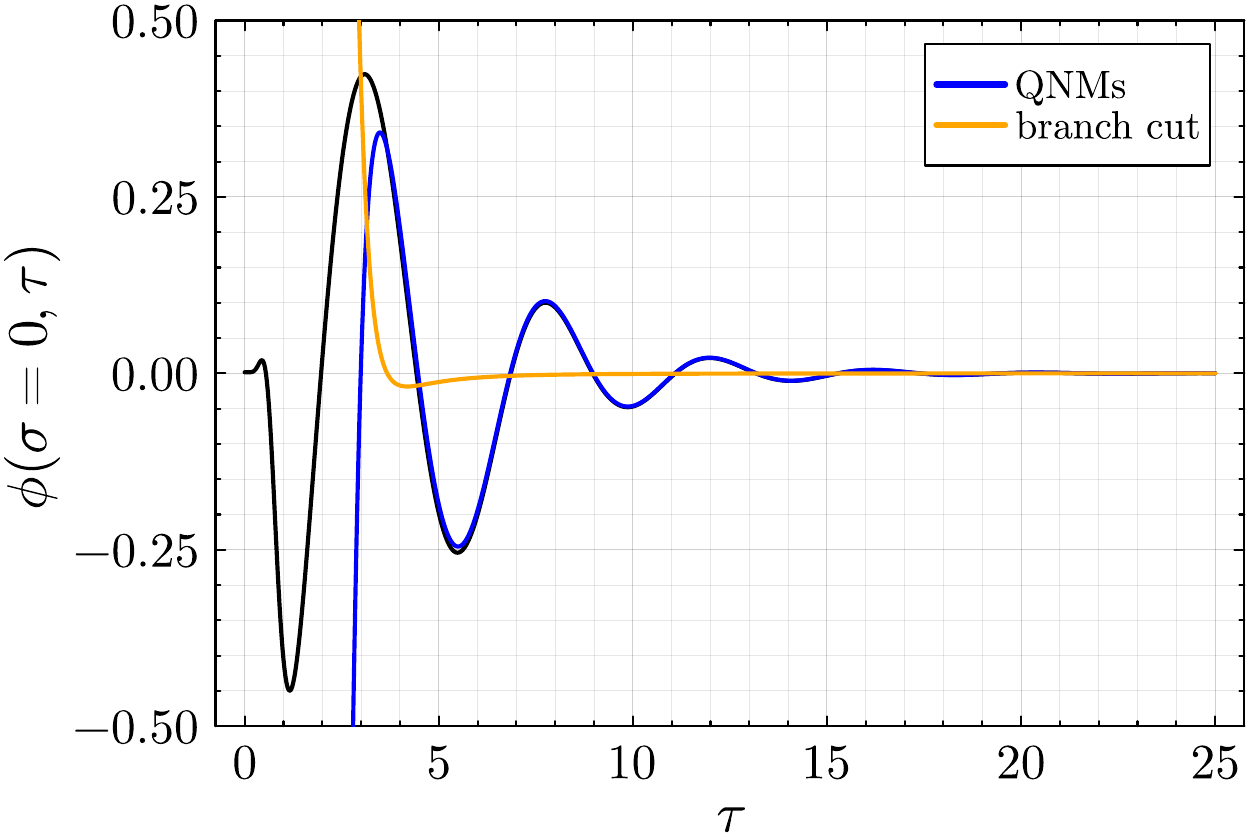}
    \label{separating_QNM_and_tail_600:waveform}
  }
  \subfloat[Eigenvalue plot]{
    \includegraphics[clip,width=0.4\columnwidth]{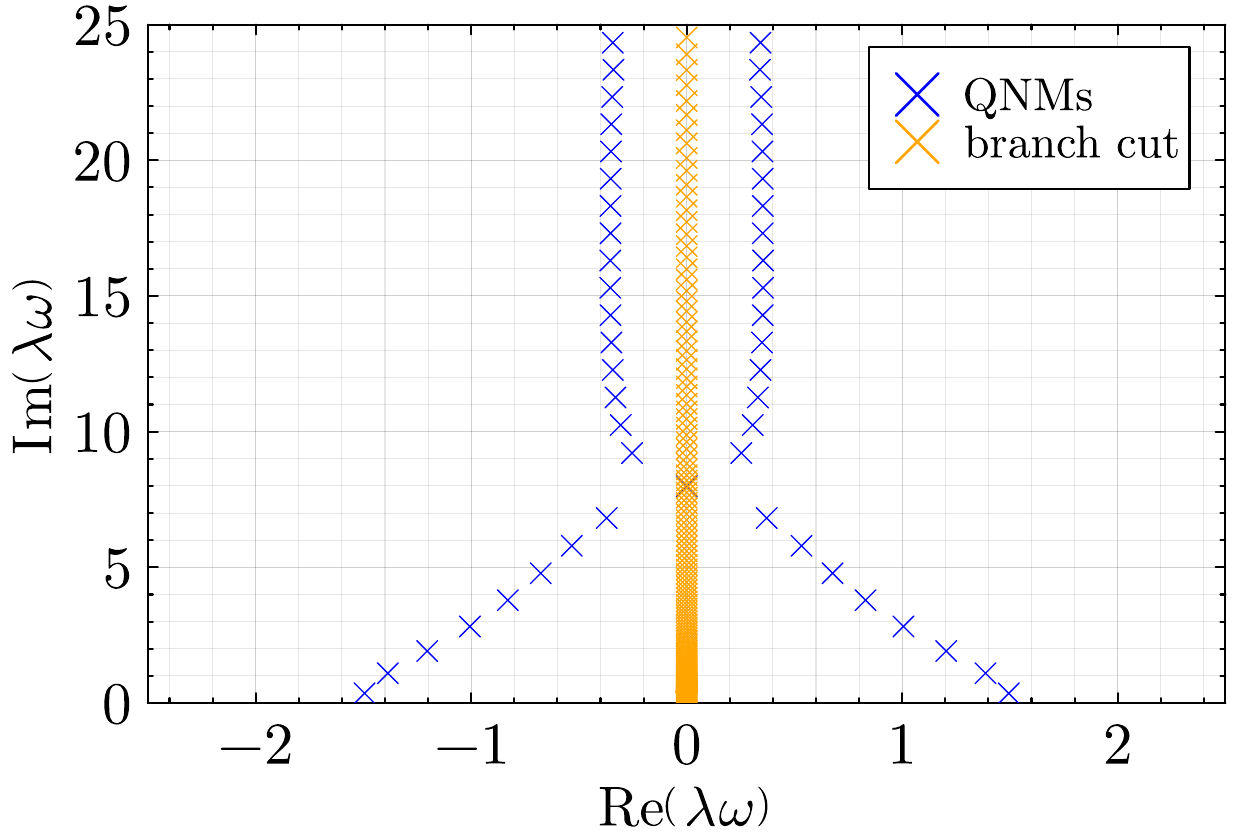}
    \label{separating_QNM_and_tail_600:eigen}
  }

  \subfloat[Waveform $\phi(\tau,\sigma)\vert_{\sigma=0}$ on a log-log plot]{
    \includegraphics[clip,width=0.6\columnwidth]{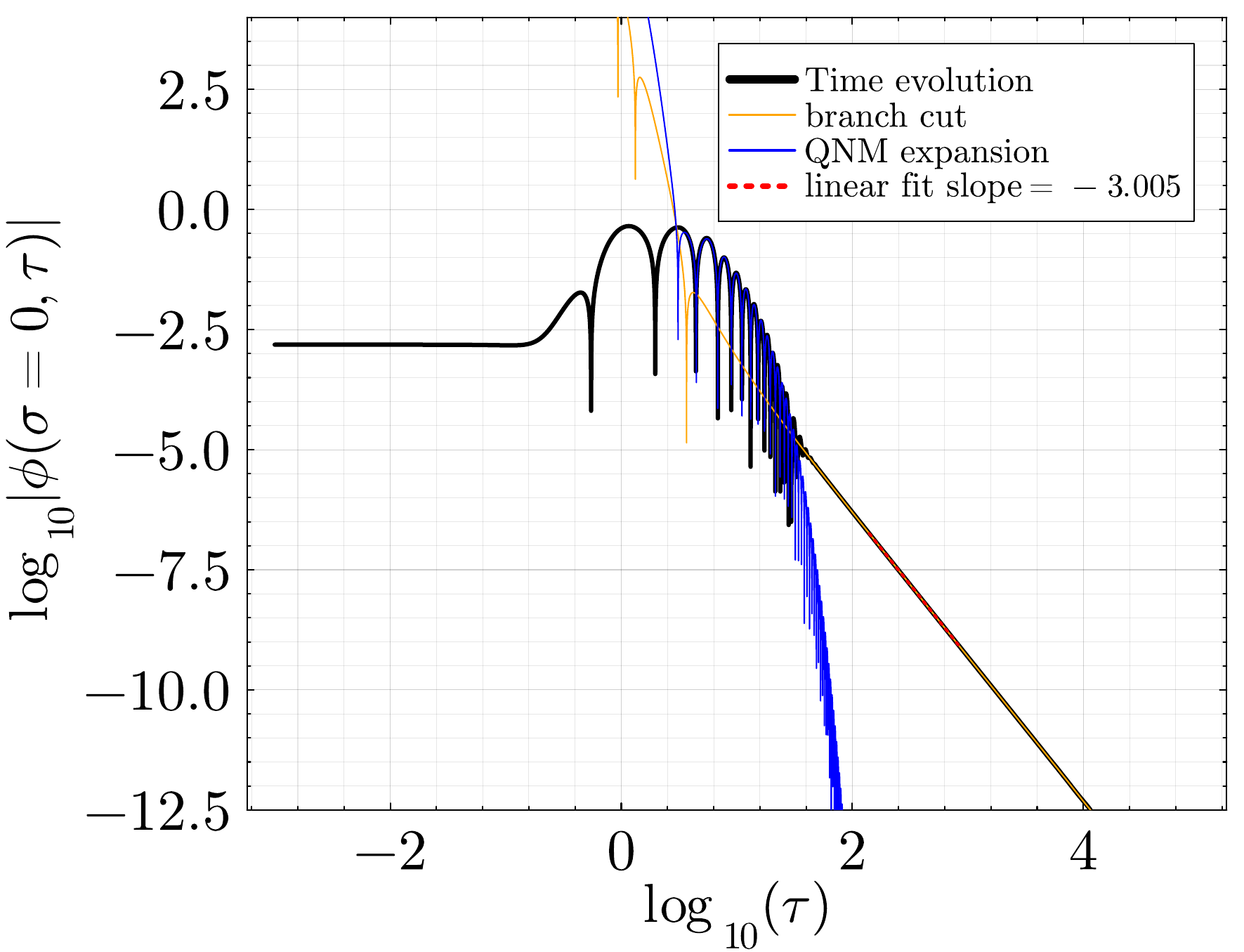}\label{separating_QNM_and_tail_600:waveform_log}
  }
  \caption{In the Schwarzschild case, with $\ell=2$ and $N=600$, we illustrate how the contribution of the branch cut and of the QNMs dominate the waveform one after the other. Initially, both of these curves contribute to the signal, then the ringdown starts at $\tau = 4$ and is dominated by QNMs, after that when the QNMs get small enough, the power-law regime dominates for a long time before it vanishes and an asymptotic exponential decay takes place due to the finite number of modes in the numerical sum. We performed a linear fit to estimate the power law tail $\tau^{\beta_{\text{fit}}}$.
  }
  \label{separating_QNM_and_tail_600}
\end{figure}
    \item[ii)] \textit{Price law.}  Figure \ref{tail_ell} presents the signal (in red) resulting from the
      sum of the QNM expansion and the signal from the branch cut
      eigenvalues. This is done for various $\ell$'s. In all cases the time-domain signal is accurately
      fitted. A robust demonstration of the good convergence behaviour of the Keldysh scheme is the
      recovery of the Price law, namely the late time decay $\phi_\ell(\tau,\sigma=0)\sim\tau^{-(\ell+1)}$.
    \item[iii)]\textit{Numerical convergence of the tail.} Fig. \ref{fig:growing_tail_with_N} shows how
      increasing the numerical resolution, by taking larger $N$'s in the Chebyshev-Lobatto's grid,
      permits to get correspondingly larger times for the tail. This provides a qualitative demonstration
      of the tail convergence in the numerical scheme.
    \end{itemize}

    \begin{figure}[htp]
      \centering
      \subfloat[$\ell=2$]{
        \includegraphics[clip,width=\sizefigmedium\columnwidth]{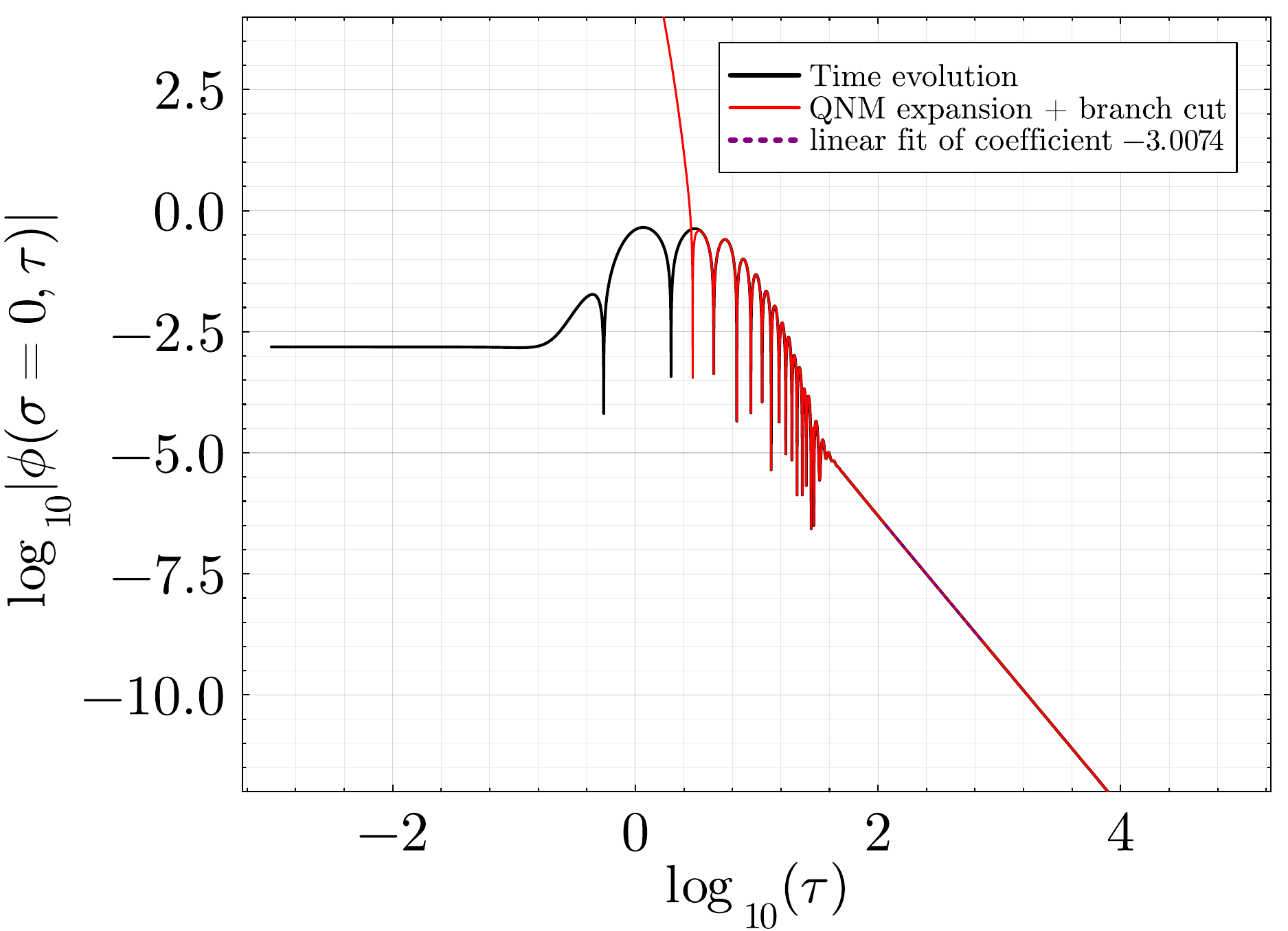}\label{tail_ell:2}
      }
      \subfloat[$\ell=3$]{
        \includegraphics[clip,width=\sizefigmedium\columnwidth]{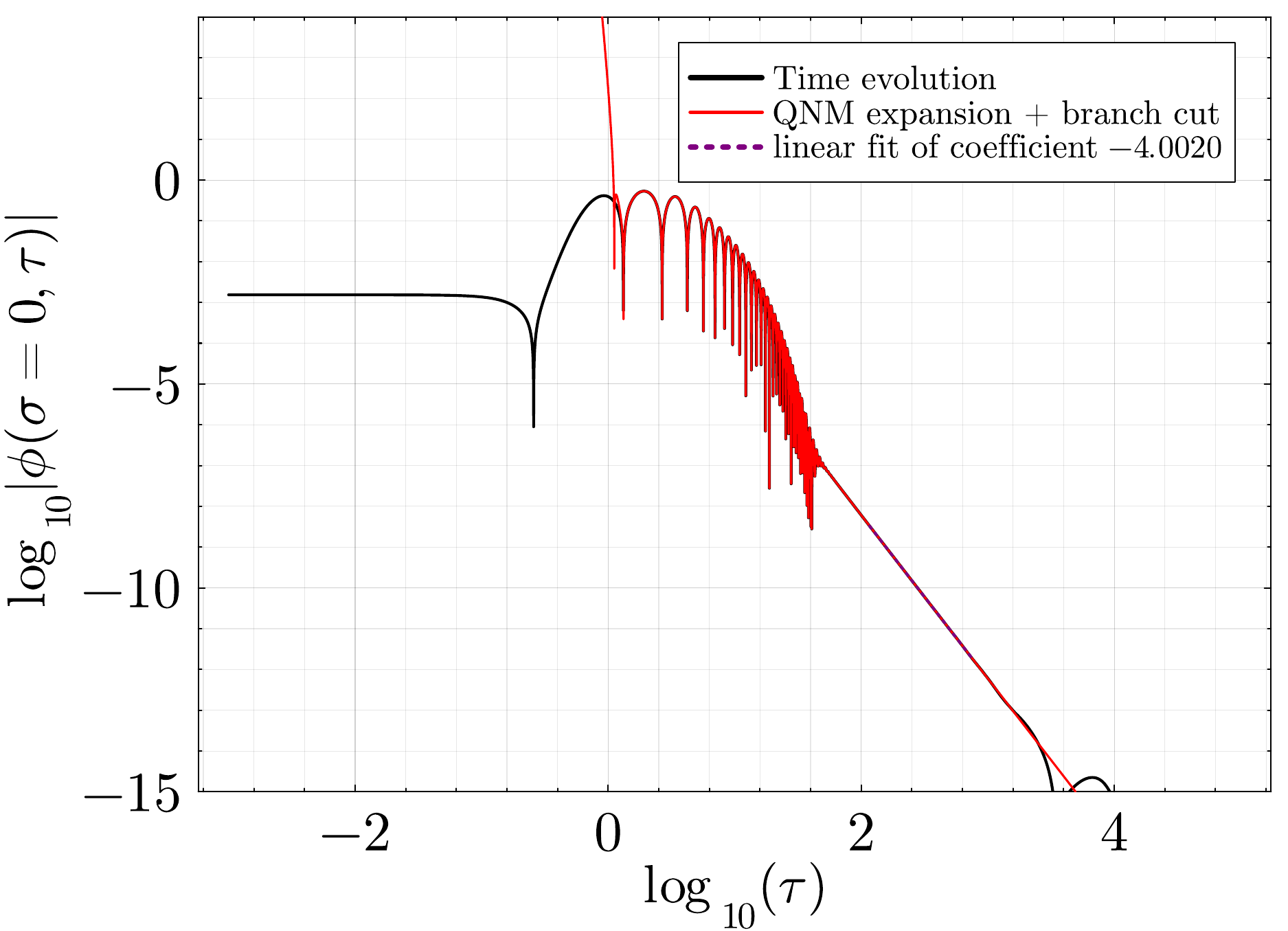}\label{tail_ell:3}
      }
    
      \subfloat[$\ell=4$]{
        \includegraphics[clip,width=\sizefigmedium\columnwidth]{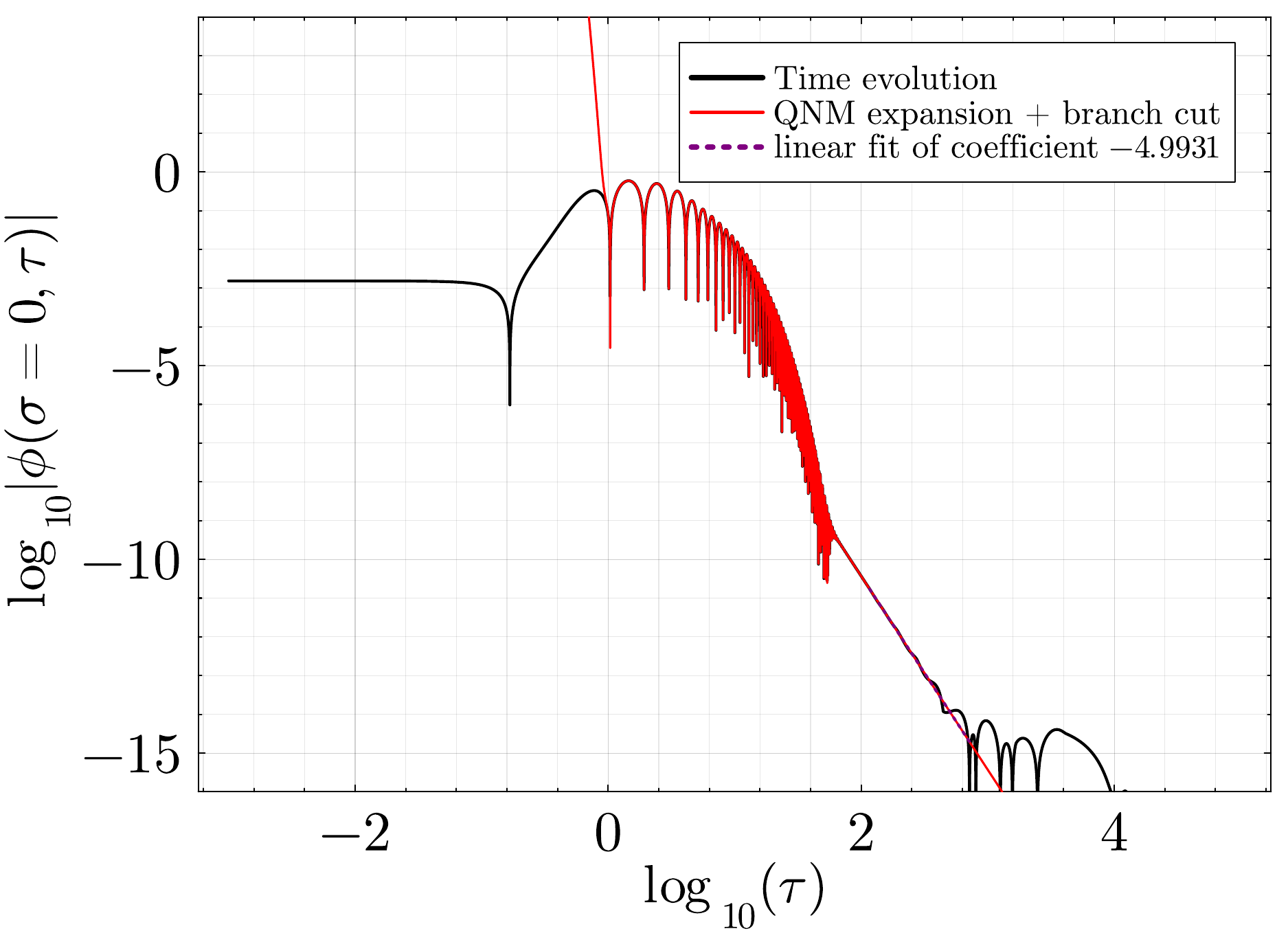}\label{tail_ell:4}
      }
      \subfloat[$\ell=5$]{
        \includegraphics[clip,width=\sizefigmedium\columnwidth]{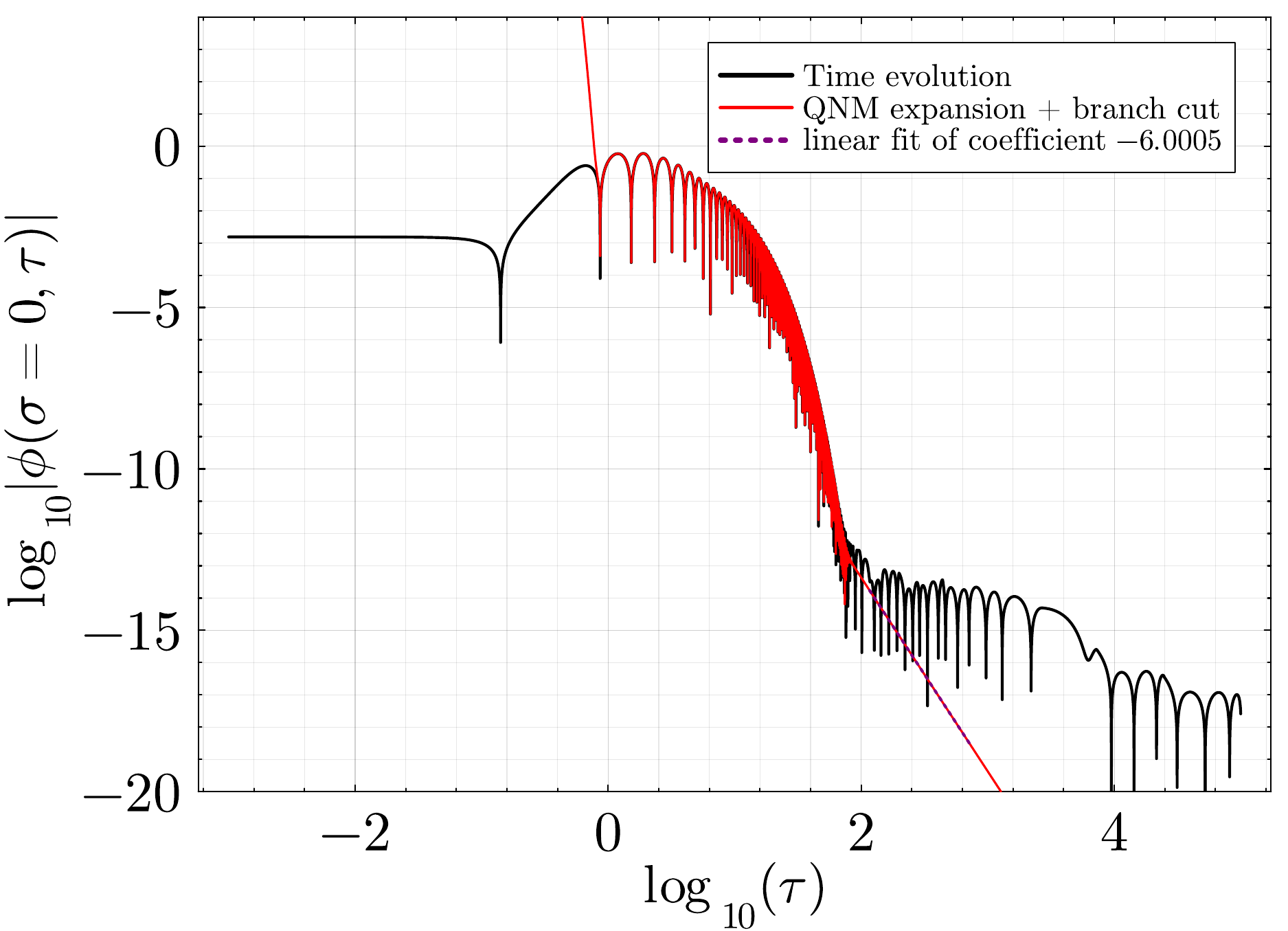}\label{tail_ell:5}
      }
      \caption{Panels \ref{tail_ell:2}, \ref{tail_ell:3}, \ref{tail_ell:4} and \ref{tail_ell:5} show the polynomial tails we get from adding the branch cut modes to the QNM expansion for different angular $\ell$'s.
        The linear fit to estimate the power law tail time decay $\sim\tau^{\beta_{\text{fit}}}$ demonstrates
        the Price law.}
      \label{tail_ell}
    \end{figure}    

    The bottom line is that the Keldysh scheme combined with the Chebyshev-pseudospectral discretisation
    provides a simple, efficient and accurate algorithm for the calculation of Schwarzschild tails. The
    Keldysh prescription provides a complete spectral account of the time-domain signal,
    providing a neat separation between the  QNM
    and tail contributions. 
    
    Furthermore, in order to give a first insight into the activation of the QNMs during the time evolution, the video \verb|Schwarzschild_mode_contribution.mp4| in supplementary material details the contribution through time of each individual "mode" (QNMs and branch-cut eigenvalues in the zone $\mathrm{Im}(\omega_n) < 25$) to the waveform at future null infinity. It shows the dominant mode at every instant $\tau\leq40$, in particular, high overtones are initially excited, then the fundamental QNM dominates at intermediate times before the activation of branch cut eigenvalues visible at the very end of the video.
    \begin{figure}[htp]
      \centering
        \includegraphics[clip,width=0.7\columnwidth]{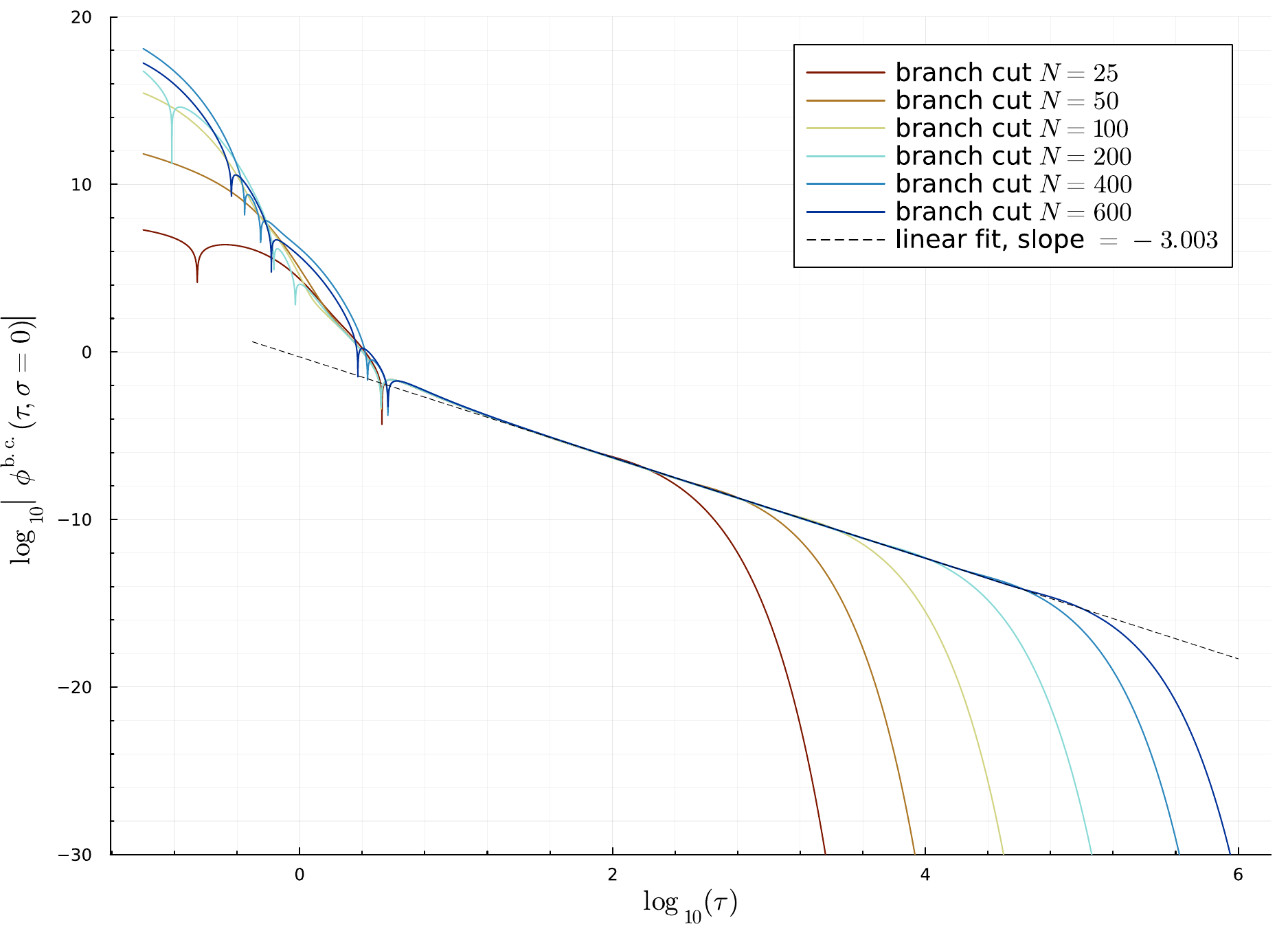}
      \caption{ We compute the expansion over the branch cut eigenvalues as a function of $\log_{10}(\tau)$ for several grid sizes $N\in\{25,50,100,200,400,600\}$ in the Schwarzschild case ($\ell=s=2$).}
      \label{fig:growing_tail_with_N}
      \end{figure}

\subsection{Early dynamics: convergence issues of the QNM expansion}
\label{s:role_overtones}
We start by rewriting the truncated  $u^{\mathrm{QNM}}(\tau,x)$ in Eq.~(\ref{e:u_truncated_sum}) as
\begin{equation}
  \label{e:u_truncated_series_v2}
  u^{\mathrm{QNM}}(\tau,x)=\sum_{n=0}^{N_{\mathrm{QNM}}} \mathcal{A}^\pm_n(x) e^{i\omega^\pm_n\tau}
  =\sum_{n=0}^{N_{\mathrm{QNM}}} a^\pm_n e^{i\omega^\pm_n\tau}v_n^\pm(x) \ .  
  \end{equation}
We can consider two natural series
from this expression, as well as
their respective notions of convergence, when taking the limit
$N_{\mathrm{QNM}}\to \infty$, namely\footnote{Taking the limit
$N_{\mathrm{QNM}}\to \infty$ in expression (\ref{e:u_truncated_series_v2}), without
fixing spatial $x_o$ or time $\tau_o$ values, provides by itself a third (or rather,
a zero-th)
``spacetime QNM Keldysh series'' that could be of interest for the
``spacetime fitting'' approach in section III.B of \cite{zhu2024challenges}.
We restrain ourselves to space-fixed and time-fixed QNM series, respectively
in subsections \ref{s:convergence_fixed_x} and \ref{s:convergence-fixed_tau},
where some additional structure permits to better control the convergence properties,
leaving the convergence of such a ` spacetime series' for future work.}:
\begin{itemize}
\item[i)] {\em Time-series at fixed $x_o$}. By making $x=x_o$ in Eq. (\ref{e:u_truncated_series_v2})
  we get
\begin{equation}
  \label{e:u_truncated_series_fixed_x}
  u_{x_o}^{\mathrm{QNM}}(\tau)=\sum_{n=0}^{N_{\mathrm{QNM}}} \mathcal{A}^\pm_n(x_o) e^{i\omega^\pm_n\tau}
  = \sum_{n=0}^{N_{\mathrm{QNM}}} c_n^{\pm}(x_o) e^{i\omega^\pm_n\tau}\ .
\end{equation}
This is the natural series to be considered in the analysis of the observed gravitational
wave signal, where $x_o=x_{\scri^+}$. The notion of convergence that we will
consider in this case is `pointwise' (uniform and non-uniform) 
convergence of function series \footnote{The series (\ref{e:u_truncated_series_fixed_x}) is said
to converge pointwise to $u_{x_o}(\tau)$ at $\tau$ when $N_{\mathrm{QNM}}\to \infty$ if
\bea
\label{e:pointwise_convergence}
\forall \varepsilon>0 \ \ \exists K_\tau \ \ \hbox{such that}, \ \ \forall N_{\mathrm{QNM}}> K_\tau,
\ \ \hbox{it holds} \ \ |u_{x_o}(\tau) - u_{x_o}^{\mathrm{QNM}}(\tau)|<\varepsilon \ .
\eea
Note that here $K_\tau$ depends explicitly on $\tau$ so, in principle, the series converge
`at different speeds' at different times. When  $K_\tau$ does not depend
on $\tau$ the series converges uniformly, i.e. the series (\ref{e:u_truncated_series_fixed_x})
is said to converge uniformly to $u_{x_o}(\tau)$ when $N_{\mathrm{QNM}}\to \infty$ if
\bea
\label{e:pointwise_convergence_uniform}
 \forall\varepsilon>0  \ \  \exists K \ \ \hbox{such that}, \ \ \forall \tau, \forall N_{\mathrm{QNM}}> K
\ \ \hbox{it holds} \ \ |u_{x_o}(\tau) - u_{x_o}^{\mathrm{QNM}}(\tau)|<\varepsilon \ ,
\eea
and therefore the rate of convergence of the function series (\ref{e:u_truncated_series_fixed_x}) can be controlled
in a uniform manner for all times $\tau$. This is crucial to preserve the analytical
properties of the partial sums  in their limit
$u_{x_o} = \displaystyle \lim_{N_{\mathrm{QNM}}\to\infty} u_{x_o}^{\mathrm{QNM}}$, so that we can do analysis with it.}.

\item[ii)] {\em Asymptotic series at fixed $\tau_o$}.  By making $\tau=\tau_o$
  in Eq. (\ref{e:u_truncated_series_v2})
  we get
\begin{equation}
  \label{e:u_truncated_series_fixed_tau}
  u_{\tau_o}^{\mathrm{QNM}}(x)
  =\sum_{n=0}^{N_{\mathrm{QNM}}} a^\pm_n e^{i\omega^\pm_n\tau_o}v_n^\pm(x)
  = \sum_{n=0}^{N_{\mathrm{QNM}}} c_n^{\pm}(\tau_o) v_n^\pm(x)\ .  
\end{equation}
This is the series of functions naturally considered, for instance, in optical cavities~\cite{LalYanVyn17}
and non-Hermitian quantum mechanics \cite{moiseyev2011}, as well as the one
underlying the QNM ``spatial fitting'' approach in \cite{zhu2024challenges}.
In this case, in addition to (uniform) pointwise convergence,
it is natural to consider the convergence in the Hilbert space ${\cal H}$ to which
the QNMs $v_n(x)$'s belong\footnote{\label{footnote:convergence_norm}Given the Hilbert space $({\cal H}, \langle\cdot,\cdot\rangle)$
with associated norm $||\cdot||$
(more generally the Banach space $({\cal H}, ||\cdot||$), the series (\ref{e:u_truncated_series_fixed_tau})
is said to converge to $u_{\tau_o}(x)\in {\cal H}$ when $N_{\mathrm{QNM}}\to \infty$ if
\bea
\label{e:norm_convergence}
\forall\varepsilon>0  \ \  \exists K \ \ \hbox{such that}, \ \ \forall N_{\mathrm{QNM}}> K
\ \ \hbox{it holds} \ \ ||u_{\tau_o}(x) - u_{\tau_o}^{\mathrm{QNM}}(x)||<\varepsilon \ .
\eea
}.
It is in this latter sense that the series (\ref{e:normal_modes}) in the
self-adjoint case of section \ref{s:normal_modes} is said to converge and, in particular,
orthonormal modes $\hat{v}_n(x)$'s are said to be a Hilbert basis. And, in our QNM setting,
this is the sense in which the QNM expansion is in general non-convergent, but only
asymptotic in the specific sense of Eq. (\ref{e:u_Keldysh_v7}).
This is also the natural convergence setting for discussing non-modal transients (see
section \ref{s:transients}).

\end{itemize}

\subsubsection{Convergence of the QNM time-series: QNM expansion at fixed $x_o$}
\label{s:convergence_fixed_x}
As discussed in section \ref{s:Time-domain_evolution}, as we add more and more QNM overtones to
the truncated QNM
expansion $u^{\mathrm{QNM}}_{x_o}(\tau)$ in Eq. (\ref{e:u_truncated_sum}),
the resulting QNM time-series starts to agree with the time-domain calculated signal at  earlier
and earlier times $\tau$. This is illustrated in Fig. \ref{role_of_overtones} for the P\"oschl-Teller case
(see below for the other cases):
the more overtones are added, the earlier the corresponding
coloured curves (passing from the red to the blue) smoothly join the time-domain signal (black curve),
supporting pointwise convergence at each $\tau$.

\begin{figure}[htp]
  \centering
  \subfloat[Adding overtones to the truncated QNM expansion]{
    \includegraphics[clip,width=0.45\columnwidth]{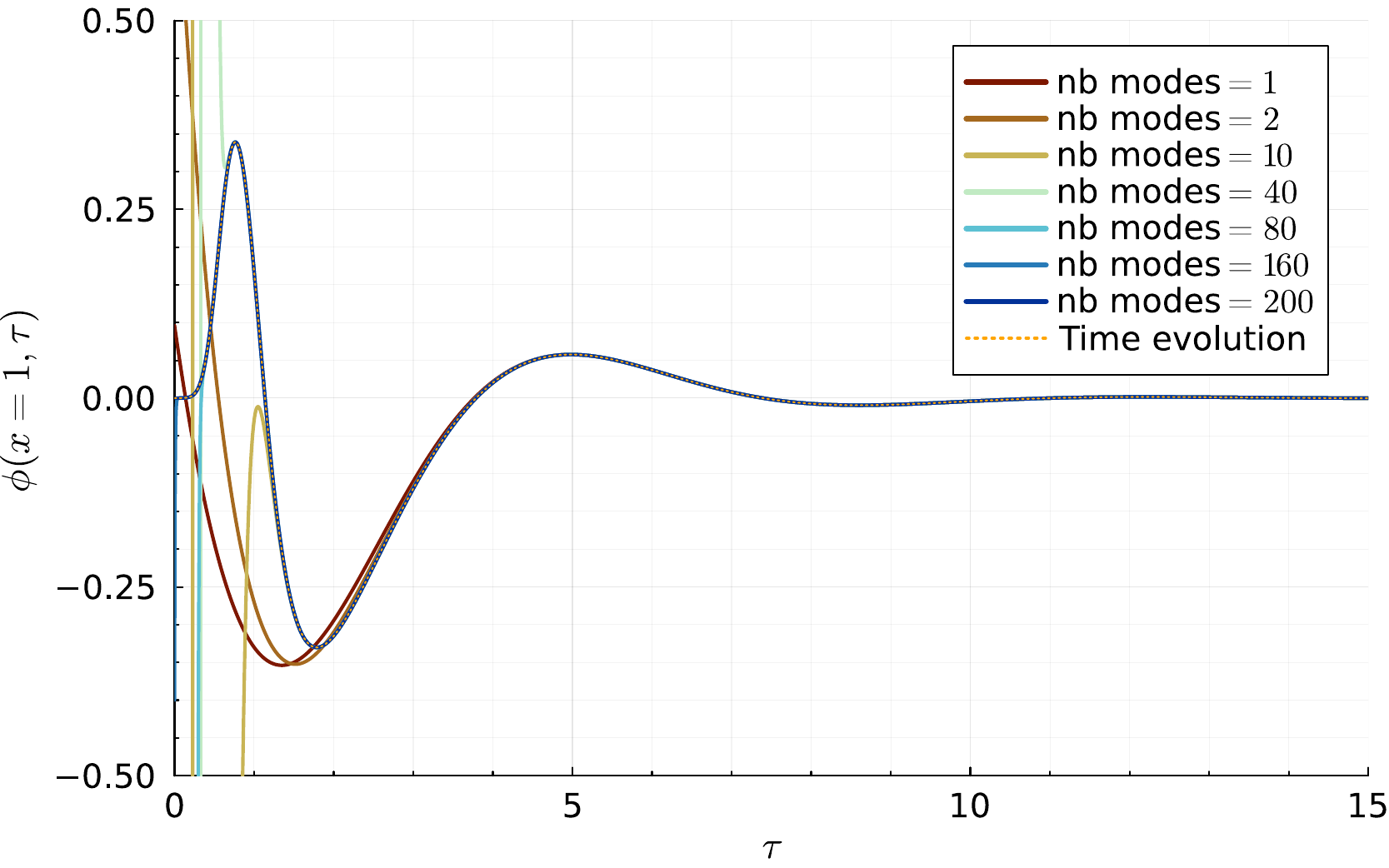}\label{role_of_overtones:full}
  }
  \subfloat[Early $\tau$ view (zoom)]{
    \includegraphics[clip,width=0.45\columnwidth]{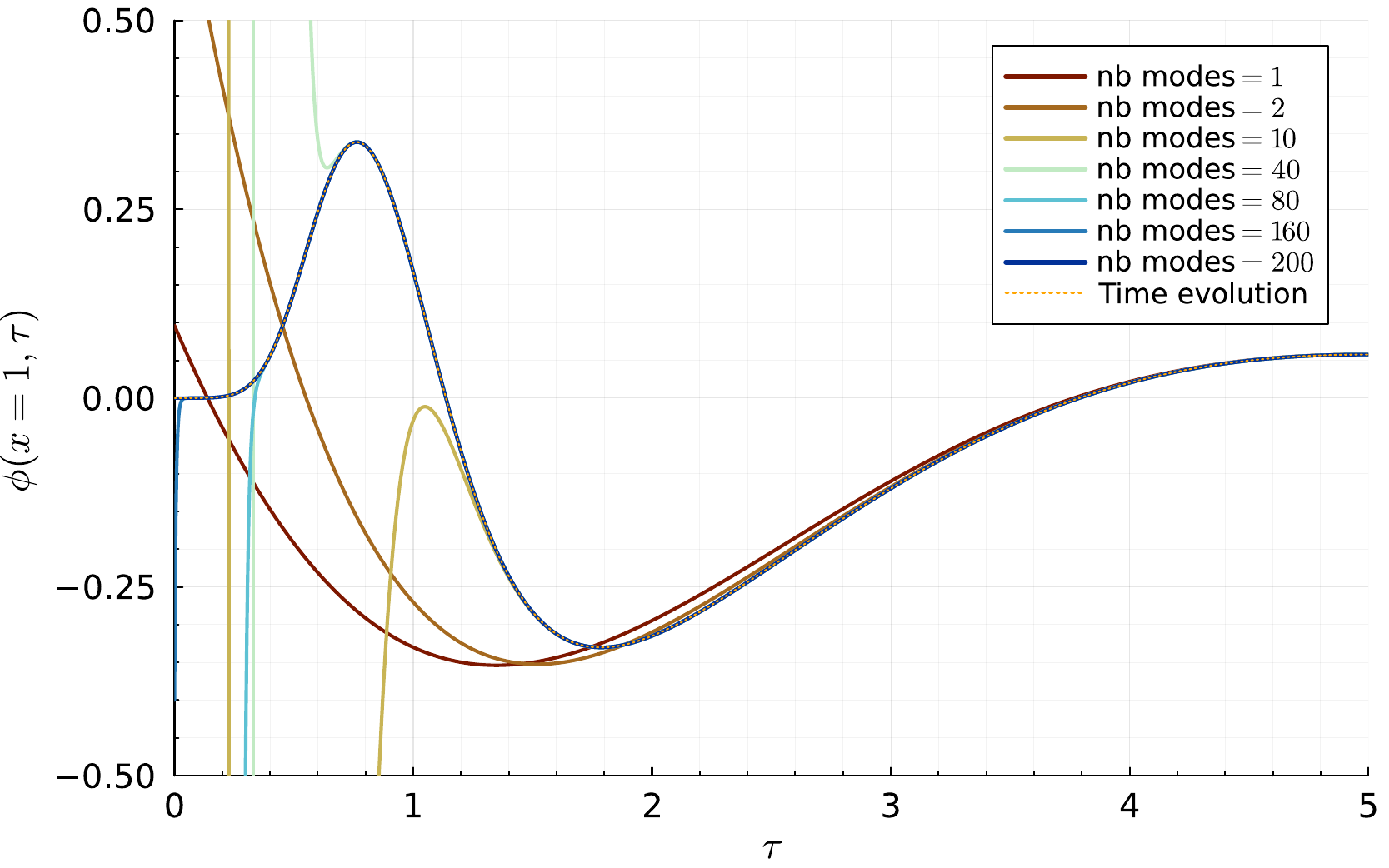}\label{role_of_overtones:early}
  }
  \caption{Panels \ref{role_of_overtones:full} show the truncated QNM expansion in the P\"oschl-Teller case with a variable number of modes. We start with only the fundamental mode and we finish with $200$ modes which corresponds to a dark blue curve hidden behind the black one.}
  \label{role_of_overtones}
\end{figure}

\paragraph{Initial time $\tau^{\mathrm{QNM}}_{\mathrm{init}}$ of the QNM time-series expansion:
Ansorg \& Macedo proposal}
    A natural question to ask from Fig. \ref{role_of_overtones} is
    whether there exists an earliest time $\tau^{\mathrm{QNM}}_{\mathrm{init}}$
    after which the QNM series converge pointwise, for all $\tau\geq \tau^{\mathrm{QNM}}_{\mathrm{init}}$,
    to the time-domain calculated signal.
    This problem has been addressed in \cite{Ansorg:2016ztf}
    leading to a conjecture in the affirmative. More specifically  Ansorg \& Macedo propose that,
    for a given initial data $u_0(x)$
    and a given time-series at fixed $x_o$, i.e. for $u_{x_o}(\tau) = u(\tau, x=x_o)$
    with $u_0(x) = u(\tau=0, x)$,
    such an initial time exists and it is given by  $\tau^{\mathrm{QNM}}_{\mathrm{init}}(x_o) = \nu_{\mathrm{QNM}}(x_o)$,
    where $\nu_{\mathrm{QNM}}(x_o)$   is the ``QNM growth rate of excitation coefficients''
    $\nu_{\mathrm{QNM}}(x_o)=\lim_{n\to\infty}\frac{\ln{\cal A}_n(x_o)}{\mathrm{Im}(\omega_n)}$.
    Interestingly, for the specific case of the Gaussian initial data $u_0$  in (\ref{e:initial_data}),
    the value of $\nu_{\mathrm{QNM}}(x_o)$ at the boundary $x_b$  is consistent with  $\nu_{\mathrm{QNM}}(x_{b})=0$
    in the P\"oschl-Teller, Schwarzschild\footnote{The 
    Schwarzschild case is however more delicate, since once must consider
    a `mutual growth rate' $\sigma(x_o)$ defined in terms of $\nu_{\mathrm{QNM}}(x_o)$ and 
    a corresponding $\nu_{\mathrm{cut}}(x_o)$ associated with  the branch cut
    (cf. definition in Eq. (110) of \cite{Ansorg:2016ztf}.} and Schwarzschild-dS cases.
    The Schwarzschild-AdS case is more difficult to assess, since the
    asymptotic convexity in Fig. \ref{Aconv:AdS} seems compatible with non-vanishing
    $\nu_{\mathrm{QNM}}(x_o)$.

    As a first step to assess the conjecture in \cite{Ansorg:2016ztf},
    we start exploring  the pointwise convergence of the QNM time-series and,
    most importantly,  if such convergence is `uniform'.

    \paragraph{Uniform convergence of the QNM time-series expansion}
     In Fig. \ref{f:uniform_convergence} we explore the convergence of
     $u_{x_b}^{\mathrm{QNM}}(\tau)$ towards $u_{x_o}(\tau)$ (with boundary $x_o=x_b$)
     by plotting
    the dependence of the error
    $|u_{x_b}(\tau) - u_{x_b}^{\mathrm{QNM}}(\tau)|$, as a function of
    the time $\tau$ and the number of QNMs $N_{\mathrm{QNM}}$,
    with a $\log_{10}$-scale in the colour bar and showing contour lines of
    constant $\epsilon=|u_{x_b}(\tau) - u_{x_b}^{\mathrm{QNM}}(\tau)|$.
    Specifically, to explore pointwise convergence at a given time
    $\tau_o$, we consider a vertical line at that $\tau_o$: given
    the structure of the Figures, if that vertical
    line crosses all $\epsilon$-contour lines, it means that for every $\epsilon>0$
    there exists an $N^o_{\mathrm{QNM}}$ such that for  $N_{\mathrm{QNM}}>N^o_{\mathrm{QNM}}$
    we have $|u_{x_b}(\tau) - u_{x_b}^{\mathrm{QNM}}(\tau)|<\epsilon$, and therefore
    pointwise convergence at that $\tau_o$. Considering our reference
    initial data in (\ref{e:initial_data}), the general
    structure of Fig.  \ref{f:uniform_convergence}, with `red colours' at late
    times, reflects the fact that less QNMs are needed at late times.
    More specifically, the contour-line structure suggests
    that pointwise convergence occurs for all $\tau$'s in P\"oschl-Teller
    and Schwarzschild-dS (with $\Lambda=0.11$), so we
    would have $\nu_{\mathrm{QNM}}(x_b)=0$. On the contrary,
    the vertical contour-lines for Schwarzschild-AdS (above $N_{\mathrm{QNM}}\sim 5$)
    in Fig. \ref{f:S-AdS_uniform_conv} would indicate that there is no actual
    pointwise convergence in this case.
    The Schwarzschild case is more delicate, since the vertical contour-lines
    are actually accounted for in terms of the tail, so adding QNMs does not
    actually diminish the error. Both Schwarzschild-AdS and Schwarzschild
    needs a more detailed study, but we already see the striking qualitative
    differences with Schwarzschild-dS-like cases.

    These differences impact directly the assessment of `uniform convergence'.
    Indeed, for the latter to occur in the interval $[\tau_o, \infty[$ it is
        enough to have contour-lines `decreasing' as functions of $\tau$ and
        such that all contour lines intersect the vertical line $\tau=\tau_o$.
        In that case, for any $\epsilon>0$, the $N^o_{\mathrm{QNM}}$
        read from the intersection between $\tau=\tau_o$ and the $\epsilon$-contour
        line provides the appropriate $\tau$-independent such that for
        all $\tau$ and for all $N_{\mathrm{QNM}}>N^o_{\mathrm{QNM}}$
        we have $|u_{x_b}(\tau) - u_{x_b}^{\mathrm{QNM}}(\tau)|<\epsilon$, so uniform
        convergence follows in  $[\tau_o, \infty[$. Fig. \ref{f:PT_uniform_conv}
            then suggests that the P\"oschl-Teller case
            in uniformly convergent from $\tau=0$, since all contour-lines
            seem to intersect the $\tau=0$ line. The Schwarzschild-dS is more
            difficult to evaluate, since it is not clear if the contour
            lines cut the $\tau=0$ line of if they asymptote to it.
            On the other hand, contour-lines in the Schwarzschild-AdS and Schwarzschild
            cases do not seem consistent with a good uniform convergent behaviour.
            However, Figs. \ref{f:uniform_convergence} only represent a first exploration and for
            a single initial data. A more systematic analysis will be presented
            in \cite{BesJarPoo24}.

            Independently of the insights regarding pointwise and uniform convergence,
            a pragmatic use of this kind of plot in Figs. \ref{f:uniform_convergence}
            is to provide an answer to the following question: given a number $N_{\mathrm{QNM}}$
            of `available' QNMs for the expansion and accepting an error $\epsilon$,
            what is the earliest time $\tau$ for the truncated sum $u_{x_b}^{\mathrm{QNM}}$
            to be valid? The answer is given by the $\tau$ at the intersection
            of the horizontal $N_{\mathrm{QNM}}$-line and the $\epsilon$-contour line.
            Such an application could be useful as an input for data analysis
            of observational gravitational wave time-signals.

\begin{figure}[htp]
    \centering    
    \subfloat[P\"oschl-Teller]{
      \includegraphics[clip,width=\sizefigmedium\columnwidth]{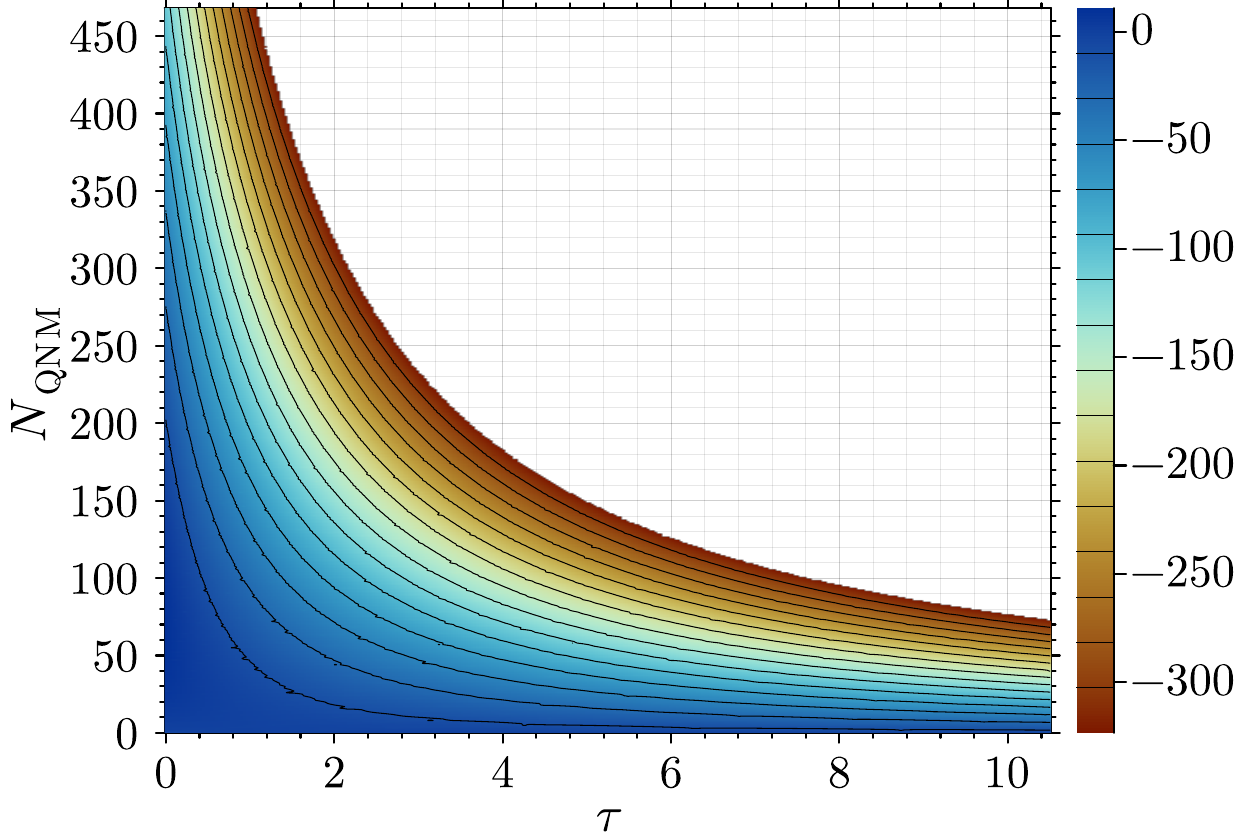}\label{f:PT_uniform_conv}
    }  
   \subfloat[Schwarzschild-de Sitter ($\Lambda=0.11$)]{
      \includegraphics[clip,width=\sizefigmedium\columnwidth]{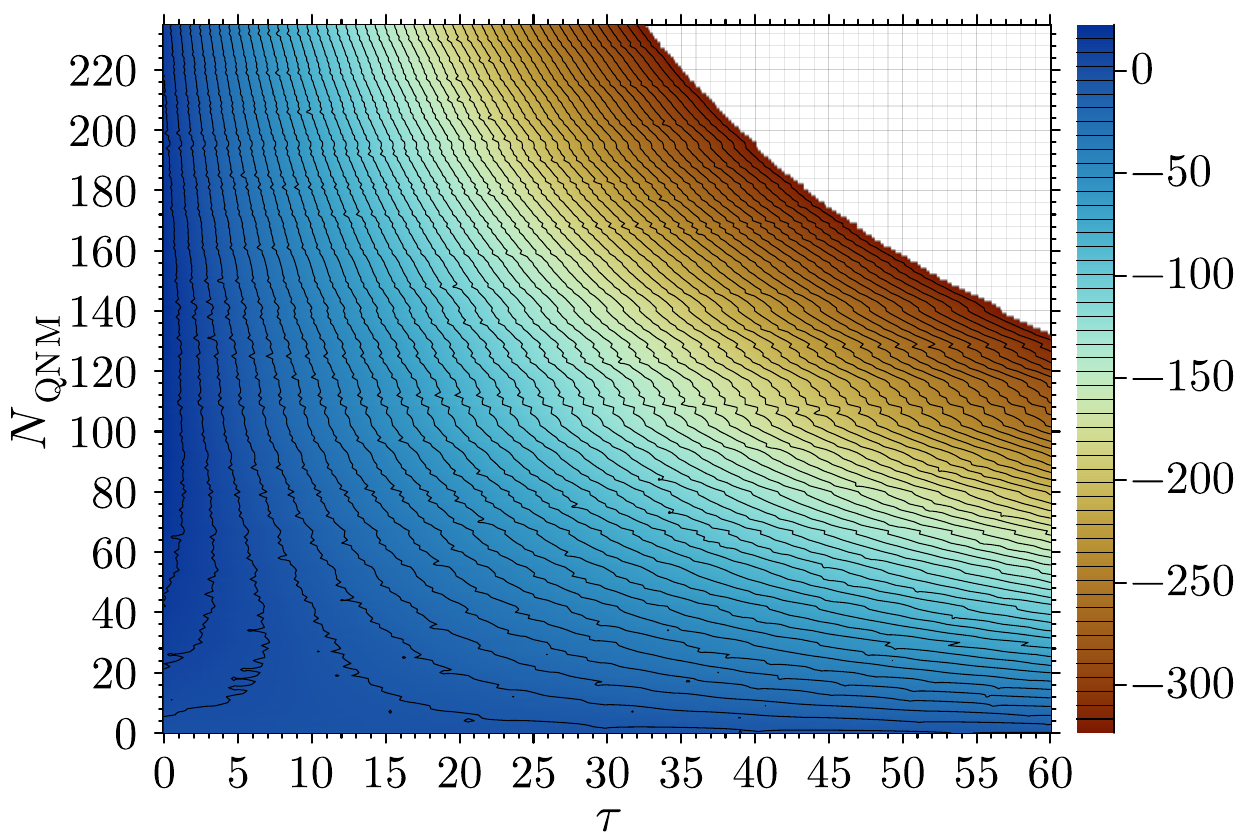}\label{f:S-dS_uniform_conv_0.11}
    }

    \subfloat[Schwarzschild-Anti de Sitter]{
      \includegraphics[clip,width=\sizefigmedium\columnwidth]{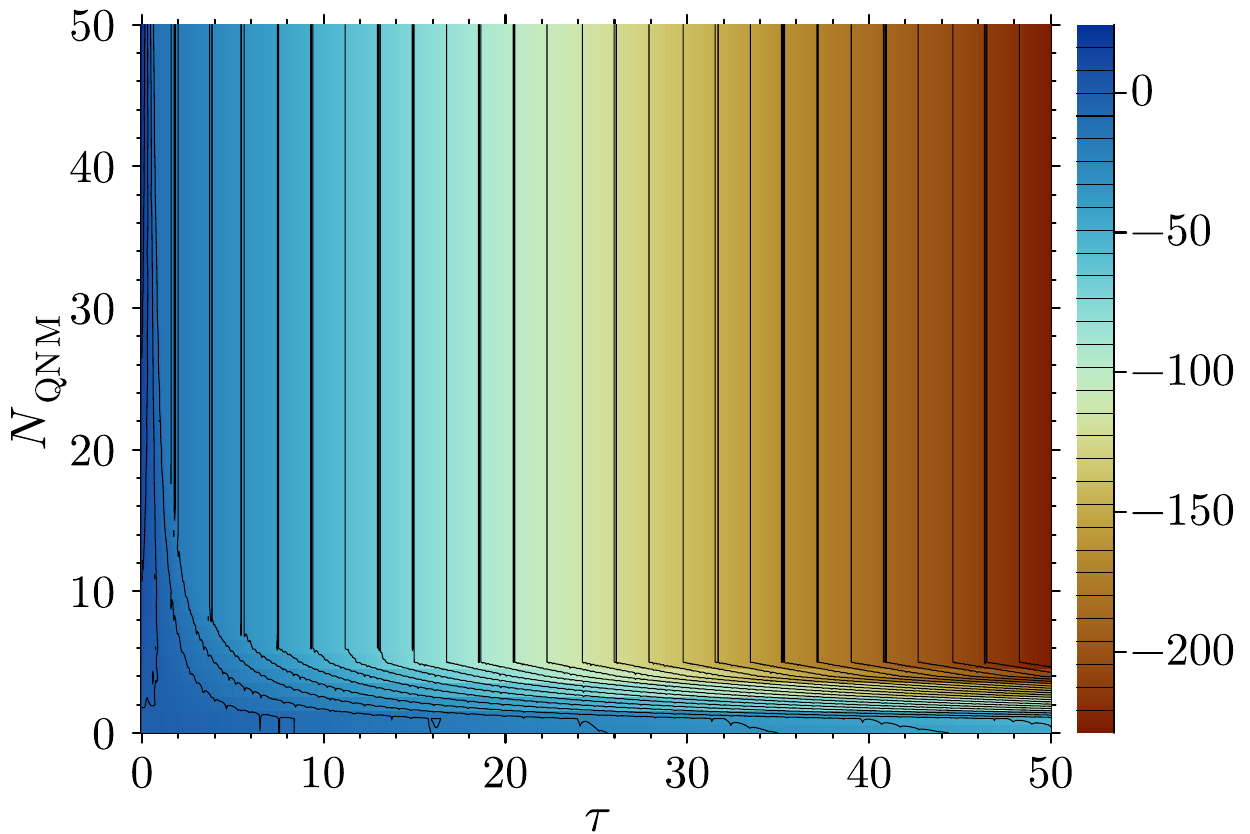}\label{f:S-AdS_uniform_conv}
    }
   \subfloat[Schwarzschild]{
      \includegraphics[clip,width=\sizefigmedium\columnwidth]{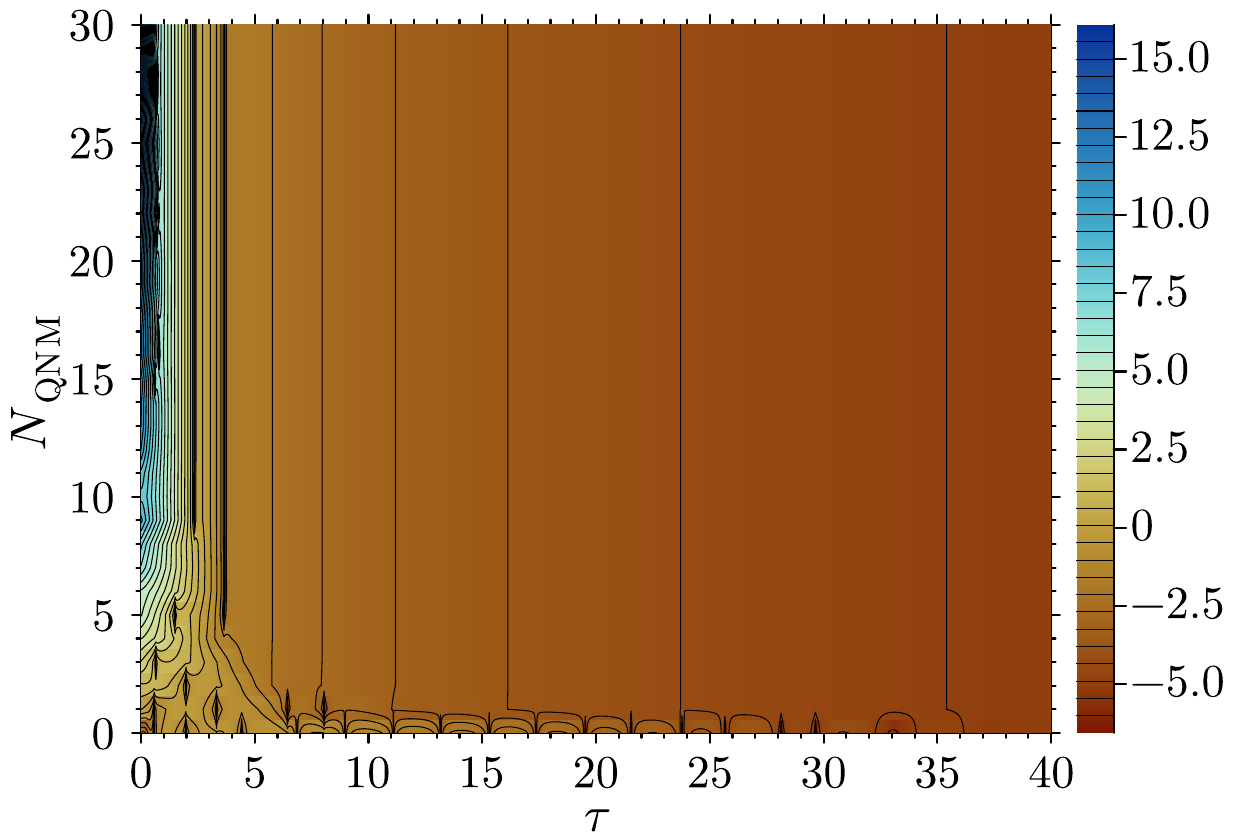}
      \label{f:S_uniform_conv}
    }
   \caption{Exploration of the pointwise and uniform convergence of the QNM
     times-series at the boundary $x_b$,
     for P\"oschl-Teller case and the Schwarzschild BHs with the different spacetime asymptotics.
     The quantity 
     $|u_{x_b}(\tau) - u_{x_b}^{\mathrm{QNM}}(\tau)|$ is plotted  as a function of
    the time $\tau$ and the number of QNMs $N_{\mathrm{QNM}}$,
    with a $\log_{10}$-scale in the colour bar.
    To estimate the pointwise convergence at a time $\tau$, we consider
    a vertical line through that $\tau$ and assess if it crosses all
    contour lines of constant $\epsilon=|u_{x_b}(\tau) - u_{x_b}^{\mathrm{QNM}}(\tau)|$.
    This suggests pointwise convergence in P\"oschl-Teller and Schwarzschild-dS
    and no convergence in the Schwarzschild-dS. The asymptotically flat Schwarzschild
    case is more difficult to assess due to the presence of tails.
    Regarding uniform convergence in the interval $[0, \infty[$, this occurs
        if the contour lines cut the $\tau=0$ vertical line. The  P\"oschl-Teller
        seem to fulfill this, whereas the Schwarzschild-dS (for $\Lambda=0.11$)
        is more difficult to assess. Finally,  given a number $N_{\mathrm{QNM}}$
        of QNMs and accepting an error $\epsilon$ in the corresponding
        QNM expansion approximation,
            the earliest time $\tau$ for the truncated sum $u_{x_b}^{\mathrm{QNM}}$
            ``to be valid'' would be given by the intersection
            of the horizontal line of constant $N_{\mathrm{QNM}}$ and the $\epsilon$-contour line.
    }
  \label{f:uniform_convergence}
  \end{figure}

\begin{figure}[htp]
  \centering
  \subfloat[$\mathrm{log}_{10}\varepsilon$ at $\tau_o=0$]{
    \includegraphics[clip,width=\sizefigmedium\columnwidth]{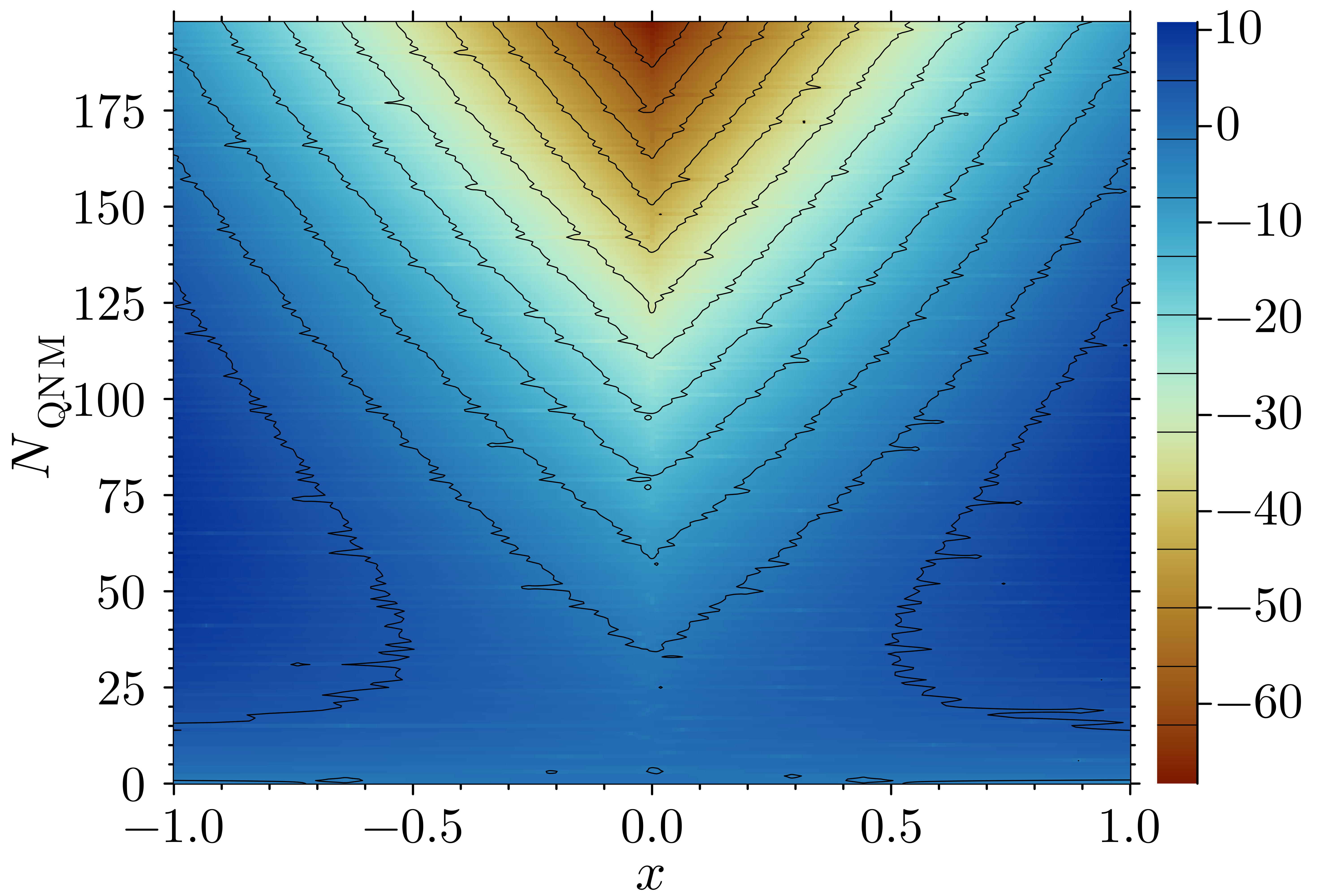}\label{f:spatial_convergence_tau_o_0}
  }
  \subfloat[$\mathrm{log}_{10}\varepsilon$ at $\tau_o=0.5$]{
    \includegraphics[clip,width=\sizefigmedium\columnwidth]{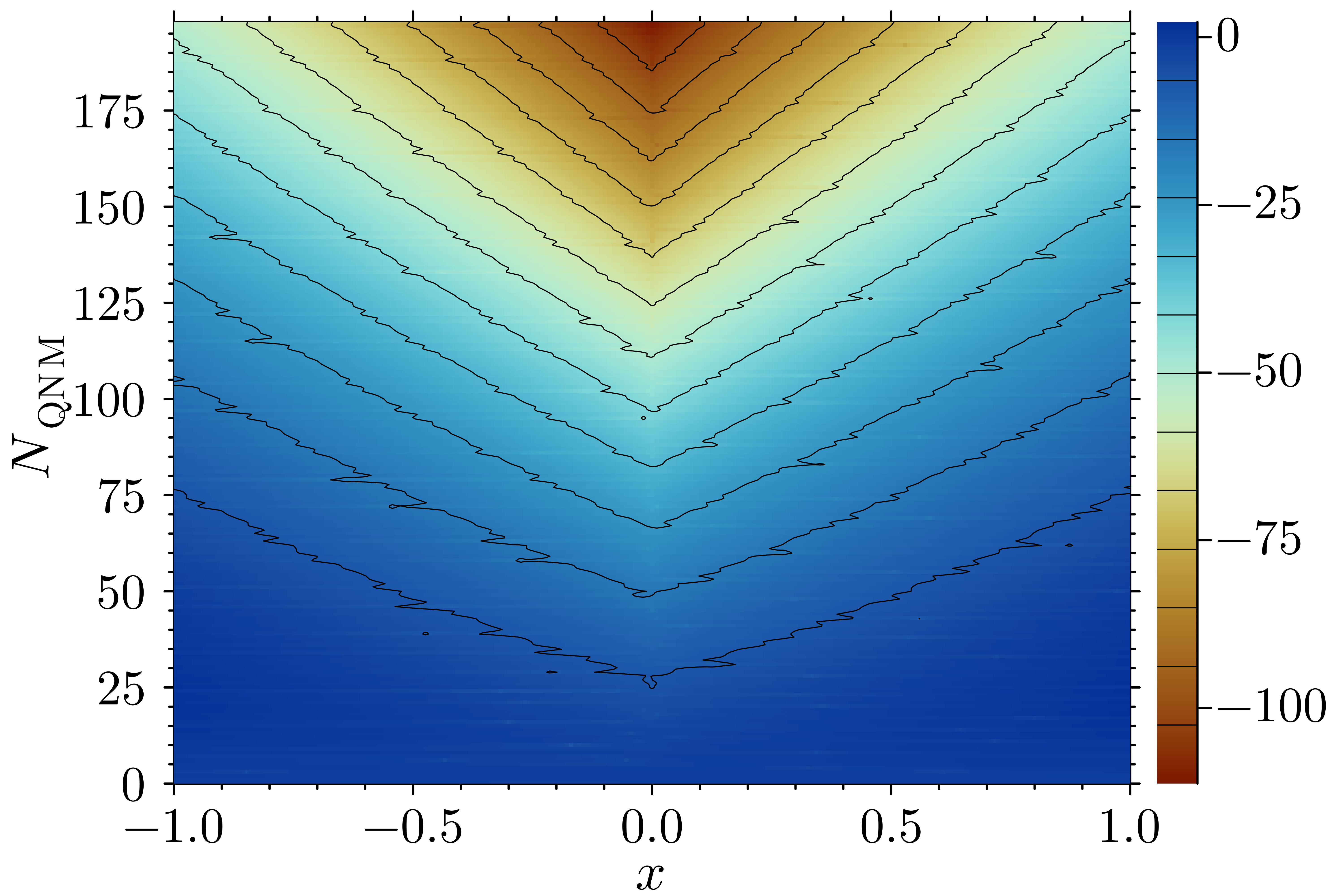}\label{f:spatial_convergence_tau_o_0.5}
  }

  \subfloat[$\mathrm{log}_{10}\varepsilon$ at $\tau_o=1$]{
    \includegraphics[clip,width=\sizefigmedium\columnwidth]{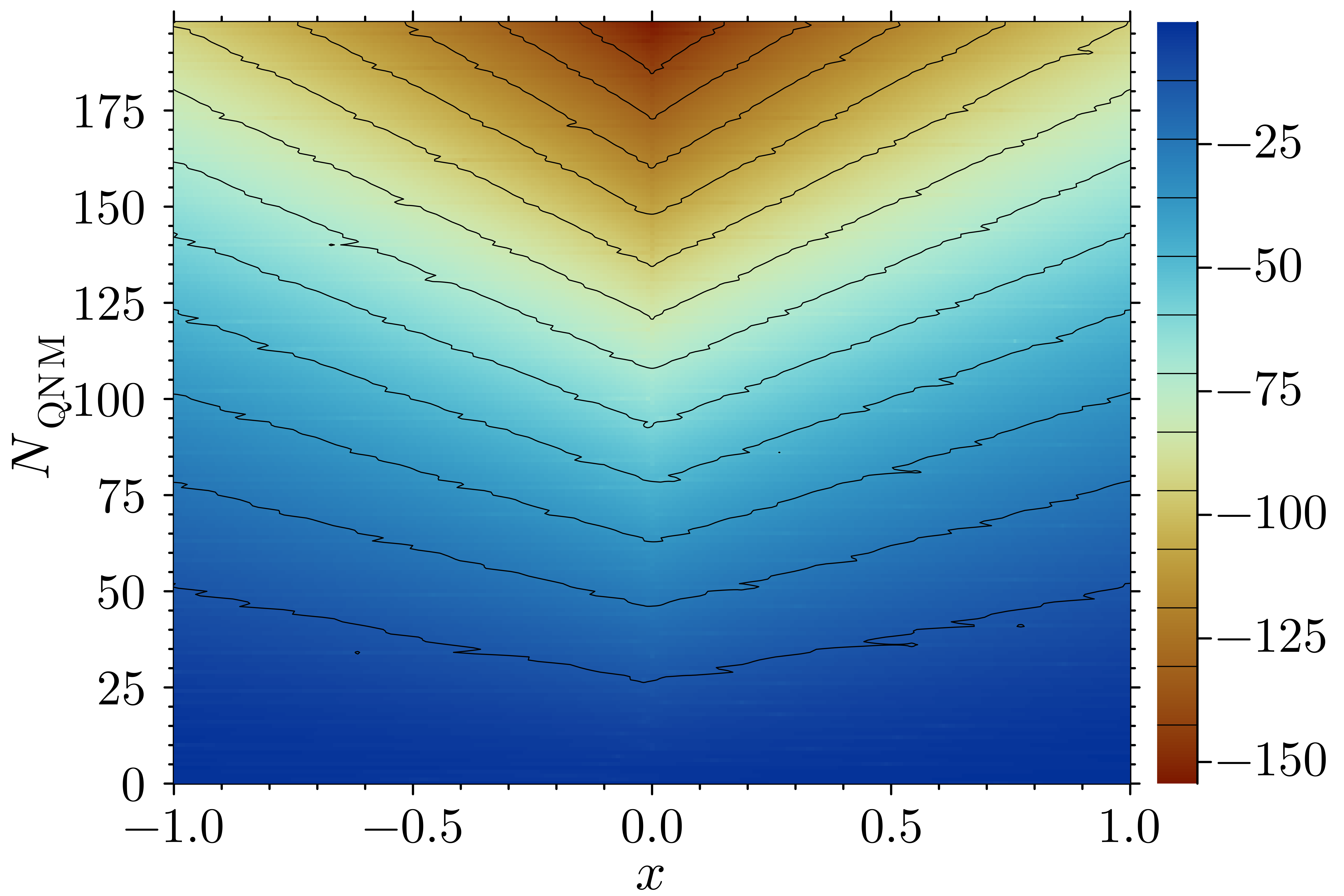}\label{f:spatial_convergence_tau_o_1}
  }
  \subfloat[$\mathrm{log}_{10}\varepsilon$ at $\tau_o=1.5$]{
    \includegraphics[clip,width=\sizefigmedium\columnwidth]{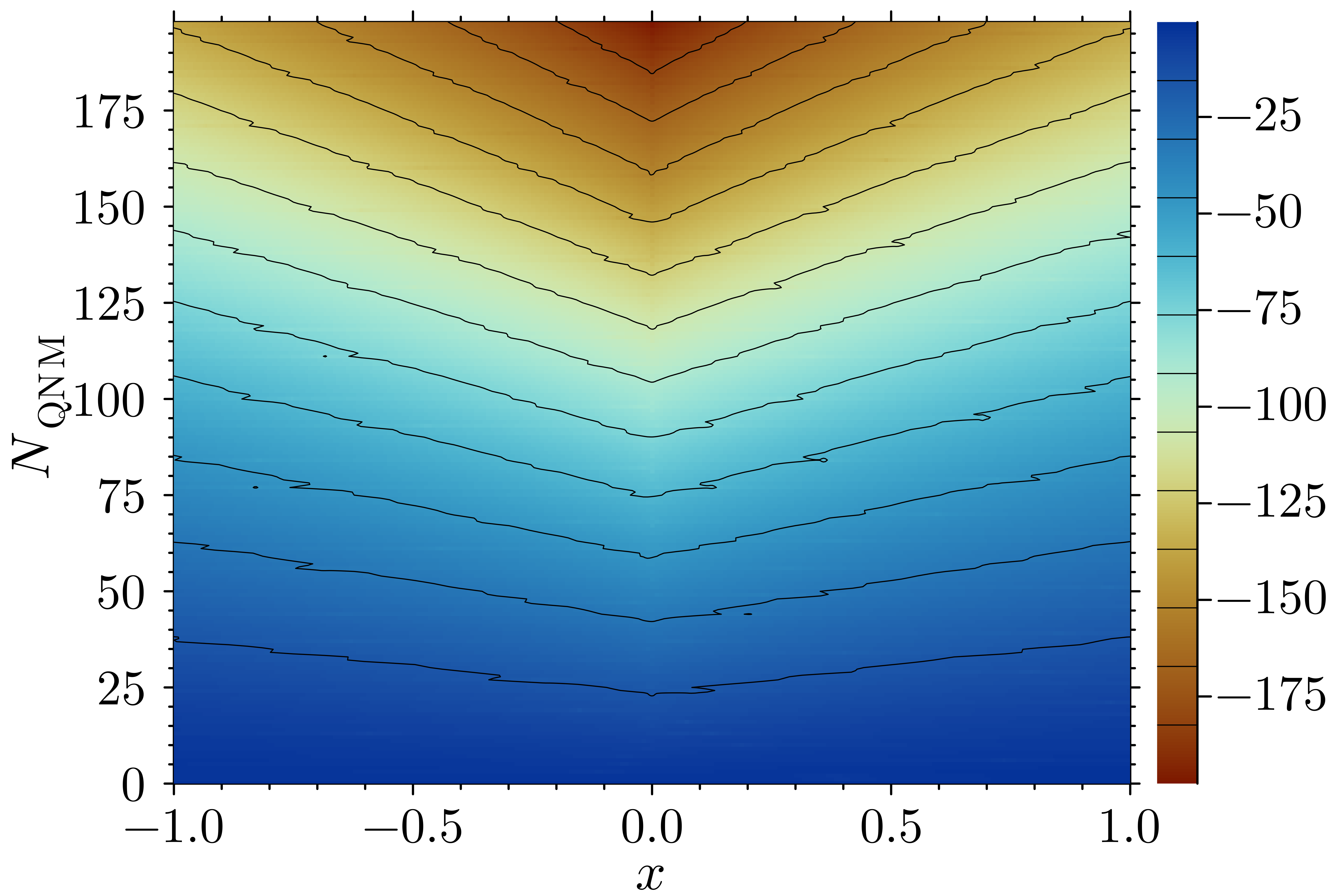}\label{f:spatial_convergence_tau_o_1.5}
  }
  \caption{Pointwise convergence in the P\"oschl-Teller case, for fixed time $\tau_o$. Illustration of the error $\varepsilon=\left|u(x,\tau_o) - u^{\mathrm{QNM}}(x,\tau_o)\right|$ as a function of $x$ at fixed times $\tau_o\in\{0,0.5,1,1.5\}$ in the P\"oschl-Teller case. The contour lines intersect the boundaries $x=\pm 1$ (the edges of the plot), which means for a given $\tau_o$ the error $\varepsilon$ can be made arbitrarily uniformly small over all the spatial domain $[-1,+1]$ by adding enough overtones, that is, we find uniform convergence in $x$. Panel \ref{f:spatial_convergence_tau_o_0} shows that adding more overtones at early times does not diminish the error systematically, in particular near the boundaries where the error shrinks only past a certain threshold $N_\mathrm{QNM}$.}
  \label{f:spatial_convergence}
\end{figure}

\subsubsection{Convergence of the asymptotic series: QNM expansion at fixed $\tau_o$}
\label{s:convergence-fixed_tau}
We consider now the convergence\footnote{In this subsection we focus on the
convergence of the series (\ref{e:u_truncated_series_fixed_tau}) in the norm
of the Hilbert space ${\cal H}$. Nevertheless, the
(uniform) pointwise convergence can also be considered, as in subsection
\ref{s:convergence_fixed_x}. Figure \ref{f:spatial_convergence} illustrates the pointwise
convergence in P\"oschl-Teller, that it is indeed uniform (since contour lines
cut the boundaries),
although pointwise convergence is faster at the center of the grid.} of the series at fixed time $\tau_o$, namely
the existence of the limit $\displaystyle \lim_{N_\mathrm{QNM}\to\infty}u_{\tau_o}^{\mathrm{QNM}}(x)$
in the norm $||\cdot||$, with $u_{\tau_o}^{\mathrm{QNM}}(x)$ given in
(\ref{e:u_truncated_series_fixed_tau}). As discussed above, in the selfadjoint case
(more generally, normal case) the spectral theorem guarantees the convergence of this series,
since the normal modes form Hilbert basis. In the non-normal case, the series is in
general non-convergent and only asymptotic in the sense of Eq. (\ref{e:u_Keldysh_v7}),
since the QNMs are not in general a basis of the functional space (see e.g.
\cite{Warnick:2022hnc}). This fact does not prevent however the convergence for a
particular initial data $u_0$. This can be useful if convergence can be shown
for an sub-ensemble of data of physical interest.

We proceed now to discuss the convergence
of the partial sums $u_{\tau_o}^{\mathrm{QNM}}(x)$, for the particular Gaussian initial
data $u_0$  in (\ref{e:initial_data}).
We first rewrite  Eq. (\ref{e:u_Keldysh_v7}) as
   \begin{eqnarray}
     \label{e:u_Keldysh_discussion_C}
     u_{\tau_o}(x) &=& u_{\tau_o}^{\mathrm{QNM}}(x) + E_{N_{\mathrm{QNM}}}(\tau_o;u_0) = \sum_{n=0}^{N_{\mathrm{QNM}}} a^\pm_n e^{i\omega^\pm_n\tau_o}v_n^\pm(x) +  E_{N_{\mathrm{QNM}}}(\tau_o;u_0) \nn \\
     \hbox{with } &&||E_{N_{\mathrm{QNM}}}(\tau_o;u_0)|| \leq C(N_{\mathrm{QNM}}, L) e^{-(\kappa N_{\mathrm{QNM}} + \mathrm{Im}(\omega_0))\tau_o}||u_0|| \ , 
   \end{eqnarray}
   where the expression
   $a_{_{N_{\mathrm{QNM}}}} = \kappa N_{\mathrm{QNM}} + \mathrm{Im}(\omega_0)$
   for $a_{_{N_{\mathrm{QNM}}}}$ in (\ref{e:u_Keldysh_v7})   is consistent with the BH QNM asymptotics
   (actually it is exact in the P\"oschl-Teller case
   on which we focus now), 
   inferred from Figs. \ref{spectres} and \ref{spectresconv} and leading to a BH Weyl law
   (see section \ref{a:Weyl_law}) that extends
   \cite{Jaramillo:2022zvf}.

   The obstacle to prove convergence of the series stems from the `a priori' lack of control
   on the growth of the constant  $C(N_{\mathrm{QNM}}, L)$ with $N_{\mathrm{QNM}}$.
   If the latter grows too strongly then no convergence can be shown. 
   On the contrary, if a `uniform bound' (i.e. not depending
   on $N_{\mathrm{QNM}}$) could be found for  $C(N_{\mathrm{QNM}}, L)$, then for every $\epsilon>0$
   one could easily use  the function  $E_{N_{\mathrm{QNM}}}(\tau_o;u_0)$ to construct the constant $K$ in
   the footnote
   \ref{footnote:convergence_norm} and convergence would follow. Actually it is enough to
   show that the growth of  $C(N_{\mathrm{QNM}}, L)$ with $N_{\mathrm{QNM}}$ is not faster than exponential.
   Interestingly, this is precisely the situation in our case.

   Figure \ref{f:C_N} presents the dependence of the constant $C(N_{\mathrm{QNM}}, L)$,
   estimated from the  particular case of the Gaussian initial data (details of the calculation
   will be presented in \cite{BesJarPoo24}). In particular, from the tangent of the curve
   when $N_{\mathrm{QNM}}\to 0$ one can estimate
   \bea
   \label{e:estimate_C}
   C(N_{\mathrm{QNM}}, L)\lesssim C \cdot e^{N_{\mathrm{QNM}}} \ ,
   \eea
   for some constant $C$. This permits to bound $||E_{N_{\mathrm{QNM}}}(\tau_o;u_0)||$
   in Eq. (\ref{e:u_Keldysh_discussion_C}) as
   \begin{eqnarray}
     \label{e:u_Keldysh_discussion_C_bound}
     ||E_{N_{\mathrm{QNM}}}(\tau_o;u_0)|| &\leq& C e^{N_{\mathrm{QNM}}}
     e^{-(\kappa N_{\mathrm{QNM}} + \mathrm{Im}(\omega_0))\tau_o}||u_0|| \ , \nn \\
      &=&C ||u_0|| e^{-\mathrm{Im}(\omega_0)\tau_o} e^{(1  -\kappa \tau_o)N_{\mathrm{QNM}}} \ .
   \end{eqnarray}
   The key remark is that if the coefficient $(1 -\kappa \tau_o)$ is negative, i.e.
   if $\kappa\tau_o>1$,
   then the error can be arbitrarily small for a sufficiently large $N_{\mathrm{QNM}}$.
   Specifically, given any $\epsilon>0$, set
   $\epsilon = C ||u_0|| e^{-\mathrm{Im}(\omega_0)\tau_o} e^{(1  -\kappa \tau_o)N_{\mathrm{QNM}}}$.
     If we consider now times $\tau_o$ satisfying
     \bea
     \label{e:tau_o_convergence_asymptotics}
   \tau_o > \frac{1}{\kappa} \ ,
   \eea
   then by taking
   \bea
   K_o =
   \frac{\left|\ln\left(\frac{\epsilon}{C||u_0|| e^{-\mathrm{Im}(\omega_0)\tau_o}}\right)\right|}{\kappa\tau_o - 1} \ ,
   \eea
   it holds that for any $N_{\mathrm{QNM}}>K_o$ we find $||E_{N_{\mathrm{QNM}}}(\tau_o;u_0)|| < \epsilon$
   and convergence follows.

   This is a suggestive result in the early dynamics context we are discussing
   in this subsection.
   Although we have considered a very special initial
   data, it indicates that convergence properties of the QNM series improve with time
   in the  decay timescale (\ref{e:tau_o_convergence_asymptotics}) naturally provided by the
   problem\footnote{The present discussion of the QNM expansion at fixed time $\tau_o$
   is akin to the discussion in the ``spatial fitting'' approach in section III.A of
   \cite{zhu2024challenges}. In particular, the requirement of convergence in the
   norm $||\cdot||$ offers a methodological guideline leading to the
   natural lower bound  (\ref{e:tau_o_convergence_asymptotics}), i.e.
   $\tau_o > 1/\kappa$, for the earliest time when the fitting should start. 
   Such a bound could prove useful in the setting of the overfitting problem
   in \cite{zhu2024challenges}.
   Of course $\kappa$
   is part of the unknown in the fitting, but it provides a first-principles
   insight into the problem. More generally, 
   finding an estimate  for $C(N_{\mathrm{QNM}}, L)$
   in the upper bound of the error function $E_{N_{\mathrm{QNM}}}(\tau_o;u_0)$,
   of the type (\ref{e:estimate_C}) but  independent of the initial data $u_o$,
   could help harnessing such an overfitting problem. This will be explored in \cite{BesJarPoo24}.}
   $\sim 1/\kappa$. If this conclusion can be extended to BH spacetimes, or
   if it is rather a peculiarity of the P\"oschl-Teller case (known for the good
   convergence properties of QNM expansions \cite{Beyer:1998nu}) or actually
   an artefact of the initial data, will be studied in \cite{BesJarPoo24}.

\begin{figure}[htp]
  \centering    
  \includegraphics[clip,width=\columnwidth]{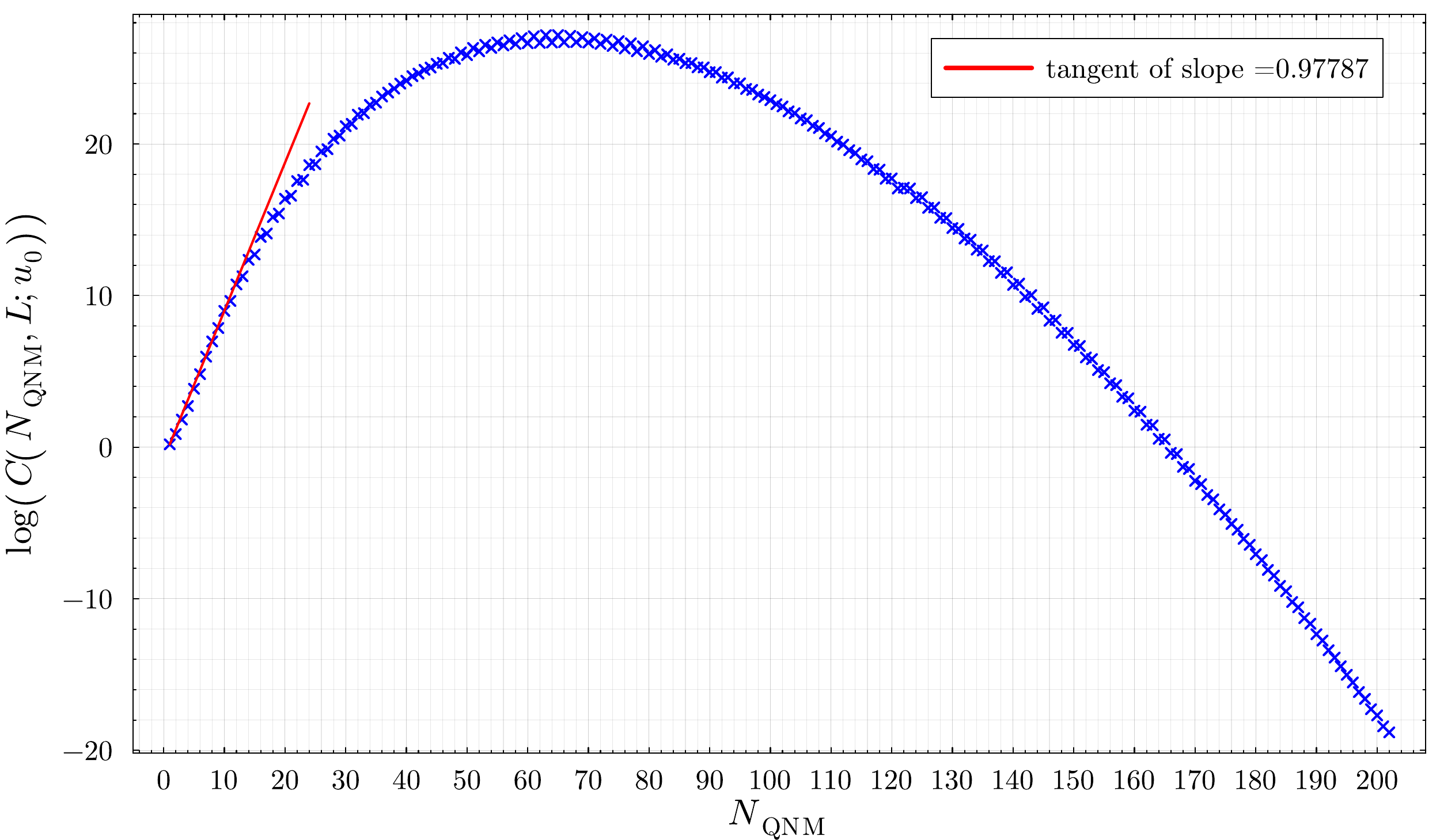}
  \caption{Coefficients $C(N_{\text{QNM}},L;u_0)$  controlling the error in the
    QNM series expansion at a fixed time $\tau_o$, 
    in the P\"oschl-Teller case and for the
    Gaussian initial data $u_0$ in (\ref{e:initial_data}). The slope of the tangent at the origin
    permits to establish the exponential bound $C(N_{\mathrm{QNM}}, L)\lesssim C \cdot e^{N_{\mathrm{QNM}}}$
    from which convergence for $\tau_o > 1/\kappa$ follows. 
} \label{f:C_N}
\end{figure}
 
    \subsection{Early dynamics: non-modal transient growths}
    \label{s:transients}
    In the discussion of the early-time dynamical behaviour of linear
    evolution problems driven by a non-selfadjoint (actually non-normal)
    infinitesimal operator it is natural to
    assess the presence of non-modal dynamical transients. Indeed,  in stark
    contrast with dynamics driven by `normal' operators, non-normal
    dynamics may present initial growth transients~\cite{TreTreRed93,trefethen2005spectra,Schmi07}.
    
    Specifically, in the setting of the dynamical equation (\ref{e:wave_eq_1storder_u_tau}),
    we consider the maximum growth function $G(\tau)$ associated with $L$ (namely
    the norm of the evolution operator $e^{i\tau L}$)
\bea
\label{e:G(t)}
G(\tau) = \sup_{u_0\neq0}\frac{||u(\tau)||}{||u_0||} = \sup_{u_0\neq0}\frac{||e^{i\tau L}u_0||}{||u_0||}
= ||e^{i\tau L}|| \  .
\eea
The function $G(\tau)$ provides the maximum possible amplification that can be attained,
at a given time $\tau$,  in the evolution $u(\tau)$ of a given some $u_0$ among all possible
initial data (note that
the `maximising' $u_0$ is generically different for distinct $\tau$'s, so the function $G(\tau)$
itself is not the evolution of any initial object). For normal time
generators $L$ with stable spectrum $\sigma(L)$ (namely $\sigma(L)$ in the upper
complex plane, in our convention)  it holds~\cite{trefethen2005spectra}
$G(\tau) = ||e^{i\tau L}||\leq 1, \forall \tau\geq 0$.
However, if $L$ is non-normal, although the late dynamics is still controlled by the
spectrum $\sigma(L)$ (modal behaviour), an initial (non-modal) {\em transient growth} characterised by
\bea
\label{e:transient_growth}
G(\tau)= ||e^{i\tau L}||>1 \ ,
\eea
can actually happen even if its spectrum  $\sigma(L)$ is stable.  The presence of such
an initial transient amplification becomes a `smoking gun' of non-normal dynamics.
`Modal analysis' focused on the spectrum $\sigma(L)$ becomes inadequate to discuss such
transient dynamics and one needs to resort to the full resolvent $R_L(L)$. Notions such as
the $\epsilon$-pseudospectrum $\sigma_\epsilon(L)$ (see section \ref{s:Hp-pseudospectrum}), the numerical range $W(L)$ and
the numerical abscissa $\omega(L)$ or
the Kreiss constant ${\cal K}(L)$ and the Kreiss matrix theorem, the growth function $G(\tau)$
(cf. further concepts and details in \cite{trefethen2005spectra,Schmi07}, also \cite{TreTreRed93})
provide a set of tools for the so-called 'non-modal analysis'. 
QNM expansions discussed here  provide (in the diagonalisable case),
a neat account of transient growths in terms of the non-orthogonality of
QNM functions. Such a mechanism is absent in the normal case due to the spectral theorem
(normal operators are unitarily diagonalisable, so normal modes are orthogonal) but the loss of
the spectral theorem in the non-normal case permits a transient constructive interference
phenomenon between QNMs at early times.

In the gravitational setting, an elementary implementation of some of such non-modal analysis
tools\footnote{\label{footnote:no_transients}The analysis in \cite{Jaramillo:2022kuv} provided
the values of the numerical abscissa and the
Kreiss constant for the class of hyperboloidal evolutions (\ref{e:wave_eq_1storder_u_tau}) with $L$ given
in Eqs.(\ref{e:L_L_1_L_2}) and (\ref{e:L_1-L_2}). This was done for arbitrary
potentials but  only for the energy norm $||\cdot||_E$ case.
The respective results were a vanishing numerical abscissa, $\omega(L)=0$,
and therefore a trivial Kreiss constant ${\cal K}(L)=1$.
This result implies that no energy transient growths can appear in this class
of problems, since $G(\tau) = ||e^{i\tau L}||_E \leq e^{i\tau \omega(L)} = 1, \ \forall \tau\geq 0$.}
was presented in \cite{Jaramillo:2022kuv}, 
applied in particular to an attempt to address binary black hole transients and, subsequently
in \cite{BoyCarDes22}, to assess transient growths in ultracompact objects, in both cases
with negative results. The pioneer work in gravity on the time-domain non-modal analysis
for the study of transients through the analysis of
QNM non-orthogonality was then presented in
\cite{Carballo_2024}, whereas the application to transients of the time-domain counterpart
of the generalised eigenvalue problem was first presented in \cite{Chen_2024}.
A recent remarkable application to the study of transient growths
in a superradiance setting is discussed in \cite{Carballo:2025ajx}.

As it is apparent in expression (\ref{e:G(t)}), the assessment of transient growths
depends on the choice of norm.  The energy norm is a natural candidate but
no transient growths can happen in this norm (cf. footnote \ref{footnote:no_transients}
and the discussion below, in particular Fig. \ref{f:transient_energy}).
On the other hand, from the discussion in
section~\ref{s:regularity_excitation_coefficients}
it is natural to consider the presence of transient growths in the context of 
$H^p$-norms, in order to assess  regularity aspects in the early evolution the solution.
We present here some first results on this problem,
in the P\"oschl-Teller case (a detailed
analysis, involving also proper black hole spacetimes, will be presented in \cite{BesBizJar25}).

\begin{figure}[htp]
    \centering
    \subfloat[Energy norm]{
      \includegraphics[clip,width=\sizefigmediumsmall\columnwidth]{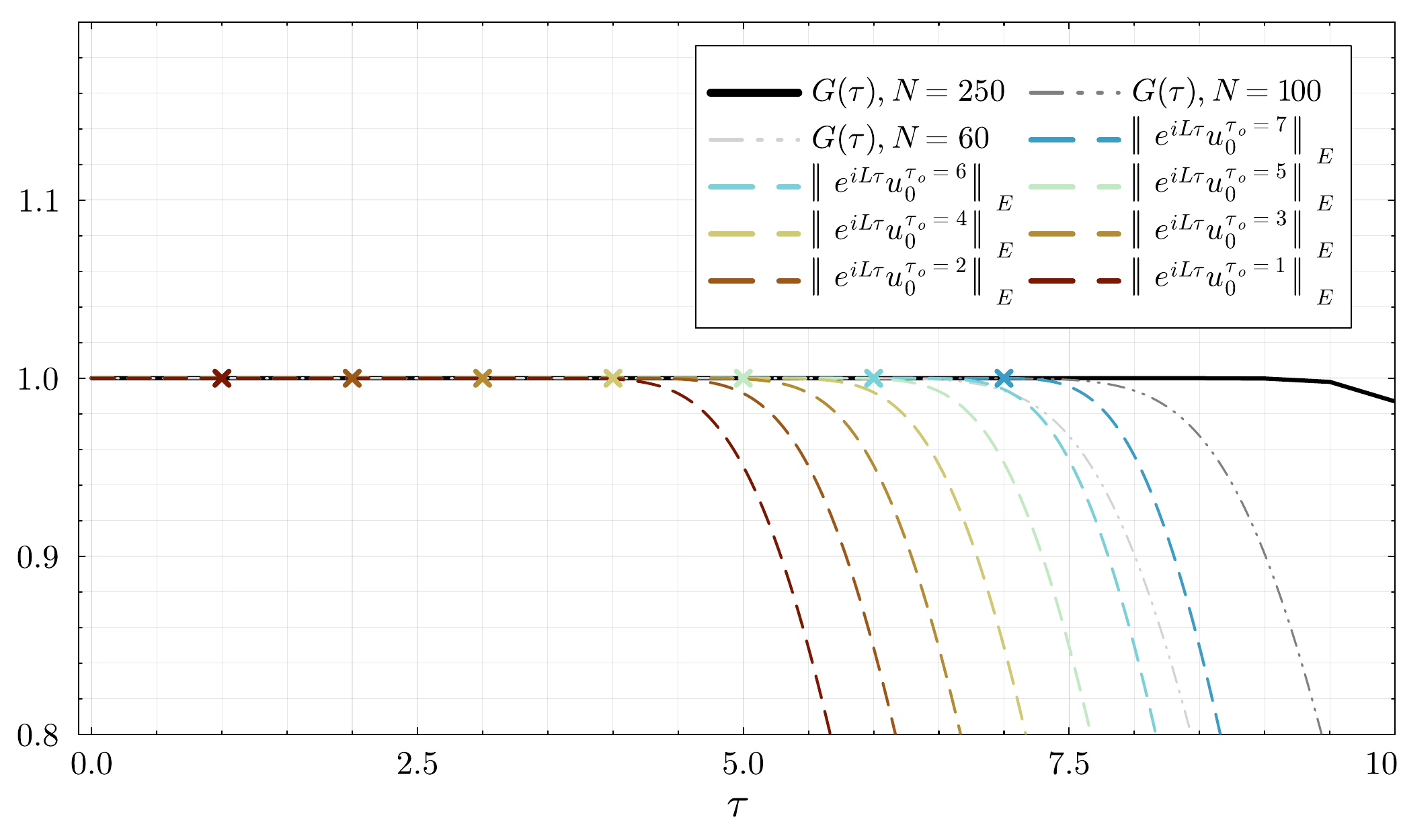}\label{f:transient_energy}
    }    
    \subfloat[Sobolev $H^5$-norm]{
      \includegraphics[clip,width=\sizefigmediumsmall\columnwidth]{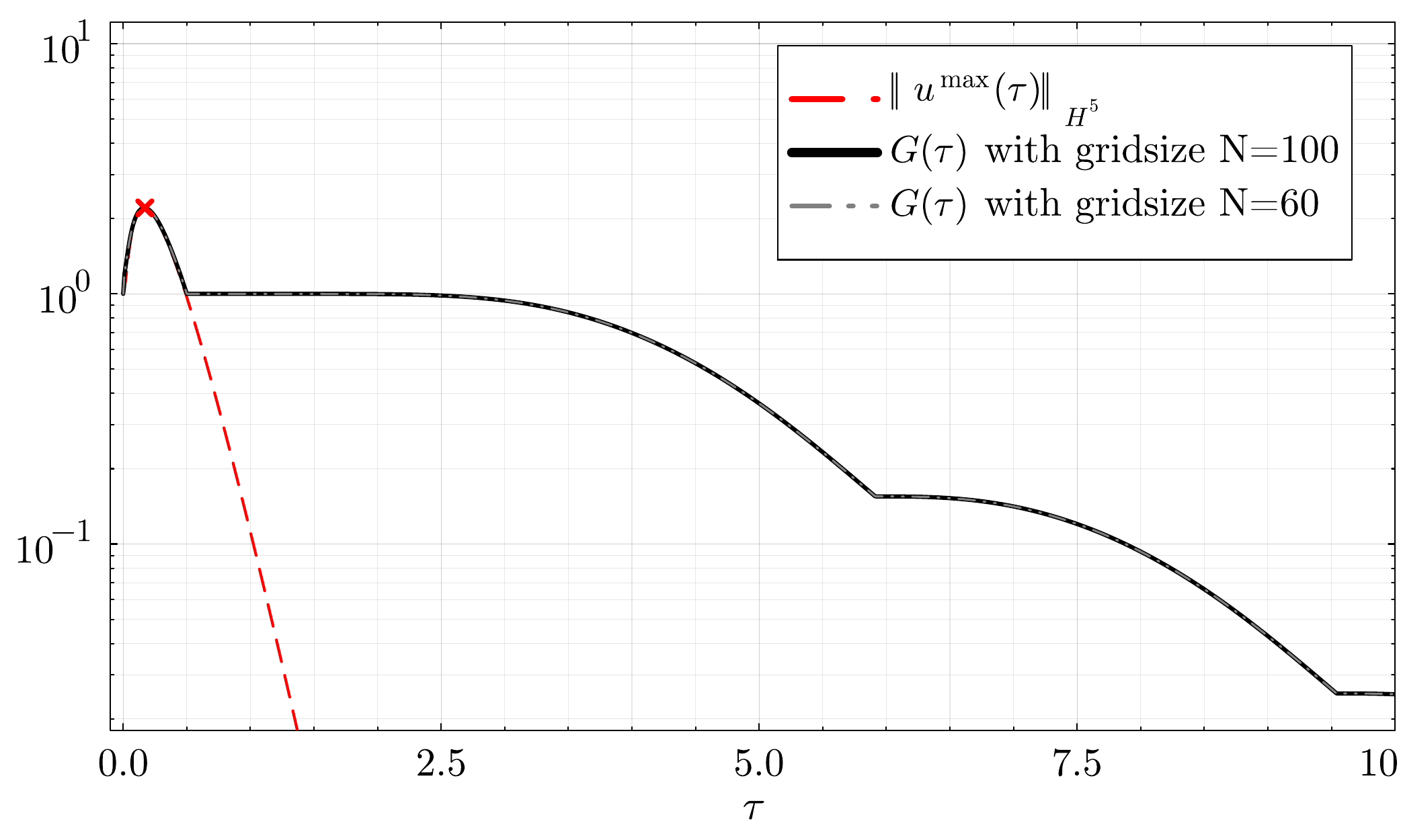}\label{f:Sobolev_H5}
    }

 \subfloat[Sobolev $H^{10}$-norm]{
      \includegraphics[clip,width=\sizefigmediumsmall\columnwidth]{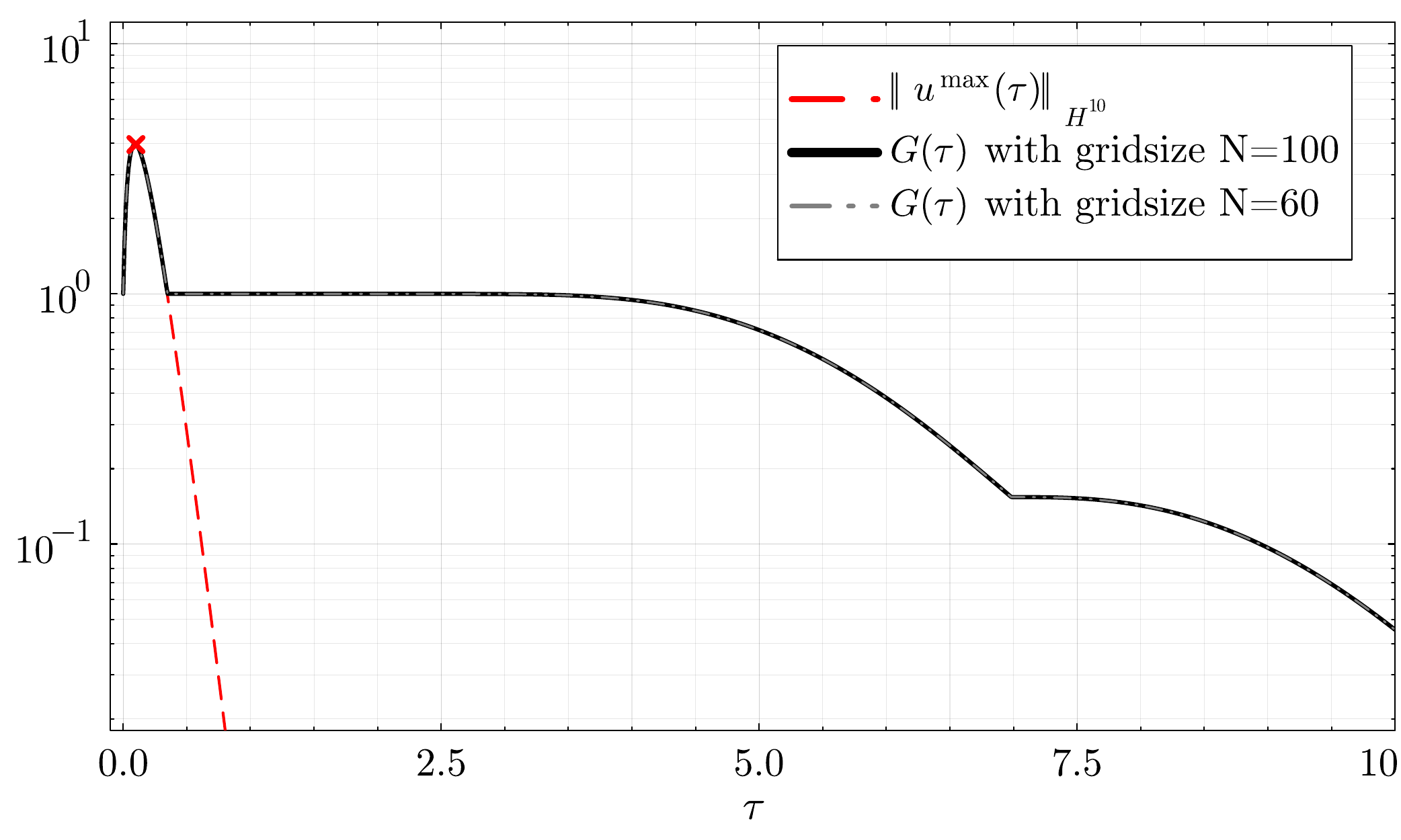}\label{f:Sobolev_H10}
 } 
    \subfloat[Sobolev $H^{15}$-norm]{
      \includegraphics[clip,width=\sizefigmediumsmall\columnwidth]{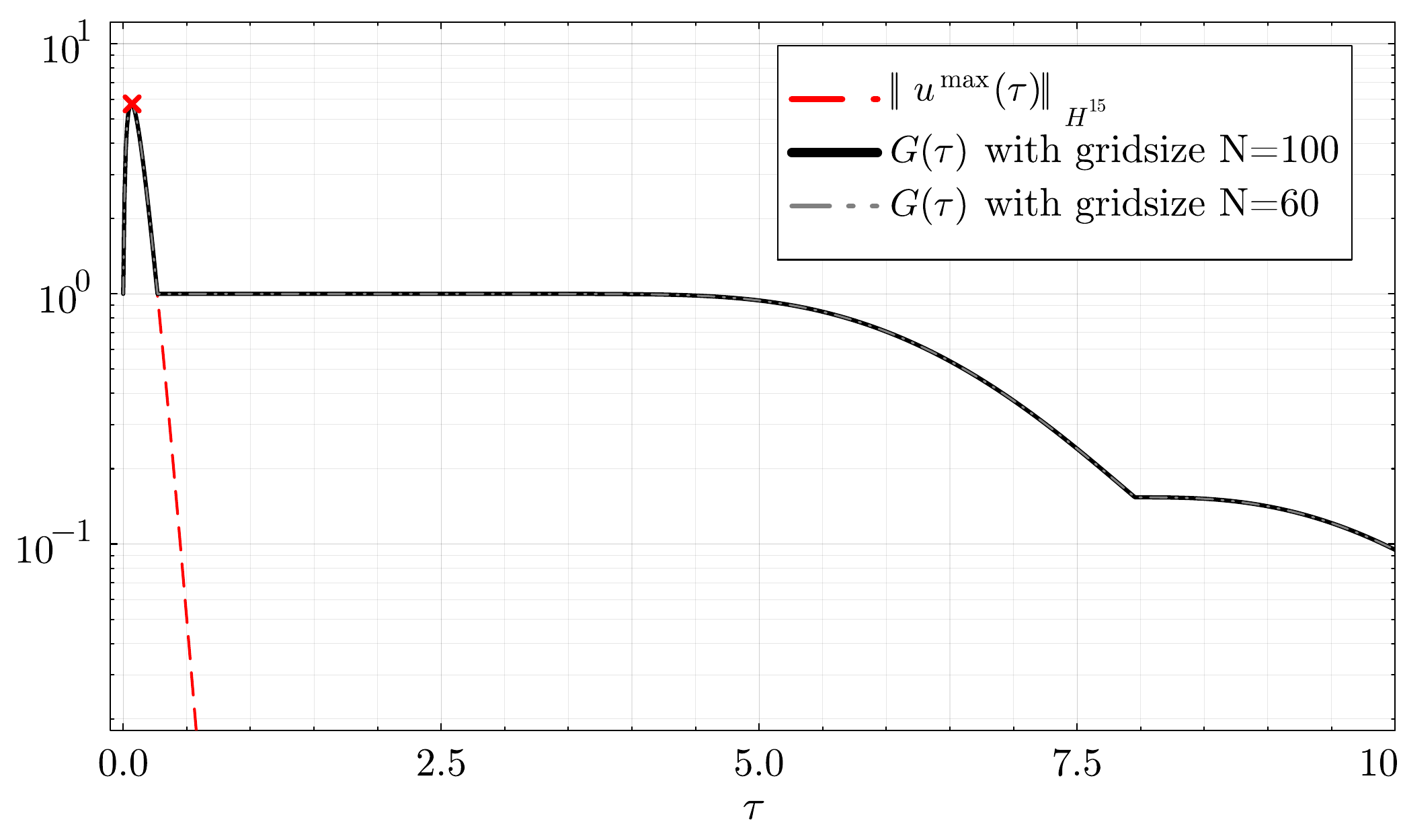}\label{f:Sobolev_H15}
    }

    \subfloat[Sobolev $H^{20}$-norm]{
      \includegraphics[clip,width=\sizefigmediumsmall\columnwidth]{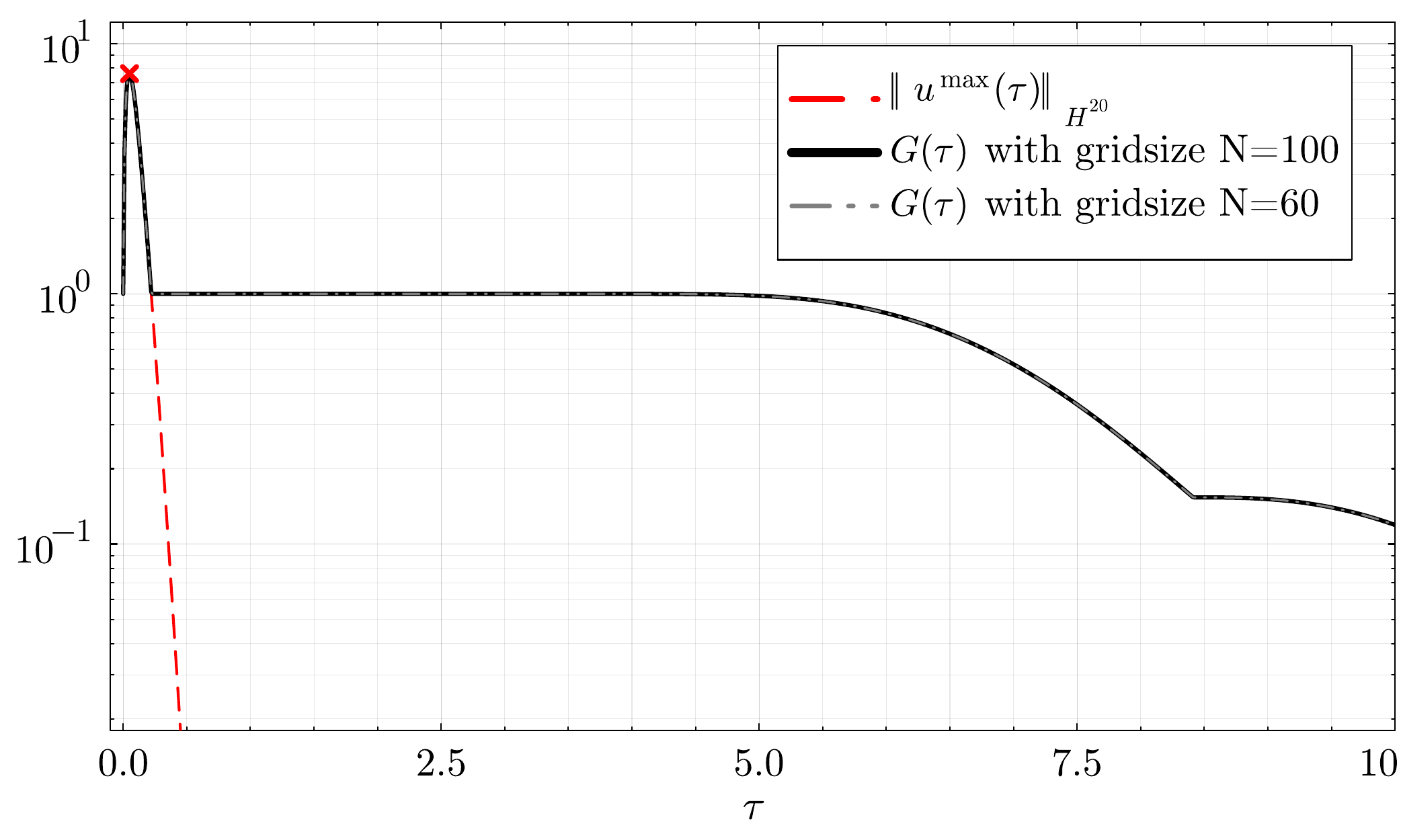}\label{f:Sobolev_H20}
    }
        \subfloat[Sobolev $H^{25}$-norm]{
      \includegraphics[clip,width=\sizefigmediumsmall\columnwidth]{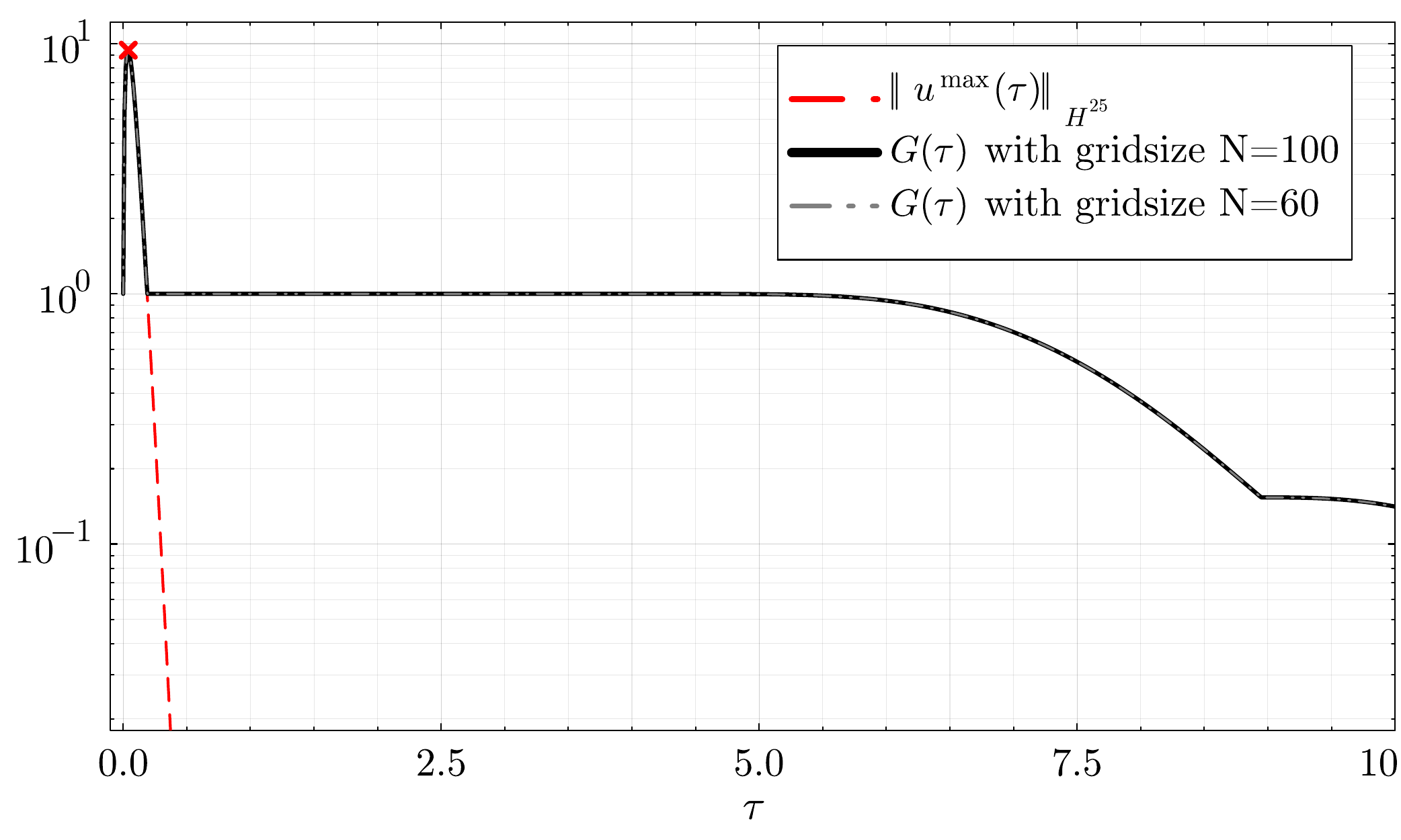}\label{f:Sobolev_H25}
    }
    \caption{Growth functions $G(\tau)$ and evolutions of `optimal perturbations/excitations' $\norm{u^\text{max}(\tau)}_{H^p}=\norm{e^{iL\tau}u_0^\text{max}}_{H^p}$ for the energy norm and five instances of  Sobolev $H^p$ norms. Panel \ref{f:transient_energy} illustrates there is no growth in the energy norm, we have instead a plateau whose length is proportional to $\log(N)$, none of the curves on this panel converge. Panels \ref{f:Sobolev_H5}, \ref{f:Sobolev_H10}, \ref{f:Sobolev_H15}, \ref{f:Sobolev_H20} and \ref{f:Sobolev_H25} display transient growths in the form of an initial (maximal) peak whose amplitude increases with $p$ and the time $\tau_{\text{max},p}$ when this peak is achieved gets closer to zero.}
    \label{f:transients}
  \end{figure}

\begin{figure}[htp]
    \centering
    \subfloat[$u_0^{\tau_o}(x)$ as a function of $x$ (energy norm)]{
      \includegraphics[clip,width=\columnwidth]{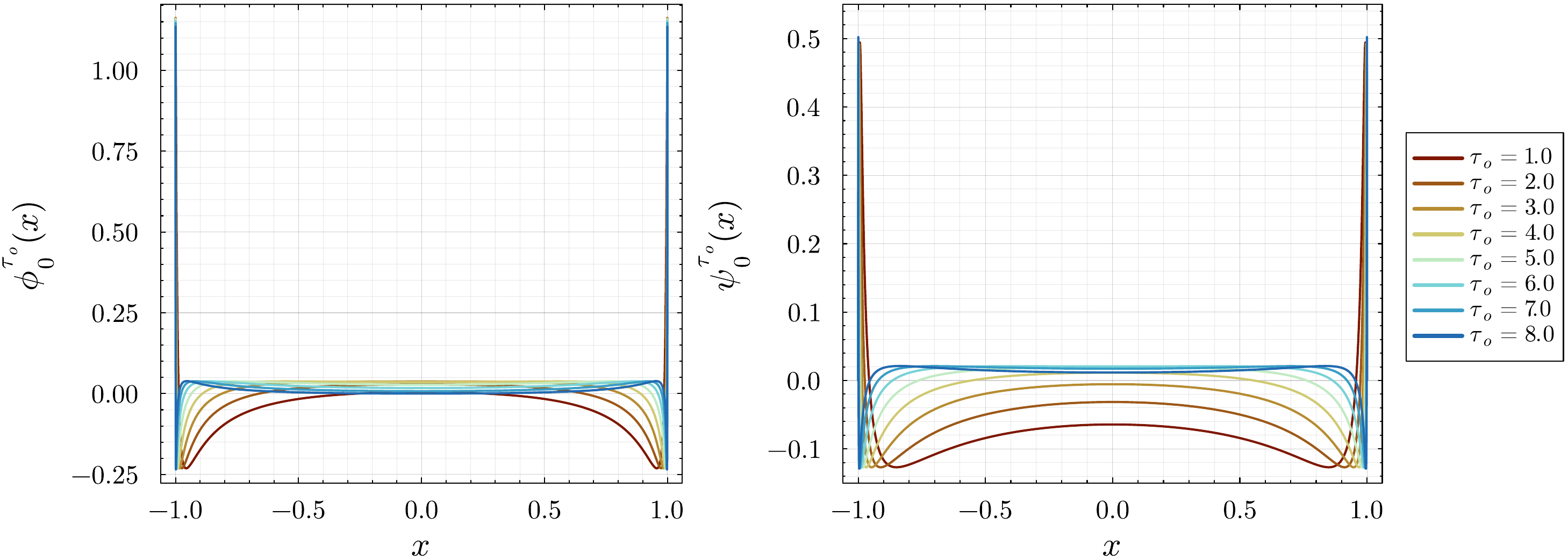}\label{f:u0_E_optimal_x}
    }    

    \subfloat[$u_0^{\tau_o}(x)$ as a function of $\arccos(x)/\pi$ (energy norm)]{
      \includegraphics[clip,width=\columnwidth]{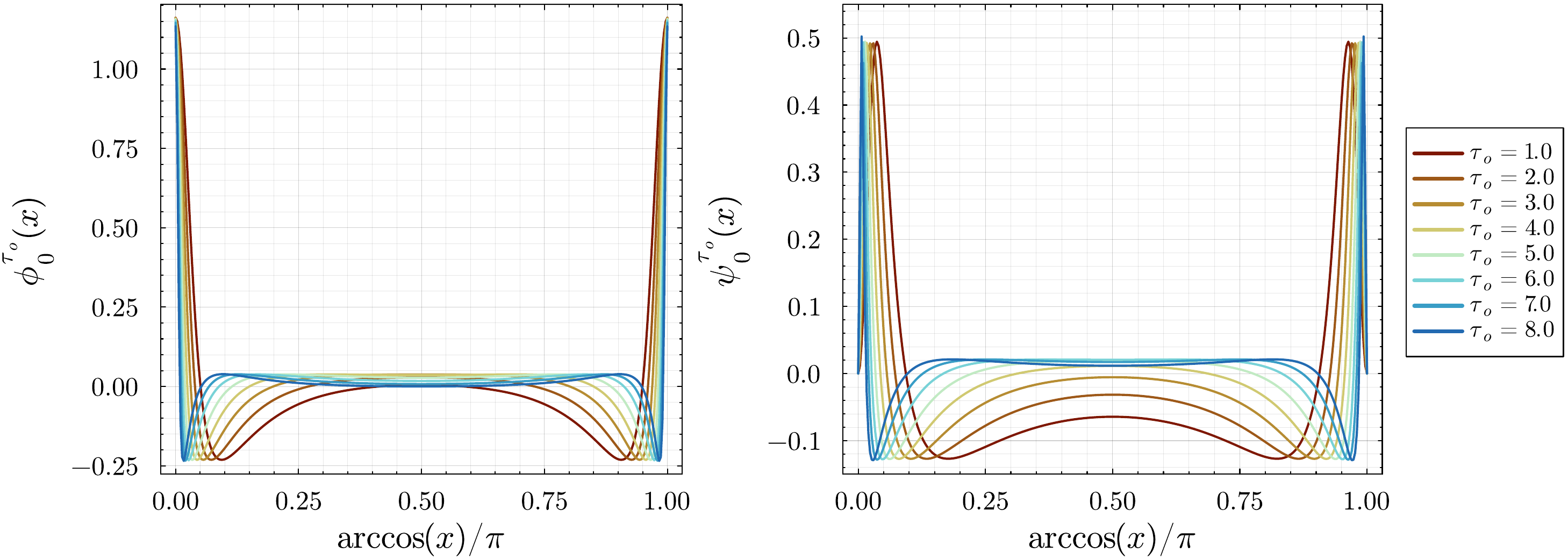}\label{f:u0_E_optimal_acosx}
    }

    \subfloat[$u_0^\mathrm{max}(x)$ as a function of $x$ for different $H^p$ norms]{
      \includegraphics[clip,width=\columnwidth]{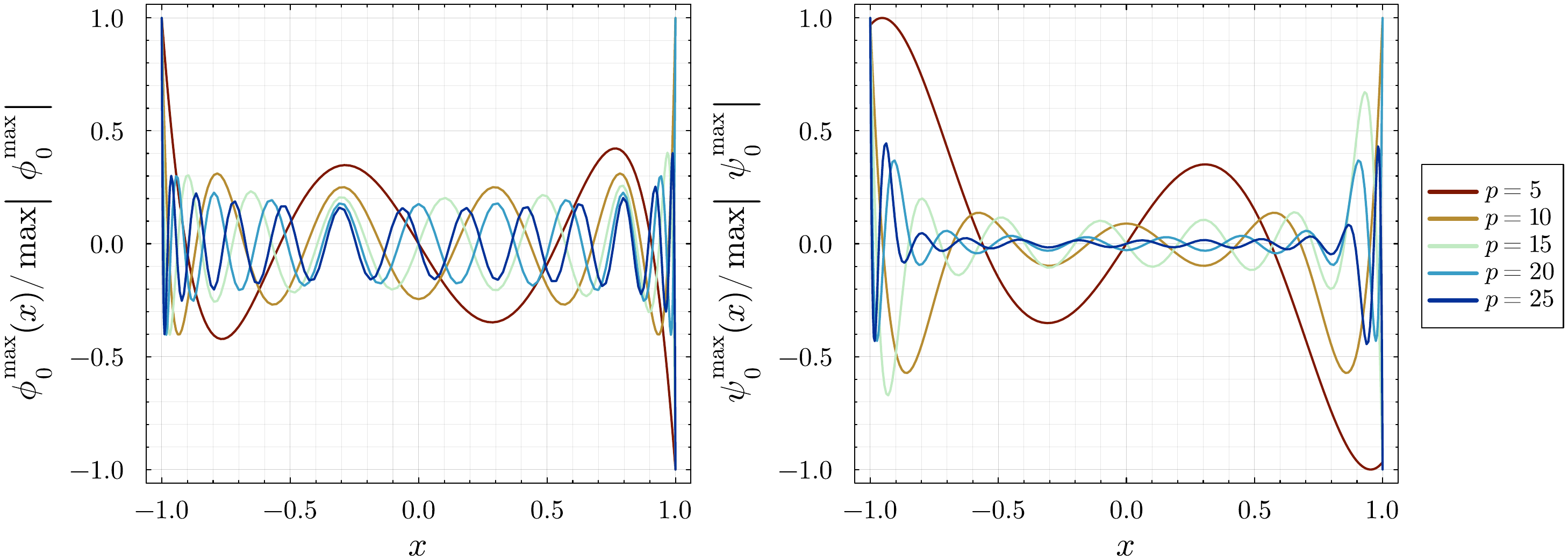}\label{f:u0_max_optimal_Hp}
    }
    \caption{We plot the optimal initial conditions for two different cases: \textit{i) the energy norm,} $u_0^{\tau_o}=(\phi_0^{\tau_o},\psi_0^{\tau_o})^t$ is plotted at different times $\tau_o$ that realise the plateau displayed in \ref{f:transient_energy}, meaning that $\norm{e^{iL\tau}u_0^{\tau_o}}_E= 1$ for $0\leq\tau\leq\tau_o$. Panel \ref{f:u0_E_optimal_x} is a view of the optimal initial conditions as a function of $x$ and \ref{f:u0_E_optimal_acosx} is a different view of the initial conditions as a function of $\arccos(x)/\pi$, this  permits to observe the behaviour of the functions at the boundaries. \textit{ii) Sobolev $H^p$ norms}, we plot the optimal initial conditions $u_0^\mathrm{max}=(\phi_0^\mathrm{max},\psi_0^\mathrm{max})^t$ that realise the peaks observed in Panels \ref{f:Sobolev_H5} to \ref{f:Sobolev_H25}, the norm of the time evolutions $\norm{e^{iL\tau}u_0^\mathrm{max}}_{H^p}$ yield the red dotted lines on these 5 panels. The grid size is $N=300$ for panels \ref{f:u0_E_optimal_x} and \ref{f:u0_E_optimal_acosx}, whereas $N=150$ in panel \ref{f:u0_max_optimal_Hp}. 
    }
    \label{f:u0_optimal}
  \end{figure}

Fig. \ref{f:transients} presents the growth function $G(\tau)$, as well as the
evolution of the norm for the `optimal perturbation' (or `optimal excitation')
$u^{\mathrm{max}}(\tau,x)=e^{i\tau L}u^{\mathrm{max}}_0(x)$ introduced in \cite{Carballo_2024}
(see also \cite{farrell1988optimal}).
For `vector norms' associated with a scalar product (as it is the case here),
$u^{\mathrm{max}}_0(x)$ can be found by solving a spectral problem. Specifically, the
corresponding induced `operator norm' is given by the square root of the (generalised) spectral radius,
namely the highest of the `generalised' singular values (with the adjoint defined by the given
scalar product, cf. e.g. \cite{Jaramillo:2020tuu}). The maximiser $u_0(x)$ at each $\tau$
in expression (\ref{e:G(t)}) is therefore obtained as
the eigenfunction of the corresponding highest generalised singular value and, in particular, the
`optimal perturbation' $u^{\mathrm{max}}_0(x)$ is the eigenfunction  at the $\tau_{\mathrm{max}}$
maximising $G(\tau)$.

In the following we discuss $H^p$-transient growths, dwelling in the P\"oschl-Teller case,
after reviewing the energy norm case:
\begin{itemize}
\item[i)] {\em Energy norm}. The growth function is constant with $G(\tau)=1$, therefore
  no transient growths appear in this case. This is consistent with the analysis
  in \cite{Jaramillo:2022kuv} (see footnote~\ref{footnote:no_transients})
  and Fig. \ref{f:transient_energy} recovers
  the curve first presented in \cite{Carballo_2024}. Specifically, $G(\tau)=1$ in the
  continuum limit $N\to\infty$ of the grid size (for finite grid resolution,
  the curve $G(\tau)$ decays at a time $\tau_{\mathrm{decay}}\sim\ln(N)$, consistently with the behaviour
  found \cite{Carballo_2024} $\tau_{\mathrm{decay}}\sim\ln(M)$ when considering a finite number $M$ of QNMs).

  Such a $G(\tau)=1$ behaviour, characteristic of a conservative system, is not in conflict with
  the dissipative nature of the dynamics, actually it is feature confirming that dissipation occur
  through the boundaries, and not in the bulk.
  Indeed, considering an arbitrary late $\tau_o$, an `optimal initial data' $u^{\tau_o}_0$ can be found
  (by enforcing sufficiently high grid resolution $N$)
  such that $||e^{i\tau L}u^{\tau_o}_0||_{_E}=1$  until $\tau\sim\tau_o$,
  when it starts to decay. This behaviour corresponds to optimal initial data
  peaked close to the boundaries whose evolution travels (in a conformal Penrose diagram
  picture) `in parallel' to such boundaries, so energy is conserved
  and  therefore $G(\tau)=1$ all the way through until they hit the opposite boundary around
  $\tau\sim\tau_o$, when they dissipate away through that second boundary.
  Optimal initial data $u^{\tau_o}_0$ get more and more peaked towards the boundary
  as $\tau_o$ grows (as illustrated in Panels \ref{f:u0_E_optimal_x} and \ref{f:u0_E_optimal_acosx} of Fig. \ref{f:u0_optimal}), their evolution therefore lasting longer before meeting the opposite boundary
  and eventually dissipating away. The optimal initial data in the
  limit $\tau_o\to\infty$
  corresponds to a distributional `Dirac-delta-like' pulse (cf. \cite{Carballo_2024,BesBizJar25})
  concentrated and propagating along the null boundary, for which no energy loss ever occurs, therefore
  leading to $G(\tau)=1, \forall \tau$.

  Such an energy conservation picture, with dissipation through the boundaries, is confirmed by setting
  initial data $u_0$ with arbitrary profiles (not necessarily optimal ones) centred at $x_o$
  close to a given boundary ($|x_o|\lesssim 1$ in in P\"oschl-Teller) and enforced to move along the
  characteristic (null) direction defined by that nearby boundary. Integrating along the
  null geodesics characterised by $\displaystyle \frac{d\tau}{dx} = \frac{\gamma\pm 1}{p}$
  (cf. notation in appendix \ref{a:Hyperboloidal approach}) one readily gets
  the estimated time  to hit the opposite
  boundary. In the in P\"oschl-Teller case this results in
  \bea
  \tau\sim\ln(2/(1 - |x_o|)) \ ,
  \eea
  that perfectly coincides with the numerically obtained decay time
  $\tau_{\mathrm{decay}}$, before which  $||e^{i\tau L}u_0||_{_E}=1$ holds
  \footnote{\label{footnote:transient_energy}It is illustrative to compare
  the analysis of this dissipation through the boundaries
  both from a purely `non-normal dynamics' and from a `geometric' perspective.
  Concerning the former, it holds for any scalar product (writing $A=iL$ and $u(\tau)=e^{\tau A}u_0$)
  \bea
  \label{e:ddt_normu}
  \frac{d}{d\tau}(||u(\tau)||) = \frac{1}{||u(\tau)||}\left\langle u(\tau), \frac{1}{2}(A + A^\dagger)u(\tau)\right\rangle \ . 
  \eea
  In the particular case of the energy scalar product in Eq. (\ref{eq_eff_en_inner_product}),
  the operator in the right-hand-side writes (cf. Eq. (31) in \cite{Jaramillo:2022kuv})
   \bea
  \label{e:A+Adagger}
 \frac{1}{2}(A + A^\dagger) 
  =
  \left(
  \begin{array}{c|c} 
    0 & 0 \\ \hline 0 &-\displaystyle \frac{\gamma}{w} \bigg(
\delta(x-a)-\delta(x-b)\bigg)
  \end{array}
  \right) \ ,
  \eea
  from which it follows ($\gamma(b)<0$ and $\gamma(a)>0$ in the hyperboloidal framework
  implementing outgoing boundary conditions)
  \bea
  \label{e:energy_flux}
  \frac{d}{d\tau}\left(\frac{||e^{i\tau L}u_0||_{_E}}{||u_0||_{_E}}\right) = \frac{1}{2}\frac{1}{||u_0||_{_E}||u(\tau)||_{_E}}
  \left(\gamma(b)|\psi(b)|^2 - \gamma(a)|\psi(a)|^2\right) \ .
  \eea
  This rather cryptic result, namely $||e^{i\tau L}u_0||_{_E}$ only changes (to decrease)
  when either $\psi(a)\neq 0$ or $\psi(b)\neq 0$,
  becomes transparent from a geometric perspective when writing the (boundary) energy flux 
  in the hyperboloidal approach (cf. Eqs. (95) and (97) in  \cite{Gasperin:2021kfv})
  \bea
  \frac{dE}{d\tau} = \gamma(b)|\partial_\tau\phi(b)|^2 - \gamma(a)|\partial_\tau\phi(a)|^2 \ ,  
  \eea
  with $||u(\tau)||_{_E}=E^{\frac{1}{2}}$ and $\psi=\partial_\tau\phi$. In summary, Fig.~\ref{f:transient_energy}
  reflects energy conservation till it dissipates through the boundaries.}.
  This discussion simply paraphrases the results found and discussed in detail
  in \cite{Carballo_2024}\footnote{In reference \cite{Carballo_2024} the
  initial plateau $||e^{i\tau L}u_0||_{_E}=1$ is referred to as a `transient'. We interpret this terminology as
  a manner of emphasising the `non-modal' character (i.e. non-fully characterised by the frequencies in the
  spectrum of the specific infinitesimal generator $L$ in the hyperboloidal approach) of the evolution of $u(\tau)$,
  in contradistinction with its late properly `modal' behaviour. This terminology has the virtue
  of stressing the non-normal character of the dynamics, but perhaps at the risk of being misleading
  regarding proper `transient growths' with $G(\tau)>1$. On the other hand, since the latter cannot occur
  when using the energy norm (cf. footnote \ref{footnote:no_transients}), the discussion in the otherwise
  remarkable work \cite{Chen_2024} must contain a flaw
  in the assessment of the numerical abscissa (that must vanish) and the boundary flux.
  In this sense, the first reference presenting a `transient growth' in gravity would be
  \cite{Carballo:2025ajx}.}.

\item[ii)] {\em Sobolev $H^p$-norm}. In contrast with the energy norm $||\cdot||_E$, the time
  derivative of the Sobolev norm $||\cdot||_{H^p}$  contains `bulk terms', and not only a boundary flux
  as in Eq. (\ref{e:energy_flux}) (see discussion in footnote \ref{footnote:transient_energy}).
  Or, equivalently, the operator $\frac{1}{2}(A + A^\dagger)$ in Eq. (\ref{e:ddt_normu}) is not
  determined in terms of distributional operators at the boundaries. As a consequence of this, actual transient growths
  can and do happen. Figs. \ref{f:Sobolev_H5}-\ref{f:Sobolev_H25} show the functions $G(\tau)$ for a
  series of Sobolev norms, as introduced in Eq. (\ref{e:Hp_norm}), showing indeed transient growths as expected.
  Actually, a first peak appears, followed by a series of secondary structures that do not
  show transient growth behaviour (namely $\displaystyle \frac{d G(\tau)}{d\tau}\leq0$).

  Focusing on the first transient peak,  two features are apparent in Fig\ref{f:transients}:
  a) the height $G_{\mathrm{max}}=G(\tau_{\mathrm{max}})$
  of the transient's peak
  increases with $p$,  and b) the timescale $\tau_{\mathrm{max}}$ for the transient peak decreases with $p$.
  Figure \ref{f:transients2} makes this quantitative, showing 
  \bea
  \label{e:peak_p_dependence}
  \tau_{\mathrm{max}} \sim \frac{1}{p} \qquad , \qquad  G_{\mathrm{max}}  \sim p \ ,
  \eea
  for sufficiently high $p$.
  In order to assess such a behaviour, we consider the optimal data $u^\mathrm{max}_0(x)$ corresponding
  to the $\tau_\mathrm{max}$ at the peak. Panel \ref{f:u0_max_optimal_Hp} of Fig. \ref{f:u0_optimal} illustrates the functions $u_0^\mathrm{max}(x)=(\phi_0^\mathrm{max}(x),\psi_0^\mathrm{max}(x))^t$ for different $H^p$ norms. On the one hand, a decomposition onto the basis of Chebyshev polynomials shows that $\phi_0^\mathrm{max}$ (corresponding to the $H^p$-norm) can be approximated by a polynomial of order $p$, the (relative) error made by this approximation is of the order $\sim 10^{-3}$ (further details will be given in \cite{BesBizJar25}). On the other hand,  performing the Keldysh decomposition introduced in section \ref{s:Keldysh}
  of such function $u^\mathrm{max}_0$,
  we find that the transient growth is fundamentally controlled by the superposition of the two
  $p+1$-th QNMs (i.e. the $p$-th overtones), namely $\hat{v}^+_{p}$ and $\hat{v}^-_{p}$,
  with a marginal  contribution
  from $\hat{v}^\pm_q$ with $q<p$
  (note that the normalization is performed with the $H^p$ norm). This is consistent with the
  polynomial nature of  $u^\mathrm{max}_0$ found through Chebyshev expansion, since
  P\"oschl-Teller eigenfunctions are indeed (Gegenbauer) polynomials [cf. Eq.
    (\ref{a:Gegenbauer})].  
  But most remarkably, this is the first overtone that lies beyond the region
  $\Delta \omega_{H^p-\mathrm{QNM}}$ in Eq.  (\ref{e:QNM_band}), where QNMs
  can be defined as discrete eigenvalues in a spectral problem
  with an $H^p$-scalar product (see \cite{Warnick:2013hba} and subsection \ref{s:Hp-pseudospectrum} below).
  In particular, this permits the account for $\tau_{\mathrm{max}}$ in Eq. (\ref{e:peak_p_dependence})
  in terms of the time decay 
  scale $\tau_p$ associated with the $p$-th overtone, given by the inverse of the
  frequency band $\Delta \omega_{H^p-\mathrm{QNM}}$
  in Eq. (\ref{e:QNM_band}), namely
  \bea
\label{e:tau_p}
\tau_p \sim \frac{1}{\Delta \omega_{H^p-\mathrm{QNM}}} \sim \frac{1}{\kappa \cdot p} \ .
\eea
  This is exact in the P\"oschl-Teller case, but it is indeed generic in the characterisation
  of QNM frequencies as eigenvalues in a $H^p$ space.
  The estimation of $G_{\mathrm{max}}$ is however more delicate, being dependent
  on the scalar products $\langle \hat{v}_p^-, \hat{v}_p^+\rangle_{H^p}$. Although its
  module (the `cosinus' of the angle between $\hat{v}_p^-$ and $\hat{v}_p^+$)
  grows with $p$ as 
  \bea
  |\langle \hat{v}_p^-, \hat{v}_p^+\rangle_{H^p}|\sim 1 - \frac{1}{p^4} \ ,
  \eea
   indicating that such $(p+1)$-ths QNMs (i.e. $p$-ths overtones) become more
   and more collinear as $p$ grows, therefore enhancing the constructive superposition,
   this is not enough to explain the height of the peak since it also
   depends on the interplay among the relative phases of $u_0$, $u(\tau)$ and $\langle \hat{v}_p^-, \hat{v}_p^+\rangle_{H^p}$.

   In any case,
   the asymptotic behaviours in Eqs. (\ref{e:peak_p_dependence}) are robust
   and imply that the quantity 
   $G_{\mathrm{max}} \cdot \tau_{\mathrm{max}}=||u^{\mathrm{max}}(\tau_{\mathrm{max}})||_{H^p}\cdot \tau_{\mathrm{max}}$ does not depend on $p$,
   suggesting that the $H^p$-transient growth $G(\tau)$, in the limit $p\to\infty$,
   has a Dirac-delta-like\footnote{This `impulsive disturbance' connects with the fundamental notion of `response function'~\cite{Chand87},
     related to the `impulse response' of the system~\cite{Schmi07},
     that
     ``...furnishes a rich and interesting picture of energy ampliﬁcations caused by external disturbances''~\cite{Schmi07}.
   } structure $\lim_{p\to\infty}G(\tau) \sim \delta(\tau)$. Such a ``singular'' behaviour in time
   has a ``loss-of-regularity'' counterpart in space, when we notice that the corresponding 
   optimal initial data $u_0^{\text{max}}(x)$ does increase its ``small-scale structure'' as $p$ grows
   (cf. panel \ref{f:u0_max_optimal_Hp} in Fig. \ref{f:u0_optimal}),
   with a loss of regularity in the (ultraviolet) limit $p\to\infty$, realising an
   infinitely-oscillating function in space as the initial data whose evolution
   gives rise to an impulsive Dirac-delta
   transient in time.

 The hyperboloidal evolution of a linear field experiences therefore an initial
  transient loss of
  regularity, whose strength increases with higher derivative in the norm\footnote{\label{footnote-Aretakis}We thank C. Warnick for pointing out
  some formal similarity of this relation between the transient-growth strength and the order $p$ of the
  considered spatial derivatives with the Aretakis instability. Although the latter occurs
  for extremal black holes,  possible structural connections may exist and will be investigated  in \cite{BesBizJar25}.}.
  A detailed analysis of such $H^p$-transient growths will be presented in \cite{BesBizJar25}.

\end{itemize}
\begin{figure}[htp]
    \centering
    \subfloat[Height $G_{\mathrm{max}}$ of the transient peak]{
      \includegraphics[clip,width=\sizefigmediumsmall\columnwidth]{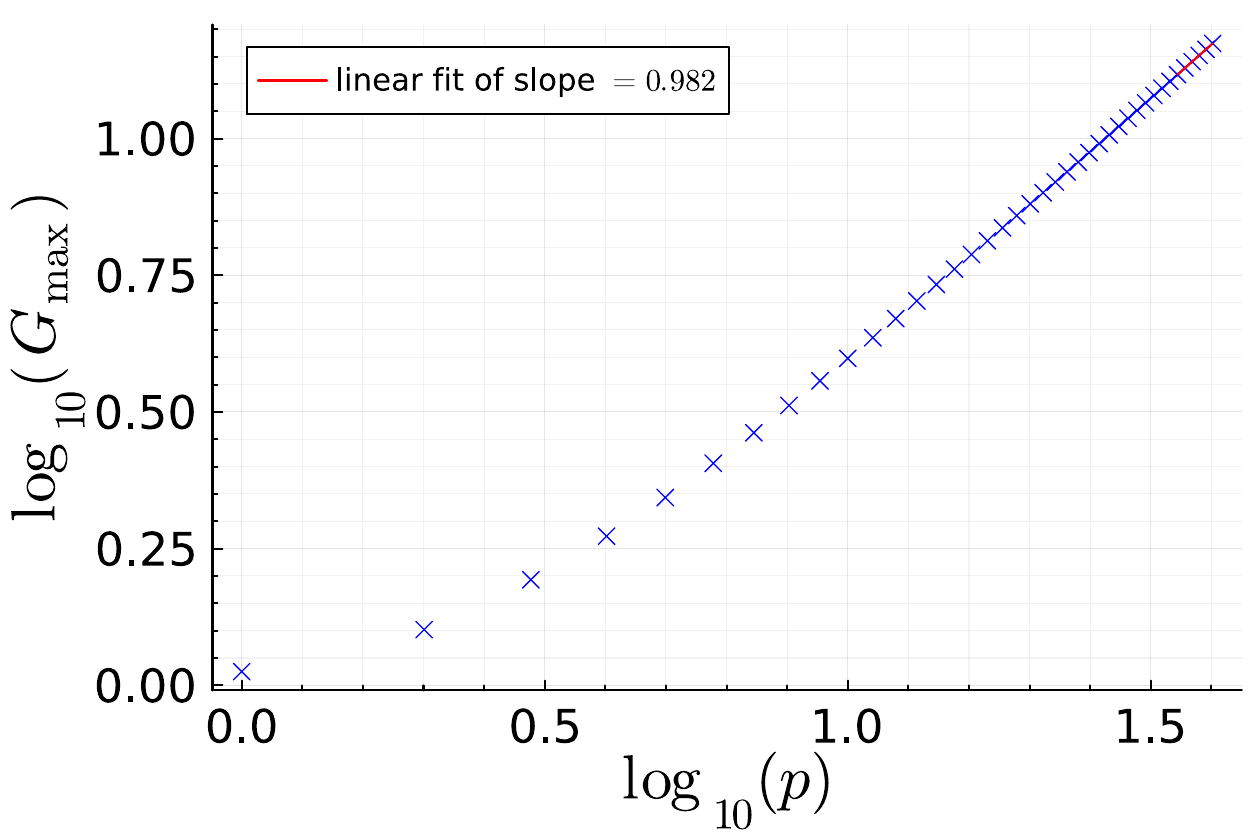}\label{f:height_peak}
    }    
    \subfloat[Time $\tau_{\mathrm{max}}$ of the transient peak]{
      \includegraphics[clip,width=\sizefigmediumsmall\columnwidth]{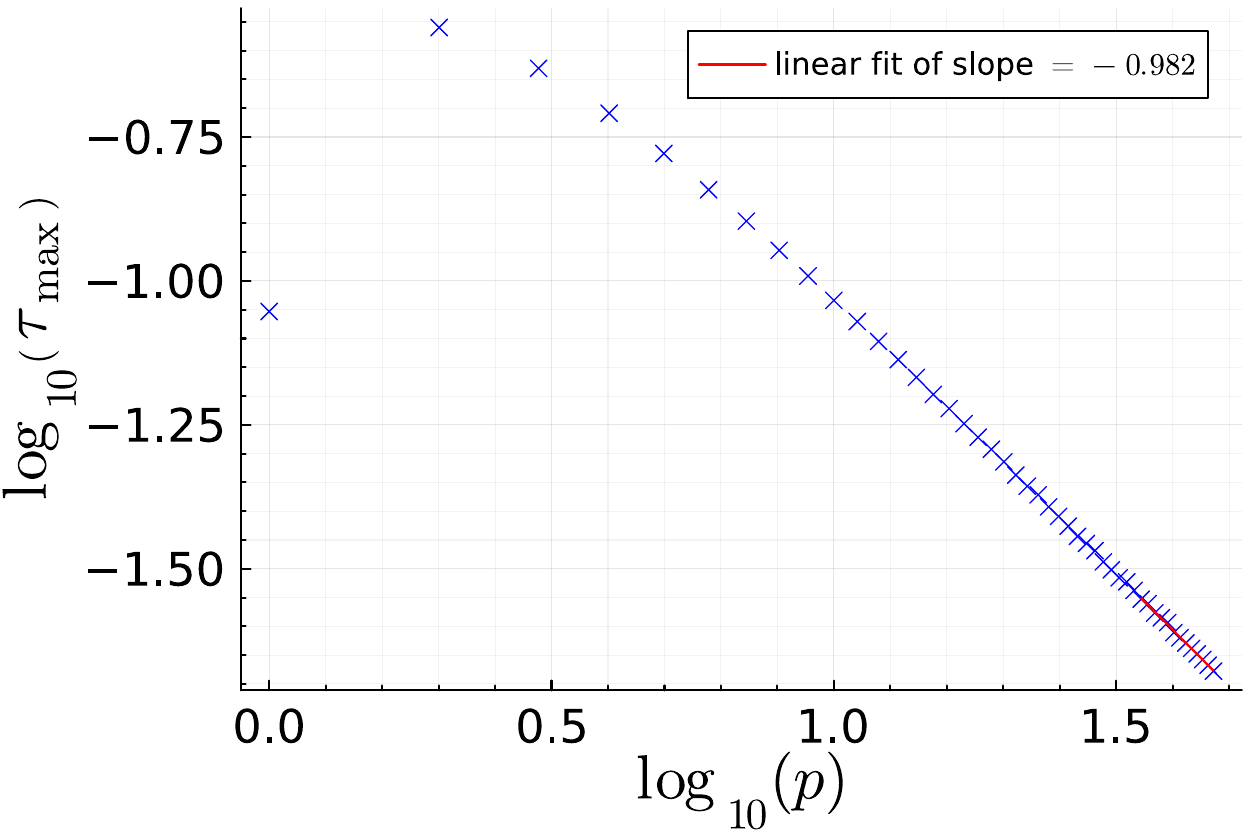}\label{f:time_peak}
    }
    \caption{Dependence in $p$ of the height $G_{\mathrm{max}}=G(\tau_{\mathrm{max}})$ and time  $\tau_{\mathrm{max}}$ of the peaks of
      the $H^p$-Sobolev transient growths in the
      P\"oschl-Teller case. The log-log panels permit to determine
      $\tau_{\mathrm{max}} \sim \frac{1}{p}$ and $G_{\mathrm{max}}  \sim p$ .
      }
    \label{f:transients2}
  \end{figure}
 
    \subsubsection{Other non-modal analysis tools: energy and $H^p$-pseudospectrum.}
    \label{s:Hp-pseudospectrum}
    As discussed in section \ref{s:transients}, in non-selfadjoint (more generally,
    non-normal) dynamics, non-modal tools are needed since the modal tools
    based on normal modes are not available and, in particular, the spectrum $\sigma(L)$
    of the time generator $L$ only controls then the late time dynamics.
    Non-modal analysis essentially resorts to the full knowledge encoded in
    the resolvent $R_L(\omega)$, providing a set of tools that capture different aspects
    of the latter. On these tools in the growth function $G(\tau) = ||e^{i\tau L}||$
    introduced in (\ref{e:G(t)}). Another important non-modal tool is the pseudospectrum
    given in terms of the norm of the resolvent, $||R_L(\omega)||$, and therefore
    in the spirit of a frequency-domain counterpart of $G(\tau)$. Specifically,
    the pseudospectrum provides a `topographic map' of the function $||R_L(\omega)||$,
    with the $\epsilon$-pseudospectrum sets defined as
\bea
\label{e:pseudospectrum_def}
\sigma^\epsilon(L) &=& \{\omega\in\mathbb{C}:
||R_L(\omega)|| = ||(L - \omega)^{-1}||>1/\epsilon\} \ .
\eea
Such a tool has been extensively employed to study spectral instability, transient growths
or pseudoresonances~\cite{TreTreRed93,Trefe97,Davies99,Davie00,trefethen2005spectra,Schmi07}
in fluid dynamics and numerical analysis and has been recently introduced in the
gravitational setting~\cite{Jaramillo:2020tuu}. In the specific setting of non-modal transient
growths,
the $\epsilon$-pseudospectra sets $\sigma^\epsilon(L)$ permit to define the $\epsilon$-pseudospectral abscissa
$\displaystyle \alpha_\epsilon(L)=-\inf_{\sigma\in \sigma^\epsilon(L)}\mathrm{Im}(\omega)$ that controls
the growth along evolution, in particular providing a lower bound to the transient peak in terms of 
the Kreiss constant $\displaystyle {\cal K} = \sup_{\epsilon>0}\frac{\alpha_\epsilon(L)}{\epsilon}$.
The $\epsilon$-pseudospectra indeed interpolate between the late modal dynamical behaviour
at small $\epsilon$'s, whose limit is the spectrum $\displaystyle \sigma(L)= \lim_{\epsilon\to 0} \sigma^\epsilon(L)$,
and the very early behaviour in the limit $\epsilon\to \infty$, specifically through the so-called numerical
abscissa that can be characterised as $\displaystyle \omega(L)= \lim_{\epsilon \to \infty}(\alpha_\epsilon(L) -\epsilon)$
and controls the initial slope of the transient growths. These latter `frequency-domain' non-modal
tools have been discussed in a gravitational setting \cite{Jaramillo:2022kuv,BoyCarDes22} and more recently
used in combination with `time-domain' non-modal tools in \cite{Carballo_2024,Chen_2024,Carballo:2025ajx}.

Therefore, for its relevance in this transient growth setting, as well as for its role in the
`definition and stability problems' of section \ref{s:coeffs_Sobolev}, we comment now
on the pseudospectra, with a special emphasis on the $H^p$-Sobolev case, where new results are presented.

\paragraph{Energy norm pseudospectrum}
The pseudospectrum depends on the choice of norm, as it is apparent
in its characterisation in Eq. (\ref{e:pseudospectrum_def}).
In Fig. \ref{PT} we present the pseudospectrum of P\"oschl-Teller QNMs
in the energy norm, whereas Fig. \ref{L2=0} presents the
pseudospectrum in the selfadjoint case obtained by setting
$L_2=0$ in the expression (\ref{e:L_L_1_L_2}) of $L$.
These figures have already been discussed in detail
in the literature (cf. \cite{Jaramillo:2020tuu}).
The reasons
to present them again here are two-fold.

On the one hand, they represent
the lowest $p$ case of the $H^p$-norm pseudospectra we discuss below,
therefore putting in context the new results on Sobolev pseudospectra.
On the other hand, these two figures illustrate two extreme patterns that
we are going to find below in  $H^p$-pseudospectra, associated with the
pseudospectrum convergence issues in the grid resolution limit $N\to\infty$.
Indeed, regarding  Fig. \ref{L2=0} the concentric contour lines
around the eigenvalues provide a neat control of the
spectrum $\sigma(L)$ that can actually be used to characterise (`define')
its eigenvalues (see~\cite{ColRomHan19}) and, crucially, the value of $||R_L(\omega)||$ converge to
finite values as $N\to \infty$ for frequencies $\omega\notin\sigma(L)$.
On the contrary, in  Fig. \ref{PT} the values $||R_L(\omega)||$ do diverge
as $N\to \infty$ if $\mathrm{Im}(\omega)\gtrsim \kappa$ (cf. \cite{Boyanov:2024}),
so the interpretation of the logarithmic patterns of the contour lines
is a delicate issue: although they capture the actual QNM
instabilities~\cite{Jaramillo:2020tuu}, they strictly do not exist in the continuum limit.
This is a problem that still needs to be elucidated.
The point to retain is that the  $||R_L(\omega)||$ divergence is not a numerical
artefact, but actually a faithful imprint of Warnick's theorem in \cite{Warnick:2013hba}, as we see below.

\begin{figure}[htp]
  \centering
  \subfloat[P\"oschl-Teller]{
    \includegraphics[clip,width=0.5\columnwidth]{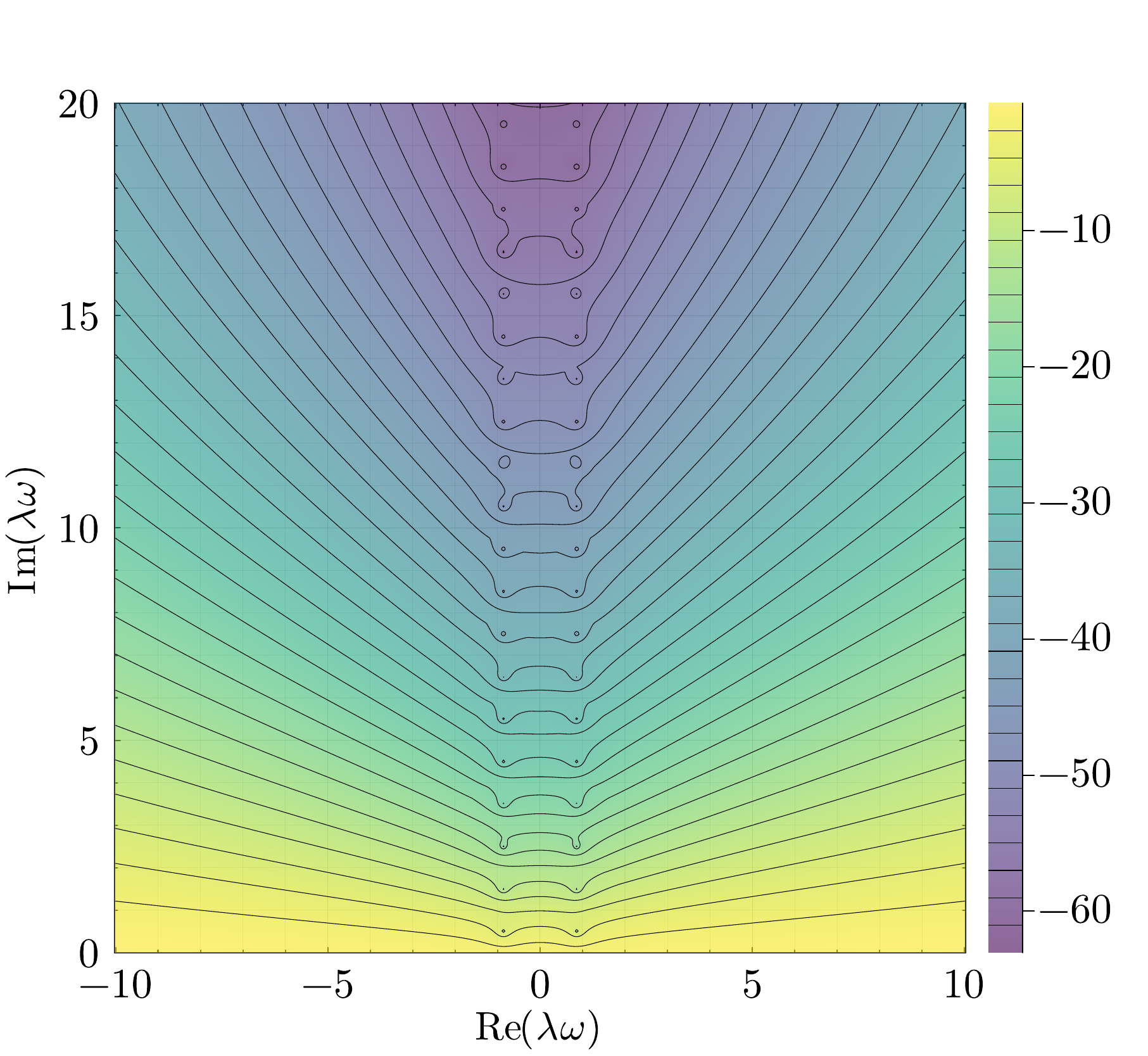}\label{PT}
  }
  \subfloat[self-adjoint case]{
    \includegraphics[clip,width=0.5\columnwidth]{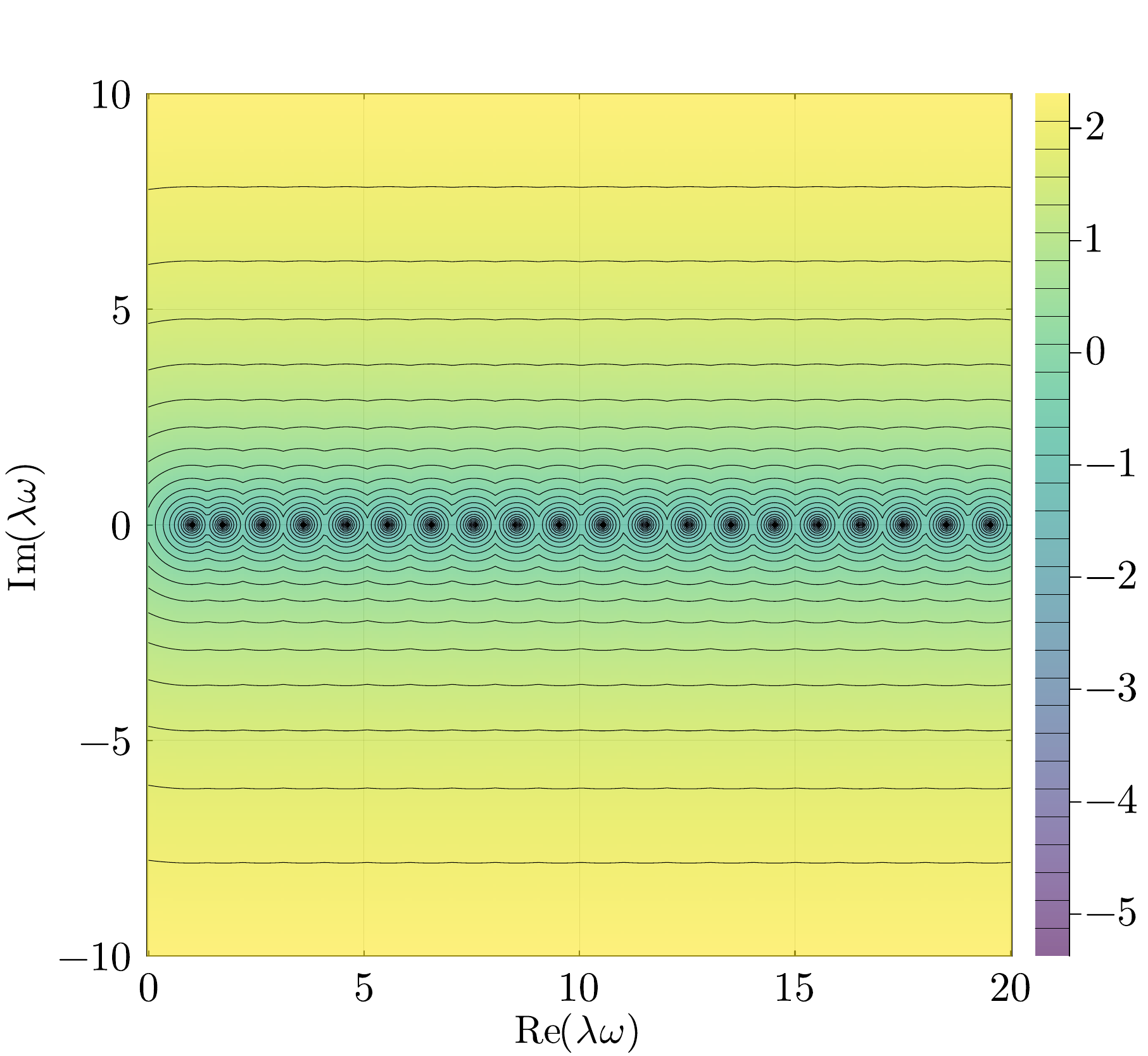}\label{L2=0}
  }
  \caption{Panel \ref{PT} shows the energy pseudospectrum in the P\"oschl-Teller case.
    Panel \ref{L2=0} shows the test provided by the self-adjoint case $L_2=0$. Note,
    in particular, the horizontal pseudospectral contours far from the
    spectrum. This is a non-trivial test indicating exactly the same stability for all eigenvalues,
    consistently with the condition number $\kappa=1$ for all eigenvalues in the
    self-adjoint case. We have set $N=40$ for both panels.}
  \label{pseudospectrum_enery_norm}
  \end{figure}

\paragraph{$H^p$-Sobolev norm pseudospectrum.}
Warnick's theorem in \cite{Warnick:2013hba} for asymptotically AdS BHs applies also
for dS asymptotics, this including the P\"oschl-Teller case.
A summary of its key elements in our context is presented
in points (i)-(iv) in section III.A of \cite{Boyanov:2024}. 
The latter can be summarised in the emergence of a horizontal
$\kappa$-band structure of the QNMs
in the $\omega$-complex plane, as follows:
\begin{itemize}

\item[i)] The restriction of the operator $L$ to $H^p$-Sobolev spaces
  permits to define the discrete QNM frequencies $\omega_n$'s as eigenvalues of $L$,
  with the respective eigenfunctions providing the QNMs, but only
  in the horizontal band of width $\Delta \omega_{H^p-\mathrm{QNM}} \sim p\cdot \kappa$ in
  Eq. (\ref{e:QNM_band}).

\item[ii)] In this $\Delta \omega_{H^p-\mathrm{QNM}}$ band, the resolvent $R_L(\omega)$ is
  a $H^p$-bounded operator, i.e. $||R_L(\omega)||_{H^p}$ is finite except at the discrete set
  corresponding to the QNM eigenvalues. 

\item[iii)] Above the $\Delta \omega_{H^p-\mathrm{QNM}}$ band, i.e. when
  $\mathrm{Im}(\omega_n)\gtrsim  p\cdot\kappa$, all frequencies $\omega$'s are proper
  eigenvalues so $||R_L(\omega)||_{H^p}$ diverges in this upper-half plane.

\item[iv)] In order to add one more $\kappa$-width band  to the region
  containing discrete QNM eigenvalues,  adding one more overtone,
  we must increase regularity by passing from $H^p$ to $H^{p+1}$. 

\end{itemize}
These properties of $H^p$-QNMs explain the structural features of
the P\"oschl-Teller QNM pseudospectra 
calculated with the $H^p$ norm, and displayed in 
in Figs. \ref{pseudospectrum_p_norm}, \ref{pseudospectrum_p_norm_conv} and
\ref{pseudospectrum_p_norm_colorbar_fixed}.
In particular, such pseudospectra present two clearly structurally separated zones:
\begin{itemize}

\item[a)] A $\Delta \omega_{H^p-\mathrm{QNM}}$ band above the real axis, where the structure
  is similar to that in Fig. \ref{L2=0}, in particular with circular contour lines
  around the spectrum controlling and permitting to defined the QNM frequencies and, crucially,
  presenting convergence  to a finite value  of $||R_L(\omega)||_{H^p}$
  in the grid resolution $N\to\infty$(except at QNMs), i.e. convergence of the $H^p$-pseudospectrum
  in the continuum limit. This is consistent with points i) and ii) above.

\item[b)] A zone with $\mathrm{Im}(\omega_n)\gtrsim  p\cdot\kappa$ similar to
  the  pseudospectrum in Fig. \ref{PT}, in particular where  $||R_L(\omega)||_{H^p}$
  diverges and all points are eigenvalues, consistently with point iii) above.

\end{itemize}
Specifically, we present in  Fig. \ref{pseudospectrum_p_norm} the $H^p$-pseudospectra for different
values of $p$ (compare with Fig.~\ref{pseudospectrum_enery_norm} for the energy norm),
calculating  straightforwardly the norm of the resolvent $R_L(\omega)$ at a given resolution
(in this case, $N=40$).

Then, we demonstrate the $H^p$-pseudospectrum convergence patterns in $\kappa$-bands
in Fig.~\ref{pseudospectrum_p_norm_conv}, showing full agreement with 
Warnick's theorem discussed above. Following this, and in an attempt to better visualise the
consistency between the calculated  $H^p$-pseudospectra and Warnick's theorem
by making directly apparent their convergence properties, we recast pseudospectra in Fig. \ref{pseudospectrum_p_norm}
again in Fig. \ref{pseudospectrum_p_norm_colorbar_fixed}, but  enforcing the
same colour scale in  $||R_L(\omega)||_{H^p}$ for all $p$ and setting as ``divergent''
all $||R_L(\omega)||_{H^p}=1/\epsilon$, with $\epsilon<10^{10}$. The resulting figure captures better the
content of Warnick's theorem\footnote{See \cite{Besson24} for an extended presentation of these
$H^p$-pseudospectra (including convergence tests). See also \cite{cai2025pseudospectrum}
for a recent application of this $H^p$-norm approach to the study of the important Kerr black hole pseudospectrum.}.

\medskip

 In summary, the construction of these
$H^p$-pseudospectra represent by themselves
a non-trivial result (see also \cite{Warnick:2024} for related
results\footnote{The pseudospectra in Fig. \ref{pseudospectrum_p_norm}
  are calculated from the resolvent $R_L(\omega)$ of $L$. Such
  resolvent $R_L(\omega)$ is a non-compact operator, a fact making of 
  Fig. \ref{pseudospectrum_p_norm} a remarkable result.
  The corresponding pseudospectra in \cite{Warnick:2024}, presenting
  the same structure in bands, are calculated from the norm
  of the inverse of the Laplace transform of the wave equation in second-order
  form.  In contrast with the fist-order version here studied,
  in that case the ``resolvent'' is compact.}).
From a structural perspective, $H^p$-pseudospectra  make apparent the
link between regularity and definition of QNMs: as we increase
$p$ we gain control on more and more overtones, as it can be seen
in the circular pseudospectral lines around growing overtones,
where QNMs are defined in horizontal bands\footnote{This structure in bands of
  width $\kappa$, with one QNM frequency per band, is
  the responsible of the writing  $a_{N_{\mathrm{QNM}}} = \kappa N_{\mathrm{QNM}}$ in
  Eq. (\ref{e:u_Keldysh_discussion_C}). It also underlies the Weyl's law discussed
  in \ref{a:Weyl_law} (see \cite{Jaramillo:2022zvf} for details).} of width $\kappa$, in such a
way that an $H^p$ norm is needed to control the first $p$ bands (see
details in \cite{Warnick:2013hba,Boyanov:2024}).
On the other hand, in our non-selfadjoint dynamics context and as commented at the beginning of this section,
the pseudospectrum
provides a fundamental `frequency-domain' non-modal analysis tool, in particular
 with applications in transient growths, from which
practical diagnostic tools as
the $\epsilon$-pseudospectral and numerical abscissa or the Kreiss constant can be defined.
More details on this `definition versus stability' problem, as well
as on $H^p$-transient growths will be presented in dedicated works \cite{BesBoyJar24}
and \cite{BesBizJar25}, respectively.

\begin{figure}[htp]
  \centering
  \subfloat[$H^5$-pseudospectrum]{
    \includegraphics[clip,width=0.5\columnwidth]{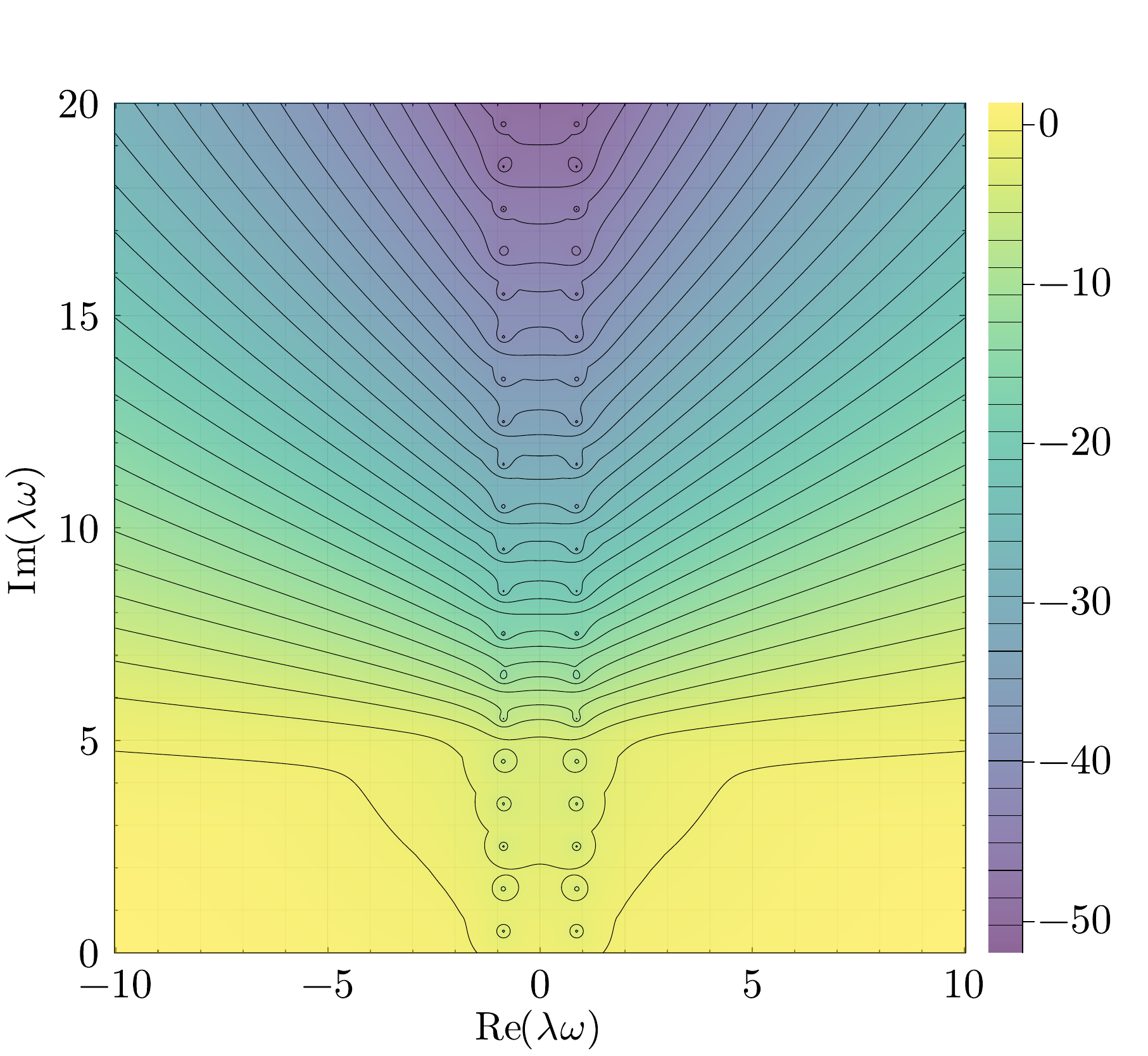}\label{b_}
   }
  \subfloat[$H^{10}$-pseudospectrum]{
    \includegraphics[clip,width=0.5\columnwidth]{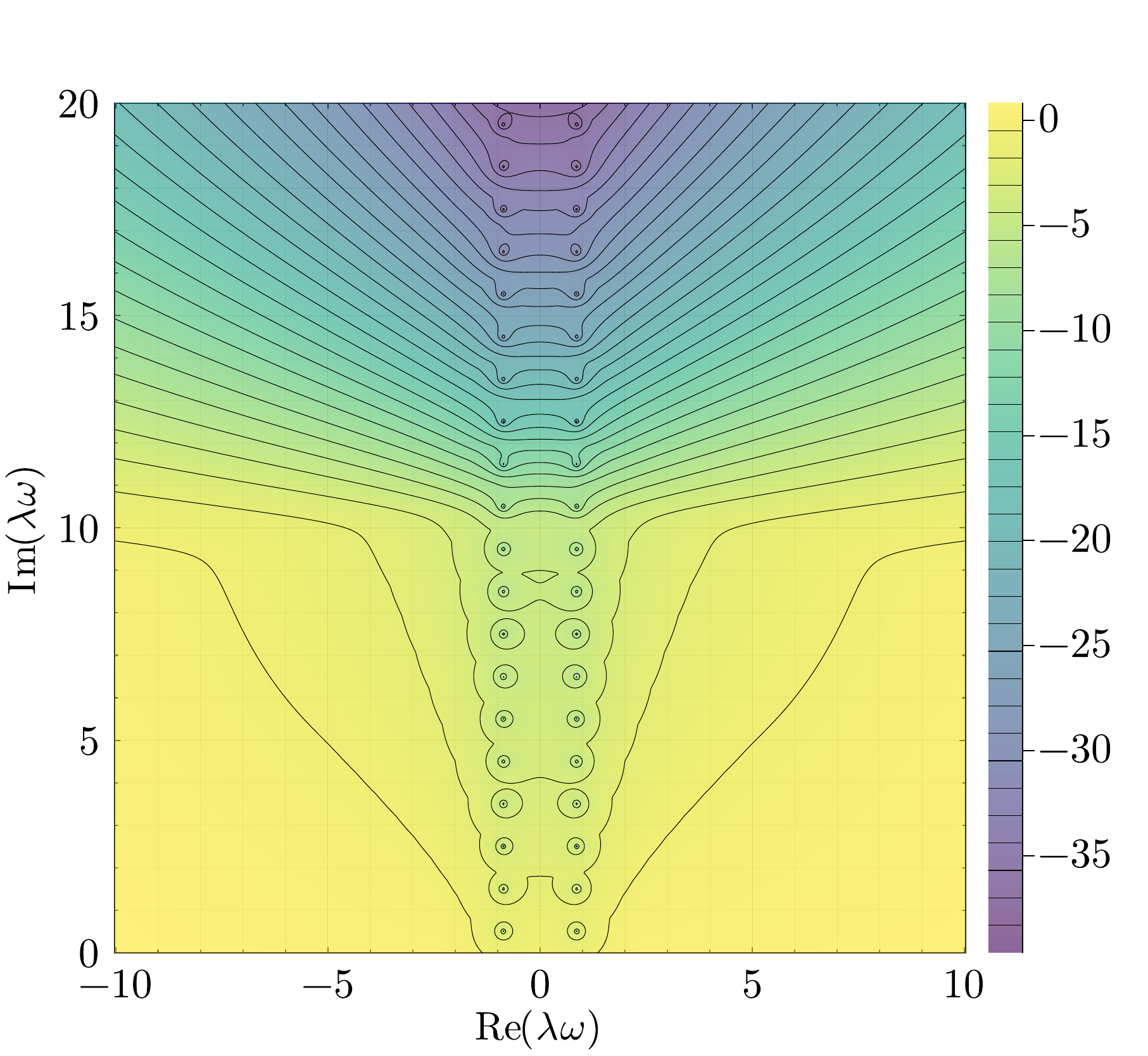}\label{c_}
  }
  
  \subfloat[$H^{15}$-pseudospectrum]{
    \includegraphics[clip,width=0.5\columnwidth]{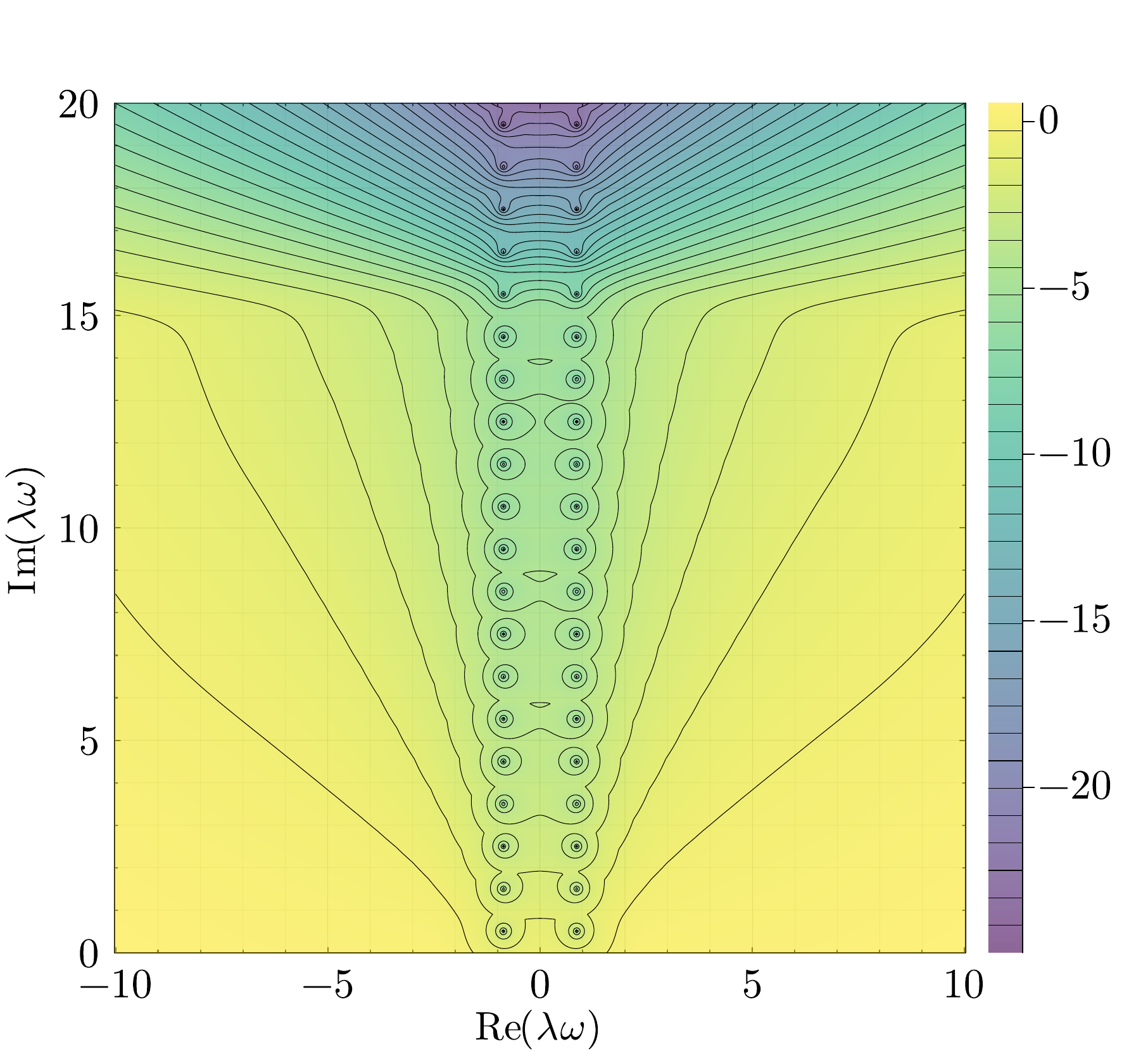}\label{d_}
  }
  \subfloat[$H^{20}$-pseudospectrum]{
    \includegraphics[clip,width=0.5\columnwidth]{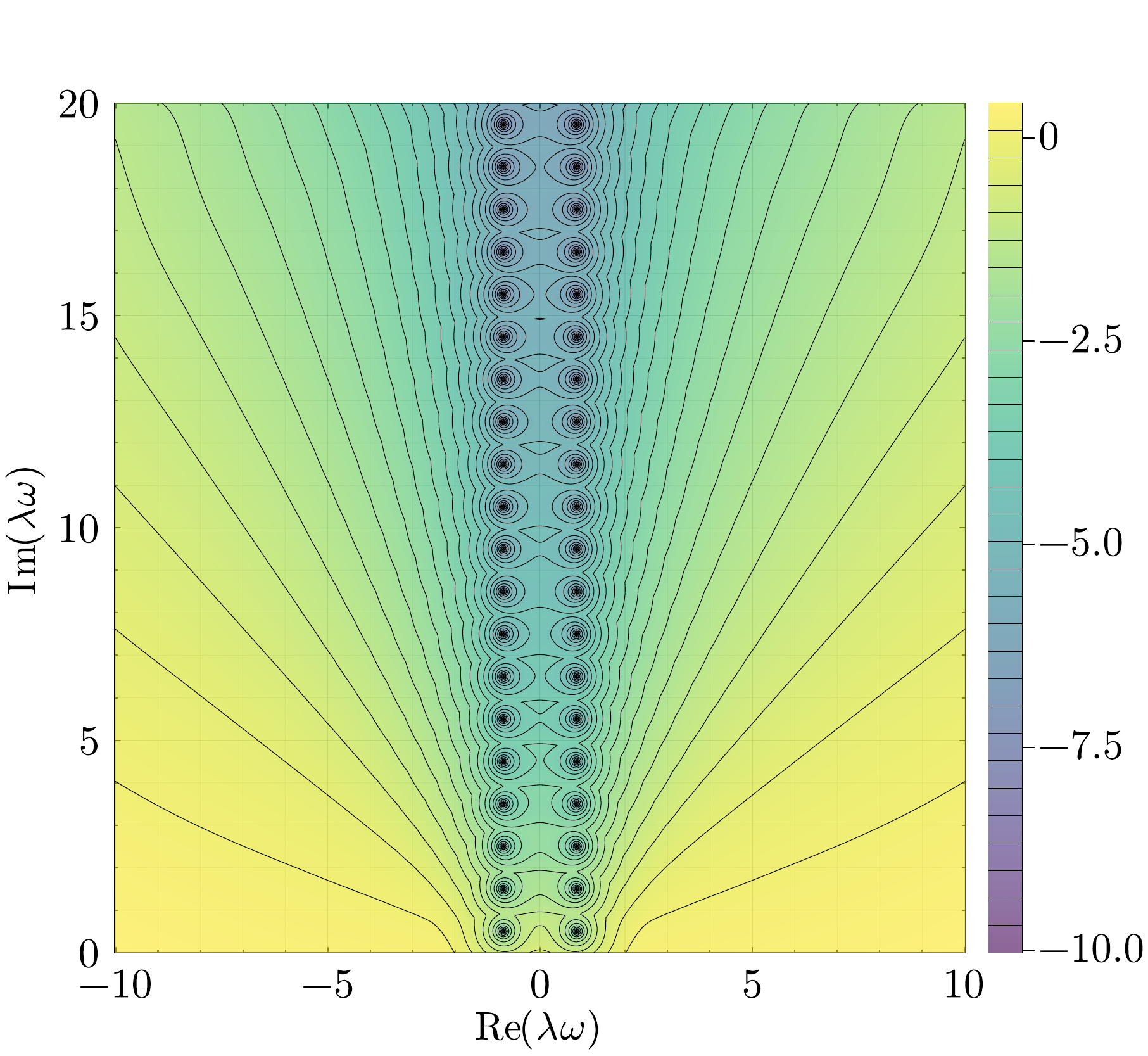}\label{e_}
  }
  \caption{Panels \ref{b_}, \ref{c_}, \ref{d_} and \ref{e_} show the $H^p$-pseudospectrum in the P\"oschl-Teller case with $p=5,10,15$ and $20$. For all these panels $N=40$.}
  \label{pseudospectrum_p_norm}
  \end{figure}

\begin{figure}[htp]
  \centering
  \subfloat[Sampling points plotted on top of the $H^{20}$-pseudospectrum]{
    \includegraphics[clip,width=0.75\columnwidth]{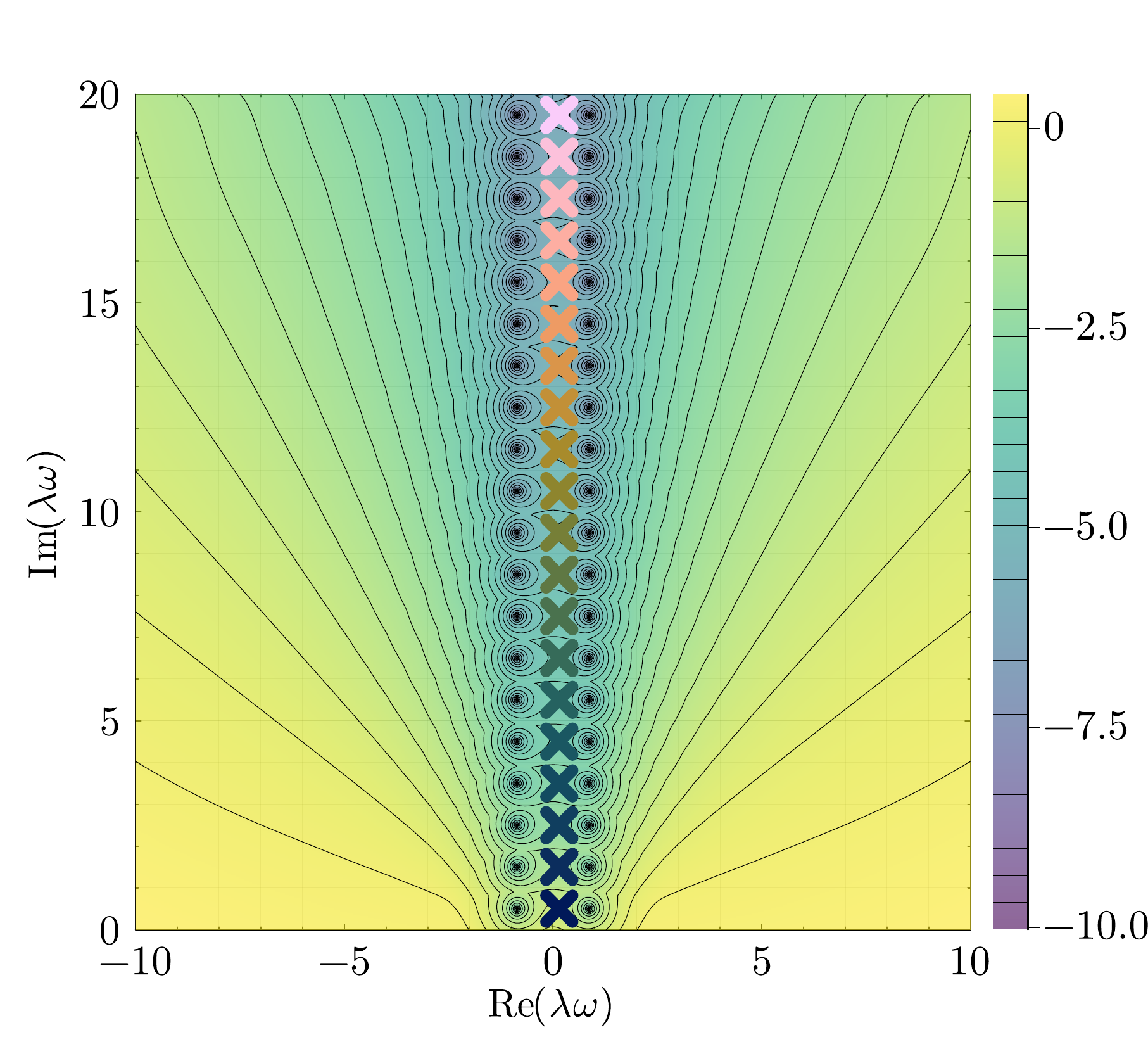}\label{pseudospectrum_p_norm_conv:sampling}
  }

  \subfloat[$H^5$-pseudospectrum convergence]{
    \includegraphics[clip,width=0.5\columnwidth]{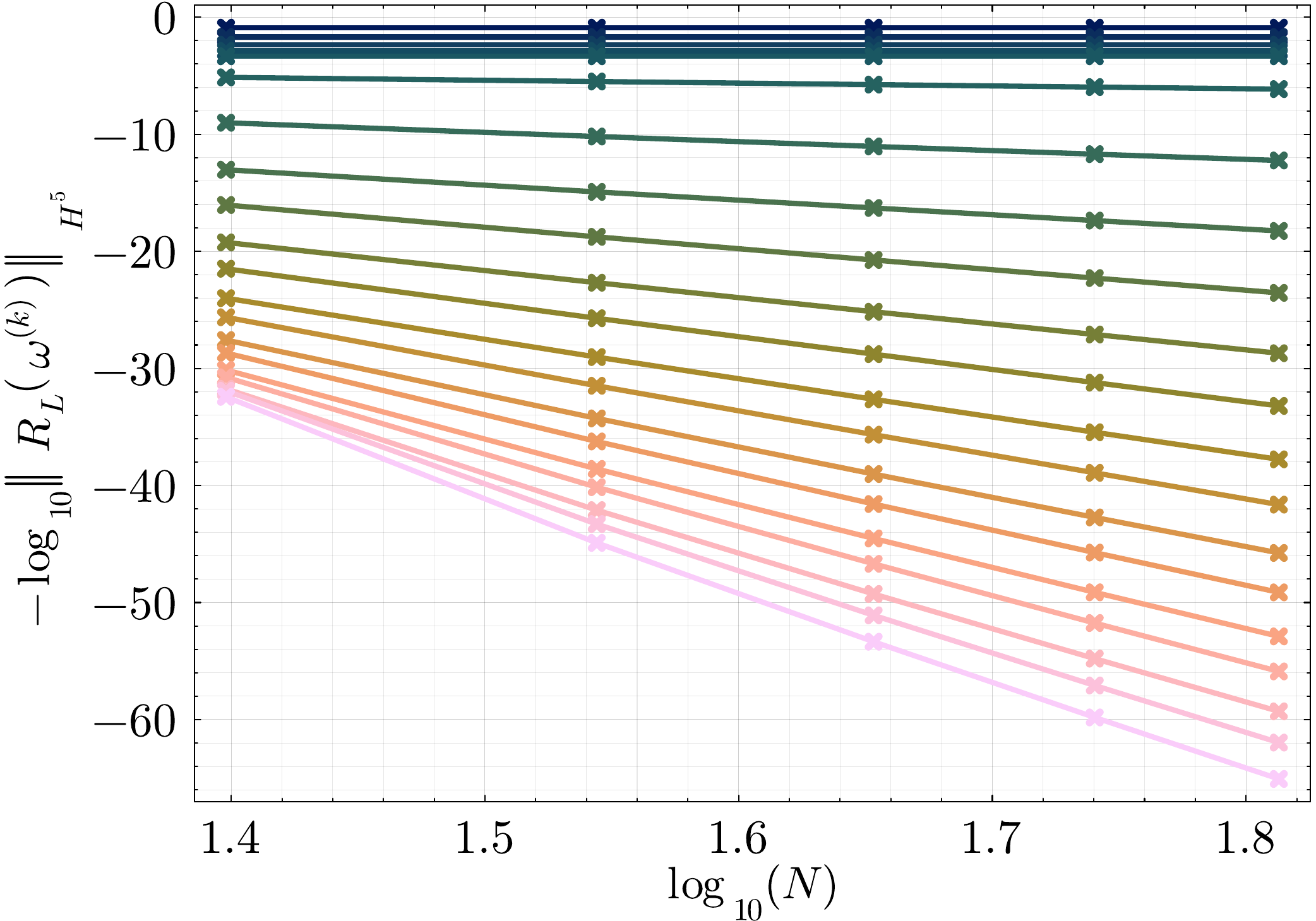}\label{pseudospectrum_p_norm_conv:b_}
  }
  \subfloat[$H^{10}$-pseudospectrum convergence]{
    \includegraphics[clip,width=0.5\columnwidth]{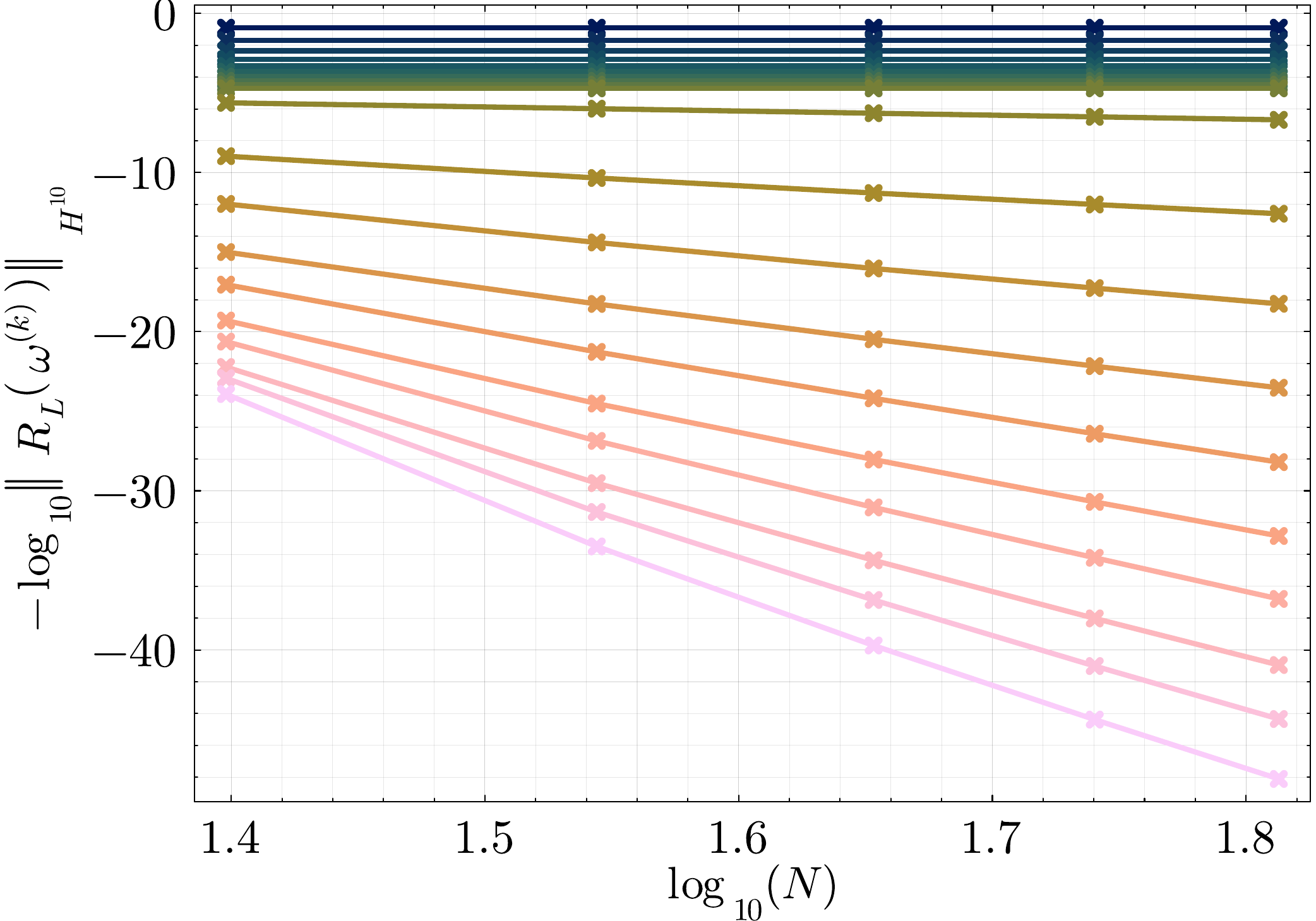}\label{pseudospectrum_p_norm_conv:c_}
  }
  
  \subfloat[$H^{15}-$pseudospectrum convergence]{
    \includegraphics[clip,width=0.5\columnwidth]{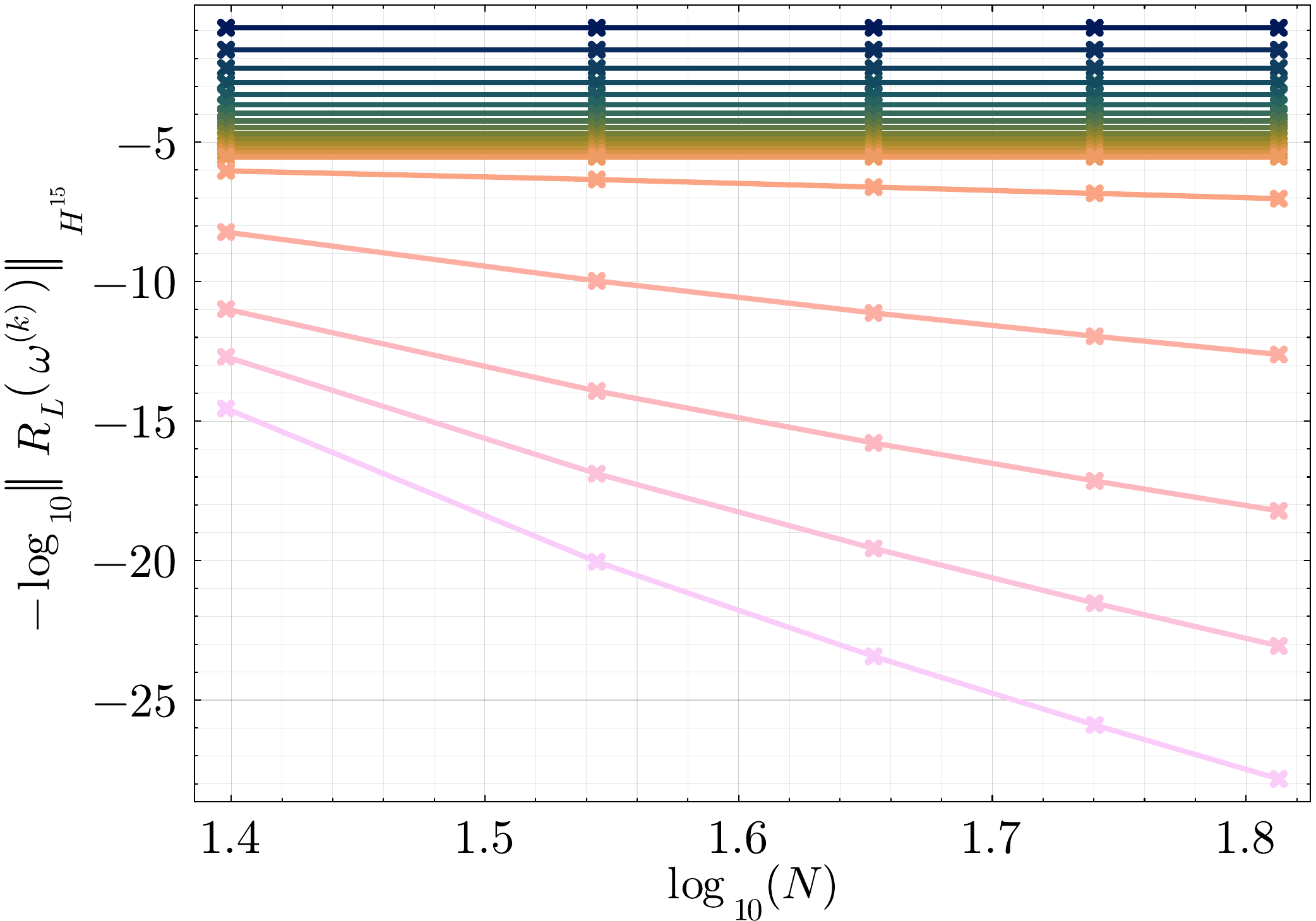}\label{pseudospectrum_p_norm_conv:d_}
  }
  \subfloat[$H^{20}-$pseudospectrum convergence]{
    \includegraphics[clip,width=0.5\columnwidth]{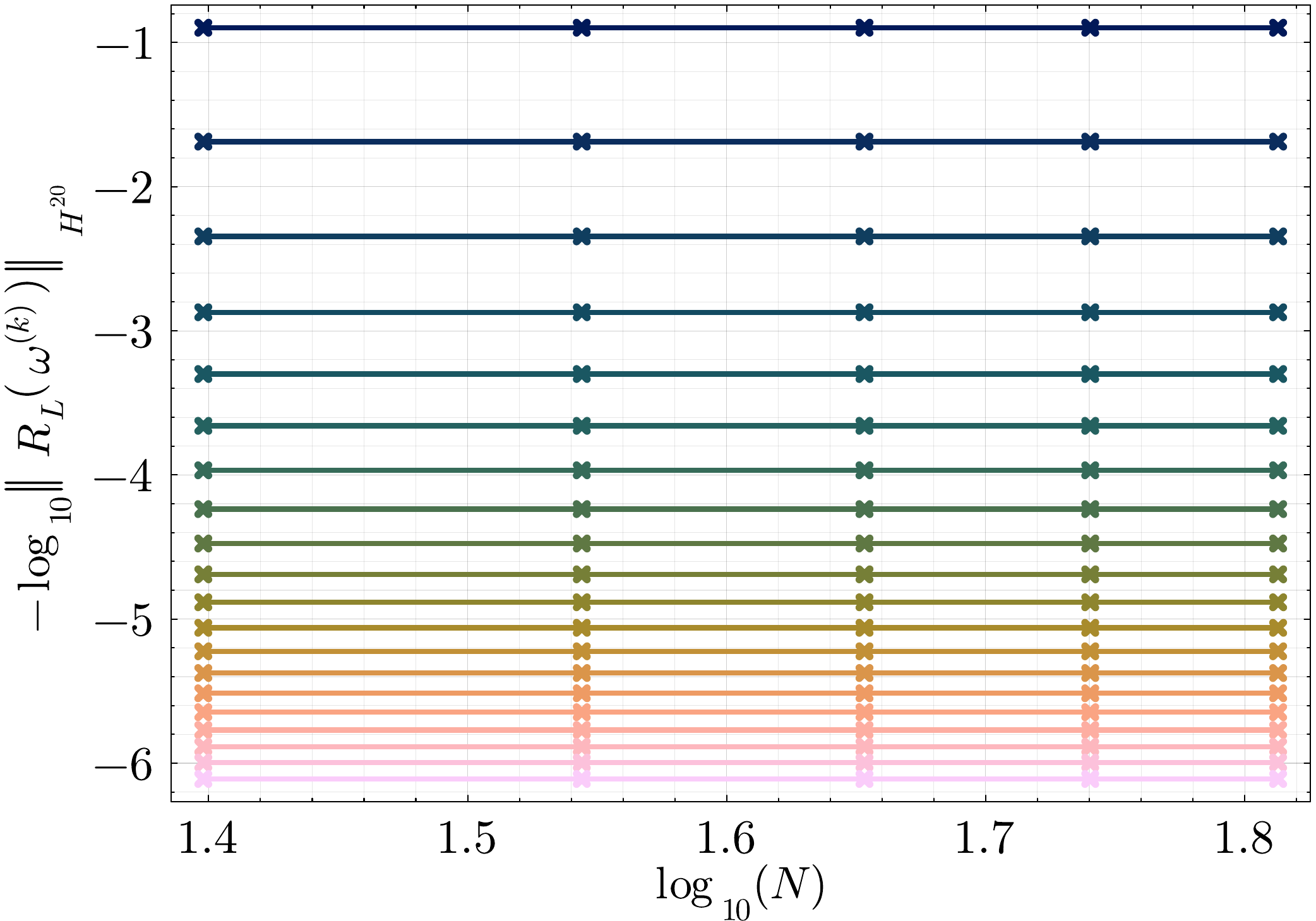}\label{pseudospectrum_p_norm_conv:e_}
  }
  \caption{Panels \ref{pseudospectrum_p_norm_conv:b_}, \ref{pseudospectrum_p_norm_conv:c_}, \ref{pseudospectrum_p_norm_conv:d_} and \ref{pseudospectrum_p_norm_conv:e_} illustrate the convergence in bands of width $\kappa$ of the $H^p$-pseudospectrum in the P\"oschl-Teller case with $p=5,10,15$ and $20$. The convergence test of the $H^p$-pseudospectrum consists in computing the $H^p$-norm of the resolvent $R_L(\omega)$ at the sampling points
    $\lambda \omega^{(k)}=0.15+i(0.5+k), k\in \{0, \ldots, 19\}$
    as a function of the Chebyshev resolution $N$. Consistently with Warnick theorem, for a given $p$, 
    convergence in the $H^p$-norm is found for all $\omega^{(k)}$ with $\mathrm{Im}(\omega^{(k)})<\kappa\cdot p$,
    whereas for those  $\omega^{(k)}$ with $\mathrm{Im}\left(\omega^{(k)}\right)>\kappa\cdot p$ it is
    found that $||R_L(\omega)||_{H^p}$ divergence (as a power-law in $N$).
    The sampling points $\lambda \omega^{(k)}$ are illustrated in panel \ref{pseudospectrum_p_norm_conv:sampling},
    setting the employed colour code.
  }
  \label{pseudospectrum_p_norm_conv}
\end{figure}

\begin{figure}[htp]
  \centering
  \subfloat[$H^5$-pseudospectrum]{
    \includegraphics[clip,width=0.5\columnwidth]{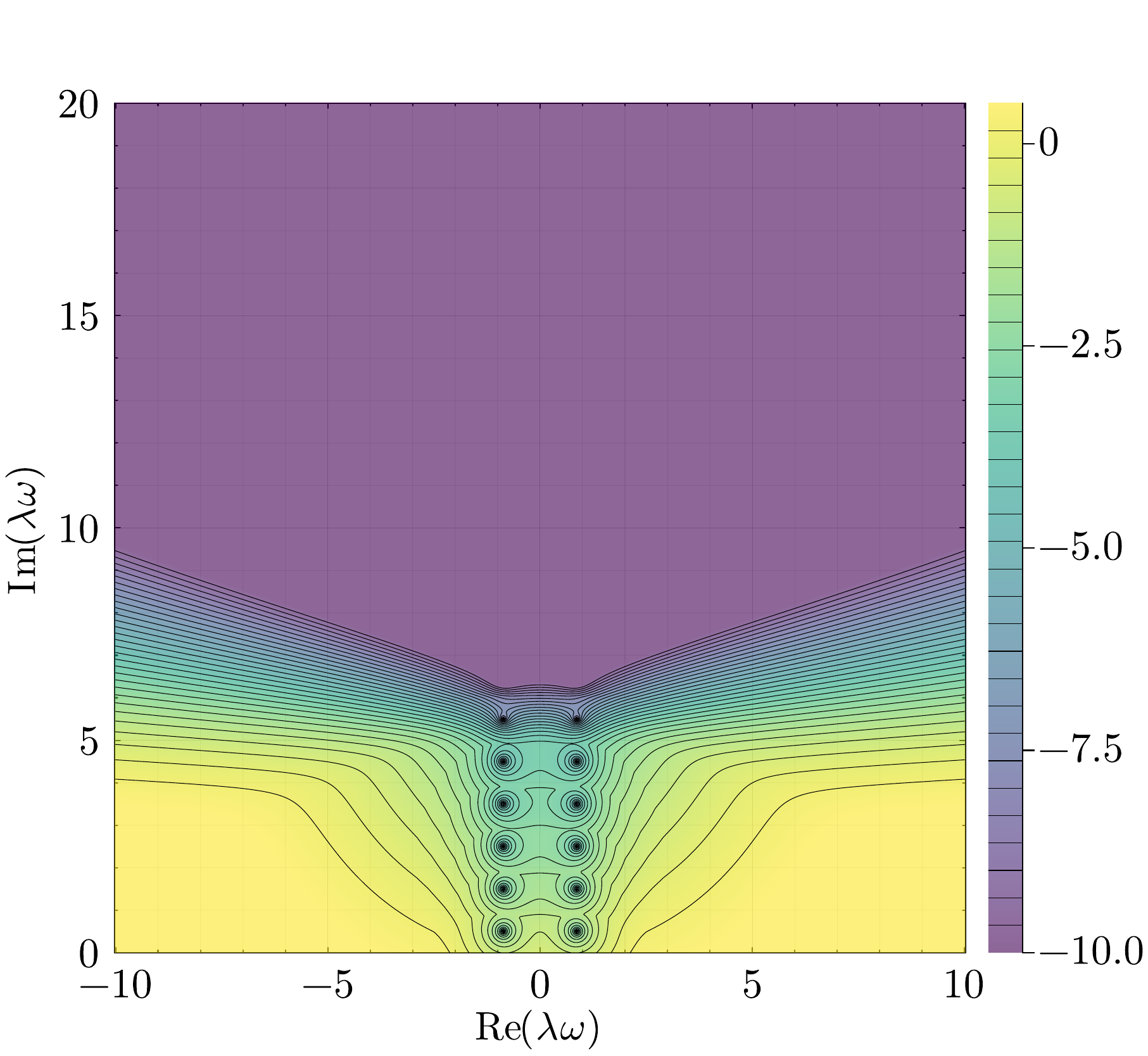}\label{pseudospectrum_p_norm_colorbar_fixed:b_}
  }
  \subfloat[$H^{10}$-pseudospectrum]{
    \includegraphics[clip,width=0.5\columnwidth]{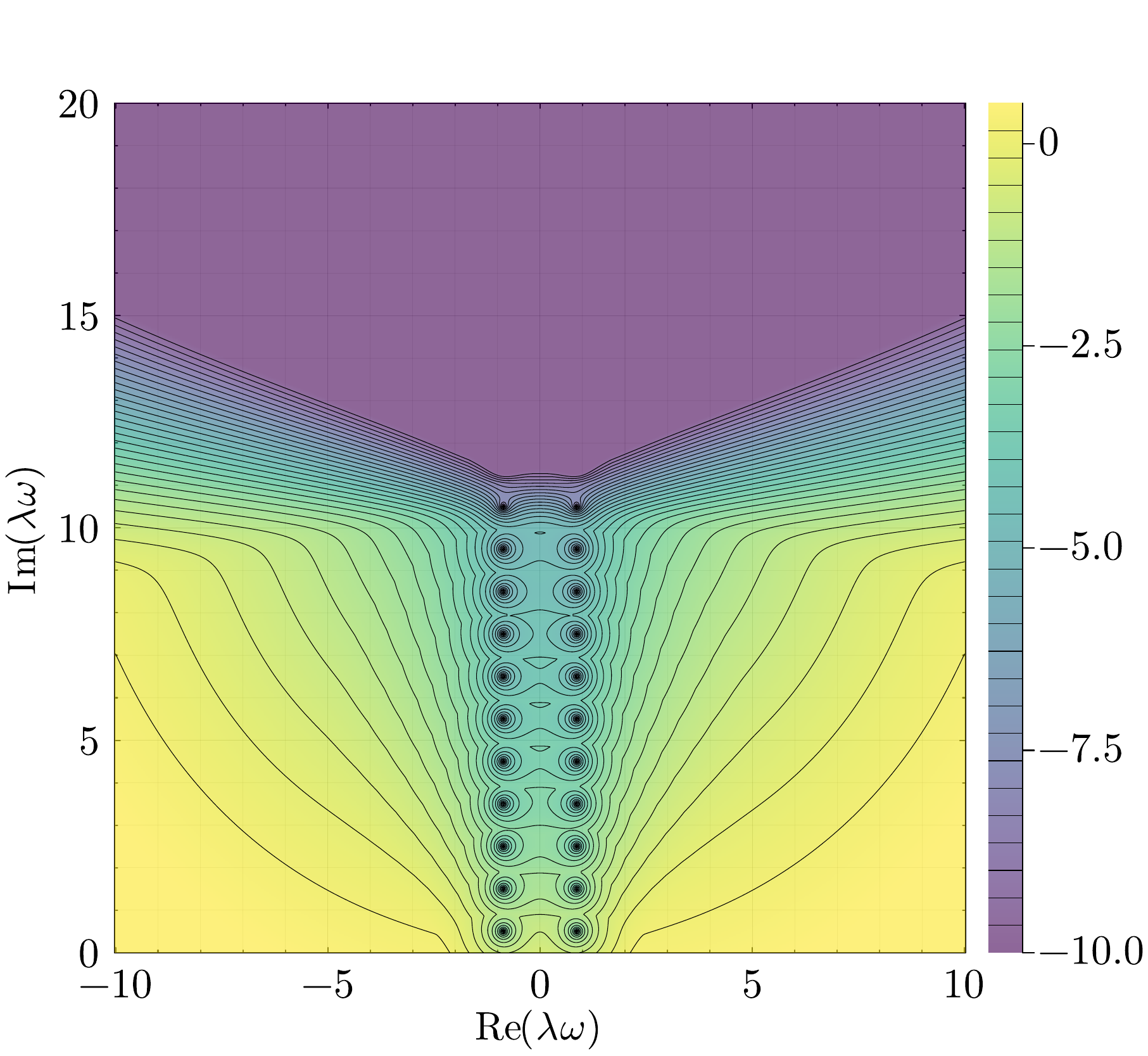}\label{pseudospectrum_p_norm_colorbar_fixed:c_}
  }
  
  \subfloat[$H^{15}$-pseudospectrum]{
    \includegraphics[clip,width=0.5\columnwidth]{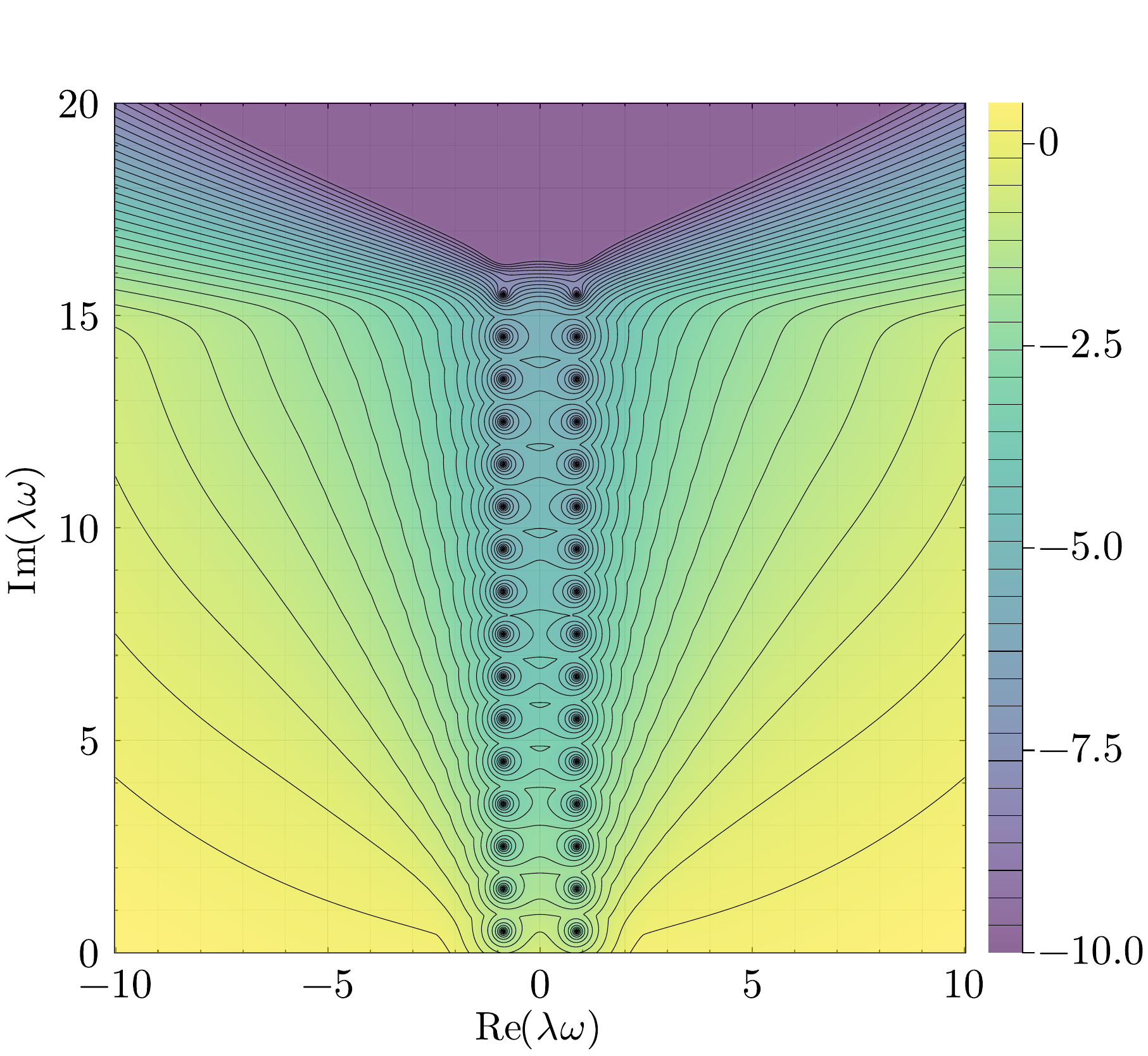}\label{pseudospectrum_p_norm_colorbar_fixed:d_}
  }
  \subfloat[$H^{20}$-pseudospectrum]{
    \includegraphics[clip,width=0.5\columnwidth]{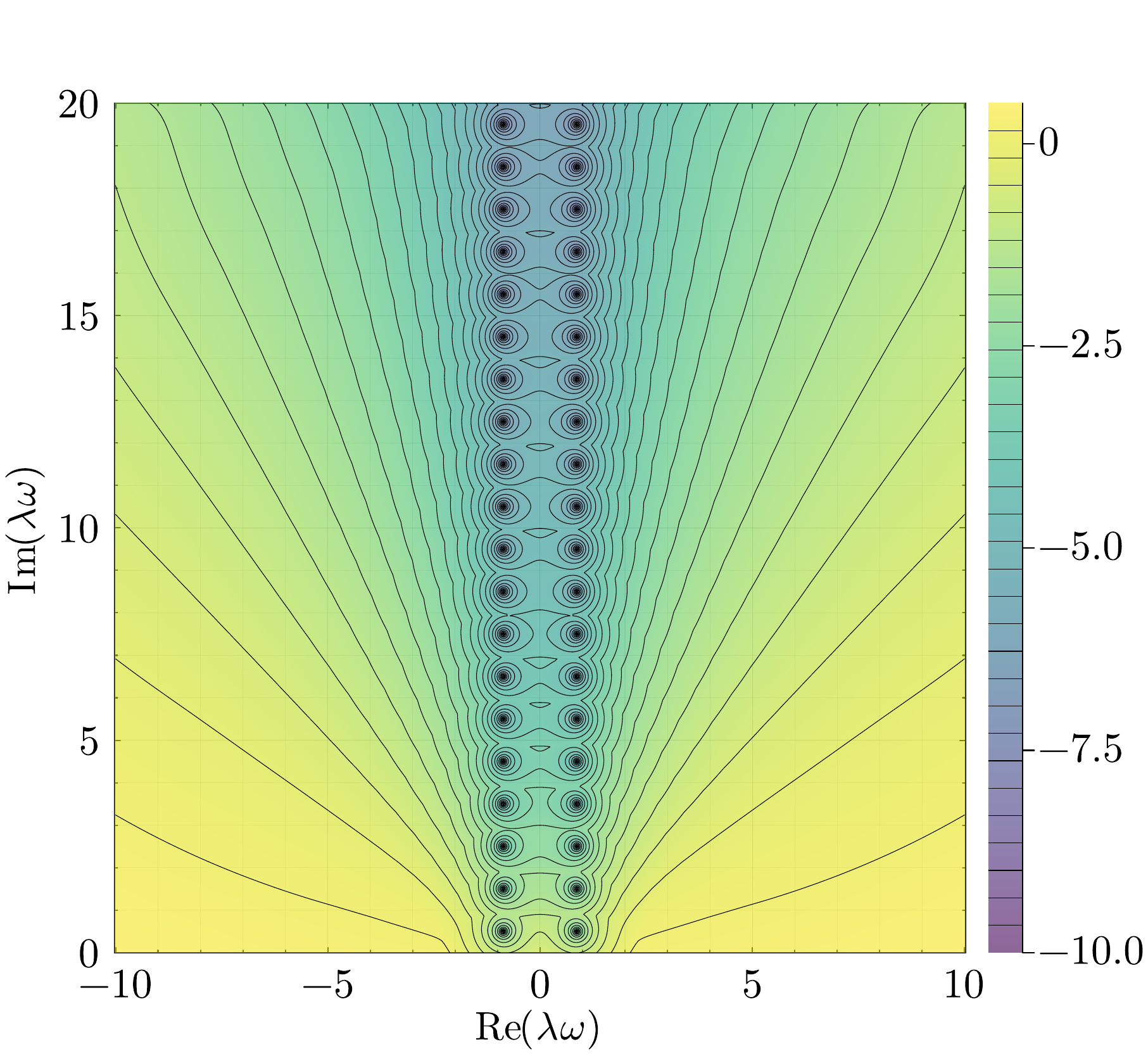}\label{pseudospectrum_p_norm_colorbar_fixed:e_}
  }

  \subfloat[$H^{20}$-pseudospectrum, zoom on the first 3 modes]{
  \includegraphics[clip,width=0.5\columnwidth]{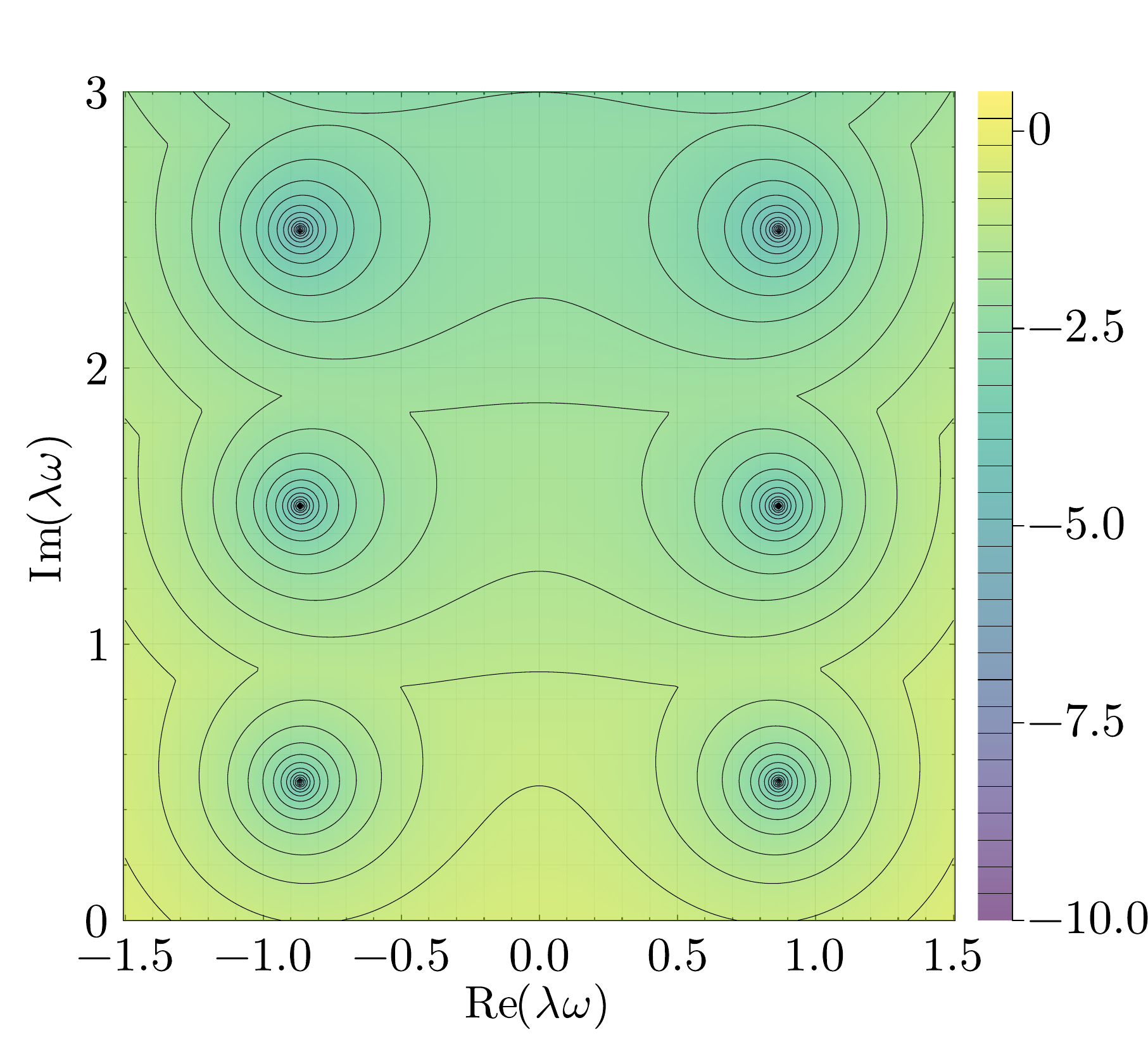}\label{pseudospectrum_p_norm_colorbar_fixed:f_}
  }
  \subfloat[$H^{20}$-pseudospectrum, zoom on the 18th to 20th modes]{
  \includegraphics[clip,width=0.5\columnwidth]{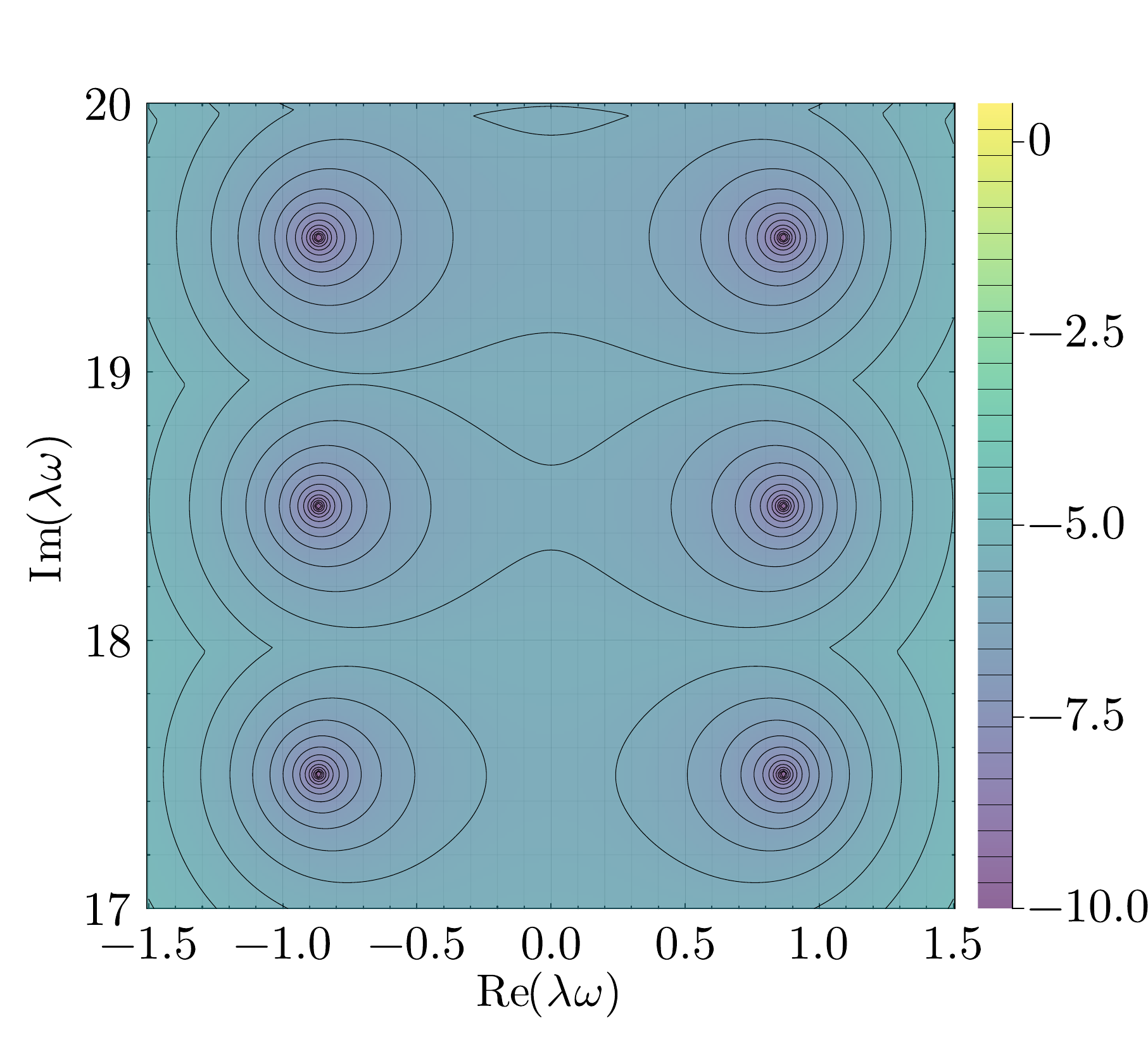}\label{pseudospectrum_p_norm_colorbar_fixed:g_}
  }
  \caption{Panels \ref{pseudospectrum_p_norm_colorbar_fixed:b_}, \ref{pseudospectrum_p_norm_colorbar_fixed:c_}, \ref{pseudospectrum_p_norm_colorbar_fixed:d_} and \ref{pseudospectrum_p_norm_colorbar_fixed:e_} show the $H^p$-pseudospectrum in the P\"oschl-Teller case with $p=5,10,15$ and $20$. For all panels $N=40$. In contrast with
    Fig.~\ref{pseudospectrum_p_norm}, the same colour bar is set for all panels. On the one hand this enables a better comparison between different $H^p$-pseudospectra, as well as it effectively sets as divergent those $||R_L(\omega)||_{H^p}=1/\epsilon$, with $\epsilon<10^{10}$. As a combination of these elements, the convergence pattern is more apparent. Panels \ref{pseudospectrum_p_norm_colorbar_fixed:f_} and \ref{pseudospectrum_p_norm_colorbar_fixed:g_} are zooms of the $H^{20}$-pseudospectrum, illustrating the circular contours around QNM frequencies for $\omega_0$, $\omega_1$ and $\omega_2$, Panel \ref{pseudospectrum_p_norm_colorbar_fixed:f_}, and $\omega_{17}$, $\omega_{18}$ and $\omega_{19}$, Panel \ref{pseudospectrum_p_norm_colorbar_fixed:g_}, demonstrating the spectral stability of
    $\omega_{n\leq 19}$ QNMs in this $H^{20}$ norm.}
  \label{pseudospectrum_p_norm_colorbar_fixed}
  \end{figure}

 \subsection{BH QNM Weyl's law in generic spacetime asymptotics}
\label{a:Weyl_law}

\begin{figure}[htp]
  \centering
  \subfloat[Schwarzschild frequencies]{
    \includegraphics[clip,width=0.5\columnwidth]{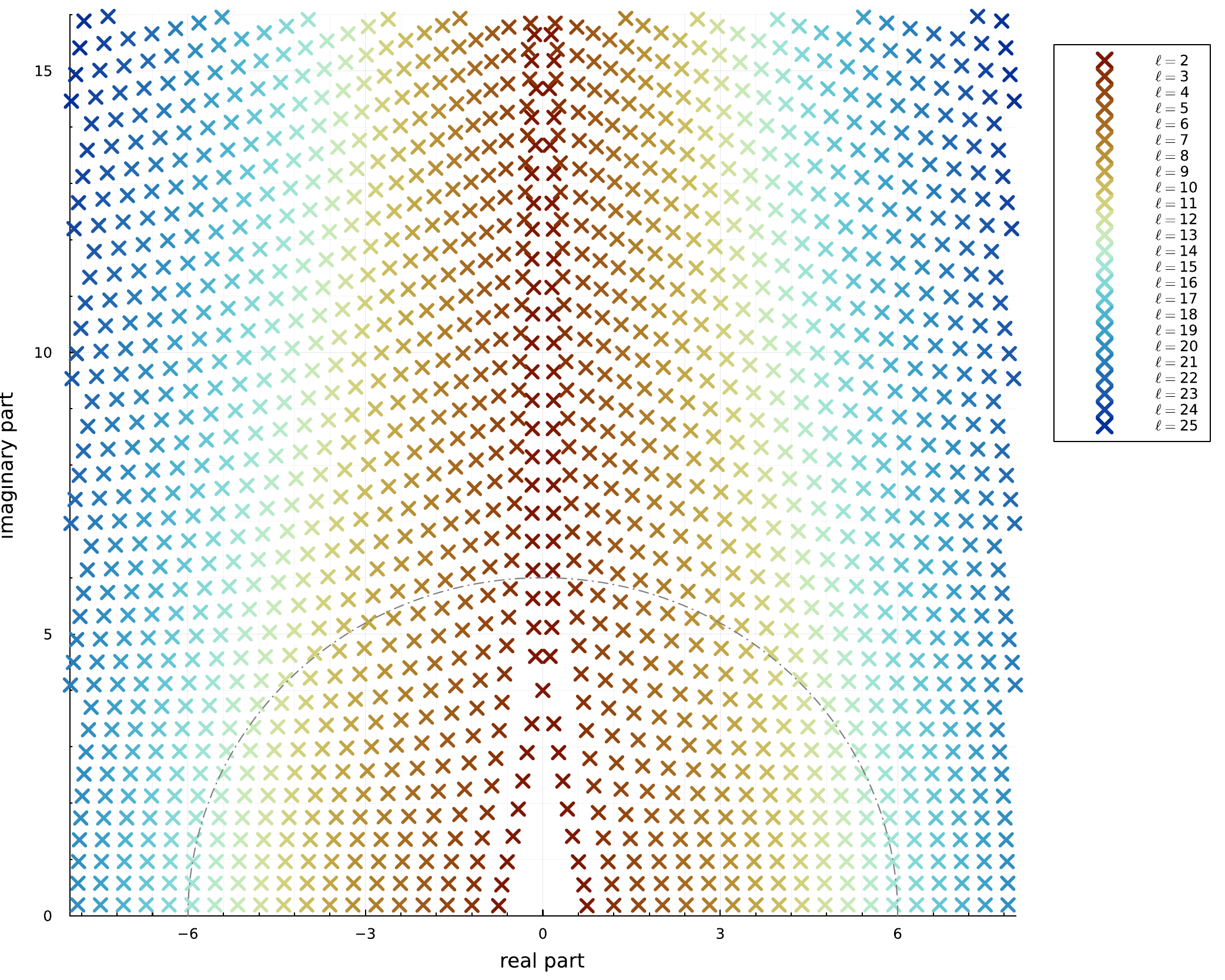}\label{QNF_S}
  }
  \subfloat[$N(R)$ for Schwarzschild]{
    \includegraphics[clip,width=0.4\columnwidth]{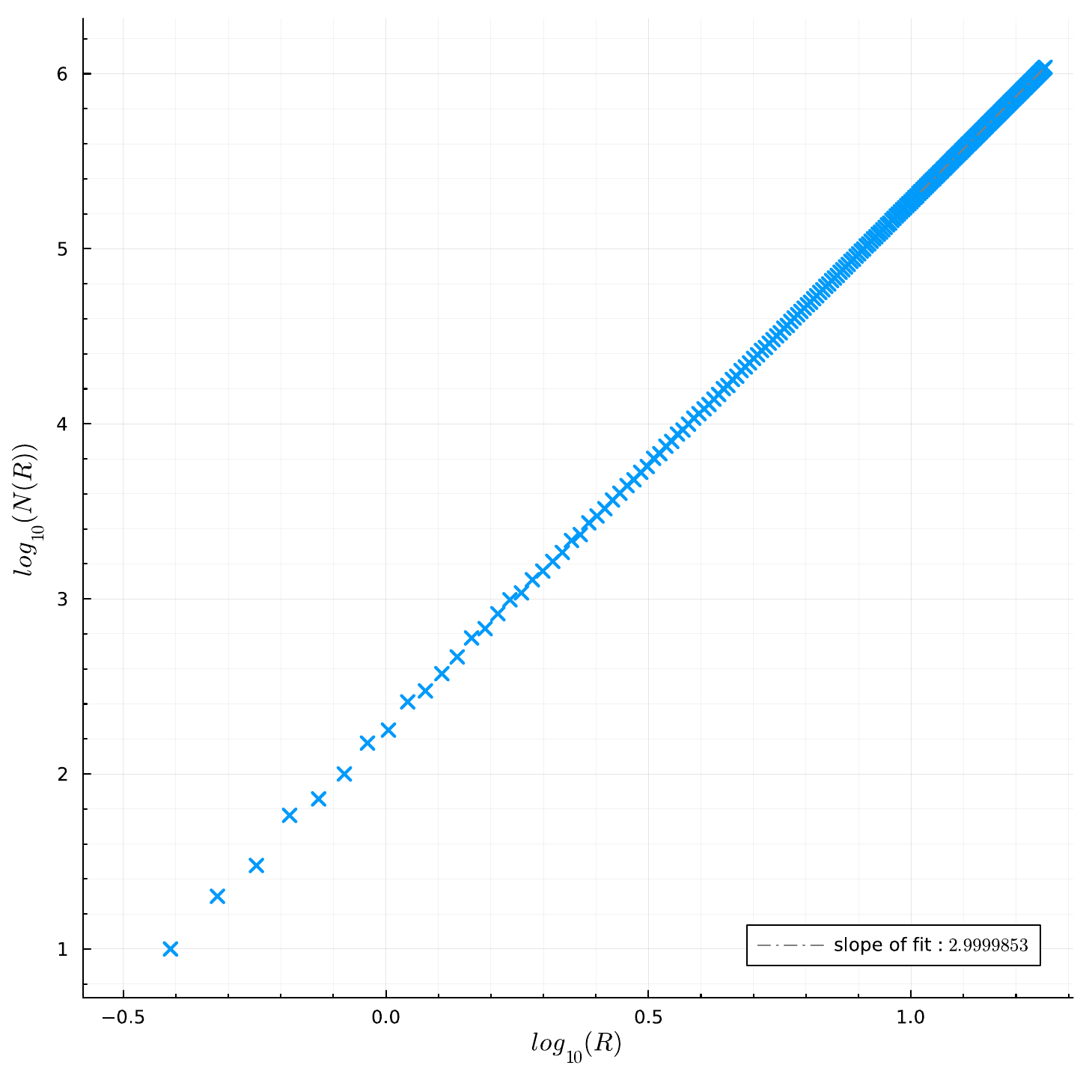}\label{weyl_S}
  }

  \subfloat[Schw.-dS frequencies frequencies]{
    \includegraphics[clip,width=0.5\columnwidth]{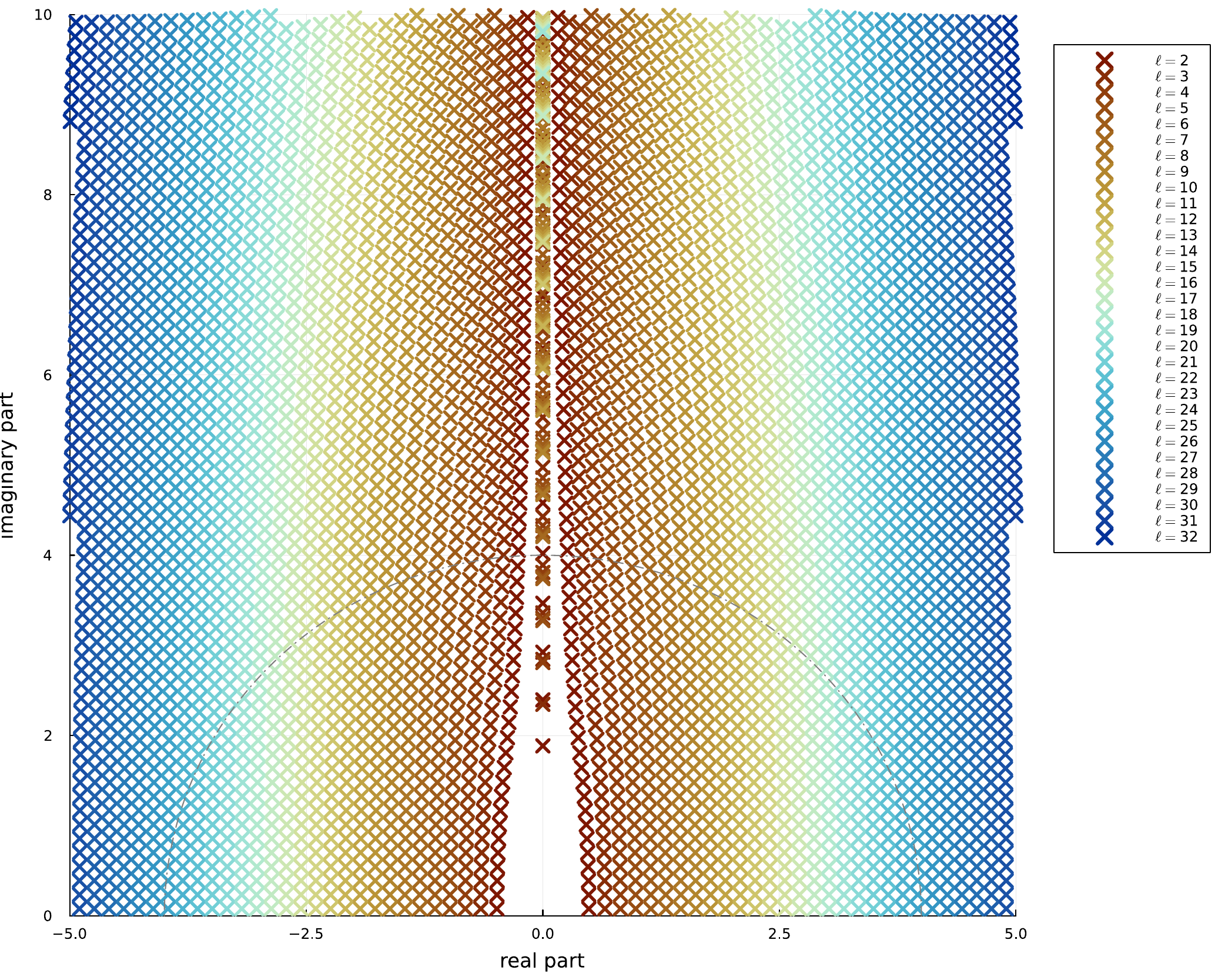}\label{QNF_dS}
  }
  \subfloat[$N(R)$ for Schw.-dS]{
    \includegraphics[clip,width=0.4\columnwidth]{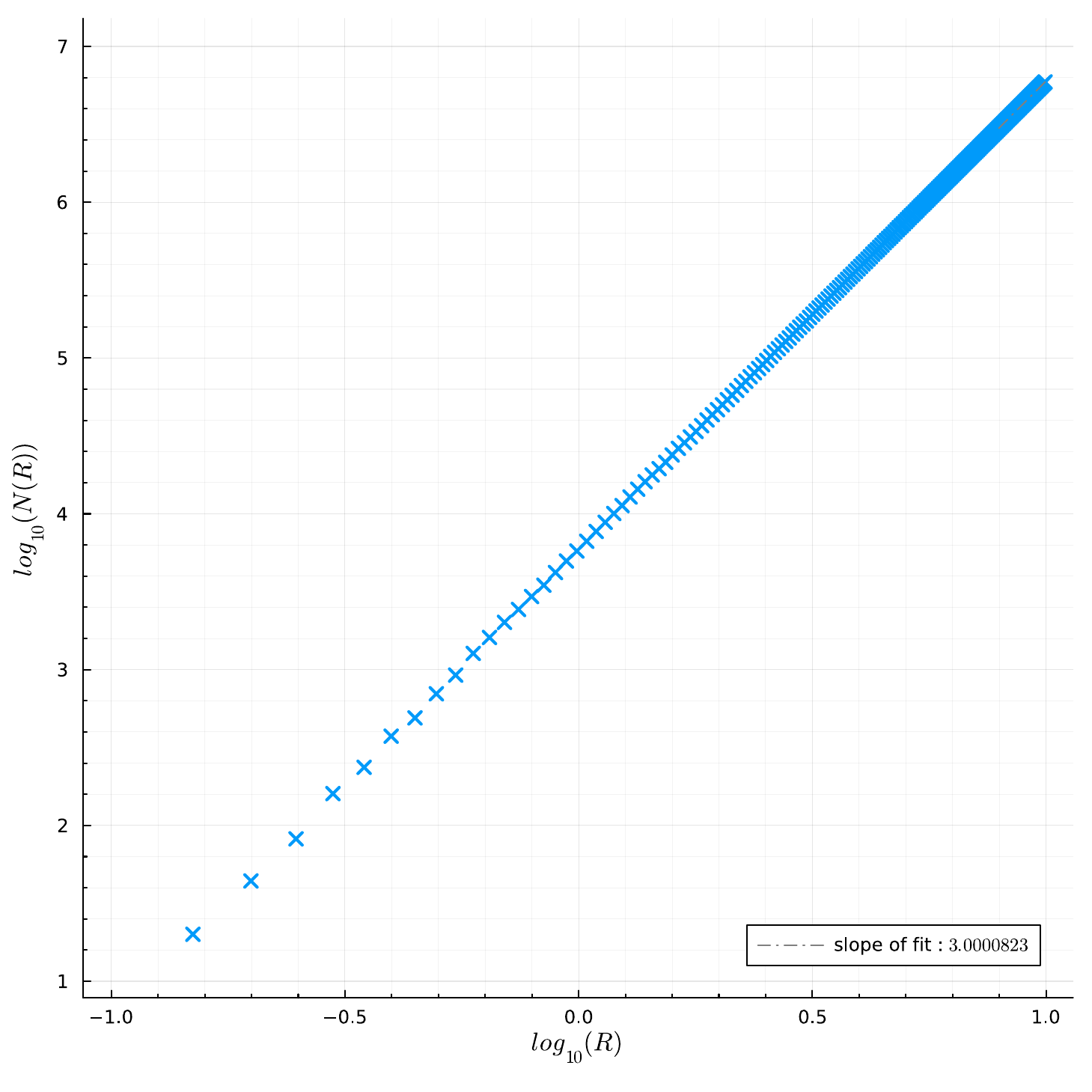}\label{weyl_dS}
  }

  \subfloat[Schw.-AdS frequencies]{
    \includegraphics[clip,width=0.5\columnwidth]{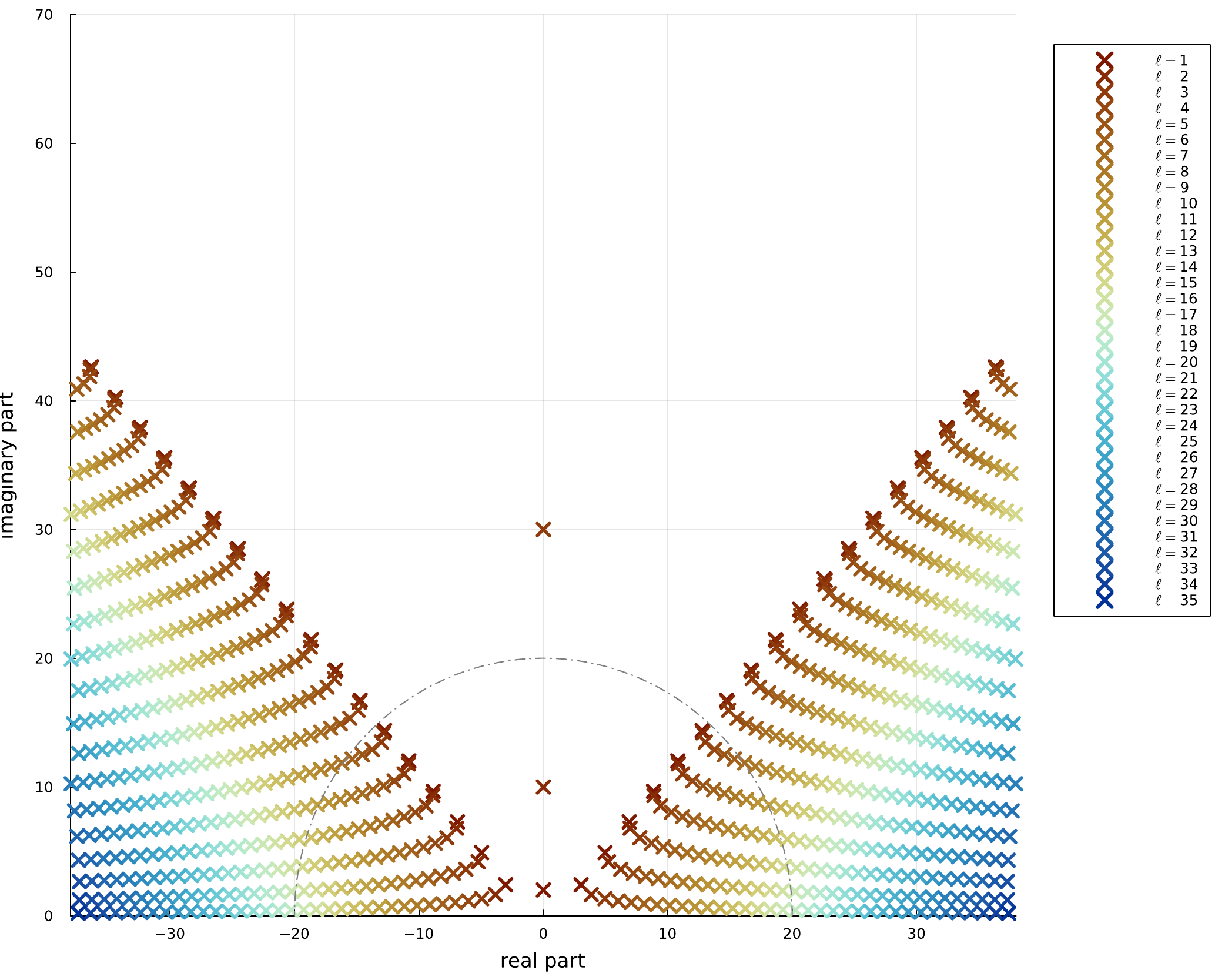}\label{QNF_AdS}
  }
  \subfloat[$N(R)$ for Schw.-AdS]{
    \includegraphics[clip,width=0.4\columnwidth]{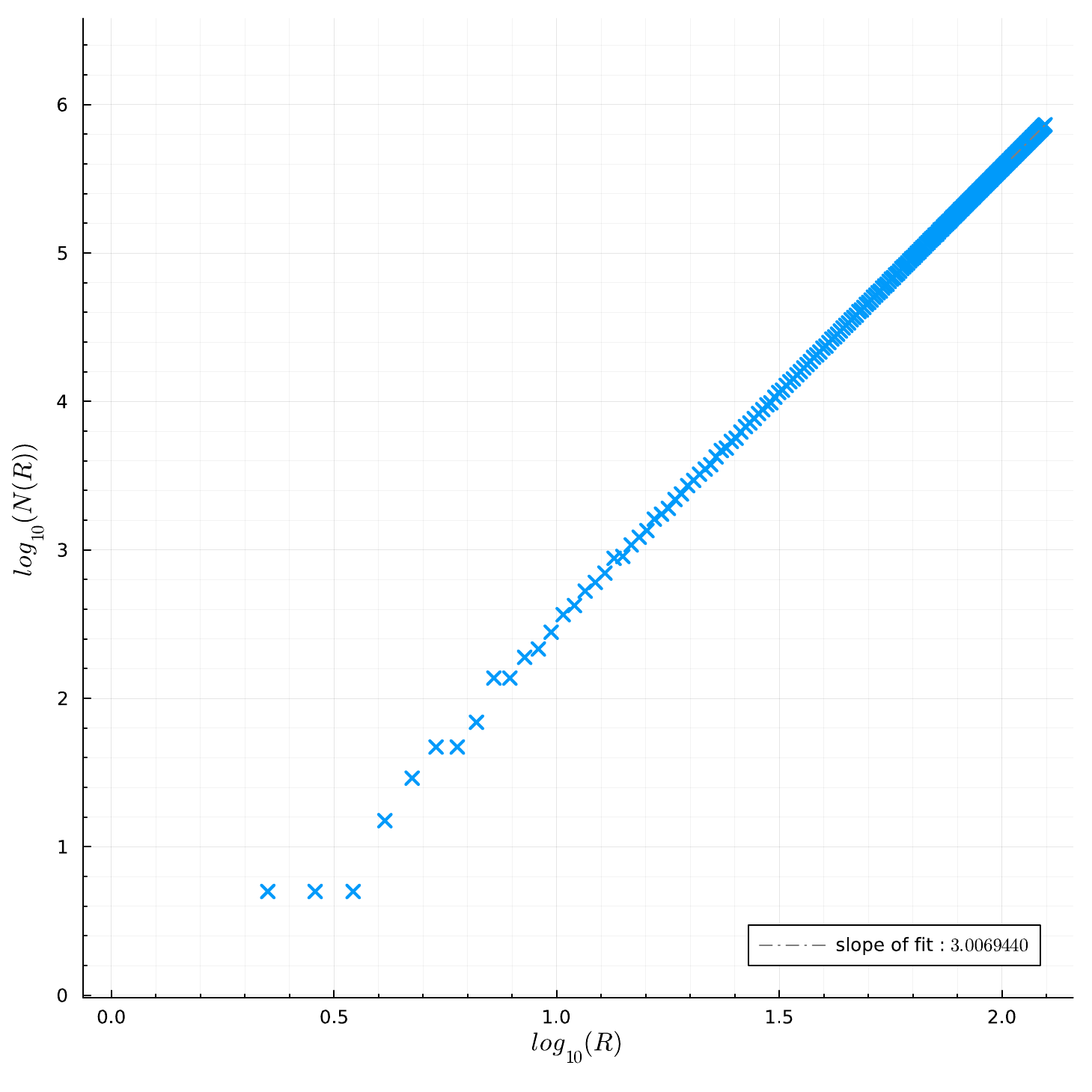}\label{weyl_AdS}
  }
  \caption{Panels \ref{QNF_S}, \ref{QNF_dS} and \ref{QNF_AdS} show the quasinormal frequencies for different quantum numbers $\ell$. Panels \ref{weyl_S}, \ref{weyl_dS} and \ref{weyl_AdS} count the number of modes $N(R)$ within a circle of radius $R$ centered at $0+0i$, these panels each have a total of 200 points. The linear fit is performed with a few dozens of points at the end of the series.}
  \label{weyl_fig}
\end{figure}
In reference \cite{Jaramillo:2022zvf} a Weyl's law is conjectured for the BH QNM overtone asymptotics.
Specifically,  given the QNM counting function $N(\omega)$, defined as
\bea
\label{e:counting_function_QNMs}
N(\omega) = \# \{ \omega_n \in \mathbb{C}, \hbox{ such that } |\omega_n|\leq \omega \} \ ,
\eea
it is conjectured that, for a $(d+1)$-dimensional BH spacetime, $N(\omega)$ satisfies
the Weyl's law
\bea
N(\omega) \sim \omega^d \quad , \quad \omega\to \infty \ ,
\eea
independently of the spacetime asymptotics. Although (semi-)analytical expressions suggest that this
behaviour is valid for generic spacetime asymptotics, the actual fact is that all numerical examples
considered in
\cite{Jaramillo:2022zvf} involve asymptotically flat spacetimes\footnote{It is worth mentioning
  that Ref. \cite{HitZwo24} provides a full proof of a related Weyl's law (involving QNM counting
  only in the angular quantum numbers) in the Schwarzschild-de Sitter case.} 
(actually Schwarzschild and Reissner-Nordstr\"om). 
In Fig. \ref{weyl_fig}  the Weyl's asymptotics $N(\omega)\sim \omega^3$
is numerically recovered for $3+1$-dimensional Schwarzschild,
Schwarzschild-de Sitter and Schwarzschild-Anti de Sitter. These results are, to our knowledge, 
the first factual evidence that the (full) conjecture holds true also in dS and AdS BH asymptotics.
Its validity strongly relies on the role of $\kappa$ in the overtone separation, something that
underlies the $\kappa$-band structure discussed in section \ref{s:Hp-pseudospectrum}
and is apparent in Figs. \ref{pseudospectrum_p_norm}. In our Keldysh discussion,
such $\kappa$-band structure justifies the
$a_{_{N_{\mathrm{QNM}}}} = \kappa N_{\mathrm{QNM}} + \mathrm{Im}(\omega_0)$ expression
for  $a_{_{N_{\mathrm{QNM}}}}$ in (\ref{e:u_Keldysh_v7}) in Eq. (\ref{e:u_Keldysh_discussion_C}).
The Weyl's law conjecture remains an open problem, but the numerical evidence here presented
strongly supports its universality with respect to
spacetime asymptotics.

\section{Conclusions}
\label{s:conclusions}
In this work we have discussed asymptotic QNM resonant expansions for
scattered fields on BH (stationary) spacetimes 
in an approach that makes key use of hyperboloidal foliations to
render the problem into a non-selfadjoint spectral setting where use can
be made of a so-called Keldysh expansion of the resolvent.

We have implemented the Keldysh approach to QNM expansions
in a set of BH spacetimes with different asymptotics and in the
toy-model provided by the P\"oschl-Teller potential. This extends
to the gravitational setting a tool that has been successfully applied
before in optics and mechanical problems~\cite{nicolet2023physically}.
The accuracy of the QNM expansions in the comparison with 
time-domain signals in this gravitational setting, where
dissipation occurs only at the boundaries, is remarkable. Results
in this article can be classified in:
\begin{itemize}

\item[i)] {\em Structural results}.  We have revisited 
the discussion of Keldysh QNM expansions in \cite{Gasperin:2021kfv},
clarifying the role of the scalar product. We conclude:
\begin{itemize}
\item[i.1)] {\em Keldysh QNM expansions only involve `dual space-pairing' notions}:
  specifically, the scalar product is not a structure needed for constructing
  Keldysh QNM expansions $u(\tau,x) \sim \sum_n {\cal A}_n(x) e^{i\omega_n\tau}$.
  In particular, no canonical choice of constant $a_n$
  in $u(\tau,x) \sim \sum_n a_n v_n(x) e^{i\omega_n\tau}$ (with $v_n$ QNM eigenfunctions) can be made, only the
  product ${\cal A}_n(x)=a_n v_n(x)$ being defined at this level.

\item[i.2)] {\em Keldysh QNM expansions generalise previous QNM expansion schemes}:
  they extend to arbitrary dimensions (and to more general spectral
  formulations of the scattering  problems) the BH QNM expansion by Ansorg \& Macedo.
  Likewise, it provides a spectral description of Lax-Phillips resonant expansions,
  adapted to the hyperboloidal scheme.

\item[i.3)] {\em Uniqueness of the QNM time-domain series at null infinity $\scri^+$}:
  the time-series $u(\tau) \sim \sum_n {\cal A}_n^\infty e^{i\omega_n\tau}$, where $\tau$ is the
  retarded time at null infinity, is
  obtained by straightforward evaluation of the Keldysh expansion at $\scri^+$, namely
  ${\cal A}_n^\infty={\cal A}_n(x_{\scri^+})$. This would correspond directly to idealised
  GW detector observational data.

\item[i.4)] {\em  Second-order QNMs in general relativity}: we have demonstrated the usefulness
  of the Keldysh expansion of the resolvent by calculating
  the correction to the QNM coefficients in second-order general relativity perturbation
  theory. This is presented with illustration purposes, being limited
  to overtones with a fixed $\ell$. The extension to the coupling of different
  $\ell$ modes is straightforward when  going beyond the one-dimensional case.

\item[i.5)] {\em Role of the scalar product to determine constant coefficients $a_n$}:
  the determination of constants $a_n$ requires a scalar product (actually, only a norm)
  to fix the ``size'' of QNM eigenfunctions $\hat{v}$. Then we can meaningfully
  write $u(\tau,x) \sim \sum_n a_n \hat{v}_n(x) e^{i\omega_n\tau}$. This is the closest
  form to standard self-adjoint normal mode expansions.

\item[i.6)] {\em Choice of scalar product in non-modal analysis: $H^p$-transient growths and $H^p$-pseudospectra}:
  the choice of scalar
  product (or, more generally, of non-degenerate
  quadratic form) is not unique and different choices can be of interest in different
  contexts. We have explicitly illustrated this feature by considering two particular tools in the non-modal
  analysis of the dynamics of non-normal systems, namely the growth function $G(\tau)$ and the
  $\epsilon$-pseudospectra, and applying them in P\"oschl-Teller to the  
  assessment of transient growths and the construction pseudospectra with different $H^p$-Sobolev norms.

\end{itemize}

\item[ii)] {\em Particular results in the black hole scattering.} We have demonstrated that
   the Keldysh approach to QNMs provides, in the BH case, an efficient and accurate
  scheme to calculate the QNM expansions and study (non-normal/non-selfadjoint) dynamics. In particular:
\begin{itemize}

\item[ii.1)] {\em Comparison of Keldysh QNM expansions and time-domain scattered field}.
  We have explicitly implemented the Keldysh expansion, studied the convergence of expansion
  coefficients $a_n$ and compared with the time-domain evolved field for a family
  of test-bed initial data. Results are remarkable even at early times.

  \item[ii.2)] {\em Keldysh recovery of Schwarzschild's tails}. As an unexpected result, we
  have found that the scheme works also beyond its `limit of validity'. Specifically,
  although the Keldysh expansion is only guaranteed to work when applied to discrete eigenvalues,
  we have found part that its `blind' application to (eigenvalues corresponding to finite-rank
  approximations of) the `branch cut' successfully construct the late power-law tails,
  correctly recovering the correct Price law.

\item[ii.3)] {\em Early evolution and overtones: convergence of the (boundary) QNM time-series and
  of (bulk) QNM asymptotic expansion}. We have clarified the role of two notions of convergence
  of QNM expansions and studied them for Gaussian test-bed initial data:
  \begin{itemize}
  \item[a)] {\em  Pointwise (uniform) convergence of the QNM time-series at fixed space $x_o$
    (e.g. the boundary)}. We have concluded that P\"oschl-Teller and dS asymptotics present good,
    indeed uniform, convergence properties from early $\tau_{\mathrm{init}}\gtrsim 0$ time,
    whereas the asymptotically flat and AdS cases are more
    difficult to assess and further research is needed to elucidate Ansorg \& Macedo conjecture
    $\tau_{\mathrm{init}}=\nu(x_o)$.
    Of potential interest for data analysis, a scheme has been sketched to choose the initial
    time for a valid QNM expansion of the observed time-series, for a given acceptable error and an available
    number of QNMs.
  \item[b)] {\em Convergence in the (bulk) norm of QNM asymptotic series at fixed time $\tau_o$}.
    Although the expansion of the scattered field in QNM functions at a given time $\tau_o$
    is generically divergent and only an asymptotic series, for particular classes of initial
    data the series can actually converge in the relevant Hilbert space. We have studied the
    case of the Gaussian initial data in the energy norm (similar results hold for $H^p$-Sobolev),
    finding the suggestive result that the (bulk) QNM series actually converge after the
    natural timescale $\tau_o\sim 1/\kappa$ of the problem.

  \end{itemize}

\item[ii.4)] {\em $H^p$-Sobolev (low-regularity) transient growths.} A main result in the considered
  non-selfadjoint dynamics setting is the existence of non-modal transient growths in the
  $H^p$-Sobolev norm. The here introduced QNM Keldysh expansion then reveals that the
  corresponding optimal excitation $u^{\mathrm{max}}_0(\tau,x)$ is built, for a given $H^p$-norm, by the
  constructive interference of the two highest $H^p$-QNMs (in the sense of Warnick) $v_p^+(x)$
  and $v_p^-(x)$. Interestingly,
  the product of the height and the time at the peak, namely
  $||u^{\mathrm{max}}_0(\tau_{\mathrm{max}})||_{H^p}\cdot \tau_{\mathrm{max}}$, does not depend on $p$, in particular
  indicating that the transient growth $G(\tau)$, when $p\to\infty$, has a delta-like structure in time,
  i.e. $\lim_{p\to\infty}G(\tau) \sim \delta(\tau)$.

\item[ii.5)] {\em High overtones and Weyl's law for BH spacetimes with different spacetime asymptotics}.
  Given the relevant role of high overtones at early times, we have provided a qualitative
  test of their distribution in the complex plane by calculating their Weyl's law.
  We have found, through a straightforward numerical evaluation,
  that the QNM counting function presents the correct power-law
  $N(\omega)\sim \omega^3$ independently of the spacetime asymptotics.
  
  \end{itemize}
\end{itemize}

\subsection{Future prospects}
\label{s:future}
In this work we have remained at a proof-of-principle level, therefore a systematic extension
and deepening of the presented work must be done. Some directions for such an extension
are along  the following points:

\begin{itemize}
\item[i)] {\em Systematic study of generic initial data}. All the results here presented
  make use of the same Gaussian initial data. One imperative need is that of studying
  systematically larger classes of initial data, in particular realistic ones in
  physical scenarios.

\item[ii)] {\em QNM expansions for generic pencils: null foliations and
quadratic pencils}. The Keldysh expansion is valid
  for a very general class of spectral problems. Here we have implemented the case
  corresponding to the standard eigenvalue problem (P\"oschl-Teller, Schwarzschild,
  Schwarzschild-dS) and the generalised one (Schwarzschild-AdS). Regarding the latter,
  the adaptation to the operators in the null foliation of \cite{cownden:2024pseudospectra,Chen_2024}
  is a natural step, before extending the null slicing treatment to other
  spacetime asymptotics. Another natural extension is to the quadratic pencils discussed
  in \cite{Warnick:2024}.

\item[iii)] {\em  Assessment of the series convergence: ``boundary'' time-series (constant $x_o$)
  and ``bulk'' QNM  expansion  (constant $\tau_o$).}
  In subsection \ref{s:role_overtones} we have discussed both the pointwise 
  convergence of the QNM time-series at a given $x_o$, with an interest in exploring
  its possible uniform convergence nature and its earliest time of validity (namely assessing
  Ansorg \& Macedo proposal $\tau_{\mathrm{init}}\sim \nu(x_o)$),
  as well as the convergence in the Hilbert space norm of the QNM series at a given $x_o$,
  where we need to control the growth $N_{\mathrm{QNM}}$ in the constant $C(N_{\mathrm{QNM}}, L)$.
  However a systematic extension of this preliminary work, in particular
  relaxing the conditions on the  choice of initial data, is needed. Details will be given in
  \cite{BesJarPoo24}.

\item[iv)] {\em Non-modal analysis: $H^p$-pseudospectra and $H^p$-transient growths}.
  Key aspects of our discussion depend on the choice of scalar product and its
  associated norm (e.g. excitation coefficients in section
  \ref{s:regularity_excitation_coefficients} or QNM convergence in subsection \ref{s:convergence-fixed_tau}).
  In particular, we have implemented $H^p$-Sobolev scalar products to probe regularity aspects
  in the non-modal analysis of the discussed non-normal dynamics. A systematic exploration
  of the `definition versus stability' QNM problem through the use of $H^p$-pseudospectra will be presented
  in \cite{BesBoyJar24}, whereas a detailed account of $H^p$-transient growths will
  be developed in \cite{BesBizJar25}.

\item[v)] {\em QNM expansions and perturbed potentials.}  QNM frequencies
  migrate to new branches in the complex plane in the presence of (ultraviolet)
  perturbations. This fact changes completely the set of QNM frequencies and eigenfunctions
  on which the QNM expansion is constructed, although the potentials are very similar, leading
  to the notion of $\epsilon$-dual QNM expansions introduced in \cite{Gasperin:2021kfv}.
  The tools here discussed permit to address this point.
  
\item[vi)] {\em QNM Keldysh expansion in the non-diagonalisable case.}
  Our discussion has been restricted to the case in which $L$ is diagonalisable.
  The Keldysh expansion extends to the non-diagonalisable case involving Jordan
  decompositions. This makes appear
  power-exponential terms in the time dependence, namely $t^k e^{i\omega_n t}$
  were $k$ runs from zero to the algebraic multiplicity of $\omega_n$ (minus one).
  This recovers the general Lax-Phillips expansion~\cite{dyatlov2019mathematical}.
  This construction is needed in situations when QNM
  branching may occur.

\item[vii)] {\em Beyond the one-dimensional case.} The Keldysh approach generalises
  to higher-dimensions the one-dimensional algorithm in \cite{Ansorg:2016ztf} for constructing
  the QNM (and branch) expansion. However, {\em all} examples here discussed involve effective
  one-dimensional problems, i.e. they can all be treated with the tools introduced
  by Ansorg \& Macedo. A genuine higher-dimensional case is needed to
  test the reach of the Keldysh method  presented here.

\item[viii)]  {\em High overtones at early times}.  A better understanding of the
  contributions of highly-damped QNMs is crucial to assess the early
  behaviour of the signal. In the considered non-selfadjoint setting this translates
  in the integration of several points: convergence issues and earliest time
  of validity of the QNM expansion, transient growths, second-order corrections
  to excitation coefficients and QNM asymptotics including the Weyl's law.

\item[ix)] {\em Structure of the eigenfunctions $v_n$'s and $\alpha_n$'s.} The {\em a priori}
  control of the Keldysh QNM expansion, in particular in its action on given initial data/external sources,
  requires a good understanding of the qualitative
  and quantitative structure of the QNM eigenfunctions.

\item[x)] {\em QNM expansion and late time BBH waveform data}. For methodological and presentation reasons,
  the present work has mostly remained at the discussion of the formal aspects of the
  problem. An application to the study of the  observational signals from BBH mergers in the setting of the
  BH spectroscopy program is the subject of current research.

\end{itemize}

\appendix
\addtocontents{toc}{\fixappendix}

\section{Explicit expressions of asymptotic Keldysh QNM expansions}
\label{a:easy_access_expressions}
In order to ease the access for the implementation to the relevant material,  we collect
here the relevant expressions for the  QNM asymptotic expansions presented in this work.

\begin{itemize}
\item[i)] {\em Keldysh asymptotic QNM expansion}:

  Given the standard spectral problems for the infinitesimal generator of time $L$
  \bea
       \label{e:eigen_L-Lt_conclusions}
       L v_n = \omega_n v_n \ \ , \ \  L^t \alpha_n = \omega_n \alpha_n
       \ \ , \ \ v_n\in{\cal H}, \alpha_n\in{\cal H}^* \ ,
       \eea
       we can write the scattered field evolved from initial data $u_0$ as
       \bea
   \label{e:Keldysh_QNM_expansion_conclusions}
   u(\tau,x) &\sim& \sum_n e^{i\omega_{n}\tau} a_n v_n(x) \ ,
   \eea
   with
   \bea
   a_n =
   \left\{\begin{array}{lcl}
   \displaystyle
   \frac{\langle\alpha_n, u_0\rangle}{\langle\alpha_n,
     v_n\rangle}  \ \  &,& \ \  \hbox{no relative scaling between } \alpha_n  \hbox{ and } v_n \\
   \langle\alpha_n, u_0\rangle \ \  &,&  \ \ \hbox{with} \
   \ \ \langle\alpha_n, v_n\rangle=1    
   \end{array}
   \right. \ ,
   \eea
   Keldysh expansion in invariant under: $v_n\to f v_n$, $a_n\to 1/f a_n$. Therefore $a_n$'s are not intrinsically
   defined.

   \medskip
   
   These expressions extend to the general eigenvalue problems 
\bea
       \label{e:generalized_eigen_L-Lt_conclusions}
       L v_n = \omega_n B v_n \ \ , \ \  L^t \alpha_n = \omega_n B^t\alpha_n
       \ \ , \ \ v_n\in{\cal H}, \alpha_n\in{\cal H}^* \ ,
       \eea
       we can write the scattered field evolved from initial data $u_0$ as
       \bea
   \label{e:Keldysh_QNM_expansion_conclusions}
   u(\tau,x) &\sim& \sum_n e^{i\omega_{n}\tau} a_n v_n(x)
   \eea
   with
   \bea
   a_n =
   \left\{\begin{array}{lcl}
   \displaystyle
   \frac{\langle\alpha_n, B u_0\rangle}{\langle\alpha_n,
     B v_n\rangle}  \ \  &,& \ \  \hbox{no relative scaling between } \alpha_n  \hbox{ and } v_n \\
   \langle\alpha_n, B u_0\rangle \ \  &,&  \ \ \hbox{with} \
   \ \ \langle\alpha_n, B v_n\rangle=1    
   \end{array}
   \right.
   \eea

 \item[ii)] {\em Hyperboloidal Lax-Phillips QNM expansion,  `\`a la Keldysh'}.

   The previous expressions can we written as:
   \bea
    \label{e:Lax_Phillips_QNM_expansion_conclusions}
   u(\tau,x) \sim \sum_n{\cal A}_n(x)  e^{i\omega_{n}\tau}  \ \ \ , \ \ \ \hbox{with} \ {\cal A}_n(x) = a_n v_n(x) \ .
   \ \ 
   \eea
   
 \item[iii)] {\em QNM time-series at null infinity}.

   Evaluating the previous expressions at null infinity:
   \bea
   \label{e:Lax_Phillips_QNM_time-series_conclusions}
    u(\tau) =  u(\tau,x_{\scri^\infty}) \sim \sum_n{\cal A}^\infty_n  e^{i\omega_{n}\tau}  \ \ \ , \ \ \ \hbox{with} \
    {\cal A}^\infty_n= {\cal A}_n(x_{\scri^\infty}) = a_n v_n(x_{\scri^\infty}) \ .
   \ \ 
   \eea

 \item[iv)] {\em Keldysh asymptotic QNM expansion, with coefficient $a^{_G}_n$ fixed by scalar
   product $\langle \cdot, \cdot \rangle_{_G}$}.

   Given the additional structure provided by scalar product $\langle \cdot, \cdot \rangle_{_G}$,
   with associated norm $||\cdot ||_{_G}$ given by $||v||^2_{_G} = \langle v, v \rangle_{_G}$,
   and considering the spectral problems of $L$ and its adjoint $L^\dagger$ in this scalar product
\bea
       \label{e:eigen_L-Ldagger_conclusions}
       L \hat{v}_n = \omega_n \hat{v}_n \ \ , \ \  L^\dagger \hat{w}_n = \bar{\omega}_n \hat{w}_n
       \ \ , \ \ \hat{v}_n, \hat{w}_n\in{\cal H} \ .
       \eea
       with  $||\hat{v}_n||_{_G} = ||\hat{w}_n||_{_G}= 1$, we can write
        \bea
       \label{e:Keldysh_scalar_product_conclusions}
       u(\tau,x) \sim
        \sum_n e^{i\omega_{n}\tau} a^{_G}_n  \hat{v}_n(x) \ ,
       \eea
       with the coefficient $a^{_G}_n$ now fully determined with this choice of scalar product
       $\langle \cdot, \cdot \rangle_{_G}$
       \bea
       a^{_G}_n =  \kappa_n \langle \hat{w}_n, u_0\rangle_{_G}
       \eea
       where the condition number $\kappa_n$ is given by
       \bea
       \kappa_n = \frac{||w_n||_{_G} ||v_n||_{_G}}{\langle w_n, v_n\rangle_{_G}} \ .
       \eea

\end{itemize}

\section{Some technical elements of Keldysh expansions: matrix case}
\label{a:Keldysh_finite_rank}
In this appendix we collect certain technical points relevant for the
discussion of the main text but dwelling in the finite
rank (matrix)  case.

\subsection{Bi-orthogonal systems and Keldysh QNM expansion}\label{s:bi-orthogonal_systems}
A bi-orthogonal system is given by two families of vectors
$\{x_1, \ldots, x_K\}$ and $\{y_1, \ldots, y_K\}$, such that $B(x_i, y_j)=\delta_{ij}$
for some bilinear map $B(\cdot, \cdot)$.
If the families $\{x_i\}$ and $\{y_i\}$ are bases, they constitute bi-orthonormal bases. The latter
play a very important role in non-Hermitian quantum mechanics  \cite{moiseyev2011}
or (quantum) optical cavities.

Bi-orthonormal systems also enter naturally
in the presented Keldysh QNM expansion construction. In the diagonalisable case, eigenvectors
of the  spectral problems (\ref{e:right-left_eigenvalues}) (or
(\ref{e:eigen_L-Lt_conclusions})) form bi-orthogonal systems
(that are not necessarily bi-orthogonal bases, due to the convergence issues
in the infinite dimensional case we have discussed). The construction can however
been extended to the non-diagonalisable case, where the eigenvalue problems
must be completed with `Jordan chains', but still an appropriate
bi-orthonormal system formed by eigenvectors and `associated vectors'
(see Theorem 1.5.9 and more generally, section 1.9 in \cite{MenMol03}) can be built.
Once the bi-orthogonal systems are constructed from the spectral
problem, the key ingredient in our scheme is their use in the  Keldysh
expansion of the resolvent in Eq. (\ref{e:gen_resolvent_F_normalized}),
which is the real starting point of our QNM expansion construction.

In order to gain an insight into this construction of bi-orthogonal systems
in terms of our spectral problems, we discuss the finite-rank diagonalisable
case.
We start by rewriting the eigenvalue problems (\ref{e:right-left_eigenvalues}) or
(\ref{e:eigen_L-Lt_conclusions}), by introducing the matrix $V$ and $A$
of right- and left-eigenvectors, respectively, as ($v_i$'s and the $\alpha_i$'s are understood as ``column'' vectors)
\bea
\label{e:VA-matrices}
V = (v_1|\ldots|v_K) \qquad , \qquad  A = (\alpha_1|\ldots|\alpha_K) \ ,
\eea
and the diagonal matrix $\Omega$ of eigenvalues $\Omega = \diag(\omega_1, \ldots, \omega_N)$,
so that the respective eigenvalue problems write (note that $V^{-1}$ exists  since $L$ is diagonalisable)
\bea
L\cdot V = V\cdot \Omega \qquad &\Longleftrightarrow& \qquad  V^{-1} \cdot L = \Omega\cdot V^{-1}   \nn \\
L^t\cdot A = A \cdot \Omega \qquad &\Longleftrightarrow& \qquad  A^t \cdot L = \Omega \cdot A^t \ . 
\label{matrix_spectral_pb_simple}
\eea
From the expressions on the right, we can write the left-eigenvector $\alpha_i$ ($i$-th line vectors in $A^t$)
as proportional to the $i$-th line vector in $ V^{-1}$, with ``normalization'' constant $m_i$.
Defining the normalisation diagonal matrix $M=\diag(m_1, \ldots, m_K)$, we can write $A^t = M\cdot V^{-1}$ so
\bea
\label{eq:matrix_M}
A^t \cdot V = M \qquad  &\Longleftrightarrow& \qquad  \langle \alpha_i, v_j \rangle = m_i \delta_{ij} \ ,
\eea
that recovers the bi-orthogonal relations  in section \ref{s:Keldysh_resolvent}, where
the $m_i$'s essentially provide the normalisation (\ref{e:normalization}) 
encoding the degree of freedom for the rescaling in (\ref{e:rescaling_alpha}).
Note also that `left-eigenvectors' $\alpha_i$'s in (\ref{e:VA-matrices}) are straightforwardly
constructed from `right-eigenvectors' $v_i$'s in  (\ref{e:VA-matrices}) through
a simple matrix inversion and transposition (modulo a trivial rescaling)
\bea
A =  (V^{-1}) ^t \cdot M \ ,
\label{eq:matrix_alphas}
\eea
that explains by itself the appearance bi-orthogonal relations.
This reasoning does not work in the
non-diagonalisable case but, as pointed out above, bi-orthogonal systems can also be constructed 
in the non-diagonalisable case for eigenvectors and their (Jordan chains) associated vectors
and, crucially, the resolvent can be written in terms of the resulting bi-orthogonal
systems~\cite{MenMol03,Beyn12,BeyLatRot12}, the key point for the construction
of resonant QNM expansions.

\subsection{Scalar product and Keldysh QNM expansions}
\label{a:Keldysh_adjoint_transpose}
Here we justify the expressions presented in section \ref{s:QNM_expansions_scalar_product}
relating the standard Keldysh expansion based on the resolvent of $L$~\cite{MenMol03,BeyLatRot12,nicolet2023physically},
formulated in the terms of the transpose operator $L^t$, and the version in terms of the adjoint
operator $L^\dagger$ when a scalar product structure is provided~\cite{Gasperin:2021kfv}. For the sake
of clarity, we dwell
again in the matrix (finite-rank) case.

The two sets of spectral problems are [cf. Eqs. (\ref{e:eigen_L-Lt}) and (\ref{e:eigen_L-Ldagger})]
 \bea
       \label{e:eigen_L-Lt_Ldagger}
       &&L v_n = \omega_n v_n \ \ , \ \  L^t \alpha_n = \omega_n \alpha_n
       \ \ , \ \ v_n\in{\cal H}, \alpha_n\in{\cal H}^* \ , \nn \\
       &&L v_n = \omega_n v_n \ \ , \ \  L^\dagger w_n = \ol{\omega}_n w_n
       \ \ , \ \ v_n, w_n\in{\cal H} \ ,
       \eea
       where now ${\cal H}$ and ${\cal H}^*$ are finite-dimensional complex spaces $\mathbb{C}^N$.
         To fix notation, we write
         \bea
         \label{e:dual_pairing_app}
         \langle \alpha, v\rangle = \alpha(v) = \alpha^t\cdot v \ ,  \qquad v\in {\cal H}, \alpha\in {\cal H}^*
         \eea
         for the natural dual pairing, where $\alpha^t$ is the row-vector transpose to the column-vector
         $\alpha$ and
         \bea
          \label{e:scalar_product_app}
         \langle w, v\rangle_{_G} = w^*\cdot G \cdot v = \ol{w}^t \cdot  G \cdot v \ ,  \qquad v, w \in {\cal H} \ ,
         \eea
         for the matrix expression of the (Hermitian) scalar product $\langle \cdot, \cdot \rangle_{_G}$,
         where  $w^*=\ol{w}^t$ is the  complex-transpose of $w$ (analogously, for a matrix, $A^*=\ol{A}^t$)
         and $G$ is a Hermitian $G^*=G$ (in particular, in the cases we discuss, a real symmetric)
         positive-definite  matrix,
         referred to as the Gram matrix
         of $\langle \cdot, \cdot \rangle_{_G}$ when a basis of ${\cal H}$ is chosen.

         \begin{itemize}

         \item[i)] {\em $G$-mapping between ${\cal H}$ and ${\cal H}^*$}. We recall that in the absence
           of scalar product (more generally, of a non-degenerate quadratic form) there is no canonical
           mapping between ${\cal H}$ and ${\cal H}^*$. The additional structure given by a scalar product
           $\langle \cdot, \cdot \rangle_{_G}$  permits to introduce the mapping $\Phi_G: {\cal H}\to  {\cal H}^*$
           where its action on $v\in {\cal H}$,  namely $\Phi_G(v)\in {\cal H}^*$, is defined by
           \bea
           \Phi_G(v)(w) = \langle v, w \rangle_{_G} = \ol{\langle w, v \rangle}_{_G} \ \ , \ \ \forall w\in {\cal H} \ .
           \eea
            In the finite-rank case this
           mapping is invertible, with inverse $(\Phi_G)^{-1}: {\cal H}^*\to  {\cal H}$, 
           in such a way that, at this matrix level, it holds
           \bea
           \Phi_G(v) = \overline{G\cdot v} \ \ , \ \ (\Phi_G)^{-1}(\alpha)
           =  G^{-1}\cdot \overline{\alpha}\ \ , \ \ 
           \forall v\in {\cal H}, \alpha\in {\cal H}^* \ .
           \eea
           Such $\Phi_G$ and $ (\Phi_G)^{-1}$ are just the standard ``musical isomorphisms'' between
           a linear space and its dual, provided by the non-degenerate quadratic form defined by $G$,
           namely
           \bea
           \Phi_G(v)=v^\flat \ \ , \ \ (\Phi_G)^{-1}(\alpha) = \alpha^\sharp \ \ , \ \ 
           \forall v\in {\cal H}, \alpha\in {\cal H}^* \ .
           \eea
           
         \item[ii)] {\em  Relation between $L^t$ and $L^\dagger$}. Given an operator $L:{\cal H}\to {\cal H}$,
           the relation between its 
           transpose $L^t:{\cal H}^*\to {\cal H}$ and its
           formal adjoint $L^\dagger:{\cal H}\to {\cal H}$, follows directly from their respective definitions, namely
           \bea
           \langle L^t\alpha , v\rangle = \langle \alpha , L v\rangle \ \ , \ \ 
           \langle L^\dagger v , w\rangle_{_G} = \langle v, L w\rangle_{_G} \ \ \forall v,w\in  {\cal H} \ ,
           \alpha\in  {\cal H}^* \ ,
           \eea
           by making use of the ``musical isomorphisms'',  so it holds
           \bea
           L^\dagger = (\Phi_G)^{-1} \circ L^t\circ  \Phi_G \ \ , 
           \eea
           or, in matrix language (cf. e.g. \cite{Jaramillo:2020tuu})
           \bea
           L^\dagger = G^{-1}\cdot L^* \cdot G \ \ .
           \eea
         \item[iii)] {\em  Relation between $L^t$ and $L^\dagger$  eigenfunctions}. For a given
           eigenvalue $\omega_n$ and its conjugate $\bar{\omega}_n$,
           the corresponding eigenvectors $\alpha_n$ and $w_n$ in
           (\ref{e:eigen_L-Lt_Ldagger}),
           respectively of $L^t$ and $L^\dagger$, relate as
           \bea
           \label{e:a_n-w_n}
           \alpha_n &=& \Phi_G(w_n) = \overline{G\cdot w_n} \nn \\
           w_n  &=& (\Phi_G)^{-1}(\alpha_n) = G^{-1}\cdot \ol{\alpha}_n \ .
           \eea
           The first relation recovers expression (\ref{e:alpha_n-w_n}) used in section
           \ref{s:QNM_aymptotic_expansions}. In order to show, for instance, the first relation
           we start from the eigenvalue problem for $L^\dagger$ in (\ref{e:eigen_L-Lt_Ldagger}), then
           \bea
           &&L^\dagger w_n = \ol{\omega}_n w_n \nn \\
           &&\left(G^{-1}\cdot \ol{L}^t \cdot G\right)\cdot w_n = \ol{\omega}_n\; w_n \nn \\
           &&\ol{L}^t \cdot (G \cdot w_n) = \ol{\omega}_n \; (G\cdot w_n) \nn \\
           &&L^t \cdot (\overline{G \cdot w_n}) = \omega_n \;  (\overline{G \cdot w_n}) \ .
           \eea
           Comparing with the eigenvalue problem for $L^t$ (and assuming simple eigenvalues for simplicity),
           we conclude $\alpha_n =\overline{G \cdot w_n}= \Phi_G(w_n)$. The other relation
           in (\ref{e:a_n-w_n}) follows.

         \item[iv)] {\em  Dual and scalar-product projections, coefficients of the Keldysh
           expansion}. Given $\alpha_n$ and $w_n$ eigenvectors of $L^t$ and $L^\dagger$, for
           $\omega_n$ and $\ol{\omega}_n$ respectively, and $v\in {\cal H}$, then it holds
           \bea
           \langle w_n, v\rangle_{_G} = \langle \alpha_n, v\rangle \ .
           \eea
           Indeed, we can write
           \bea
           \langle w_n, v\rangle_{_G} &=& \ol{w}_n^t\cdot G \cdot v
           = \ol{(G^{-1}\cdot\ol{\alpha}_n)}^t\cdot G \cdot v\nn \\
           &=& \alpha_n^t\cdot \left(\ol{G}^t\right)^{-1}\cdot G \cdot v 
           =  \alpha_n^t\cdot (G^*)^{-1}\cdot G \cdot v \nn \\
           &=& \alpha_n^t\cdot G^{-1}\cdot G \cdot v
           = \alpha_n^t\cdot v = \langle \alpha_n, v\rangle \ ,
           \eea
           where in the second equality we have used relation (\ref{e:a_n-w_n}) between
           $w_n$ and $\alpha_n$ and in the fifth equality we have used the Hermitian
           nature of $G$. This relation recovers expression  
           (\ref{e:scalar_product-dual}) in section \ref{s:QNM_aymptotic_expansions},
           thus permitting to conclude the validity of the two expressions
           for $a_n$ in (\ref{e:coeff_an_scalar-dual_projections}) and therefore the
           equivalence between the standard Keldysh expansion (defined purely in terms
           of duality notions) and the version in \cite{Gasperin:2021kfv} (using a scalar
           product).

         \end{itemize}

\subsection{Dynamics from the (exponentiated) evolution operator}\label{a:Keldysh_evol_op}
In the literature \cite{BoyCarDes22,destounis2024black,Destounis:2021lum,sarkar23} it is found
the formal solution $u(\tau, x)$ of the time evolution problem (\ref{e:wave_eq_1storder_u}),
in terms of the (formal) evolution operator $e^{iL\tau}$, expressed as
\begin{equation}
u(\tau, x)=e^{iL\tau}u_0(x) \ .
\end{equation}
However, this expression raises the issue (cf. e.g. \cite{trefethen2005spectra}) underlying the meaning
of the exponentiation of the operator $L$. If $L$ is a bounded operator we can make sense of
the exponential as a convergent power series. On the contrary,  if $L$ is unbounded,
we must resort to the theory of semigroups to define such exponential. In the footsteps
of appendix \ref{s:bi-orthogonal_systems}, in this appendix 
we consider instead the simpler matrix  case corresponding to the finite-rank approximants
for the operator $L$. In particular we show that, in the diagonalisable case we have
focused on along the whole discussion (but this point can be generalised),
the spectral treatment of the evolution
problem straightforwardly recovers the very same expressions in the Keldysh scheme,
but now applied to all eigenvalues
of the  finite-rank approximant, not only to those ones converging to QNM frequencies in
the $N\to\infty$ limit.

In other words, the forthright  application of the
expressions in the Keldysh scheme to {\em all} the eigenvalues of the matrix approximant of $L$
does recover the dynamics from the evolution operator.
This fact justifies the presence of polynomial tails when applying the scheme
to the appropriate subset of eigenvalues (cf. section \ref{s:S_tails}).
We discuss this finite-rank case.

As indicated above, we assume the matrix $\w L$ to be diagonalisable and,
following the notation in appendix \ref{s:bi-orthogonal_systems}, we
write $\w L=\w V \cdot \w \Omega \cdot\w V^{-1}$  with $\w \Omega =\text{diag}(\omega_1,...,\omega_{K})$,
where  the set $\{\omega_i\}$ is now formed by all the eigenvalues of the finite-rank approximant of
$L$, and not only those converging to QNM frequencies.
The finite-rank approximation to the solution of the evolution problem can then be written
(as a ``column vector'') as
\begin{equation}
  \label{e:solution_expL}
  \w u(\tau)=e^{i\w L\tau} \cdot \w u_0 \ ,
\end{equation}
 in terms of the evolution operator $e^{i\w L\tau}$ and, using the diagonalisability of $L$, we have
 \begin{equation}
  \w u(\tau)=\w V \cdot e^{i \w \Omega \tau} \cdot  \w V^{-1} \cdot  \w{u_0} \ .\label{evol_op_arg}
\end{equation}

In the notation of  appendix \ref{s:bi-orthogonal_systems}, assuming without loss of generality
(through appropriate rescalings) that
$\langle \w \alpha_i,\w v_j \rangle =\w \alpha_i^t\cdot\w v_j= \delta_{ij}$, i.e.
$M=\mathrm{I}_{K}$ in Eqs. (\ref{eq:matrix_M}) and (\ref{eq:matrix_alphas}), we write
\bea
\label{e:VA-matrices_Id}
V = (v_1|\ldots|v_K) \ , \qquad  A =\left(\w V^{-1} \right)^t = (\alpha_1|\ldots|\alpha_K) \ ,
\eea
where, as in Eq. (\ref{e:VA-matrices}), $\w v_i$ and $\w \alpha_i$ are respectively (column) eigenvectors of
$\w L$ and $\w L^t$, but again not restricted to QNM eigenvalues.
Then it follows by direct calculation (see below)
\begin{equation}
  u(\tau)= \w V \cdot e^{i\w \Omega \tau} \cdot \w V^{-1} \cdot \w{u_0} = \sum_{n=1}^{K} \frac{\langle \w \alpha_n,\w{u_0}\rangle}{\langle \w \alpha_n,\w{v_n} \rangle}e^{i\omega_n\tau}\w{v_n} \label{evol_res} \ ,
\end{equation}
that exactly coincides with the expression in the Keldysh QNM expansion
(\ref{e:Keldysh_QNM_expansion_u_scale_invariant}),
but now not corresponding to an asymptotic (infinite) series involving only QNM eigenvalues, but to a finite
sum over {\em all} eigenvalues of the matrix approximant to the operator $L$.

For the sake of clarity we present now the calculation details.
First, the normalization condition arises from the characterisation (\ref{e:VA-matrices_Id})
of $A$ in terms of the inverse of $V$, that is
\bea
\label{e:eq_just1}
\delta_{ij} &=& (\w{V^{-1}} \cdot \w{V})_{ij} = \sum_{k=1}^K (\w V^{-1})_{ik}(\w V)_{kj}
= \sum_{k=1}^K \left(\left(\w V^{-1}\right)^t\right)_{ki}(\w V)_{kj}\nn \\
&=& \sum_{k=1}^K \left(\w A\right)_{ki}(\w V)_{kj} = \sum_{k=1}^K (\w{\alpha}_i)_{k}(\w{v}_j)_{k}
=  \w{\alpha}^t_i \cdot \w{v}_j = \langle \w{\alpha}_i,\w{v}_j \rangle \ ,
\eea
thus recovering the bi-orthogonal relations in the chosen normalisation. 

On the other hand 
\bea
(\w V^{-1} \cdot \w{u_0})_i &=& \sum_j \left(\w V^{-1}\right)_{ij}(\w{u_0})_j
=\sum_{j=1}^K \left(\left(\w V^{-1}\right)^t\right)_{ji}(\w{u_0})_j
=\sum_{j=1}^K (\w{\alpha_i})_{j}(\w{u_0})_j \nn \\
&=&\w{\alpha_i}^t \cdot \w{u_0}
=\langle \w{\alpha_i},\w{u_0} \rangle \ ,
\eea
and then, acting in this with $e^{i\w \Omega\tau}$, we have 
\bea
\left(e^{i\w \Omega\tau}\w \cdot V^{-1} \cdot \w{u_0}\right)_k = \sum_{i=1}^K \left(e^{i\w \Omega\tau}\right)_{ki}(\w V^{-1} \cdot \w{u_0})_i
=\sum_{i=1}^K e^{i\omega_i\tau}\delta_{ki}\langle \w{\alpha_i}, \w{u_0}\rangle
=e^{i\omega_k\tau}\langle \w{\alpha_k}, \w{u_0}\rangle \,
\eea
and finally we can write
\bea
\label{e:eq_last}
  (\w V  \cdot e^{i\w \Omega\tau} \cdot \w V^{-1} \cdot \w{u_0})_\ell &=& \sum_{=1}^K (\w V)_{\ell k}(e^{i\w \Omega\tau} \cdot \w V^{-1} \cdot \w{u_0})_k
=\sum_k (\w{v_k})_\ell (\w{\alpha_k}^t \cdot \w{u_0}) e^{i\omega_k\tau} \nn \\
&=&\sum_k e^{i\omega_k\tau}\langle \w{\alpha_k}, \w{u_0}\rangle \left(\w{v_k}\right)_\ell 
=\sum_k e^{i\omega_k\tau}\frac{\langle\w{\alpha_k}, \w{u_0}\rangle}{\langle \w{\alpha_k},\w{v_k} \rangle}\left(\w{v_k}\right)_\ell \nn \\
&=&\sum_k e^{i\omega_k\tau}\w{\mathcal{A}}_k(x_\ell) \ .
\eea
These relations lead  to Eq. (\ref{evol_res}) (in the
second equality of the second line in (\ref{e:eq_last}) we have
reintroduced the denominator to allow for an arbitrary rescaling
in the final expression).

\paragraph{Evolution problem associated with the generalised eigenvalue problem.}
We briefly comment on the same question, but in the context of the more
general evolution problem in Eq. (\ref{e:wave_eq_1storder_u_tau_generalised})
and, in particular, the discussion in point v) of the Remarks in
section \ref{s:QNM_aymptotic_expansions}. Specifically we consider the evolution problem
with a ``mass'' matrix $B$
\begin{equation}
  B\partial_\tau u =iLu\ , \qquad u(\tau=0,x)=u_0(x) \,\ ,
 \label{dynamical_pb_evolution_operator}
\end{equation}

at the matrix level, that leads to the generalised eigenvalue problems
\bea
\label{spectral_gen_decomp}
\left\{
\begin{array}{rcl}
    \w L \cdot \w v_n &=& \omega_n \w B \cdot \w v_n\\
    \w L^t \cdot \w \alpha_n &=& \omega_n \w B^t \cdot \w \alpha_n \ ,
\end{array}
  \right. 
  \eea
  that we rewrite as
  \bea
  \label{e:eq1}
  \left\{
  \begin{array}{rcl}
  L\cdot V &=& B \cdot V\cdot \Omega \nn \\
  L^t\cdot A &=& B^t \cdot A \cdot \Omega
  \end{array}
  \right. 
  \eea
  Then the solution to (\ref{dynamical_pb_evolution_operator})
  can still be written, in the diagonalisable case, as in Eq. (\ref{evol_op_arg}) 
  \bea
  \label{e:solution_B}
  \w u(\tau)= u(\tau)=  e^{iV\cdot\Omega\cdot V^{-1}\tau}\cdot \w{u_0} =\w V \cdot e^{i \w \Omega \tau} \cdot  \w V^{-1} \cdot  \w{u_0} \ ,
  \eea
  though note that now $V\cdot\Omega\cdot V^{-1}\neq L$, so Eq. (\ref{e:solution_expL})
  does not hold. We check that (\ref{e:solution_B}) solves indeed the evolution problem
  (\ref{dynamical_pb_evolution_operator})
  by direct evaluation
  \bea
  B\cdot\partial_\tau u(\tau) = i(B\cdot V\cdot\Omega)\cdot V^{-1}\cdot
          (e^{iV\cdot\Omega\cdot V^{-1}\tau}\cdot u_0) =  iL\cdot u(\tau) \ ,
  \eea
  where we have used $L\cdot V=B\cdot V\cdot\Omega$  in   Eq. (\ref{e:eq1}).  Finally,
  by direct evaluation on can show
  \bea
  u(\tau)= \w V \cdot e^{i\w \Omega \tau} \cdot \w V^{-1} \cdot \w{u_0} = \sum_{n=1}^{K} \frac{\langle \w \alpha_n,\w B \cdot \w{u_0}\rangle}{\langle \w \alpha_n,\w B \cdot \w{v_n} \rangle}e^{i\omega_n\tau}\w{v_n} \ .\label{gen_evol_res}
\eea
  where, as in Eq. (\ref{evol_res}), the sum runs over all eigenvalues and not only QNMs.
  The calculation is however more subtle than in the case of Eqs. (\ref{e:eq_just1})-(\ref{e:eq_last})
  in the generic case in which $B$ is not invertible\footnote{Indeed, in such a case of non-invertible
    mass matrix  $B$, some lines and columns
    in the matrix must be eliminated to deal with ``diverging'' eigenvalues in  $\Omega$, but this can
    be done self-consistently. This is actually the situation in the
    asymptotically AdS case (see Eq.~(\ref{eq:dynamical_system_AdS}) in section~\ref{a:SAdS}), where the mass matrix $B$ is diagonal and singular (due to a single zero diagonal element). Since we have a pencil $(L,B)$ with $B$ singular , there is one infinite eigenvalue and $K-1$ generalised eigenvalues and eigenvectors. The equality (\ref{gen_evol_res}) is changed so that it excludes the single infinite eigenvalue.}. Some insight can be gained
  by looking however to the case with invertible $B$. In this case Eqs. (\ref{e:eq1}) can be rewritten as
  \bea
  \label{e:eq2}
  \left\{
  \begin{array}{rcl}
    V^{-1} \cdot B^{-1} \cdot L &=&  \Omega \cdot V^{-1} \nn \\
     A^t \cdot L &=& \Omega \cdot A^t \cdot B
  \end{array}
  \right. 
  \eea
  from which it follows
  \bea
  B^t \cdot A = \left(V^{-1}\right)^t \ ,
  \eea
  namely the analogue to $A =\left(\w V^{-1} \right)^t$ in Eq. (\ref{e:VA-matrices_Id}),
  that leads to the bi-orthogonal relation
  \bea
  A^t\cdot B\cdot V =(B^t \cdot A)^t \cdot V = V^{-1}  \cdot V = \mathrm{I}_K \ ,
  \eea
  from which expression (\ref{gen_evol_res}) readily follows.

  \paragraph{A ``shortcut'' for the projection algorithm.}
  Besides justifying why the tails are recovered in the finite rank case, another by-product of (\ref{evol_res}) is a convenient computational shortcut for a rather tedious numerical aspect of the projection algorithm described in footnote \ref{note:indexing} of section \ref{s:calculation_keldysh}. The Keldysh expansion is fully characterized by the frequencies $\omega_n$ and the coefficients $\mathcal{A}_n(x_k)$ using
\begin{equation}
\w{\mathcal{A}}_n(x_k)=(\w V)_{kn}( \w V^{-1} \w{u_0})_n \ , \label{shortcut_A}
\end{equation}
where $\{x_k\}_{1\leq k\leq N+1}$ are the collocation points on the Chebyshev-Lobatto grid and $K=2N+2$. As a consequence, our numerical expansions are greatly simplified and reduce to these 3 steps : (i) find the eigenvalues and eigenvectors of $\w L$, (ii) inverse the matrix $\w V$ of the eigenvectors (directly; no normalisation nor indexation is needed) and (iii) evaluate (\ref{shortcut_A}).

\section{Hyperboloidal approach to scattering}
\label{a:Hyperboloidal approach}
In this appendix we give the details of the hyperboloidal approach to
scattering in section \ref{s:Time-domain_evolution}.

\subsection{The evolution problem: perturbations on spherically symmetric black holes}
We start by considering the spherically symmetric line element 
\bea
ds^2 = -f(r)dt^2 + f(r)^{-1} dr^2 + r^2\Omega_{AB}dx^Adx^B \ , 
\eea
with $\Omega_{AB}$ the metric of the unit sphere and $r\in ]r_{\mathrm{H}}, \infty[$ with $r_{\mathrm{H}}$
the coordinate radius of the BH event horizon. We introduce the tortoise coordinate $r_*$, satisfying
$dr/dr_* = f(r)$, with range $r_*\in ]-\infty, \infty[$.
    Making use of the spherical symmetry scalar, electromagnetic or gravitational field
    perturbations can be written in terms of scalar master functions $\Phi$ that, once rescaled
    $\Phi=\phi/r$ and decomposed
    in spherical harmonics components $\phi_{\ell m}$, satisfy 
    \bea
\left(\frac{\partial^2}{\partial t^2} - \frac{\partial^2}{\partial r_*^2} + V_\ell \right)\phi_{\ell m}=0 \ ,
\label{wave_equation_tortoise}
    \eea
    where the expression $V_\ell$ depends on the BH background and on the nature of the field (its spin $s$).
    Boundary conditions for Eq. (\ref{wave_equation_tortoise}) in our scattering problem and in this
    Cauchy formulations are purely outgoing at (spatial) infinity ($r_*\to+\infty$)
    and purely ingoing at the horizon ($r_*\to-\infty$). For convenience, we introduce
    the  dimensionless quantities
\bea
\bar t = \frac{t}{\lambda} \ \ , \ \  \bar x = \frac{r_*}{\lambda}
\ \ , \ \  \hat V_\ell &= \lambda^2 V_\ell,
\eea
in terms of a length scale $\lambda$ appropriately chosen in each problem, so Eq. (\ref{wave_equation_tortoise})
for perturbations on spherically symmetric BHs is cast in the form of the wave equation (\ref{e:wave_equation_generic}) 
\bea
\label{e:wave_equation_Cauchy}
\left(\frac{\partial^2}{\partial \overline{t}^2} - \frac{\partial^2}{\partial \overline{x}^2} + \hat{V}_\ell \right)\phi_{\ell m}=0
\,.
\eea

\subsection{Hyperboloidal scheme: outgoing boundary conditions and non-selfadjoint infinitesimal
  generator $L$}
\label{a:hyperboloidal_scheme}
Massless perturbations (in odd space-dimensions) propagate along null characteristics reaching the wave
zone modelled by null infinity $\scri^+$ at far distances and traversing the event horizon
when propagating in a black hole spacetime. Considering first asymptotically
flat spacetimes, outgoing boundary conditions are imposed at these respective outer and inner spacetime
null boundaries. A natural manner of adapting the evolution problem to this propagation behaviour at the spacetime
boundaries consists in choosing a spacetime foliation $\{\Sigma_\tau\}$ that is transverse to $\scri^+$ and the BH
event horizon. From a geometric perspective, null cones are outgoing at the intersection
between the slices $\Sigma_\tau$ at the spacetime boundaries, enforcing in a geometric
manner the outgoing boundary conditions for physical fields\footnote{\label{footnote:AdS-dS_BCs}In Anti-de Sitter null infinity asymptotics,
we rather impose homogeneous Dirichlet conditions at $\scri^+$, a timelike hypersurface in this case. In the case
of de Sitter there are two natural possibilities for the outer boundary ``asymptotics'': i) the cosmological
horizon, namely a null hypersurface, or ii) null infinity $\scri^+$, a spacelike surface in this case.
Although the former is perhaps preferred from a physical perspective,
since it restraints the study to the patch of spacetime accessible to the observer (see
e.g. \cite{albrychiewicz2021scattering}), from the mathematical perspective both are natural and
have been considered in literature, for instance \cite{daude2010inverse} chooses  i), whereas
  \cite{friedrich1986existence,hintz2024asymptotically} opt for ii). In our case,
 we choose the cosmological
 horizon, i.e. i),  for a ``technical'' reason: the need to deal with a stationary (actually static
 in our case) patch of spacetime,
  namely having an (asymptotically) timelike Killing, to define QNMs. Notice that exactly the
  same questions is posed in the black hole interior. In summary we choose 
 outgoing boundary conditions to be imposed at the cosmological horizon, a null hypersurface,
  leading to a similar boundary problem to that of the  
   asymptotically flat case, although asymptotic decay conditions are faster,
   making the outer (null) boundary more regular in the de Sitter case when a compactification is considered.}.
Hyperboloidal foliations,  interpolating between the BH horizon and $\scri^+$, are therefore
a natural setting in our scattering problem (cf. the excellent discussions
in~\cite{panosso2024hyperboloidal,zenginouglu2024hyperbolic,macedo2024hyperboloidal}).

On the other hand, the choice of such hyperboloidal foliations is particularly
interesting in the setting of the Keldysh expansion of the resolvent discussed
in section~\ref{s:Keldysh}. Indeed the outgoing boundary conditions entail a loss
of {\em energy}, namely a decrease of the (energy) norm of the scattered field
in the slice $\Sigma_\tau$ and, therefore, the evolution
cannot be conservative (unitary) in such foliations (see more details in section
\ref{s:transients}).
In other words, under the  choice of hyperboloidal  foliations to describe the time evolution,
the enforcement of outgoing boundary conditions in the dynamical problem implies 
the non-selfadjoint nature of the infinitesimal time generator $L$, when Eq. (\ref{e:wave_equation_Cauchy})
is written in the  first-order form (\ref{e:wave_eq_1storder_u}).

Hyperboloidal foliations therefore provide a natural setting to apply the Keldysh expansion in scattering problems,
namely for the resolvent of the  infinitesimal time-generator non-selfadjoint operator $L$.
We sketch now the basic elements to cast the evolution
problem in a hyperboloidal slicing (we follow closely the notation in \cite{Jaramillo:2020tuu}, see~\cite{panosso2024hyperboloidal,zenginouglu2024hyperbolic,macedo2024hyperboloidal} and references therein for a more extensive discussion).
First we consider the coordinate change
\bea
\label{e:hyperboloidal_change}
\begin{cases}
  	\bar{t} = \tau - h(x)\\
  	\bar{x} = g(x)
\end{cases} \ .
\label{compactified_hyperboloidal}
\eea
The height function $h$ implements the hyperboloidal foliation, in such a way that
$\tau=\mathrm{const.}$ slices are spacelike hypersurfaces penetrating the horizon
and extending to future null infinity $\scri^+$. The function $g$ defines a compactification
mapping of $\Sigma_t$ that brings (null) infinity at $\bar{x}\to +\infty$
and the BH horizon at  $\bar{x}\to -\infty$ to a finite interval $]a, b[$, namely
  \bea
  g \colon ]a,b[  &\to& ]-\infty, +\infty[ \nn\\
  x &\mapsto& \overline x=g(x) \ .
\eea
  Adding the points $a$ and $b$ implements the spatial compactification, allowing to incorporate null infinity $\scri^+$
  and the BH horizon in the spatial domain $[a, b]$. Inserting the change of coordinates (\ref{e:hyperboloidal_change}) into the
  wave equation (\ref{e:wave_equation_Cauchy}), we get (we drop the indices $\ell$ and $m$)
  \bea
  \label{e:evol_equation_2nd_order}
-\partial^2_\tau \phi + L_1 \phi + L_2 \partial_\tau\phi = 0 \ ,
\label{ODE_L1_L2}
\eea
where the expression of the operators $L_1$ and $L_2$ are given by
\bea
  \label{e:L_1-L_2}
  L_1 &=& \frac{1}{w(x)}\left( \partial_x(p(x)\partial_x) -  q(x) \right) \nn \\
  L_2 &=& \frac{1}{w(x)}\left(2\gamma(x)\partial_x +\partial_x\gamma(x)\right)
  = \frac{1}{w(x)}\left(\gamma(x)\partial_x +\partial_x(\gamma(x)\cdot)\right) \ ,
\eea
with
\bea
\label{e:defs_in_L_1-L_2}
p(x) = \frac{1}{|g'|}, \quad  q(x) = \lambda^2 |g'(x)| V_\ell,
\quad w(x) = \frac{h'^2-g'^2}{|g'|},  \quad \gamma(x)=\frac{h'}{|g'|} \ .
	\label{pqwg}
        \eea
        Here $L_1$ is a singular Sturm-Liouville operator, with $p(x)$ vanishing at the boundary of the interval,
        i.e. $p(a)=p(b)=0$.  
        The key consequence of the singular character of $L_1$ is that, if we require sufficient regularity
        on the solutions, then no boundary conditions are allowed to be enforced.
        Actually, boundary conditions
        are now encoded in the wave equation itself, corresponding to the evaluation of Eq. (\ref{e:evol_equation_2nd_order})
        at the boundaries $x=a$ and $x=b$.
        Regarding the $L_2$ operator, it is a dissipative term  characterising and enforcing the
        (outwards) leaking at the boundary, the energy flux of the field at the boundary being proportional
        to $\gamma$ (cf.~\cite{Gasperin:2021kfv}). In brief, boundary conditions are in-built in the $L$ operator,
        the singular character of $L_1$ recasting their explicit enforcement
        into a demand of regularity of the solutions,
        whereas the dissipative character of $L_2$ guarantees their `outgoing' nature.
        This is the analytic counterpart of the geometric enforcing of outgoing boundary conditions in the hyperboloidal scheme.
       
        As a final step to get the appropriate operator $L$ for the Schr\"odinder-like equation,
        we perform a first
        order reduction in time of the equation (\ref{e:evol_equation_2nd_order})
        by introducing the fields
        \bea
        \psi = \partial_\tau \phi \ \ , \  \ u = \begin{pmatrix} \phi\\ \psi \end{pmatrix} \ .
        \eea
        This permits the evolution equation (\ref{e:evol_equation_2nd_order}) to be cast in the form
        (\ref{e:wave_eq_1storder_u}), namely
        \bea
        \partial_\tau u = i L u \ ,
\label{eq:ODE}
\eea
with the infinitesimal time generator identified as
\bea
\label{e:L_L_1_L_2}
	 L = \frac{1}{i}
	\left(\begin{array}{c|c}
    0 & 1\\ \hline
    L_1 & L_2
	\end{array}\right)
        \eea
        with $L_1$ and $L_2$ given in (\ref{e:L_1-L_2}). For non-vanishing $L_2$
        the operator $L$ is non-selfadjoint and this is the starting point
        for the application of the Keldysh expansion in our hyperboloidal setting.
        Given its role in the discussion in section \ref{s:transients},
        in particular in footnote \ref{footnote:transient_energy}, we
        write the formal adjoint $L^\dagger$ of $L$, in the energy scalar product
        (\ref{eq_eff_en_inner_product}) (cf. discussion in \cite{Jaramillo:2020tuu})
        \bea
        \label{e:formal_adjoint_Gas}L^\dagger
=\frac{1}{i}\!  \left(
  \begin{array}{c|c}
    0 & 1 \\ \hline L_1 & L_2+ L_2^\partial
  \end{array}
  \right) \! \ ,
  \eea
  where $L_2^\partial$ is a purely distributional Dirac-delta-like term
  \bea\label{e:L2_adjoint_pert}
  L_2^\partial = 2\frac{\gamma}{w} \bigg(\delta(x-a)-\delta(x-b)\bigg) \ .
  \eea

\subsection{QNMs as eigenvalues of the non-selfadjoint operator $L$}
\label{a:ANM_eigenvalues}
The characterization of QNMs can be formally addressed by considering Eq. (\ref{e:wave_equation_generic}), by
taking the Fourier transform in time $\overline{t}$ and imposing `outgoing' boundary conditions,
that leads
to 
\bea
\label{e:BH_QNM_eigenvalue_Cauchy}
\left(-\frac{\partial^2}{\partial \overline{x}^2} + \hat{V}\right)\phi=
  \omega \phi \ \ , \ \ \hbox{plus `outgoing' boundary conditions} \ .
\eea
The resulting  QNM functions $\phi(\omega)$ associated with each harmonic
$e^{i\omega \overline{t}}$ mode have however a singular
behaviour (non-bound oscillations) at the boundaries (cf. e.g. the discussion in \cite{macedo2024hyperboloidal}).
Actually no natural Hilbert space is associated with those (generalised) `eigenmode' solutions.
In stark contrast with this, the hyperboloidal framework provides a proper `eigenvalue problem',
where eigenfunctions are regular functions at the boundaries. More specifically, eigenfunctions
belong actually to an appropriate Hilbert space where the infinitesimal time generator $L$
is a proper non-selfadjoint operator~\cite{Warnick:2013hba,gajic2024quasinormal}.
Therefore, we are in the natural setting to apply the
Keldysh expansion of the resolvent of $L$.

Concretely, taking the Fourier transform in Eq. (\ref{eq:ODE}) with respect to the hyperboloidal
time $\tau$  (with convention $\phi(\tau, x) \sim e^{i\omega \tau}$, as in \cite{Jaramillo:2020tuu})
we get the eigenvalue problem 
\bea
\label{e:BH_QNM_eigenvalue_hyperboloidal}
L u = \omega u \ ,
\eea
where boundary conditions are encoded in $L$ as long as solutions are enforced to belong to the 
appropriate regularity class of functions.
Eigenfunctions of this spectral problem are the QNMs.
This eigenvalue approach to QNMs has been introduced in \cite{Zenginoglu:2011jz,Warnick:2013hba,Ansorg:2016ztf}.

It is worthwhile to note that
the QNM eigenfrequencies $\omega_n$ of this hyperboloidal (proper) eigenvalue problem are the same that those
in the (formal) eigenvalue problem (\ref{e:BH_QNM_eigenvalue_Cauchy}), obtained in the Cauchy formulation.
The reason is that both the Cauchy time $\overline{t}$ and the hyperboloidal time $\tau$
are (up to the constant $\lambda$) `affine' parameters of the stationary Killing vector $t^a$, since
\begin{equation}
  \label{e:t_Killing}
t^a=\partial_t = \frac{1}{\lambda} \partial_{\overline t} = \frac{1}{\lambda} \partial_\tau \ ,
\end{equation}
so it holds $\lambda t^a(\overline t)=\lambda t^a(\tau)=1$.
This is a key consequence of the specific form of the first equation (\ref{e:hyperboloidal_change}), defining
the height function $h(x)$.

The eigenvalue approach to QNMs has been subject of mathematical relativity studies
aiming at characterising the proper Hilbert space for the eigenfunctions
\cite{Warnick:2013hba,Gajic:2019qdd,Gajic:2019oem,Bizon:2020qnd,galkowski2021outgoing,gajic2024quasinormal,Warnick:2024}, as well of numerical investigations~\cite{Ansorg:2016ztf,Ammon:2016fru,PanossoMacedo:2018hab}, the latter with a particular
focus on spectral stability issues by taking advantage of concepts adapted from the spectral theory of non-selfadjoint operators, such as the notion of the pseudospectrum~\cite{Jaramillo:2020tuu,Jaramillo:2021tmt,alsheikh:tel-04116011,Destounis:2021lum,BoyCarDes22,sarkar23,Destounis:2023nmb,Arean:2023pseudospectra,cownden:2024pseudospectra,Boyanov:2024,Carballo_2024,cao2024pseudospectrum,Chen_2024,cao2024stability,cai2025pseudospectrum}.
In the present work, the hyperboloidal QNM spectral problem
(\ref{e:BH_QNM_eigenvalue_hyperboloidal}), together with its associated transpose problem, matches the
standard spectral problem (\ref{e:right-left_eigenvalues_v2}), so we can apply directly the
 Keldysh expansion of the resolvent discussed in section \ref{s:Keldysh_resolvent}.

\section{Elements of the hyperboloidal formulation of black holes: spherically symmetric case}
\label{a:cases of study}
We provide here the  definitions and basic elements for the explicit expression of the hyperboloidal slicing and the
differential operator $L$ for the four considered cases of study. 

\subsection{A toy model : the P\"oschl-Teller case}
\label{a:PT}
We follow  \cite{Jaramillo:2020tuu} to study the  P\"oschl-Teller toy model, corresponding
exactly with the Klein-Gordon equation in the static patch of de
Sitter spacetime studied in \cite{Bizon:2020qnd}. This case is integrable and
can be explicitly solved, providing analytic expressions for its QNMs.
Given $V_0$ (corresponding to the mass-squared term $m^2$ in \cite{Bizon:2020qnd}) and $b>0$, the potential is
given by 
\begin{alignat}{2}
V(\bar x)&=V_0 \sech^2(\bar x/b) \ , \qquad \bar x \in \mathbb{R} \ .
\label{PT_potential}
\end{alignat}
Adopting the hyperboloidal foliation corresponding to the Bizo\'n-Mach coordinates $(\tau,x)$
\begin{equation}
  \displaystyle
\begin{cases}
\overline t = \tau - \frac{b}{2}\log(1-x^2)\\
\overline x = b \arctanh(x)
\end{cases} \ ,
\end{equation}
and injecting these coordinates into the expression (\ref{compactified_hyperboloidal}), we have
\begin{alignat}{4}
p(x)&=\frac{1-x^2}{b}, \qquad w(x)&=b,\qquad \gamma(x)&=-x, \qquad q(x)&=bV_0 \ ,
\end{alignat}
which translates into the following differential operators 
\begin{align}
L_1&=\partial_x \left(\frac{1-x^2}{b^2}\partial_x \right)- V_0\\
L_2&=-\frac{1+2x\partial_x}{b} \ ,
\end{align}
where we have used the identity $\sech^2(\arctanh(x)) = 1-x^2$. With these expressions of $L_1$ and $L_2$, we develop the equations (\ref{ODE_L1_L2}), producing then an equation for the field $\phi(x)$. Upon expanding the field into a power series, it can be shown (see appendix of \cite{Jaramillo:2020tuu}) that enforcing the appropriate regularity condition
($\phi(x)$ is actually a polynomial in this case) we recover the modes $\phi_n(x)$ and the two branches of quasinormal frequencies
\begin{equation}
    \omega^\pm_n=\frac{i}{b}\left(\frac{1}{2}+n \pm \sqrt{\frac{1}{4}-b^2V_0}\right) \ .
\end{equation}
If the expression inside the square root is positive the eigenfrequencies are purely imaginary, whereas if it is negative
the two branches $\omega_n^\pm$ are parallel to the imaginary axis with respectively positive ($\omega_n^+$)
and negative ($\omega^-_n$) real part (cf. also  expressions in \cite{Beyer:1998nu}) 
\begin{align}
  \label{e:PT_QNMs}
  \displaystyle
  \omega^{\pm}_n = \begin{cases} \frac{i}{b}\left(\frac{1}{2}+n \pm \sqrt{\frac{1}{4}-b^2V_0}\right)
    &\text{ if }b^2V_0<\frac{1}{4}\\
    \frac{1}{b}\left(\pm\sqrt{b^2V_0-\frac{1}{4}} + i\left(\frac{1}{2}+n\right) \right) &\text{ if }b^2V_0\geq\frac{1}{4} 
    \end{cases} \ .
\end{align}
In these coordinates, the eigenfunctions (namely, the QNMs) are given in terms of
the Jacobi polynomials $P_n^{(\alpha, \beta)}$,
specifically as $\phi_n^\pm(x) = P_n^{(ib\omega_n^\pm,ib\omega_n^\pm)}(x)$ (cf.~\cite{Jaramillo:2020tuu}).
This special case of Jacobi polynomials corresponds to
the Gegenbauer polynomials $C_n^{(\lambda)}(x) = c_n P_n^{(\lambda-1/2,\lambda-1/2)}(x)$, with $c_n$
a constant factor that depends on $n$, so we conclude
\begin{equation}
  \label{a:Gegenbauer}
\phi^{\pm}_n(x)=C_n^{(ib\omega_n^\pm+1/2)}(x) \ ,
\end{equation}
(note that the term $1/2$ exactly cancels the $1/2$ in $(\ref{e:PT_QNMs})$).
Then, from $C_n^{(\lambda)}(-x)=(-1)^n C_n^{(\lambda)}(x)$, it follows that the eigenfunction $\phi^{\pm}_n(x)$ is an even (odd) function of $x$ if $n$ is even (respectively odd). The knowledge of the properties of the eigenfunctions is particularly useful when studying the effect of a symmetric initial data on the evolution problem (\ref{eq:ODE}).

\subsection{Black hole spacetimes}
For spherically symmetric black hole spacetimes, we work
with $\sigma\in [0,1]$ instead of $x \in [-1,1]$. We start with a line element 
\begin{equation}
ds^2 = -f\left(r\right)dt^2 + \frac{1}{f(r)} dr^2 +r^2(d\theta^2+sin^2\theta d\varphi^2) \ .
\label{line_element_BH}
\end{equation}
Using the tortoise coordinate and of the compactification
function~\cite{Ansorg:2016ztf,PanossoMacedo:2018hab,Jaramillo:2021tmt} we have
\begin{equation}
\overline{\sigma} = \frac{r_*}{\lambda}= \frac{1}{\lambda} \int^{r(\sigma)} \frac{dr}{f(r)} = g(\sigma) \ ,
\end{equation}
with $\lambda$ an appropriate scale in the problem.
Then the radial wave equation then reads
\begin{equation}
\left(\frac{\partial^2}{\partial \overline{t}^2} - \frac{\partial^2}{\partial \overline{\sigma}^2} + \lambda^2 V_\ell\right)\phi = 0 \ .
\end{equation}

We point out that the expression of $q(x)$ in (\ref{e:defs_in_L_1-L_2}) is simplified due to the presence of a factor $f(r)$ for all these potentials
\begin{equation}
q(\sigma)=\lambda^2 \lvert g'(\sigma) \rvert V_\ell(r(\sigma))=\lambda \left\lvert \frac{dr}{d\sigma} \right\rvert \frac{V_\ell(r(\sigma))}{f(r(\sigma))} \ .
\end{equation} 
The potential $V_\ell$ depends on the black hole case and the type of the perturbation listed in table \ref{table_potential_tiny}.
In the following sections we provide $h(\sigma)$, the scale $\lambda$ and $r(\sigma)$ for each
black hole case. Once we have the expressions for the height function $h(\sigma)$, the compactification function $g(\sigma)$, the scale factor $\lambda$ and the potential $V_\ell$ then a straightforward computation gives $p(\sigma)$, $w(\sigma)$, $q(\sigma)$ and $\gamma(\sigma)$ contained within the differential operators $L_1$ and $L_2$.
\begin{table}[h]
    \footnotesize{\begin{tabular}{|c||c|c|c|c|c|}
    \hline
    \textbf{Cases} & $f(r)$
      & \multicolumn{3}{c|}{potential for $s=0, 1$ or $2$ axial perturbations}& potential for $s=2$ polar perturbations \rule{0pt}{9pt} \\[2pt]\hline
         P\"oschl-Teller  & \multicolumn{5}{c|}{$V_0 \sech^2\left(\ol x/b\right)$} \rule{0pt}{10pt} \\[3pt] \hline
           Schw              & $1-\frac{2M}{r}$                   &
    \multicolumn{3}{c|}{$f(r)\left(\frac{\ell(\ell+1)}{r^2}+(1-s^2)\frac{2M}{r^3} \right)$} & $f(r)\frac{2}{r^3}\frac{9M^3+3c^2Mr^2+c^2(1+c)r^3+9M^2cr}{(3M+cr)^2}$ \rule{0pt}{13pt} \\[6pt]\hline
  Schw-(A)dS              & $1-\frac{2M}{r}-\frac{\Lambda r^2}{3}$                      
    & \multicolumn{3}{c|}{$f(r)\left[\frac{\ell(\ell+1)}{r^2}+(1-s^2)\left(\frac{2M}{r^3}-\frac{2-s}{3}\Lambda\right)\right]$} & $\frac{2f(r)}{r^3}\frac{9M^3+3c^2Mr^2+c^2(1+c)r^3+3M^2(3cr-\Lambda r^3)}{(3M+cr)^2}$ \rule{0pt}{13pt} \\[5pt]\hline
    \end{tabular}}
    \caption{Expressions for the potential in the four  cases we consider. Note
      that the cosmological constant $\Lambda$ may be positive or negative
      corresponding, respectively, to Schw-dS and Schw-AdS.  We denote $c=\frac{(\ell-1)(\ell+2)}{2}$.}\label{table_potential_tiny}
  \end{table}

The table \ref{table_hyperboloidal_cases} provides a view of the hyperboloidal approach for all of our cases of study.
\begin{table}[htp!]
  \footnotesize{\begin{tabular}{|c||c|c|c|c|}
  \hline
  \textbf{Cases} & $r(\sigma)$ & height function $h$ & compactification function $g$ &
      scale $\lambda$ \rule{0pt}{9pt} \\[2pt] \hline \rule{0pt}{13pt}
  \Centerstack[c]{ P\"oschl-Teller\\ $V_0=b=1$}  &  
      & $\frac{1}{2}\log(1-x^2)$& $\tanh^{-1}(x)$& $1/\sqrt{V_0}$ \rule{0pt}{9pt} \\[6pt] \hline
  \Centerstack[c]{ Schw.\\ $M=1$}              & $2M/\sigma$ 
      &  $\frac{1}{2}\left( \log\sigma + \log(1-\sigma) -\frac{1}{\sigma}\right)$& $\frac{1}{2}\left( \frac{1}{\sigma} + \log(1-\sigma) -\ln\sigma\right)$& $4M$  \rule{0pt}{12pt} \\[6pt]\hline
  \Centerstack[c]{Schw.-dS\\   $M=1$\\ $\Lambda=0.07$}     & $r_{+}\sigma+r_c(1-\sigma)$       & 
      \Centerstack[c]{ $\frac{1}{4\kappa_+r_+}\log(1-\sigma)+\frac{1}{4|\kappa_c|r_+}\log(\sigma) +$\\\\ $\frac{1}{4\kappa_0r_+}\log\left( 1+\frac{r_c}{r_0}+\sigma\frac{r_+-r_c}{r_0}\right)$}  & 
      \Centerstack[c]{ $\frac{1}{4\kappa_+r_+}\log(1-\sigma)-\frac{1}{4|\kappa_c|r_+}\log(\sigma) +$\\\\ $\frac{1}{4\kappa_0r_+}\log\left( 1+\frac{r_c}{r_0}+\sigma\frac{r_+-r_c}{r_0}\right)$} &$2r_+$  \rule{0pt}{20pt}\\[18pt]\hline
  \Centerstack[c]{Schw.-AdS\\   $r_h=R=1$ \\($\alpha=1$)}     & $r_h/\sigma$       & 
            $\frac{1}{3\alpha^2+1}\log(1-\sigma)$
                      & \Centerstack[c]{  $-\frac{\log(\alpha^2(\sigma^2+\sigma+1)+\sigma^2)-2\log(1-\sigma)}{2(3\alpha^2+1)}$\\\\\\$-\frac{(6\alpha^2+4)\tan^{-1}\left(\frac{\alpha^2(2\sigma+1)+2\sigma}{\alpha\sqrt{3\alpha^2+4}}\right)}{2\sqrt{3\alpha^2+4}\alpha(3\alpha^2+1)}$} &$r_h$  \rule{0pt}{25pt} \\[25pt]\hline
  \end{tabular}}
  \caption{Expressions for the height $h$ and compactification $g$ functions, as well as
    the scale $\lambda$ in all of our cases of study. We also present the expression of $r(\sigma)$ when applicable.}\label{table_hyperboloidal_cases}
\end{table}

\subsubsection{The Schwarzschild case.}
\label{a:S}
In the Schwarzschild coordinates we have $f(r)=1-\frac{2M}{r}$, following \cite{Ansorg:2016ztf,PanossoMacedo:2018hab,Jaramillo:2021tmt} the compactified coordinate is $r(\sigma)=\frac{2M}{\sigma}$ such that $\sigma=1$ at the horizon and $\sigma=0$ at null infinity and the hyperboloidal slicing associated to the change of coordinates (\ref{compactified_hyperboloidal}) is 
\begin{equation}
\begin{cases}
h(\sigma)=\frac{1}{2}\left( \log\sigma + \log(1-\sigma) -\frac{1}{\sigma}\right)\\
g(\sigma)=\frac{1}{2}\left( \frac{1}{\sigma} + \log(1-\sigma) -\ln\sigma\right)
\end{cases} \ .
\end{equation}
The scale $\lambda=4M=1/\kappa$ has been chosen. One can check the correct implementation of the boundary conditions~\cite{Bizon:2020qnd} by verifying $h(\sigma) \sim -g(\sigma)$ near the horizon and $h(\sigma) \sim g(\sigma)$ at future null infinity, so spatial $\tau=\mathrm{const}$ slices asymptote to outgoing null
directions. Functions in Eq. (\ref{e:defs_in_L_1-L_2}) can then be computed and are given by
\begin{alignat}{4}
p(\sigma)&=2\sigma^2 (1-\sigma), \quad w(\sigma)&=2(1+\sigma),\quad \gamma(\sigma)&=1-2\sigma^2, \quad q(\sigma)&=2(\ell(\ell+1)-3\sigma) \ ,
\end{alignat}
which yields the following differential operators 
\bea
L_1&=&\frac{1}{2(1+\sigma)}[\partial_\sigma (2\sigma^2(1-\sigma)\partial_\sigma)- 2(\ell(\ell+1)-3\sigma)] \nn\\
L_2&=&\frac{1}{2(1+\sigma)}[2(1-2\sigma^2)\partial_\sigma - 4\sigma] \ .
\eea
A key difference with the  P\"oschl-Teller toy model is the power-law decay of the potential $V_\ell$
at infinity.
This feature translates into the factor $\sigma^2$ in $p(\sigma)$ vanishing quadratically at null infinity, in contrast
with the term $(1 - \sigma)$ vanishing linearly at the horizon. 
This structure leads to  continuous part of the spectrum along the imaginary axis (corresponding to the `branch cut',
absent in P\"oschl-Teller) in addition to actual eigenvalues (QNM frequencies).
As a consequence, it appears a power-law time decay of the scattered field (at late times) at future null infinity, in contrast to the late exponential time decay when only QNMs are present.

\subsubsection{The Schwarzschild-de Sitter case.}
\label{a:SdS}
Following \cite{Griffiths_Podolsky_2009}, we write the Schwarzschild asymptotically de Sitter metric, with cosmological constant $\Lambda$, as
\begin{equation}
  \label{e:f_SdS}
f(r) = 1-\frac{2M}{r}-\frac{\Lambda r^2}{3} =-\frac{\Lambda}{3r}(r-r_{+})(r-r_c)(r+r_0) \ ,
\end{equation}
where $r_{+}$, $r_c$  are respectively the black hole and the cosmological event horizon (satisfying $r_{+}<r_c$) and $r_0=r_c+r_+$. We introduce the compactified radial coordinate $\sigma$ that maps the event horizon $r_{+}$ to $1$ and the cosmological horizon\footnote{We choose the cosmological horizon as the outer
boundary in order to guarantee that the Killing $t^a$ is a timelike vector. This permits to use its affine
parameter $\tau$ as the dual time variable to the frequency variable $\omega$ in which the
 QNM frequencies $\omega_n$'s are invariantly defined (cf. Eq. (\ref{e:t_Killing})).
This choice of stationary (actually static, in our spherically symmetric context) patch 
is natural from a physical perspective and mathematical perspective, but it is not the only choice
in other settings, cf. footnote \ref{footnote:AdS-dS_BCs}.} $r_c$ to $0$, namely
\begin{equation}
r(\sigma)=r_{+}\left(\sigma+\frac{r_c}{r_{+}}(1-\sigma)\right)\in[r_+,r_c] \ .
\end{equation}
Following \cite{sarkar23} and using the scale $\lambda=2r_+$, the hyperboloidal slicing is chosen to be
\begin{equation}
  \begin{cases}
  h(\sigma)=\frac{1}{4\kappa_+r_+}\log(1-\sigma)+\frac{1}{4|\kappa_c|r_+}\log(\sigma)+\frac{1}{4\kappa_0r_+}\log\left( 1+\frac{r_c}{r_0}+\sigma\frac{r_+-r_c}{r_0}\right)\\
  g(\sigma)=\frac{1}{4\kappa_+r_+}\log(1-\sigma)-\frac{1}{4|\kappa_c|r_+}\log(\sigma)+\frac{1}{4\kappa_0r_+}\log\left( 1+\frac{r_c}{r_0}+\sigma\frac{r_+-r_c}{r_0}\right)
  \end{cases} \ ,
  \label{e:SdS_slicing}
\end{equation}
that are  expressed in terms of the three ``surface gravity'' expressions
\begin{equation}
  \begin{cases}
  \kappa_+=\frac{\Lambda}{6r_+}(r_c-r_+)(2r_++r_c)\\
  \kappa_c=\frac{\Lambda}{6r_c}(r_+-r_c)(2r_c+r_+)\\
  \kappa_0=\frac{\Lambda}{6r_0}(r_c+2r_+)(r_++2r_c)
  \end{cases} \ .
\end{equation}
It is only a difference of sign in the $\log \sigma$ term what distinguishes the height and the compactification functions $h$ and $g$, thus ensuring $h(\sigma) \sim -g(\sigma)$ near the BH horizon and $h(\sigma) \sim g(\sigma)$ at the de Sitter cosmological horizon. The slicing (\ref{e:SdS_slicing}) yields then for  (\ref{e:defs_in_L_1-L_2})
\bea
    p(\sigma)&=&\frac{{2r_+} {\left({r_c} - {r_{+}}\right)}}{{L_{dS}}^{2}}\frac{{\left(({r_c} - {r_{+}} )\sigma - 2 \, {r_c} - {r_{+}}\right)}  {\left(\sigma - 1\right)} \sigma}{({r_c} - {r_{+}} )\sigma - {r_c} } \nn \\
    \gamma(\sigma)&=& \frac{1}{2 \, {r_c} + {r_{+}}}\frac{2{r_c}( {r_c} - {r_{+}}) \sigma^{2} - (4 \, {r_c}^{2} + {r_c} {r_{+}}+ {r_{+}}^{2}) \sigma + {r_c}(2 \, {r_c} + {r_{+}})}{({r_c} - {r_{+}} )\sigma - {r_c} }\nn \\
    w(\sigma) &=& -\frac{2{L_{dS}}^{2} {r_c}}{({r_c} - {r_{+}}) {r_+} {\left(2 \, {r_c} + {r_{+}}\right)}^{2} }\frac{ r_c(r_c - {r_{+}} )\sigma - {r_c}^{2} - {r_c} {r_{+}} - {r_{+}}^{2} }{{({r_c} - {r_{+}}) \sigma - {r_c}}}\nn \\
    q_\ell(\sigma)&=& -{2r_+} {\left({r_c} - {r_{+}}\right)}\left(\frac{\ell(\ell+1)}{r(\sigma)^2}-\frac{6M}{r(\sigma)^3}\right) \ .
  \eea
Unlike the Schwarzschild case, instead of the `branch cut', actual QNM eigenvalues corresponding to
de Sitter modes are found along the imaginary axis and do not manifest themselves in a power-law tail.
In particular, in the discretised version of the operator, the corresponding
eigenvalues are convergent, in contrast with the `eigenvalues' corresponding to the `branch cut'. These features are consistent with $p(\sigma)$ vanishing linearly at the boundaries $\sigma=0$ and $1$. We do not use analytical formulas for the parameters $r_c$, $r_+$ and $r_0$, we determine the latter numerically as the roots of the polynomial $rf(r)$. 

\subsubsection{The Schwarzschild-Anti-de Sitter case.}\label{a:SAdS}
QNMs of asymptotically AdS spacetimes, characterised as proper eigenvalues of a non-selfadjoint
operator, have been fully discussed in \cite{Warnick:2013hba}. In the particular case of Schwarzschild-AdS QNMs,  their spectral stability have been studied in \cite{Boyanov:2024,cownden:2024pseudospectra,Arean:2023pseudospectra}. In contrast with the Schwarzschild asymptotically flat or de Sitter cases, AdS null infinity is a timelike hypersurface that acts like a boundary box that, when choosing homogeneous Dirichlet conditions, confines the field in a conservative manner.
Dissipation happens only at the event horizon. The function $f(r)$ writes in this case as
(following \cite{Boyanov:2024})
\begin{equation}
f(r)=1-\frac{r_s}{r}+\frac{r^2}{R^2}=\left(1-\frac{r_h}{r}\right)\left(1+\alpha^2\left(1+\frac{r}{r_h}+\frac{r^2}{r_h^2}\right)\right) \ ,
\end{equation}
with $r_h$ the event horizon radius,  $\alpha=r_h/R$ and $r_s=r_h(1+\alpha^2)$. 
We chose $\sigma=\frac{r_h}{r}$ that maps $r_h$ to $\sigma=1$ and $r\to\infty$ to $\sigma=0$. 
Upon choosing the scale factor $\lambda=r_h$ the hyperboloidal foliation becomes  
\bea
  g(\sigma)&=&-\frac{\log(\alpha^2(\sigma^2+\sigma+1)+\sigma^2)-2\log(1-\sigma)}{2(3\alpha^2+1)}-\frac{(6\alpha^2+4)\tan^{-1}\left(\frac{\alpha^2(2\sigma+1)+2\sigma}{\alpha\sqrt{3\alpha^2+4}}\right)}{2\sqrt{3\alpha^2+4}\alpha(3\alpha^2+1)} \nn\\
h(\sigma)&=&\frac{1}{1+3\alpha^2}\log(1-\sigma) \ .
\eea
 The expression of the differential operator $L_1$ and $L_2$, namely (\ref{e:L_1-L_2}), follows from the following expressions for the functions in (\ref{e:defs_in_L_1-L_2})
 \bea
  p(\sigma)&=&{\left((1 + {\alpha}^{2}) \sigma^{2} + {\alpha}^{2}( \sigma + 1)\right)} {\left(1-\sigma\right)} \nn\\
  \gamma(\sigma)&=& -\frac{(1 + {\alpha}^{2}) \sigma^{2} + \alpha(1 + {\sigma}^{2})}{1 + 3 \, {\alpha}^{2}}\nn \\
  w(\sigma) &=& \frac{{\left((1 + {\alpha}^{2}) \sigma^{2} + {\alpha}^{2} \sigma + 1 + 4 \, {\alpha}^{2}\right)} {\left((1  + {\alpha}^{2}) \sigma + 1 + 2 \, {\alpha}^{2}\right)}}{{\left((1 + {\alpha}^{2}) \sigma^{2} + \alpha^2 (\sigma+1)\right)} {\left(1 + 3 \, {\alpha}^{2}\right)}^{2}}\nn \\
  q_\ell(\sigma)&=& \ell(\ell+1) -3(1+\alpha^2)\ .
  \eea
The function $p(\sigma)$ vanishes only at the horizon (and actually linearly), thus encoding the outgoing boundary conditions only at $\sigma=1$. In order to impose homogeneous Dirichlet boundary conditions at $\sigma=0$, we introduce the rescaling
$u(\sigma,\tau) = \sigma \widetilde u(\sigma,\tau) = \left(\begin{array}{c}
  \sigma \widetilde \phi\\
  \sigma \widetilde \psi 
\end{array}\right)$
that forces $u(\sigma,\tau)$ to vanish at this point, if regularity is enforced on $\widetilde u(\sigma)$.
The spectral problem is rewritten as a generalised eigenvalue problem
\begin{equation}
\widetilde L \widetilde u = \lambda B \widetilde u \ ,
\label{eq:dynamical_system_AdS}
\end{equation}
with
\renewcommand\arraystretch{1.5} 
\begin{alignat}{2}
 \widetilde L &= \frac{1}{i}
	\left(\begin{array}{c|c}
    0 & 1\\ \hline 
    \widetilde L_1 & \widetilde L_2
	\end{array}\right) \ , &\quad\quad B=\left(\begin{array}{c|c}
    \ 1 \, \, & \ 0 \ \\ \hline
   \ 0 \ & \ \sigma \
	\end{array}\right) \ ,
\end{alignat}
\renewcommand\arraystretch{1}
and
\bea
	\widetilde L_1 &=& \frac{1}{w(\sigma)}\left( \sigma p(\sigma)\partial_\sigma^2 + [\sigma \partial_\sigma p + 2p(\sigma)]\partial_\sigma + \partial_\sigma p - \sigma q(\sigma) \right)\nn \\ 	
  \widetilde L_2 &=& \frac{1}{w(\sigma)}\left(2\gamma(\sigma)\sigma\partial_\sigma +2\gamma(\sigma)+\sigma\partial_\sigma\gamma(\sigma)\right) \ .
\eea

\section{Numerical method}
\subsection{(Chebyshev) Pseudospectral methods.}
\label{s:pseudospectral_methods}
Most of the numerical (pseudo-spectral) methods we use here are presented in \cite{Jaramillo:2021tmt}
and references therein. In the footsteps of these works, we use Chebyshev interpolation, namely we approximate a function $f(x)$, with $x\in[-1,1]$ by the
Chebyshev's interpolant $f^N(x)$
\begin{equation}
  f(x)\approx f^N(x)=\frac{c_0}{2}+\sum_{i=1}^N c_i T_i(x)= \frac{c_0}{2}+\sum_{i=1}^N c_i \cos(i\arccos(x)) \ ,
\end{equation}
where the $T_i(x)$'s are the Chebyshev's polynomials, and the coefficients $c_i$
are determined by requiring $f(x_i) = f^N(x_i)$, over the collocation points $x_i$ of the Chebyshev-Lobatto\footnote{This can be generalised to other collocation grids, namely
Chebyshev-Gauss or Chebyshev-Radau (left/right),  
cf. e.g. \cite{alsheikh:tel-04116011}.},
collocation grid $x_i=\cos\left(\frac{\pi i}{N}\right)\in [-1,+1]$ for $0\leq i\leq N$ that includes the endpoints $-1$ and $+1$. An affine map $\mu \colon [-1,1]  \to [a,b]$ is used to sample the space interval $[a,b]$, namely the domain of the compactification function $g$, in which the compactified coordinate lies. 

Thus the discrete counterparts of the scattered field $\phi( x,\tau)\rvert_{\tau=\mathrm{const}}$ and
its time derivative $\psi(x,\tau)\rvert_{\tau=\mathrm{const}} = \partial_\tau\phi(x,\tau)\rvert_{\tau=\mathrm{const}}$ are $N+1$ vectors, whereas the first-order reduced scattered field
$u(x,\tau)\rvert_{\tau=\mathrm{const}}$
and the eigenfunctions $v_n(x), w_n(x)$ and $\alpha_n(x)$ are
$2N+2$ vectors with complex entries.
Likewise, the interpolant of the differential operator $L$ is
a $(2N+2)\times (2N+2)$ matrix.
The interpolant of the derivative operator is obtained by left multiplication by the differentiation matrix $\mathbb{D}$ of the
form (cf. \cite{Jaramillo:2020tuu})
\begin{equation}
  (\mathbb{D})_{i,j}=
  \begin{cases}
    \vspace{0.2cm}-\frac{2N^2+1}{6}\qquad i=j=N\\
    \vspace{0.2cm}\frac{2N^2+1}{6}\qquad i=j=0\\
    \vspace{0.2cm}-\frac{x_j}{2(1-x_j)^2}\qquad 0<i=j<N\\
    \frac{\xi_i}{\xi_j}\frac{(-1)^{i-j}}{x_i-x_j}\qquad i\neq j
  \end{cases}
\end{equation}
with 
\begin{equation}
  \xi_i=
  \begin{cases}
    \vspace{0.2cm}2 \qquad i\in \{0,N\}\\
    1 \qquad i\in \{1,...,N-1\}
  \end{cases}
\end{equation}
Scalar products are discretised through the construction of the Gram matrix,  denoted $\Gram$, corresponding to a given (continuum) scalar product $\langle \cdot, \cdot \rangle_{_G}$. It is, in full generality, a positive-definite Hermitian matrix (in our case, actually a positive-definite  real-symmetric) whose expression relies on an interpolation of the functions $p$, $w$ and $q$ and derivative operator $\mathbb{D}$ on a thinner grid of size $2N+2$ (see Appendix C.3
of \cite{Jaramillo:2021tmt} for details in the particular case of the so-called ``energy scalar product'').
The discrete energy scalar is then calculated as
\begin{equation}
\bkps{u_1}{u_2} = \w{u_1}^* \cdot \Gram \cdot\w{u_2} \ , \label{scalar_product_Gram}
\end{equation}
where the star sign $(\cdot)^*$ stands for the matrix Hermitian conjugate $\ol{(\cdot)}^t$.
Following section \ref{s:Keldysh}, this notation should be
distinguished from the one for the dual pairing 
\begin{equation}
  \langle\w{\alpha},\w{v}\rangle = {\w{\alpha}}^t \cdot \w{v} \ , \label{dual_pairing}
\end{equation}
which has no subscript. The interpolant of the (formal) adjoint of $L$ with respect to the
scalar product  $\langle \cdot, \cdot \rangle_{_G}$, namely $\w{L^\dagger}$, satisfies
$\bkps{u_1}{L \cdot u_2}=\bkps{L^\dagger \cdot u_1}{u_2}$ for any vectors $\w{u_1}$ and $\w{u_2}$. From this it follows
\begin{equation}
\w L^\dagger = \w \Gram^{-1} \cdot \w L^* \cdot\w \Gram \label{adjoint_Gram} \ .
\end{equation}

\subsection{Method of lines.}
\label{a:method_lines}
Chebyshev pseudospectral methods commented above are employed in our different
numerical calculations, both in frequency-domain (eigenvalue
calculation and pseudospectrum construction)
and time-domain evolutions. Regarding the latter, we have implemented two schemes.
The first one, purely spectral, has been described in 
appendix \ref{a:Keldysh_evol_op} in the diagonalisable case.
Here we comment on another evolution scheme, the method of lines.

As commented in the precedent section, we discretise the space interval
using Chebyshev pseudospectral methods,
representing a field by a vector whose entries are the values of the field
at the Chebyshev collocation points $x_i$. Then, 
we replace spatial derivatives by their numerical approximations obtained by
acting with $\mathbb{D}$ on the components of the discretised field. This leaves us
with one continuous parameter, namely the (hyperboloidal) time variable $\tau$, and
the partial differential equation
becomes an evolution system of ordinary differential equations (ODE) in $\tau$ for the
values of the field at the collocation points. Specifically
\begin{equation}
  \w u(\tau)=(\phi_0(\tau), \phi_1(\tau),...,\phi_{N-1}(\tau),\phi_N(\tau), \psi_0(\tau), \psi_1(\tau),..., \psi_{N-1}(\tau),\psi_N(\tau))^t \ ,
\end{equation}
denotes the column vector and, then, the time evolution problem (\ref{e:wave_eq_1storder_u_tau})
is translated into the
following vector ODE for $\w u(\tau)$, with its initial condition $\w{u_0}$. That is
\begin{alignat}{2}
  &\frac{d\w u(\tau)}{d\tau}  = i\w L \cdot\w u(\tau) \ , \quad \quad &\w u(\tau=0)=\w{u_0} \ . \label{ODE_sol}
\end{alignat}
We use finite difference methods on the time coordinate $\tau$ (though other methods can be
considered in this evolution part\footnote{In particular, we bring attention to works \cite{Macedo:2014bfa,Ansorg:2016ztf,PanossoMacedo:2018hab}, not only pseudospectral in space  but, most remarkably, also in time.}). Since this method is a cornerstone of numerical simulations, we have chosen to use the already highly optimized libraries available and, for the purpose of our work, we use a Julia library as a black box that requires the choice of a discretisation scheme algorithm as a previous step to set the ODE system. The use of Julia allows us to control the accuracy (namely "tolerance") of the numerical solution (see table \ref{table:table_parameters}).

The method of lines is a general framework and not a detailed recipe, the choice of the discretisation method of the operator $L$ is arbitrary and further tests/checks are needed to validate the numerical solution, in particular, one can see in supplementary material (referred to in section \ref{s:hyperboloidal evolutions}) how the outgoing or reflective boundary conditions manifest themselves in the numerical solution.

We note that the time evolution problem in the asymptotically AdS spacetime, namely
\begin{alignat}{2}
  &\w B \cdot \frac{d\widetilde{\w u}(\tau)}{dt} = i\widetilde{\w L}\cdot \widetilde{\w u}(\tau) \ , \quad \quad &\widetilde{\w u}(\tau=0)=\widetilde{\w{u}_0} \ , \label{DAE_sol}
\end{alignat}

where $\w B$ is singular, takes the form of a mass matrix differential-algebraic equation (DAE). $\w B$ is sometimes called the mass matrix for historical reasons because it represents the mass of vibrating structures in (generalized) second order problems in mechanics.

\begin{table}[h]
        \begin{tabular}{|c|c|}
            \hline
            \textbf{Sector} & \textbf{Parameter} \rule{0pt}{9pt} \\[2pt] \hline
            Chebyshev-Lobatto grid size & $N$\\
            \hline
            arbitrary decimal precision & precision\\
            \hline
            \multirow{3}{*}{ODE/DAE solver (numerical time evolution)}
            & $dt$ increment\\
            & tolerance\\
            & algorithm\\
            \hline
        \end{tabular}
    \caption{Parameters to tune in the present numerical evolutions using Julia.
    The decimal precision controls the arithmetic precision of real or complex floating numbers. The solver's $dt$ is fixed to $10^{-7}$. The precision of the ODE solver is controlled by a parameter named "tolerance" (we assimilate the relative and the absolute tolerance parameters to a single tolerance parameter). Furthermore, the solver requires an algorithm that corresponds to the discretisation scheme of the time derivative. We have chosen to exploit Julia's automatic stiffness detection feature and we figured out that the best choice in terms of accuracy and execution time is (probably) AutoVern9(Rodas5P()).}\label{table:table_parameters}
  \end{table}

\subsubsection{Computational issues.}
A key element for the numerical resolution of the spectral problem (and, consequently, for the spectral construction  of the QNM resonant expansion as well) is the computation of eigenvalues with arbitrary precision numerics.

Julia is endowed with a library \textit{OrdinaryDiffEq} that provides a toolbox to solve differential equations of the form $\frac{d\w y}{dt}=\w f(\w y,t)$ where $\w y$ and $\w f(\w y,t)$ are real valued column vectors with an initial condition $\w y(t=0)=\w{y_0}$.
This Julia library also provides a similar DAE solver for differential problems of the type $\w B\cdot\frac{d\w y}{dt}=\w f(\w y,t)$, it uses adaptive time stepping as we can see on the panel \ref{fit_log:AdS} of Figure \ref{fit_log} and panel \ref{diff:AdS} of Figure \ref{diff} that show the time series sampling worsens when the amplitude becomes very small. 
Although its precision is easy to tune using a parameter called "tolerance", the drawback of this method is its computation time. Table \ref{table:table_parameters} shows the main parameters that drive the accuracy of the time evolution and the Keldysh (spectral) expansion. We will consider the ODE solver as a black box that yields a numerical solution. 
Although the ODE/DAE solver is the privileged way to compute time evolutions in this work, the appendix \ref{a:Keldysh_evol_op} addresses how the Keldysh scheme can be applied over all the eigenvalues of the finite rank approximant of $L$ to yield the full dynamics of the solution of (\ref{ODE_sol}) and (\ref{DAE_sol}).
This approach is fast compared to the ODE solver and its accuracy isn't limited by a tolerance parameter, it only depends on the decimal precision and the grid size $N$. We use this "fast" approach for (only) 2 figures: Figures \ref{f:C_N} 
and \ref{f:uniform_convergence} that necessitate very high precision.

\subsection{Initial data test.}
\label{e:Gaussian_ID}
Throughout this work we have fixed a reference initial condition that depends on the compactified space coordinates $x$ (for P\"oschl-Teller) and $\sigma$ (for the black hole cases). We define it on the interval $[-1,+1]$ of the Chebyshev-Lobatto grid as follows
\begin{equation}
  \label{e:initial_data}
  u_0^\text{ref.}(x_j)=\left(\begin{array}{c}
    e^{-a(x_j+b)^2}\\
    0
  \end{array}\right),\quad \forall x_j=\cos\left(\frac{\pi j}{N}\right)\in[-1,+1] \ ,
\end{equation}
with the specific choice $a=8$ and $b=0.1$.
Figure \ref{u0} shows this initial condition on $[-1,+1]$, in order to sample the interval $[0,1]$, we use $x_j=2\sigma_j-1$ for $j\in\{0,1,...,N-1,N\}$. The ODE scheme employed in the P\"oschl-Teller, Schwarzschild and Schwarzschild-de Sitter cases makes a direct use of the initial condition $u_0^\text{ref.}$ unlike the AdS case which uses $\widetilde{u_0}=u_0^\text{ref.}$ before rescaling the whole field according to $u(\tau,\sigma)=\sigma\widetilde{u}(\tau,\sigma)$.

As we have commented in the introduction of section \ref{s:Time-domain_evolution},
we have studied other initial data. Results are qualitatively the same.
We will study  systematically generic initial data in \cite{BesJarPoo24}.

\begin{figure}[htp]
  \centering
  \includegraphics[clip,width=0.7\columnwidth]{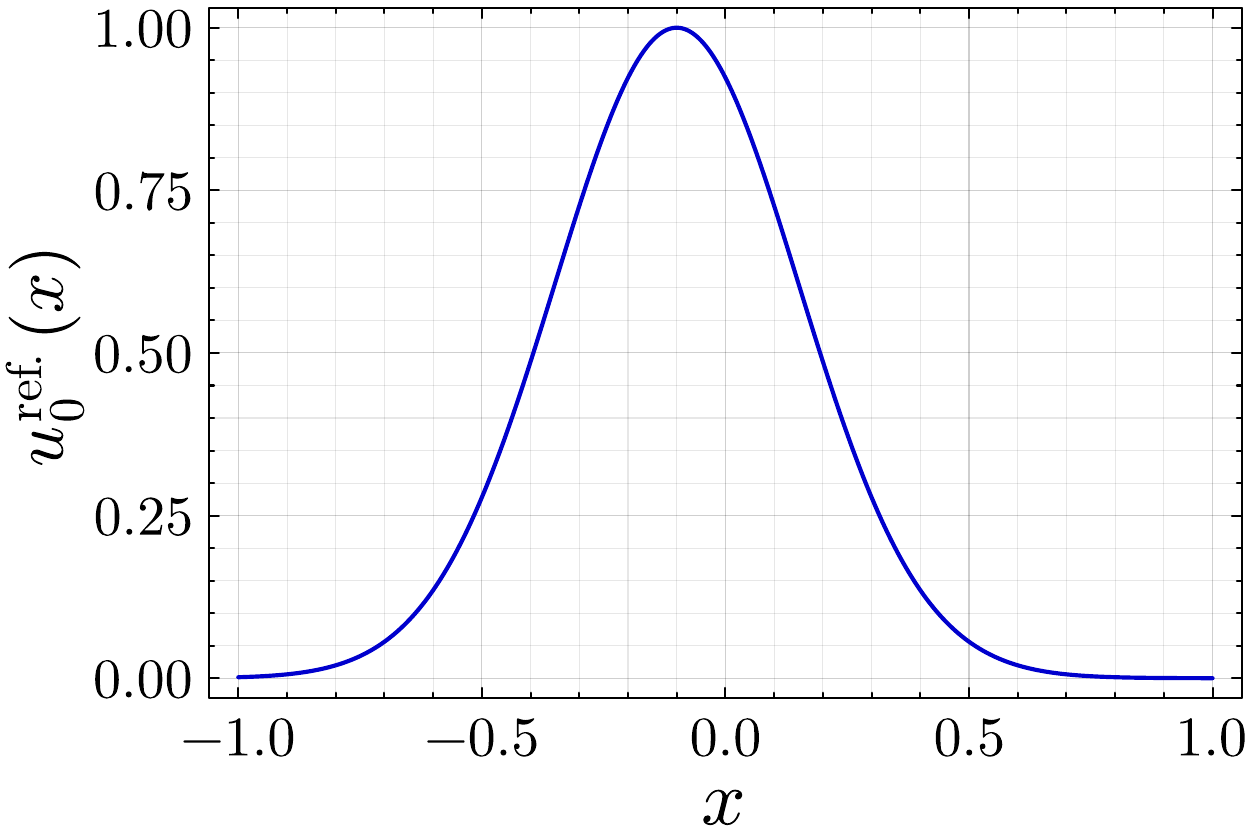}\label{u0:u0}
  \caption{Initial condition $u_0^\text{ref.}$ depicted on a Chebyshev-Lobatto grid with $500$ points.}
  \label{u0}
\end{figure}

\bigskip

\section*{Acknowledgments}
The authors would like to warmly thank Lamis Al Sheikh for the extensive discussions
and for her generous sharing of ideas.
We acknowledge the organisers of the ``Infinity on a Gridshell'' workshop held in Copenhagen, 10-13 July 2023, where part of the material contained in this manuscript was presented.
We would like to thank also Marcus Ansorg, Piotr Bizo\'n, Anne-Sophie Bonnet-BenDhia,
Valentin Boyanov, Javier Carballo, Lucas Chesnel, Edgar Gasper\'\i n, Christophe Hazard, Maryna Kachanovska,
Badri Krishnan, 
Rodrigo P. Macedo, Oscar Meneses-Rojas, Zo\"\i s Moitier, Christiana Pantelidou,
Daniel Pook-Kolb, Adam Pound, Andrzej Rostworowski,
Juan A. Valiente-Kroon, Corentin Vitel, Claude Warnick, Benjamin Withers
and An\i l Zengin\u{o}glu for enriching discussions.
 We acknowledge support
from the PO FEDER-FSE Bourgogne 2014/2020
program and the EIPHI Graduate School (contract ANR-17-EURE-0002) as
part of the ISA 2019 project. We also thank the project QuanTEdu-France (22-CMAS-0001), the ``Investissements
d'Avenir'' program through project ISITE-BFC (ANR-15-IDEX-03), the ANR
``Quantum Fields interacting with Geometry'' (QFG) project
(ANR-20-CE40-0018-02), and the Spanish FIS2017-86497-C2-1 project
(with FEDER contribution).

\bigskip

\bibliography{sn-article_resub.bib}
\end{document}